\documentclass[final,3p,times]{elsarticle}

\usepackage{amssymb}
\usepackage{amsmath}

\immediate\write18{%
  makeindex -s nomencl.ist -o \jobname.nls -t \jobname.nlg \jobname.nlo%
}
\usepackage{nomencl}
\usepackage{etoolbox}
\renewcommand\nomgroup[1]{%
  \item[\bfseries
  \ifstrequal{#1}{G}{Greek symbols}{%
  \ifstrequal{#1}{D}{Subscripts}{%
  \ifstrequal{#1}{U}{Superscripts}{}}}%
]}
\usepackage{multicol}
\makenomenclature
\renewcommand*\nompreamble{\begin{multicols}{2}}
\renewcommand*\nompostamble{\end{multicols}}

\usepackage{mdframed}

\usepackage{booktabs}
\usepackage{placeins}
\usepackage{makecell}
\usepackage{subfig}

\usepackage{hyperref}
\usepackage{cleveref}

\usepackage{tikz}
\usetikzlibrary{arrows.meta,positioning,shapes.geometric,shapes,arrows,calc}

\tikzset{
  >={Latex[length=2.2mm]},
  block/.style={rectangle, rounded corners, draw, align=center,
                minimum width=47mm, minimum height=10mm},
  smallblock/.style={rectangle, rounded corners, draw, align=left,
                     minimum width=53mm, minimum height=16mm},
  decision/.style={diamond, draw, aspect=2, align=center,
                   inner xsep=4mm, inner ysep=2mm},
  line/.style={draw, -{Latex}},
  note/.style={rectangle, draw, align=left, rounded corners,
               font=\small, inner sep=3mm},
  base/.style = {draw=black,
                 minimum width=2cm, minimum height=0.5cm,
                 text centered, font=\normalfont},
  activityStarts/.style = {base, rectangle, fill=white},
  startstop/.style = {base, rectangle, rounded corners, fill=white},
  activityRuns/.style = {base, rectangle, rounded corners, fill=white},
  process/.style = {base, rectangle, rounded corners, minimum width=2.2in,
                    fill=white, font=\normalfont},
  StartEnd/.style = {base, ellipse, fill=white},
  Decide/.style = {base, diamond, fill=white, aspect=4, text width=5em}
}

\tikzset{
  chart/.style={
    legend label/.style={font={\scriptsize},anchor=west,align=left},
    legend box/.style={rectangle, draw, minimum size=5pt},
    axis/.style={black,semithick,->},
    axis label/.style={anchor=east,font={\tiny}},
  },
  bar chart/.style={
    chart,
    bar width/.code={
      \pgfmathparse{##1/2}
      \global\let\bar@w\pgfmathresult
    },
    bar/.style={very thick, draw=white},
    bar label/.style={font={\bf\small},anchor=north},
    bar value/.style={font={\footnotesize}},
    bar width=.75,
  },
  pie chart/.style={
    chart,
    slice/.style={line cap=round, line join=round, thin,draw=black},
    pie title/.style={font={\bf}},
    slice type/.style 2 args={
      ##1/.style={fill=##2},
      values of ##1/.style={}
    }
  }
}

\journal{}
\begin{document}
\begin{frontmatter}
\title{Density-based topology optimization for turbulent fluid flow using the standard $k$-$\varepsilon$ RANS model with wall-functions imposed through an implicit wall penalty formulation}

\author[label1]{Amirhossein Bayat}
\author[label1]{Hao Li}
\author[label1]{Joe Alexandersen\corref{cor1}}
\cortext[cor1]{Corresponding author}
\ead{joal@sdu.dk}
\affiliation[label1]{organization={Institute of Mechanical and Electrical Engineering, 
University of Southern Denmark},
            addressline={Campusvej 55},
            city={Odense},
            postcode={DK-5230},
            country={Denmark}}

\begin{abstract}
Turbulent flows have high requirements for very fine meshes near the boundary to ensure accuracy. In the context of topology optimization (TO), such fine meshes become unrealistic and common approaches are hampered by low accuracy and overestimation of boundary layer thickness. Wall-functions are a natural way to ease the computational requirements, but they are not naturally imposed in density-based TO due to the diffuse design parametrization. We propose an implicit wall-function formulation for the Reynolds-Averaged Navier--Stokes (RANS), standard \(k\)–\(\varepsilon\) model that extracts wall-normal information directly from the gradient of the design variable and enables a penalty-based formulation for imposing wall-functions to the RANS equations, without the need for body-fitted meshes. The method provides a reliable route to high Reynolds number turbulent topology optimization, delivering boundary-layer accuracy comparable to explicit-wall body-fitted analyses, while retaining the flexibility of density-based TO. Furthermore, because wall effects are modeled using wall-functions, accurate solutions are obtained on substantially coarser meshes, leading to significant reductions in computational cost. The approach is validated on three canonical benchmarks over Reynolds numbers up to $\text{Re} = 2\times10^5$: a pipe--bend; a U-bend; and a Tesla--valve. Across all cases, the proposed method accurately recovers near-wall velocity profiles, closely matching verification simulations on body-fitted meshes with explicit wall-functions. In contrast, a conventional turbulent TO formulation, without the proposed wall-function treatment, mispredicts boundary-layer development and yields sub-optimal results.
\end{abstract}


%
%
%
%
%
%

\end{frontmatter}

\mbox{}

\nomenclature[A]{TuTO}{Turbulent topology optimization}
\nomenclature[A]{TO}{Topology Optimization}
\nomenclature[A]{RANS}{Reynolds--averaged Navier--Stokes equations}

\nomenclature[N]{$\boldsymbol{u} = [u_1, u_2]^\top$}{Velocity vector}
\nomenclature[N]{$\boldsymbol{x} = [x_1, x_2]^\top$}{Coordinate vector}
\nomenclature[N]{$p$}{Pressure}
\nomenclature[N]{$h_{max}$}{Maximum allowable element size}
\nomenclature[N]{$P_{con}$}{Control factor}
\nomenclature[N]{$\operatorname{Re}$}{Reynolds number}
\nomenclature[N]{$\boldsymbol{n}$}{Unit normal vector}
\nomenclature[N]{$\boldsymbol{w}$}{Test function vector}
\nomenclature[N]{$\gamma$}{Design variable}
\nomenclature[N]{$\alpha$}{Inverse permeability}
\nomenclature[N]{$\rho$}{Mass density}
\nomenclature[N]{$\nu$}{Kinematic viscosity}
\nomenclature[N]{$\nu_T$}{Turbulent viscosity}
\nomenclature[N]{$P_k$}{Production term of turbulent kinetic energy}
\nomenclature[N]{$\tau_w$}{Tangential part of the viscous stresses on the wall}
\nomenclature[N]{$y^+$}{Turbulent local Reynolds number}
\nomenclature[N]{$\delta^+_w$}{Wall lift-up}
\nomenclature[N]{$u_\tau$}{Friction velocity}
\nomenclature[N]{$\kappa$}{Von K\'arm\'an constant}
\nomenclature[N]{$\beta'$}{Empirical constant}
\nomenclature[N]{$y$}{Distance to the wall in normal direction}
\nomenclature[N]{$L$}{Characteristic length}
\nomenclature[N]{$U$}{Characteristic velocity}
\nomenclature[N]{$\sigma$}{Viscous stresses}
\nomenclature[N]{$r_{1}$}{PDE filter radius}
\nomenclature[N]{$r_{2}$}{Second PDE filter radius}
\nomenclature[N]{$\beta_0$}{Projection steepness parameter}
\nomenclature[N]{$\eta$}{Projection threshold}
\nomenclature[N]{$\varphi$}{Filtered-projected design variable}
\nomenclature[N]{$\Gamma_{\text{s,f}}$}{Fluid-solid interface}
\nomenclature[N]{$\Omega$}{Domain}
\nomenclature[N]{$\text{G}$}{Gradient normalization parameter}
\nomenclature[N]{$h$}{Characteristic mesh element size}
\nomenclature[N]{$N$}{Number of nodes}
\nomenclature[N]{$\text{V}$}{Volume}
\nomenclature[N]{$q_a$}{Brinkman interpolation exponent}
\nomenclature[N]{$\psi$}{Wall intensity}
\nomenclature[N]{$\psi_p$}{Production term wall intensity}
\nomenclature[N]{$\chi$}{Coupling parameter in $k$-$\varepsilon$ equations}
\begin{table*} 

	\begin{mdframed}
		\printnomenclature
	\end{mdframed}
\end{table*}


\section{Introduction}\label{Section: Introduction}
Turbulent flows are widely recognized to be significantly more effective than laminar flows at transporting heat, mass, and momentum, primarily due to the enhanced mixing associated with ``eddy viscosity'' in turbulence~\cite{Pope2001}. This distinctive characteristic makes the study of turbulent flows crucial and too important to be ignored, especially as many contemporary industrial processes and scientific challenges are constrained by the limitations of heat, mass, and momentum transfer. 

There are four main approaches to simulating turbulent flows. The most accurate, yet computationally demanding, is direct numerical simulation (DNS), which resolves all turbulence scales and requires extremely fine meshes, especially near solid boundaries~\citep{fox1972numerical,moin1998direct}. The second approach is large eddy simulation (LES), which applies a spatial filter to the Navier-Stokes equations, directly resolving the energetic large-scale eddies while modeling the smaller ones~\cite{lesieur1996new}. While LES is less computationally expensive than DNS, it still demands significant resources. Reynolds-Averaged Navier-Stokes (RANS) models, which compute the statistical averages of turbulent quantities, offer a significant reduction in computational cost while providing sufficient accuracy for a wide range of engineering applications~\cite{xiao2019quantification}. Finally, Hybrid LES/RANS  models combine LES in free-shear regions with RANS modeling near walls, optimizing accuracy and efficiency by leveraging LES in the main flow and RANS in the near-wall region based on the law of the wall~\cite{frohlich2008hybrid}. In this work, we have selected the standard $k$-$\varepsilon$ turbulence model with wall-functions due to its favorable balance between accuracy, computational efficiency, and computational robustness. 

Topology optimization (TO) is a powerful computational method for generating efficient and innovative designs. The concept was originally introduced for solid mechanics by Bendsoe and Kikuchi~\cite{bendsoe1988generating}, later extended to stokes flow problems by Borrval and Petersson~\cite{borrvall2003topology}, and subsequently applied to laminar flows by Gersborg et al.~\cite{gersborg2005topology}. However, for turbulent fluid flows, TO remains far from mature and requires substantial further research and development~\cite{alexandersen2020review}. In the context of TO, RANS models are the most practical choice due to the iterative process involving hundreds of simulations.

\subsection{Motivation}

To apply TO to RANS turbulence models, wall-functions are highly advantageous since they:
\begin{enumerate}[(i)]
    \item enable the simulation of turbulent flows on coarse meshes. In density-based TO, the fluid-solid interface is not explicit; any point in the design domain may act as a potential wall. Consequently, resolving near-wall turbulence would otherwise require extremely fine meshes throughout the domain to capture the viscous sub-layer and logarithmic region~\cite{dewan2010tackling,aliabadi2022turbulence}, which is computationally infeasible.
    \item approximate the near-wall behavior, simplifies the solution of the flow equations
    \citep{ciofalo2022thermofluid}, and therefore avoids the slow convergence associated with
    resolving the near-wall region in low-Reynolds-number $k$-$\varepsilon$ models
    \citep{craft2004new}. As a result, employing wall-functions substantially reduces computational cost in TO, where the forward and adjoint problems must be solved hundreds of times during the optimization.
    \end{enumerate}

Integrating wall-functions with density-based TO is, however, non-trivial because the fluid-solid interface is not explicitly represented, but rather embedded in a diffuse design field. While the diffuse domain provides flexibility for large geometric changes during the optimization~\cite{alexandersen2023detailed}, these same gray regions complicate the accurate imposition of boundary conditions at the fluid-solid interface. This makes the consistent formulation of RANS models involving wall-functions particularly challenging.

\subsection{Related works}
\subsubsection{Turbulent topology optimization (TuTO)}

Turbulent topology optimization (TuTO) was first demonstrated using the Spalart--Allmaras (SA)~\cite{spalart1992one} turbulence model \cite{zymaris2009continuous,papoutsis2011constrained,yoon2016topology,sa2021topology,holka2022density}. The SA model is a one-equation RANS closure model, that transports a working variable related to eddy viscosity, and is widely used in turbomachinery and aerospace applications \cite{Rodrguez2019RANSTM}. However, for free-shear, jet flows, the baseline SA model is often less accurate~\cite{yusuf2020short}.
Afterwards, TuTO has been extended to two-equation turbulence models such as the $k$–$\varepsilon$, $k$–$\omega$, or similar models~\cite{papoutsis2016continuous,dilgen2018topology,dilgen2018density,yoon2020topology,alonso2022topology,sun2023topology,zhang2023topology,butler2025experimental,Wu2024}. Using two-equation turbulent models is more accurate, provide greater generality and reduce dependence on empirical assumptions \cite{harrison2008comparison}, albeit at the cost of a higher computational cost. At the upper end of the fidelity spectrum, Nobis et al.~\cite{nobis2023modal} conducted TuTO directly with DNS, providing a modeling-error–free reference at a substantially higher computational cost.

Calculation of RANS adjoints adds yet another source of time-consuming and memory-intensive computations. This has motivated the development of various approximations to obtain gradients that are ``good enough'' for design~\cite{dwight2006effect}. Among these, the frozen-turbulence approach is the most common. In this approach, the adjoint is derived by considering only the mean-momentum equations, while the turbulent viscosity $\nu_t$ is held fixed and derivatives of the turbulence model are neglected~\cite{othmer2008continuous,pietropaoli2017design}. Although computationally cheap, this simplification reduces accuracy and can significantly distort adjoint gradients, especially in adverse-pressure-gradient regions~\cite{kontoleontos2013adjoint,dilgen2018topology}. In the present work, the discrete adjoint method has been implemented in COMSOL Multiphysics \cite{comsol_v6-3} without any such simplifications.

To avoid the mathematical complexity and computational cost of RANS turbulent models, some algebraic turbulent methods and simplifications are used for TO as lightweight alternatives, which are useful in large-scale or three-dimensional TuTO.
Høghøj~\cite{hoghoj2023topology} presented a so-called zero-equation model, based on Prandtl's mixing length model. Although there is no additional equation for turbulent quantities, it still requires additional equations (one reaction-diffusion equation and one Poisson equation) to compute the distance to the nearest wall.
Lundgren et al.~\cite{lundgren2025large} further simplified this approach by assuming simple wall distance functions independent of the actual design and applied the algebraic Prandtl's mixing length model for three-dimensional flow--heat--structural TO.

To reduce the need of extremely fine mesh near to the wall, the use of wall-functions are necessary.
Kubo et al.\cite{Kubo2021}, uses one equation law of the wall and modified the turbulent viscosity on the wall directly from normal gradient of velocity, within an immersed boundary framework combined with level-set-based topology optimization.
Galanos et al.\cite{galanos2024cut} used the solution of the Hamilton–Jacobi equation to compute the distance to the wall and apply wall-function to the SA model in TO using cut cells.

To the best of the authors’ knowledge, incorporating implicit wall-functions into density-based TuTO remains unexplored. Mature wall-function formulations exist for the \(k\!-\!\varepsilon\) RANS model, and adopting them can reduce computational cost, since they allow meshes with \(y^+\sim 10\text{--}300\) instead of the \(y^+\approx 1\) required for fully-resolved near-wall modeling. It should be mentioned that for most of the works listed above, the $y^+$ requirements of the underlying models are rarely fulfilled due to computational cost. Thus, incorporating wall-functions is of paramount importance in topology optimization.

\subsubsection{Implicit walls in density-based TO}
In density-based topology optimization, the topology is controlled by a continuous material distribution, and as a result, there is no explicit solid-fluid interface. To address wall-related conditions in this context, several methods have been developed that rely on the spatial gradient of the design variable.
To address coated structures, \citet{clausen2015topology,clausen2017topology} introduced a second filter allowing interfaces of controlled thickness to be defined.
This approach has been extended for controlling walls in a variety of engineering applications
\cite{luo2019topology,chu2020multiscale,hu2024topology,hoghoj2020topology}.
To impose boundary conditions on the implicit walls, density jumps (akin to gradients) have been employed in convective heat transfer~\cite{alexandersen2011topology,zhou2016industrial}, in fluid--structure interaction (FSI)~\cite{hederberg2025fluid}, and in modeling visco-thermal losses~\cite{mirpourian2025arbitrary}.

Sometimes it is not sufficient to define only an implicit wall, but the wall-normal direction can also be required. In convective heat transfer, \citet{lazarov2014topology} introduced the use of implicit normal vectors to impose a normal heat-flux wall condition in density-based topology optimization.
The same idea was used by Wang and Xiaoping~\cite{wang2020density}, to treat design-dependent loads.
In additive manufacturing, where manufacturability constraints are central, wall-normal vectors have been used to restrict overhang~\cite{qian2017undercut} and to control surface roughness~\cite{wang2019boundary}.

Most recently, \citet{bayat2025density} employed a second filter and its spatial gradient to determine the wall representation and the corresponding normal vector in density-based TO for implementing free-slip boundary conditions. This is the methodology that we extend to imposition of turbulent wall-functions herein.

\subsection{Contributions}

In this work, a robust and relatively easy-to-implement approach is introduced, that integrates the standard $k$-$\varepsilon$ $\text{RANS}$ model with wall-functions into a density-based topology optimization framework, referred to as implicit wall-function method. The proposed implicit wall-function method accurately captures near-wall physics, while maintaining tractable computational cost, thereby introducing a novel formulation for coupling $\text{RANS}$ with topology optimization.

The pipe-bend, the U-bend, and the Tesla valve are investigated as three important benchmarks to illustrate the robustness of the proposed method in complex physics and geometries. The first two benchmarks show that the proposed implicit wall-function method generates designs outperforming the ``conventional'' TuTO approach. Recent studies have advanced the understanding of Tesla valve performance with respect to geometry and operating conditions, but primarily within laminar and creeping--flow regimes~\cite{zhou2021simulation,dong2021novel,bohm2022highly,ni2023performance,feng2024multi,sasaki2024topology,ren2025multi}. To the best of our knowledge, topology optimization of Tesla valves in the turbulent, inertia-dominated regime has not yet been reported, where accurate boundary-layer modeling is essential for reliable prediction and optimization. The Tesla valve naturally produces sharp bends and complex geometries, which in turn generate highly turbulent flow. In this work, we investigate the Tesla valve at high Reynolds numbers to emphasize inertia-dominated effects on the optimized topology. Furthermore, we conduct a comparative study across a wide range of Reynolds numbers, spanning from laminar flows (under both free-slip and no-slip boundary conditions) to fully turbulent regimes.

\subsection{Paper Layout}
The article consists of the following parts:
\begin{itemize}
  \renewcommand\labelitemi{--}
  \setlength{\itemsep}{0.2em}

  \item Section~\ref{Section: Formulation} introduces the turbulent $k$--$\varepsilon$ formulation, the TO variables, and both explicit and implicit wall-function representations, and presents the final weak form and details of the solution method.

\item Section~\ref{Section: validation} applies the proposed turbulent topology optimization method, termed the implicit wall-function method, to the backward-facing step benchmark and compares the results with the experimental data of \citet{kim1978investigation}.

\item Section~\ref{Section: Test cases} applies the implicit wall-function method to several topology-optimization test cases (pipe bend, U-bend, and Tesla valve), includes a literature comparison and LES verification for the U-bend, and compares a laminar Tesla-valve design (free-slip vs.\ no-slip walls) with the turbulent optimized Tesla-valve design.

\item Section~\ref{Section: Conclusion} presents concluding remarks.

\end{itemize}

\section{Formulation of proposed implicit wall-function method}\label{Section: Formulation}

\subsection{Reynolds-Averaged Navier–Stokes (RANS) turbulence models} \label{Section: RANS}
Reynolds developed a statistical approach by decomposing the velocity field into two components, an approach known as Reynolds decomposition: the mean velocity and the fluctuating part. He further introduced additional non-linear stress terms into the Navier-Stokes equations, now referred to as Reynolds stresses or turbulent stresses. This decomposition is expressed as~\cite{pope2001turbulent}:
\begin{equation}
\boldsymbol{u} = \boldsymbol{\tilde{u}} + \boldsymbol{u'}
\end{equation}
where $\boldsymbol{\tilde{u}}$ is the mean (time-averaged) velocity and $\boldsymbol{u'}$ is the fluctuating velocity component.
The time average of the fluctuating part is equal to zero.

The standard $k-\varepsilon$ RANS model is introduced to simulate the fluid flow behavior (for a detailed review, see~\cite{ Moin_Chan_2024}). The steady flow of an incompressible fluid governed by the RANS equations for velocity $\boldsymbol{u}= [u_1, u_2]^\top$, and pressure $p$ is:
\begin{subequations} \label{eq:govequ_ns}
\begin{align}
\left(  \boldsymbol{u} \cdot \nabla \right) \boldsymbol{u}  +\nabla p - \nabla \cdot \left( (\nu + \nu_T)[\nabla \boldsymbol{u} + \nabla \boldsymbol{u}^\top] \right)+ \alpha\boldsymbol{u} &= 0
\\
\nabla \cdot \boldsymbol{u} &= 0 
\end{align}
\end{subequations}
where $\nu$ is the kinematic viscosity, which depends on the fluid itself, and $\nu_t$ is the turbulent eddy viscosity, which mimics the effect of the fluctuation part of the velocity, $\boldsymbol{u'}$, and $\alpha$ is the inverse permeability or Brinkman penalty factor. The Brinkman penalty factor is set to a large number in the solid domain, to ensure negligible fluid flow through the solid.

To calculate the turbulent eddy viscosity $\nu_t$, we need to solve two additional convection-diffusion-reaction equations known as the $k$ and $\varepsilon$ equations, which represent the turbulent kinetic energy and the dissipation rate, respectively. The turbulent eddy viscosity is calculated as $\nu_t = C_\mu \frac{k^2}{\varepsilon}$, where $k$ and $\varepsilon$ are the solution of the following equations:
\begin{subequations} \label{eq:govequ_keps}
    \begin{align}
\nabla \cdot \left( k\boldsymbol{u} - \frac{\nu_T}{\sigma_k} \nabla k \right) - P_k + \varepsilon &= 0
\label{eq: k_equation} \\
\nabla \cdot \left( \varepsilon \boldsymbol{u} - \frac{\nu_T}{\sigma_\varepsilon} \nabla \varepsilon \right) 
- \frac{\varepsilon}{k} \left( C_1 P_k - C_2 \varepsilon \right) + \alpha\varepsilon 
&= 0 \label{eq:epsilon-equation}
    \end{align}
\end{subequations}
where $\alpha$ is the same Brinkman penalty factor used in \Cref{eq:govequ_ns}, and $P_k$ is the production term of turbulent kinetic energy, defined as:
\begin{equation}
P_k = \nu_T \left( \nabla \boldsymbol{u} : \left( \nabla \boldsymbol{u} + (\nabla \boldsymbol{u})^\top \right) \right)
\label{eq: prodution_kinetik_energy_turbulence}
\end{equation}
The constant parameters are set to $C_\mu=0.09$, $C1=1.44$, $C2=1.92$, $\sigma_k=1.0$, and $\sigma_\epsilon=1.3$~\cite{kuzmin2007implementation}.
The last term of \Cref{eq:epsilon-equation} is introduced to force $\varepsilon$ to be zero in the solid domain --- similar to the Brinkman term in \Cref{eq:momentum_equation}.

\begin{figure}
    \centering
    
    \subfloat[Explicit wall\label{fig:picture1}]{
        \includegraphics[width=0.4\textwidth]{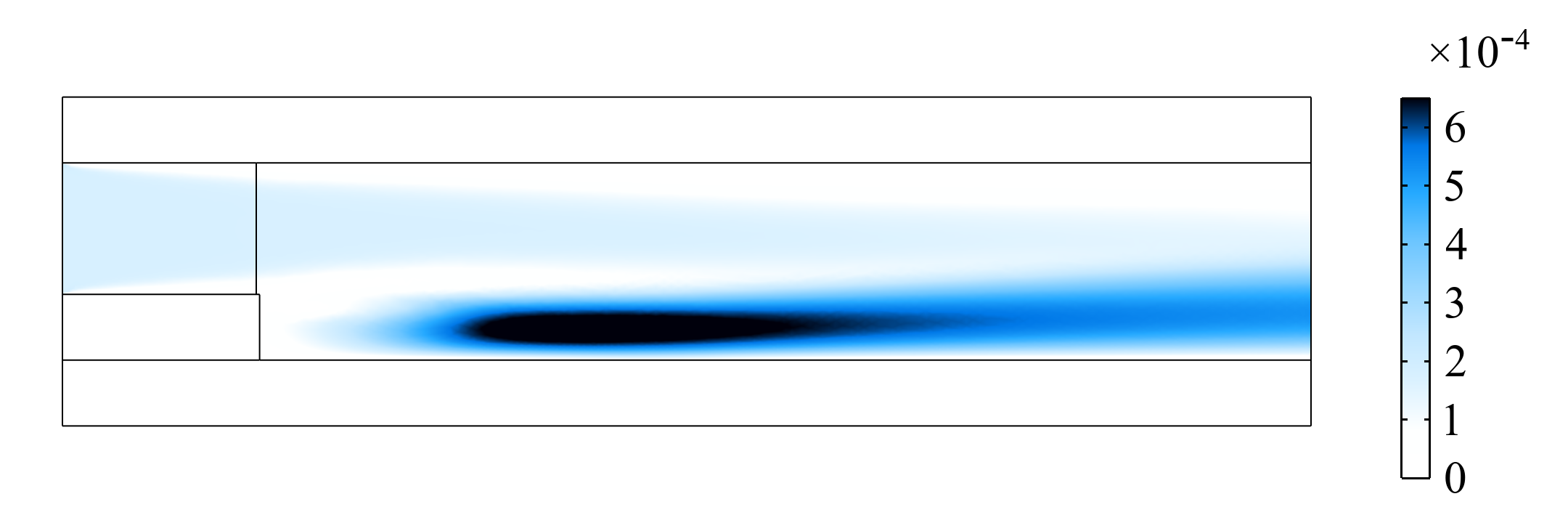}
    }\\[0.5em]
    
    \subfloat[Implicit wall without penalty in the $k$-equation\label{fig:picture2}]{
        \includegraphics[width=0.4\textwidth]{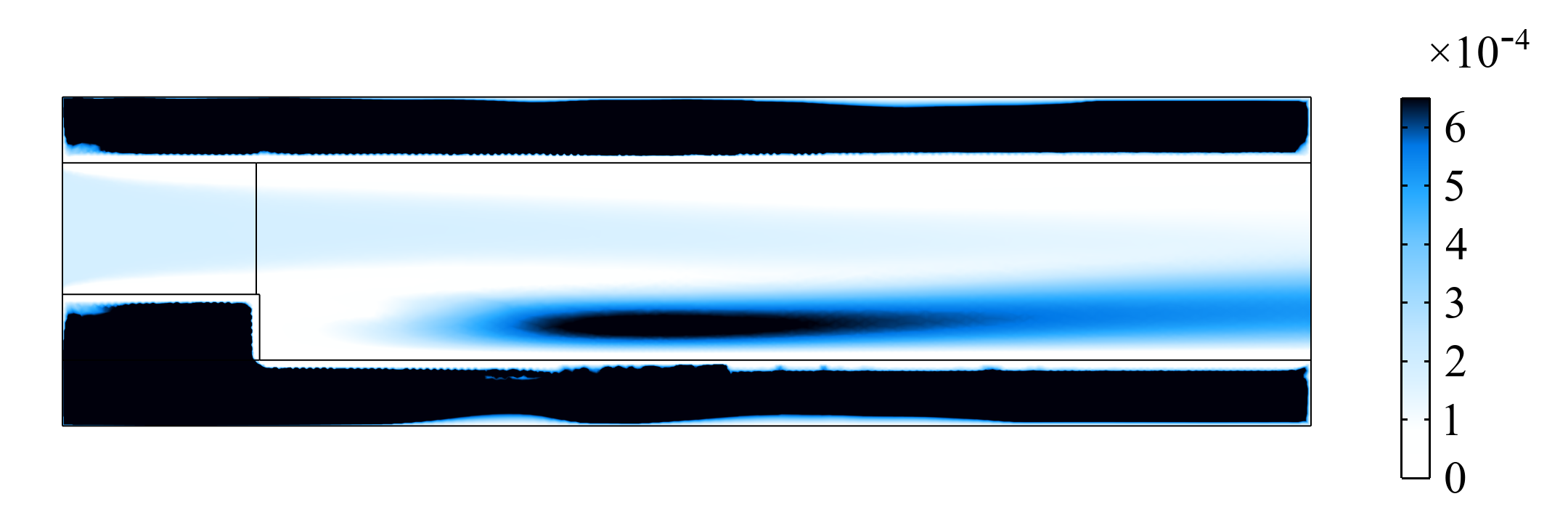}
    }\\[0.5em]
    
    \subfloat[Implicit wall with penalty in the $k$-equation\label{fig:picture3}]{
        \includegraphics[width=0.4\textwidth]{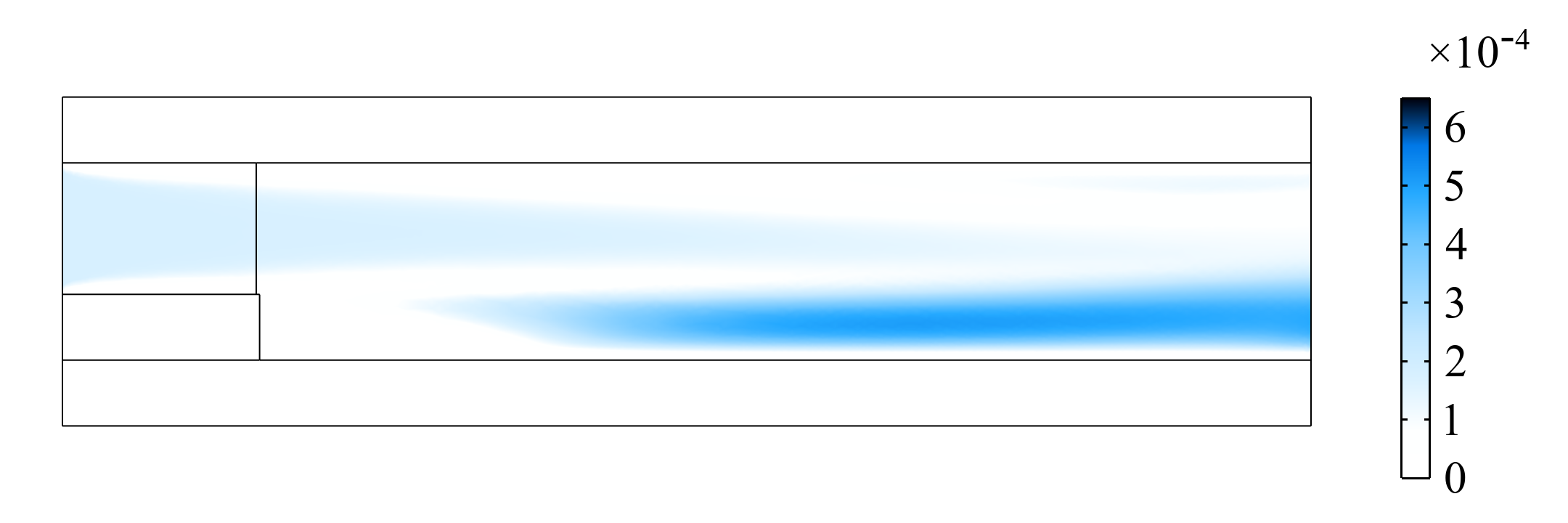}
    }
    
    \caption{Turbulent viscosity field $\nu_t$ for three cases: 
    (a) explicit wall (reference), 
    (b) implicit wall without penalty in the $k$-equation, and 
    (c) implicit wall with penalty in the $k$-equation. 
    The use of an implicit wall without penalty in the $k$-equation provides a closer match to the explicit wall, and the increased $\nu_t$ enhances the stability of the turbulence model and facilitates convergence.}
    \label{fig:turbulent_viscosity}
\end{figure}

As introduced by \citet{butler2025experimental}, we do not apply a Brinkman penalty term in the $k$ equation in contrast to previous work \cite{yoon2020topology, picelli2022topology}. Initial investigations showed a closer agreement of the final turbulent properties with explicit wall reference cases, when not applying a Brinkman penalty term to the $k$ equation. 
As shown in \Cref{fig:turbulent_viscosity}, applying the penalty term also to the $k$ equation drives the turbulent viscosity to zero within the solid region, but reduces accuracy in predicting turbulent viscosity in the bulk flow. In contrast, omitting the penalty in the $k$ equation yields turbulent viscosity predictions that closely match the explicit wall case. It allows the turbulent viscosity to increase within the solid domain, thereby enhancing the viscous terms in the turbulence equations and resemble more of a solid part.

\subsection{Explicit wall-functions}
In the vicinity of the wall, the mean velocity gradient becomes extremely steep, which requires extremely fine meshes for accurate resolution \cite{rodi1997comparison,rogallo1984numerical}. In practical applications, it is preferable to avoid the high computational cost of directly resolving these near-wall regions. Instead, appropriate boundary conditions, known as wall-functions, can be imposed at a certain distance from the wall, enabling a substantial reduction in computational expense while maintaining decent solution accuracy. The wall-function accounts for the viscous stresses at the wall and also the $k$-$\varepsilon$ equations~\cite {kuzmin2007implementation}.

\subsubsection{Viscous shear stresses at the wall}\label{sec:wallfunctions_viscous}
The viscous stresses in the fluid domain is defined as:
\begin{equation}
    \boldsymbol{\sigma} = \nu \left[ \nabla \boldsymbol{u} + (\nabla \boldsymbol{u})^\top \right]
    \label{eq: wall_viscous_stresses}
\end{equation}
and the tangential part of the viscous stresses on the wall is equal to:
\begin{equation}
\begin{aligned}
    \boldsymbol{\tau}_w = \boldsymbol{\sigma} {\boldsymbol{n}} - \left( \boldsymbol{n}^\top{\boldsymbol{\sigma}} \boldsymbol{n} \right)\boldsymbol{n} 
    \label{eq: tangential_wall_stresses}
\end{aligned}
\end{equation}
In this expression, $\boldsymbol{\sigma} {\boldsymbol{n}}$ represents the viscous traction vector on the wall, while $\left( \boldsymbol{n}^\top{\boldsymbol{\sigma}} \boldsymbol{n} \right)$ denotes its normal component. By subtracting the normal component from the total traction, the tangential component of the viscous stress on the wall can be isolated.
Then the logarithmic law of the wall can be used to estimate the boundary condition for the model to prescribe the tangential viscous stresses on the wall \cite{kuzmin2007implementation}:
\begin{equation}
\boldsymbol{\tau}_w = -\,\frac{u_\tau}{y^+}\,\mathbf{u},
\label{eq: tangential_turbu_stresses}
\end{equation}
in which \(y^{+}\) is calculated from:
\begin{equation}
  y^+ =  \frac{y u_{\tau}}{\nu}
\end{equation}
and $u_\tau$ is the friction velocity and it can be estimated as  \(u_\tau = \frac{|\boldsymbol{u}|}{y^+}\) through the domain and \(u_\tau =C_\mu^{0.25} \sqrt{k}\) on the wall~\cite{grotjans1998wall}. One can calculate it on both domain and wall as~\cite{kuzmin2007implementation}:
\begin{equation}
u_\tau = \max\left\{ C_\mu^{0.25} \sqrt{k}, \frac{| \mathbf{u} |}{ \delta_w^+} \right\}
\end{equation}
where $y$ is defined as the distance from the wall in the normal direction (coordinates are labelled $x_1$ and $x_2$).
In the context of finite element analysis, we estimate the distance to the wall, \(y\), using the half of the element size, \((h/2)\), \cite{dilgen2018topology} and compute the wall lift-up \cite{comsol_kepsilon_6_3} using: 
\begin{equation}
  \delta_w^+ = \max\left( \frac{h_w C_\mu^{0.25} \sqrt{k}}{2\nu},\, 11.06 \right)
\end{equation}
Closest to the wall, in the viscous sub-layer, the linear relation $y^+=\frac{\lvert \boldsymbol{u}\rvert}{u_{\tau}}$ holds. Further out, in the logarithmic sub--layer, the logarithmic law of the wall, \(
\frac{\lvert \mathbf{u} \rvert}{u_{\tau}} = \frac{1}{\kappa} \log y^{+} + \beta
\),
 holds. The laws of the logarithmic and linear sub--layers meet at \(y_*^+=11.06\) (assuming \(\kappa\) (von K\'arm\'an constant) \(=0.41\) and \(\beta'\) (empirical constant) \(=5.2\)), that is the reason why minimum $\delta_w^+$ in logarithm region is equal to 11.06. The value \(\delta_w^+\) should lie between 11.06 and 300 for the logarithmic law of the wall to be satisfied --- this is controlled by adjusting the mesh size in the numerical implementation.

Combining~\Cref{eq: tangential_wall_stresses,eq: tangential_turbu_stresses}, we will have:
\begin{equation}
       (\boldsymbol{\sigma} \boldsymbol{n}) 
- (\boldsymbol{n}^\top \boldsymbol{\sigma} \boldsymbol{n}) \boldsymbol{n}  + \frac{u_\tau}{\delta^+_w } \boldsymbol{u} = 0
       \quad \text{on } \Gamma_{s,f}
    \label{eq: wall_funcion_viscous}
\end{equation}
which represents the viscous wall-function boundary condition, which must be satisfied when using the standard $k$–$\varepsilon$ turbulence model.

\subsubsection{Boundary conditions for $k$ and $\varepsilon$  on the wall}\label{sec:wallfunctions_keps}

In addition to the viscous wall-function boundary condition, we also apply boundary conditions on $k$ and $\varepsilon$ at the wall, as well as modify the production term $P_k$.

Kuzmin et al.\cite{kuzmin2007implementation} suggested to apply Neumann boundary conditions for $k$ and $\varepsilon$ :
\begin{subequations}
\begin{align}
\mathbf{n}\!\cdot\!\nabla k &= -\frac{\partial k}{\partial y} = 0 & \text{on } \Gamma_{\text{wall}} \label{eq: k_boundary} \\
\mathbf{n}\!\cdot\!\nabla \varepsilon &= -\frac{\partial \varepsilon}{\partial y}
= \frac{{u_\tau}^{3}}{\kappa\, y^{2}} = \frac{\varepsilon}{y} & \text{on } \Gamma_{\text{wall}}
\end{align}
\end{subequations}
where it is important to highlight that $y$ is the distance from the wall in the normal direction.
One can alternatively apply a Dirichlet condition on $\varepsilon$ \cite{grotjans1998wall,comsol_kepsilon_6_3}:
\begin{equation}
  \varepsilon_w = \frac{{u_\tau}^4}{\kappa \delta^+_w  \nu} 
\label{eq: epsilon_on_the_wall}
\end{equation}
for which it is important to then ensure that the production term on the boundary, $P_{kw}$, is in equilibrium with the dissipation rate on the boundary ($\varepsilon_w$)  \cite{kuzmin2007implementation}: 
\begin{equation}
    P_{kw} = \varepsilon_{w} = \frac{{u_\tau}^4}{\kappa \delta^+_w \nu}
    \label{eq: production_wall}
\end{equation}
We will use this approach and enforce \Cref{eq: epsilon_on_the_wall,eq: production_wall} within the implicit wall identified, while leaving the $k$ equation without any imposed boundary conditions at the wall or in the solid (see \Cref{Section: RANS} and \Cref{fig:turbulent_viscosity} for reasoning.)

\subsection{Design variable and filtering}\label{Design variable}
In density-based TO, the design variable 
$\gamma(\boldsymbol{x})$ is introduced to distinguish between solid and fluid regions. All sensitivities and differentiations are performed with respect to this variable. The design variable takes continuous values ranging from 0, representing the solid domain, to 1, representing the fluid domain, as defined below:
\begin{equation}
\gamma(\boldsymbol{x}) =
\begin{cases}
1 & \text{if } \boldsymbol{x} \in \Omega_{f} \\[6pt]
0 & \text{if } \boldsymbol{x} \in \Omega_{s}
\label{eq:design_variable}
\end{cases}
\end{equation}

A reaction-diffusion PDE filter \cite{lazarov2011filters} is employed to smooth the design variable $\gamma$ and regularize the problem, given as:
\begin{equation}
  -\frac{{r_{1}}^2}{12} \, \Delta \hat{\gamma} + \hat{\gamma} = \gamma
  \label{eq:filtered}
\end{equation}
where $r_{1}$ is the filter radius introducing an approximate minimum length scale of this size.
While using \Cref{eq:filtered} helps to achieve mesh-independent designs, it results in a gray-scale design variable. To obtain a more binary distribution of the design variable (but at the loss of strict length scale control), 
a Heaviside projection function is employed~\cite{wang2011projection}:
\begin{equation}
  \varphi  =
  \frac{\tanh(\beta_0 \eta) + \tanh\!\big(\beta_0(\hat{\gamma}-\eta)\big)}
       {\tanh(\beta_0 \eta) + \tanh\!\big(\beta_0(1-\eta)\big)} 
       \label{eq:projection}
\end{equation}
where $\beta_0$ is the steepness parameter controlling the sharpness of the transition 
(a higher $\beta_0$ makes the design variable resemble a step function) 
and $\eta$ is the threshold that determines the location of the projection transition.

\subsection{Brinkman penalty}\label{Brinkman penalty} 
To enforce vanishing velocity and $\varepsilon$ fields inside the solid region, a volumetric penalty term is introduced \cite{borrvall2003topology,gersborg2005topology,alexandersen2023detailed}. This term, commonly referred to as the Brinkman penalty or inverse permeability term, is expressed as a function of the projected design variable $\varphi$:  
\begin{equation}
  \alpha(\varphi) = \alpha_{\text{max}} \, \frac{1 - \varphi}{1 + q_\alpha \varphi} 
  \label{eq:brinkman}
\end{equation}  
where $\alpha_{\text{max}}$ denotes the maximum penalty value and $q_\alpha$ is a tuning parameter controlling the transition sharpness. For a more detailed discussion, the reader is referred to~\cite{alexandersen2020review,alexandersen2023detailed}.

\subsection{Implicit wall and normal vectors}\label{sec:implicit_wall_and_normal} 

\begin{figure}
    \centering
    \includegraphics[width=0.7\linewidth]{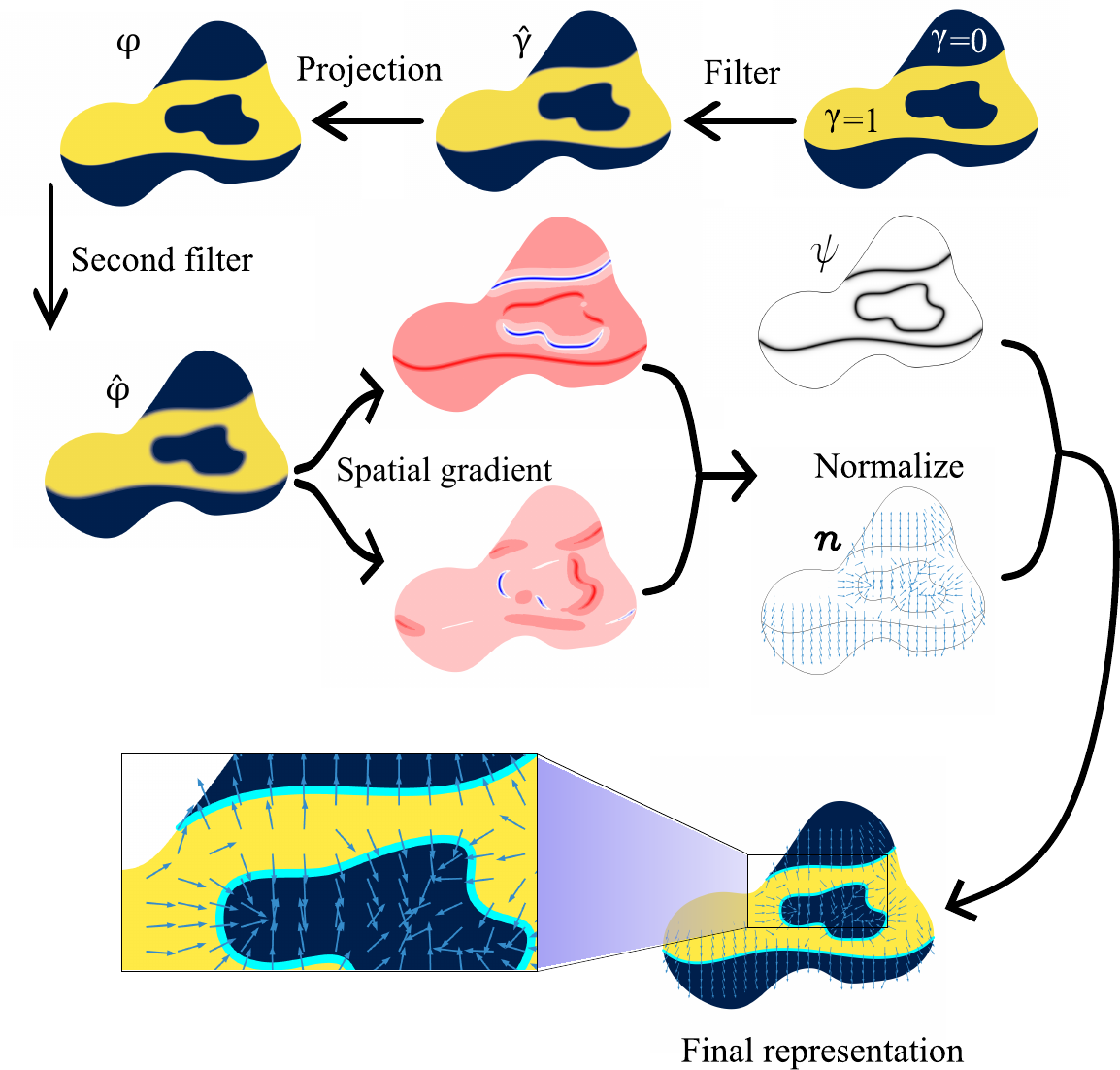}
    \caption{Schematic illustration of the process for identification of the wall region and computation of normals.}
    \label{fig:Comparison_old_new_methods_schem}
\end{figure} 
In density-based TO, it is common to employ filtering and projection techniques for the design variable as introduced in~\Cref{Design variable}. However, in order to identify and control the width of implicit walls and to compute the associated normal vectors in a smooth and continuous fashion, it is advantageous to introduce an auxiliary secondary filter \cite{clausen2015topology,bayat2025density}.
This approach provides two main benefits: (i) it separates wall-related parameters from other regularisation, thereby allowing the use of a different filter radius for wall definition and offering control over wall thickness; (ii) it smoothes the projected design variable, $\varphi$, giving better quality derivatives for normal vector computation.

The second filter uses the same reaction-diffusion PDE filter \cite{lazarov2011filters}:
\begin{equation}
  -\frac{{r_{2}}^2}{12} \, \Delta \hat{\varphi} + \hat{\varphi} = \varphi 
  \label{eq: secound_filter}
\end{equation}  
but where $r_{2}$ is the second filter radius and the output is the second filtered design variable, $\hat{\varphi}$. This variable is subsequently employed for defining the solid–fluid interface and the associated normal vectors according to the work of \citet{bayat2025density}. The solid–fluid interface is described by the wall intensity, $\psi$, for the momentum and $k$-$\varepsilon$ equations, defined as :
\begin{equation}
\psi = \psi_{\text{max}}  \left(\frac{\lVert \nabla \hat{\varphi} \rVert_2}{G_{\text{max}}}\right)^{P_{con}}
\label{eq: wall_intensity}
\end{equation}
where $\psi_{\text{max}}$ is the maximum wall intensity, $P_{con}$ is a control factor, and $G_{\text{max}}$ represents the maximum gradient magnitude derived by \citet{clausen2015topology}:
\begin{equation}
G_{\text{max}} = \frac{\sqrt{3}}{r_{2}}
\end{equation}
Similar to previous works \cite{lazarov2014topology,qian2017undercut,bayat2025density}, the normal vector $\boldsymbol{n}$ $= [n_1,n_2]^\top$ is simply calculated from
 \begin{equation}
\boldsymbol{n}(\hat\varphi) = -\frac{\nabla \hat\varphi}{|\nabla \hat\varphi|}
\label{eq:wall_normal}
\end{equation}

Regarding the production term, a separate wall intensity, $\psi_p$, will be introduced that plays a different role than the other wall intensities used for enforcing viscous and $k$-$\varepsilon$ boundary conditions. In the previous cases, the wall intensities are multiplied by large dimensional coefficients, $\psi_{\text{max}}$, (e.g.\ $\mathcal{O}(10^3$-$10^4)$) in order to create strong penalty terms that dominate the local residual and effectively enforce the desired boundary behavior in a diffuse manner near the implicit wall. But in the case of the production term, since it acts as a reaction term in the equation and modifies the magnitude of the turbulent production term, we only need to turn this modification ``on'' at the wall and ``off'' in the bulk flow, without changing its scale. For this reason, $\psi_p$ is defined as a dimensionless activation factor bounded between $0$ and $1$, rather than as a large penalty parameter. To achieve this, we employ a smooth Heaviside projection to control whether the wall-function correction to the production term is active or inactive in a given location, expressed as:
\begin{equation}
  \psi_p  =
  \frac{\tanh\!\left(\beta_p \eta_p\right) + \tanh\!\left(\beta_p \left( \frac{\psi}{max(\psi)} -\eta_p \right) \right)}
       {\tanh\!\left(\beta_p \eta_p\right) + \tanh\!\left(\beta_p(1-\eta_p)\right)} 
       \label{eq:projection_psip}
\end{equation}
Here, we fix $\eta_p = 0.5$ and $\beta_p = 64$ for all simulations and optimization runs. The relatively large value of $\beta_p$ yields a very sharp but continuous, transition of the Heaviside projection between $0$ and $1$.

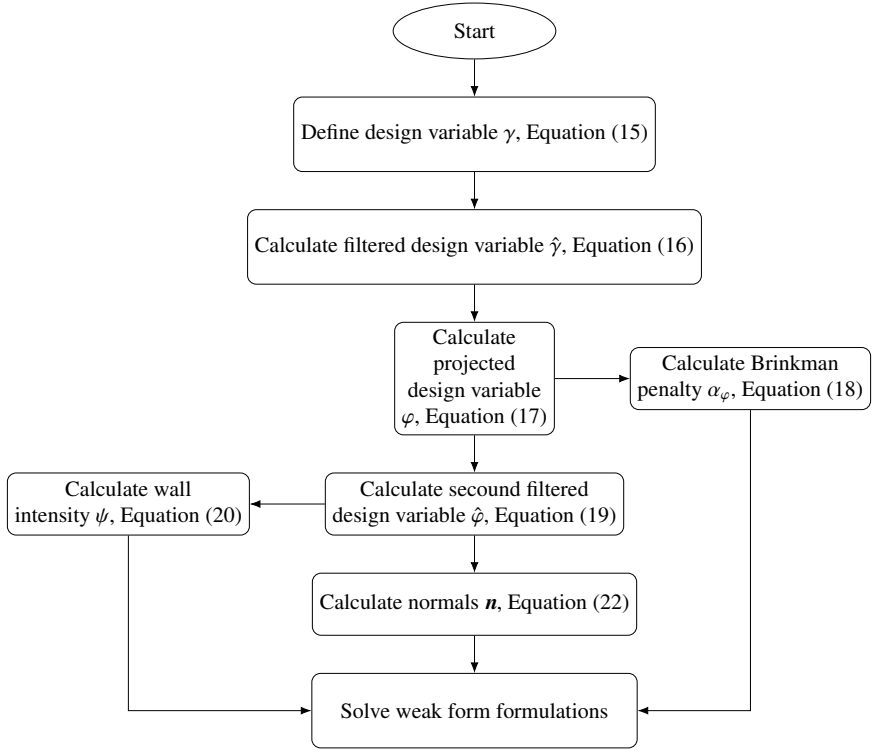
\begin{figure}
\centering
\resizebox{0.7\textwidth}{!}{
\begin{tikzpicture}[node distance=10mm and 18mm, >=Latex]
  \tikzset{
    line/.style={-Latex},
    block/.style={rectangle, rounded corners, draw, align=center,
                  minimum width=52mm, minimum height=12mm},
    smallblock/.style={rectangle, rounded corners, draw, align=center,
                       minimum width=32mm, minimum height=10mm},
    conn/.style={circle, draw, inner sep=1.4pt}
  }
  \node[ellipse, draw, align=center, minimum width=26mm, minimum height=9mm] (start) {Start};
  \node[block, below=6mm of start] (top) {Define design variable $\gamma$, \Cref{eq:design_variable}};
    \node[block, below=6mm of top] (filter) {Calculate filtered design variable $\hat{\gamma}$, \Cref{eq:filtered}};
  \node[block, below=6mm of filter, minimum width=14mm] (main) {Calculate \\projected \\design variable\\ $\varphi$, \Cref{eq:projection}};
  \node[smallblock, right=12mm of main] (rightb) {Calculate Brinkman \\penalty $\alpha_{\varphi}$, \Cref{eq:brinkman}};
  \node[smallblock, below=6mm of main] (mid) {Calculate secound filtered\\ design variable $\hat\varphi$, \Cref{eq: secound_filter}};
   \node[smallblock, below=6mm of mid] (normal) {Calculate normals $\boldsymbol{n}$, \Cref{eq:wall_normal}};
   
    \node[smallblock, left=12mm of mid] (leftb) {Calculate wall \\intensity $\psi$, \Cref{eq: wall_intensity}};
  \node[block, below=6mm of normal] (bottom) {Solve weak form formulations};
  \draw[line] (start) -- (top);
  \draw[line] (top) -- (filter);
  \draw[line] (filter) -- (main);
  \draw[line] (main) -- (mid);
  \draw[line] (mid) -- (normal);
    \draw[line] (normal) -- (bottom);
  \draw[line] (mid.west) -- ++(-8mm,0) |- (leftb.east);
  \draw[line] (leftb.south) |- (bottom.west);
  \draw[line] (main.east) -- ++(8mm,0) |- (rightb.west);
  \draw[line] (rightb.south) |- (bottom.east);
\end{tikzpicture}
}
\caption{Flowchart illustrating how to incorporate solid–fluid interface wall (implicit wall) definition into the density-based topology optimization framework.}
\label{fig: wall_flowchart}
\end{figure}

The overall procedure for the wall calculation is summarized in \Cref{fig: wall_flowchart,fig:Comparison_old_new_methods_schem}.

\subsection{Non-linear coupled solver approach}
The governing equations of \Cref{eq:govequ_ns,eq:govequ_keps} are tightly coupled and highly non-linear. Whereas the molecular viscosity $\nu$ is a material property of the fluid, the turbulent eddy viscosity $\nu_t$ depends on the turbulent fields. The turbulent fields, $k$ and $\varepsilon$, in turn depend on both the flow field and each other. Finding the steady-state solution can be very difficult for this highly non-linear system.

Thus, we use pseudo-transient continuation~\cite{comsol_pseudo_time_stepping_6_3}. An artificial time-derivative in terms of the pseudo-time $\tau$ is added to each equation and the pseudo–time step $\Delta \tau$ is gradually increased until the residuals converge. Convergence is reached as $\Delta \tau$ is ramped up to $10^{4}$, where the solution will be close to the steady-state. The pseudo-transient version of the momentum equation is:
\begin{equation}
\frac{\partial\boldsymbol{u}}{\partial \tau}+ \left(\boldsymbol{u} \cdot \nabla \right) \boldsymbol{u}  +\nabla p - \nabla \cdot \left( (\nu + \nu_T)[\nabla \boldsymbol{u} + \nabla \boldsymbol{u}^\top] \right)+ \alpha\boldsymbol{u}=0
\label{eq:momentum_equation}
\end{equation}

To solve the coupled equations, we choose to use a segregated approach. One approach is to introduce the coupling parameter \cite{kuzmin2007implementation}:
\begin{equation}
    \chi = \frac{\varepsilon}{k}
\end{equation}
which decouples the \( k \) and \( \varepsilon \) equations and also linearizes both equations. The full system is then solved iteratively until convergence is achieved \cite{comsol_kepsilon_6_3}.

Using the pseudo-transient approach, then the linearized turbulent $k$-$\varepsilon$ equations become \cite{kuzmin2007implementation,ceze2013pseudo,comsol_pseudo_time_stepping_turbulent_6_3}:
\begin{equation}
\frac{\partial k}{\partial \tau} + \nabla \cdot \left( k\boldsymbol{u} - \frac{\nu_T}{\sigma_k} \nabla k \right) - P_k + \chi k = 0
\label{eq: linearized k_equation}
\end{equation}
\begin{equation}
\frac{\partial \varepsilon}{\partial \tau}+\nabla \cdot \left( \varepsilon \boldsymbol{u} - \frac{\nu_T}{\sigma_\varepsilon} \nabla \varepsilon \right) 
- \chi \left( C_1 P_k - C_2 \varepsilon \right) + \alpha\varepsilon
= 0
\label{eq:linearized epsilon-equation}
\end{equation}
These equations are then solved using the value of \(\chi\) and $\nu_t$ from the previous outer iteration, until the solution converges. At convergence, the obtained \(k\) and \(\varepsilon\) fields correspond to the solutions of~\Cref{eq: k_equation,eq:epsilon-equation}.

\subsection{Weak forms with implicit wall-functions}

In order to introduce the implicit wall-functions described in \Cref{sec:wallfunctions_viscous,sec:wallfunctions_keps} using the approach described in \Cref{sec:implicit_wall_and_normal}, we introduce volumetric penalty terms in the weak forms using the wall intensity as the penalty field to concentrate it at the implicit walls.

In practice, we multiply the wall intensity, $\psi$, by the test function and the relevant boundary condition, integrating over the domain to obtain the additional terms for the weak forms. For the viscous stresses from \Cref{eq: wall_funcion_viscous}, we will have:
\begin{equation}
 \int_{\Omega} \psi 
\left( (\boldsymbol{\sigma} \boldsymbol{n}) 
- (\boldsymbol{n}^\top \boldsymbol{\sigma} \boldsymbol{n}) \boldsymbol{n} +  \frac{u_\tau}{\delta^+_w } \boldsymbol{u} \right) \cdot \boldsymbol{w} \, \mathrm{d}\Omega = 0
\label{eq: weak_form_viscous_wall_fumction}
\end{equation}
where $\boldsymbol{w}$ is the test function vector for the velocity field.
For the $\varepsilon$ condition from \Cref{eq: epsilon_on_the_wall}, we will have:
\begin{equation}
\int_{\Omega} \psi\left( \varepsilon - \varepsilon_{w} \right) w_{\varepsilon} \, \mathrm{d}\
\label{eq: wall-function weak form}
\end{equation}
where $w_\varepsilon$ is the test function for the $\varepsilon$ field.

In contrast, for the production term that enters in the $k$ equation, we modify production term according to \Cref{eq: production_wall} in every iteration using:
\begin{equation}
P_k = P_k +  \psi_p \left(\varepsilon_w-  P_k \right)
\end{equation}

The final full weak forms for the proposed implicit wall-function method are:
\begin{equation}
\begin{aligned}
&\textbf{Navier-Stokes equations:} \\
&\int_{\Omega} \frac{\partial \boldsymbol{u}}{\partial \tau}\, \boldsymbol{w} \, d\Omega
+\int_{\Omega} \big( (\boldsymbol{u}\!\cdot\!\nabla)\boldsymbol{u} \big)\!\cdot\!\boldsymbol{w}\, \mathrm{d}\Omega
- \int_{\Omega} p\, \nabla\!\cdot\!\boldsymbol{w}\, \mathrm{d}\Omega\\
&\quad
+ \int_{\Omega} \,\left( (\nu + \nu_T)[\nabla \boldsymbol{u} + \nabla \boldsymbol{u}^\top] \right) : \nabla \boldsymbol{w}\, \mathrm{d}\Omega
+ \int_{\Omega} \alpha \, \boldsymbol{u} \cdot \boldsymbol{w} \, \mathrm{d}\Omega
-\int_{\Omega} q \, (\nabla \cdot \boldsymbol{u}) \, \mathrm{d}\Omega\\
&\quad
+ \int_{\Omega} \psi 
\left( \boldsymbol{\sigma} \boldsymbol{n}
- ( \boldsymbol{n}^\top \boldsymbol{\sigma} \boldsymbol{n}) \boldsymbol{n} + \frac{u_\tau}{\delta^+_w } \boldsymbol{u}\right) \cdot \boldsymbol{w} \, \mathrm{d}\Omega + \mathcal{S}_{u,p} = 0
\end{aligned}
\label{eq:momentum-continuity-weak}
\end{equation}

\begin{equation}
\begin{aligned}
&\textbf{\(k\)-equation:} \\
&\int_{\Omega} \frac{\partial k}{\partial \tau}\, w_k \, d\Omega +\int_{\Omega} \boldsymbol{u} \cdot \nabla k \, w_k \,  \mathrm{d}\Omega
+ \int_{\Omega} \frac{\nu_t}{\sigma_k} \, \nabla k \cdot \nabla w_k \, \mathrm{d}\Omega
\\
&\quad
+ \int_{\Omega} \chi\, k \, w_k \, \mathrm{d}\Omega 
+ \int_{\Omega} \left( P_k +  \psi_p \left( P_{kw} - P_k \right) \right) w_k \, \mathrm{d}\Omega  + \mathcal{S}_{k}=0
\\
\label{eq:k-weak}
\end{aligned}
\end{equation}

\begin{equation}
\begin{aligned}
&\textbf{\(\varepsilon\)-equation:} \\
&\int_{\Omega} \frac{\partial \varepsilon}{\partial \tau}\, w_{\varepsilon} \, d\Omega +\int_{\Omega} \boldsymbol{u} \cdot \nabla \varepsilon\, w_{\varepsilon} \, \mathrm{d}\Omega
- \int_{\Omega} \frac{\nu_t}{\sigma_{\varepsilon}}\, \nabla \varepsilon \cdot \nabla w_{\varepsilon} \, \mathrm{d}\Omega\\
&\quad
- \int_{\Omega} \chi\, C_{1}  \left( P_k +  \psi_p \left( P_{kw} - P_k \right) \right) w_{\varepsilon} \, \mathrm{d}\Omega \\
&\quad
+ \int_{\Omega} \chi\, C_{2}\,\varepsilon \, w_{\varepsilon} \, \mathrm{d}\Omega
+ \int_{\Omega} \alpha\, \varepsilon \, w_{\varepsilon} \, \mathrm{d}\Omega
+ \int_{\Omega} \psi \, (\varepsilon - \varepsilon_w) \, w_{\varepsilon} \, \mathrm{d}\Omega  + \mathcal{S}_{\varepsilon}= 0 .
\end{aligned}
\label{eq:eps-weak}
\end{equation}

In contrast to our previous work on free-slip boundary conditions \cite{bayat2025density}, the stabilization contributions ($\mathcal{S}_{u,p}$, $\mathcal{S}_{k}$, $\mathcal{S}_{\varepsilon}$) are merely the default COMSOL implementations of pressure, streamline and discontinuity stabilization \cite{comsol_stabilization_6_3}.

\subsection{Summary of the implicit wall-function topology optimization}

\begin{figure}
\centering
\resizebox{0.75\textwidth}{!}{%
\begin{tikzpicture}[scale=1.02,every node/.style={scale=1}, node distance=10mm and 18mm]

\node[font=\Large\bfseries] (title) {};
%
\node[coordinate, below=25mm of title] (mid) {};

\node[ below=8mm of title] (start) {};

\node[block, left=8mm of mid , yshift=15mm] (init) {%
\text{Initial guess for } $(\boldsymbol{u},p,k,\varepsilon)$\\
$k_0=\left(\frac{\nu}{L_0}\right)^2,\qquad
\varepsilon_0=C_\mu\,\dfrac{k_0^{3/2}}{L_0}$\\
$\boldsymbol{u}= 0,\qquad p=0$
};

\node[block, right=8mm of mid , yshift=15mm] (bcs) {%
\text{Boundary conditions}\\[2pt]
\text{Inlet:} $\ \boldsymbol u=\text{constant}$\\[-1pt]
\hspace*{6mm}$k=C_{bc}\,\lVert\boldsymbol u\rVert^{2}$
\hspace*{6mm}$\varepsilon=C_\mu\,\dfrac{k^{3/2}}{L}$\\[2pt]
$0.003 \le C_{\mathrm{bc}} \le 0.01$\\[2pt]
\text{Outlet:} do-nothing
};

\node[note, below=37mm of start.south] (mom) {Solve momentum weak form \Cref{eq:momentum-continuity-weak}\\
\hspace*{20mm} Derive $\boldsymbol{u}$ and p};

\node[note, below=5mm of mom.south] (kep) {Solve $k$-$\varepsilon$ weak form \Cref{eq:k-weak,eq:eps-weak}\\
using $\nu_t$ and $\beta$ from the previous iteration};

\node[note, below=5mm of kep.south] (keppos) {Check the values of $k$ and $\varepsilon$ to be positive};

\node[diamond, fill=orange!10, draw, aspect=2, below=5mm of keppos.south, align=center] (cond) {$\delta t \geq 10000$,\\
$\nu_t^{i+1}- \nu_t^{i}  \leq \zeta$
};

\node[note,left=10mm of cond] (nuup) {Update $\nu_t$, and $P_k$\\ 
\hspace*{5mm}Increase $\delta t$\\
\hspace*{8mm}i = i+1};

\node[note, below right=-12mm and 20mm of cond] (mma){Update $\gamma$,\\
using MMA\\ 
algorithm.\\
$n = n+1$};

\node[note, below right=12mm and 33mm of cond] (opup){Next\\ continuation\\
parameters.\\ 
$j = j+1$};

\node[note,below=5mm of cond] (adj) {Calculate the derivative of the objective\\ function with respect to 
the design variale $\gamma$\\ using adjoint method
and automatic differentiation};

\node[diamond, fill=orange!10, draw, aspect=2, below=5mm of adj.south, align=center] (cond2) {$n = n_{max}$, or \\
$Tol \leq \zeta_2$};

\node[diamond, fill=orange!10, draw, aspect=2, below=5mm of cond2.south, align=center] (cond3) {$j= j_{max}$};

\node[ellipse, draw, left=18mm of cond3] (End) {End};

\node[note, above=3mm of init.north , yshift=3mm] (labInit) {Initial guess};
\draw[line] (labInit.south) -- (init.north);

\node[note, below=8.5mm of mid] (labStart) {Calculate wall intensities ($\psi$,$\psi_\varepsilon$,$\psi_p$), $\boldsymbol{n}$, and $\alpha(\varphi)$};

\node[note, above=3mm of bcs.north] (labBC) {Boundary condition};
\draw[line] (labBC.south) -- (bcs.north);

\coordinate (merge) at ($(labStart) - (0,-8mm)$);
\coordinate (merge2) at ($(opup) - (0,-70mm)$);
\draw[line] (labStart.south) -- (mom);
\draw[line] (init.south) -- (merge);
\draw[line] (bcs.south) -- (merge);
\draw[line] (merge) -- (labStart.north);
\draw[line] (mom.south) -- (kep);
\draw[line] (kep.south) -- (keppos);
\draw[line] (keppos.south) -- (cond);
\draw[line] (cond.west) -- node[above]{No} (nuup.east);
\draw[line] 
  (nuup.north) -- ++(0,10mm)  
  |- (mom.west);
  \draw[line] (cond.south) -- (adj);
\draw[line] (cond2.east) -- ++(22mm,0) node[above]{No} -- ++(0,25mm) -> (mma.south);
\draw[line] (adj.south) -- (cond2);
\draw[line] 
  (mma.north) -- ++(0,15mm)  
  |- (merge2);
\draw[line] (cond3.east) -- ++(40.75mm,0) node[above]{No} -- ++(0,20mm) -> (opup.south);
\draw[line] 
  (opup.north) -- ++(0,10mm)  
  |- (merge2);
\draw[line] (cond2.south) -- (cond3);
\draw[line] (cond3.west) -- (End);
\draw[line] (merge2) |- (labStart.east);

\end{tikzpicture}
}
\caption{This flowchart presents the step-by-step procedure of the proposed method for density-based turbulent topology optimization with wall-functions defined directly within the design domain, including the specific boundary conditions and continuation scheme used.}
\label{fig: TuTO flowchart}
\end{figure}
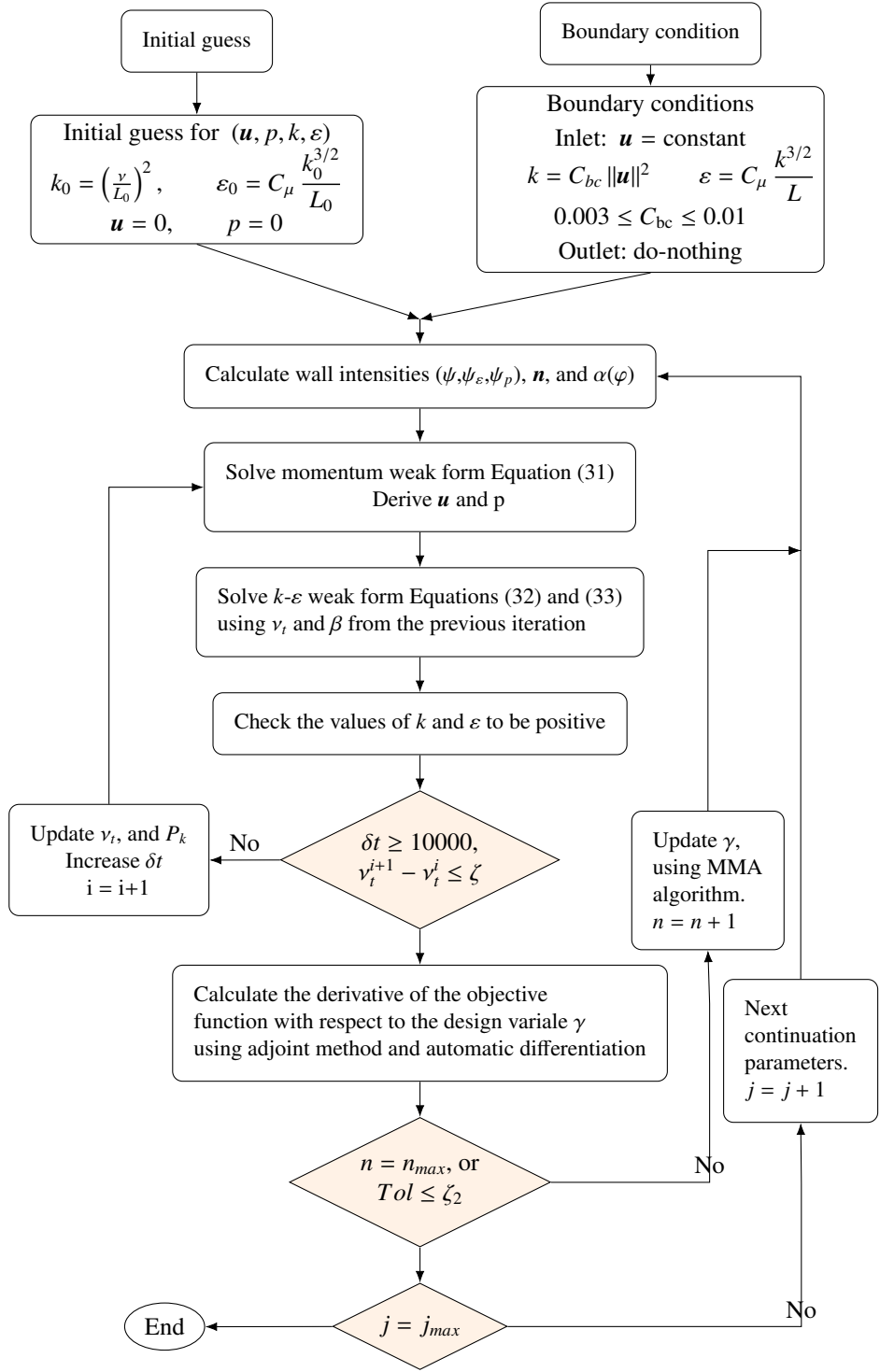

    The proposed implicit wall-function method implementation details are summarized in \Cref{fig: TuTO flowchart}. As illustrated, the wall normals and wall intensity calculated based on \Cref{fig: wall_flowchart} are incorporated into the $k$–$\varepsilon$ weak form equations. These quantities, together with the initial guesses and boundary conditions, are used in the weak formulation of the governing equations, which are solved using the finite element method with linear discretization. To obtain a stable solution of the $k$–$\varepsilon$ turbulence model, a pseudo-transient approach combined with linearization of the $k$ and $\varepsilon$ equations is employed. The pseudo-transient solution progressively converges to the steady-state solution as the time step is increased up to 10,000. The sensitivities are subsequently evaluated through a discrete adjoint formulation combined with automatic differentiation in COMSOL. The stopping criteria for the MMA is then when the relative change of all, scaled by their initial magnitudes, design variables; falls below the optimality tolerance $10^{-2}$;
or objective stagnation, $\frac{\lvert o_{k+1}-o_k\rvert}{\max\!\bigl(1,\lvert o_k\rvert\bigr)} < \zeta_2$ \text{for five consecutive iterations},
in which $o$ is the objective function
with $\zeta$ a small tolerance; or
at last when a safeguard of at most 25 MMA iterations (Pipe-bend and U-bend cases) and 30 MMA iterations (for Tesla-valve) per continuation step. Once the stopping criteria are satisfied, the procedure advances to the next continuation step, where the topology optimization parameters are updated and the process is repeated. We utilize a continuation scheme, starting with a high $q_a$ (i.e., a lower penalty) and then reducing it continuously. This helps the fluid access most regions, mitigates the influence of the initial Brinkman penalty on the design evolution, and is beneficial for non-convex problems \Citep{alexandersen2023detailed}. Furthermore, we also gradually increase the projection steepness parameter, $\beta_0$.

\section{Validation of the proposed implicit wall-function method}\label{Section: validation}

\begin{figure}
    \centering
    
    \subfloat[Explicit domain\label{subfig:schematic_explicit}]{
        \includegraphics[width=0.75\linewidth]{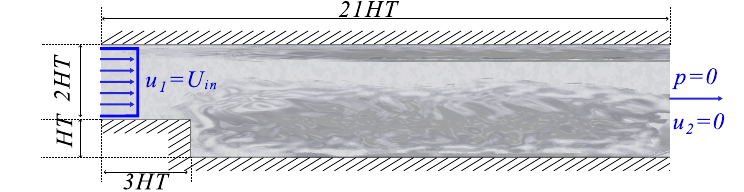}
    }\\[0.5em]
    
    \subfloat[Diffuse domain\label{subfig:schematic_implicit}]{
        \includegraphics[width=0.75\linewidth]{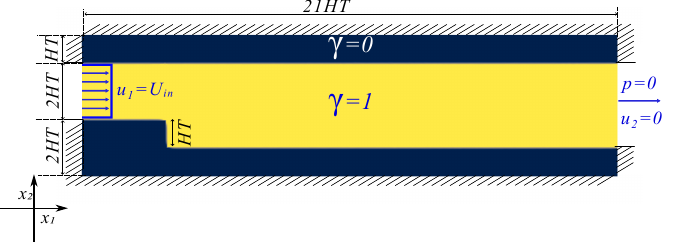}
    }
    
    \caption{Schematics of the backward-facing step benchmark geometry, employed to validate the proposed implicit wall-function method. The explicit wall case is shown in (a), while the diffuse domain used for implicit wall case, is presented in (b). $\gamma=1 $ represent the fluid domain, and $\gamma= 0$ represent the solid domain.}
    \label{fig:schematics_comparison}
\end{figure}

\begin{figure}
    \centering
    \includegraphics[width=0.6\linewidth]{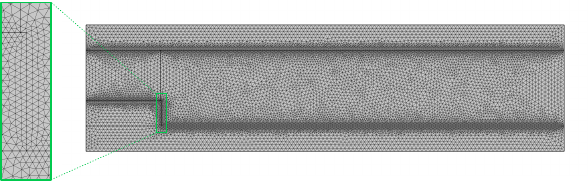}
    \caption{Schematic of the computational mesh employed to validate the proposed implicit wall-function method. The mesh is refined with an element size of $h_{max}=$ $0.002$ at the solid-fluid interface and coarsened to $h_{max}=$ $0.006$ in the free stream region and inside the solid domain ($h_{max}$ is the maximum allowable element size). In total the mesh has 24,668 triangles and 87,486 DOFs. }
    \label{fig: Schematic_kim_mesh}
\end{figure}

To validate the implementation of the implicit wall-function method in density-based TuTO, we compare it with the experimental data from the classic backward-facing step benchmark case by Kim~\cite{kim1978investigation}. This problem is widely used to evaluate separation and reattachment behavior in turbulent flows and is illustrated in \Cref{fig:schematics_comparison}.
The Reynolds number is calculated based on the momentum thickness of the boundary layer at the separation point ($x_1 = 0$), yielding a value of approximately $1.3 \times 10^3$, consistent with the experimental setup reported by Kim et al.\cite{kim1980investigation}. The Reynolds number based on the momentum thickness is defined as:
\begin{equation}
\text{Re}_\theta = \frac{U_0 \, \theta}{\nu}
\label{eq:Reynolds_Momentum}
\end{equation}
where $U_0$ denotes the free-stream velocity, $\nu$ is the kinematic viscosity of the fluid, and $\theta$ is the momentum thickness, computed as:
\begin{equation}
\theta = \int_0^{\infty} \frac{u(y)}{U_0} \left( 1 - \frac{u(y)}{U_0} \right) dy
\label{eq:momentum_thickness}
\end{equation}
The given momentum thickness Reynolds number is roughly equivalent to an inlet Reynolds number of 89,600. The simulations are performed with a kinematic viscosity of 
$\nu = 8.5\times 10^{-7}\, \textrm{m}^2\cdot\textrm{s}^{-1}$ 
and a step height of 
$HT = 0.0381\,\textrm{m}$, 
while the inlet velocity is prescribed as a uniform 
$U_{\mathrm{in}} = 1\, \textrm{m}\cdot\textrm{s}^{-1}$. Parameters used for the backward-facing step validation benchmark are summarized in \Cref{tab:backward--facing step_parameters}, and the mesh is shown in \Cref{fig: Schematic_kim_mesh} with a maximum allowable element size of $h_{max}=0.002$ on the wall and $h_{max}=0.006$ in the free stream.
It should be noted, that a relatively coarse mesh is used on purpose, to illustrate the fact that wall-functions allow this.
This benchmark provides a physically meaningful scenario for assessing the behavior of the proposed method in capturing wall effects. 

\begin{table}[]
    \caption{Parameters used for the backward-facing step benchmark validation.}
    \centering
    \label{tab:backward--facing step_parameters}
    \begin{tabular}{@{}l l@{}}
        \toprule
        \textbf{Parameter} & \textbf{Value} \\
        \midrule
        $HT$        & 0.0381 \\
        $\nu$          & $8.5\times 10^{-7}$\\
        $\psi_{\max}$  & 5000 \\
        $P_{con}$ & 4 \\
        $r_{1}$ & 4$h_{max}$\\
        $r_{2}$ & 2$h_{max}$\\
        $\alpha_{max}$ & 100 \\
        $U_{\mathrm{in}}$ & 1 \\
        $Re_{in}$ & 89,600 \\
        \bottomrule
    \end{tabular}
\end{table}

The schematic of the backward-facing step used for validating the proposed implicit wall-function method is presented in~\Cref{fig:schematics_comparison} for both the: (a) normal geometric representation with explicit walls; (b) density-based ``diffuse domain'' representation with implicit walls. Simulations are carried out for three different cases:
\begin{itemize}
    \item Case~\#1 employs an explicit wall representation with wall-functions and is considered the reference accurate solution;
    \item Case~\#2 applies the proposed implicit wall-function method;
    \item Case~\#3 follows the ``conventional''  density-based TuTO flow modeling approach without wall-functions, incorporating Brinkman-type penalization for all variables (velocity, $k$ and $\varepsilon$) \cite{yoon2020topology} to push $k$ and $\varepsilon$ towards zero in solid domains.
\end{itemize}
A relatively coarse mesh is used to emphasise that wall-functions allow for this and to illustrate that the ``convection'' TuTO model is inaccurate on commonly-used coarse meshes.

\begin{figure}
    \centering
    \includegraphics[width=0.6\linewidth]{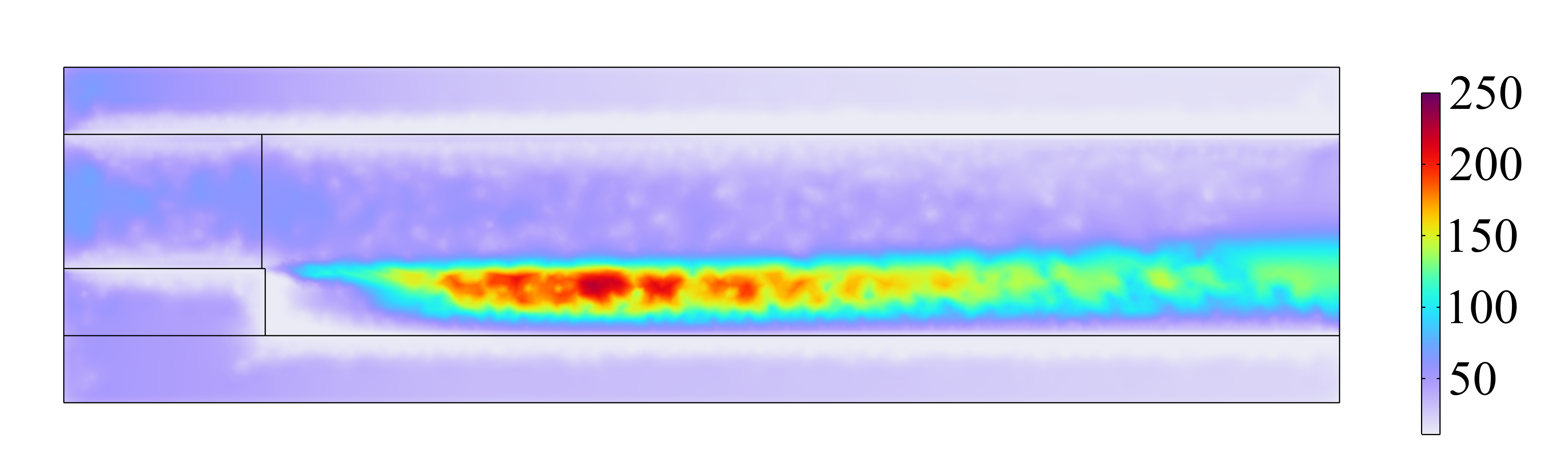}
    \caption{Implicit $y^{+}$ distributions for the backward-facing step benchmark geometry, obtained using the proposed implicit wall-function method. In the vicinity of the walls, where the wall intensity field, $\psi$, is large, $y^{+}$ remains between 11.06 and 300, which satisfies the logarithmic law of the wall.}
    \label{fig: Validation_y_plus}
\end{figure}
The proposed implicit wall-function method employs an implicit $y^+$ field that is defined throughout the entire domain, in contrast to explicit wall-function approaches where $y^+$ is defined only on the wall. The volumetric field is illustrated in \Cref{fig: Validation_y_plus}, where the values are only practically used near the boundary where the wall intensity field, $\psi$, is large.

\begin{figure}
    \centering
    \subfloat[$x_1/HT = -1$]{%
        \includegraphics[width=0.4\textwidth]{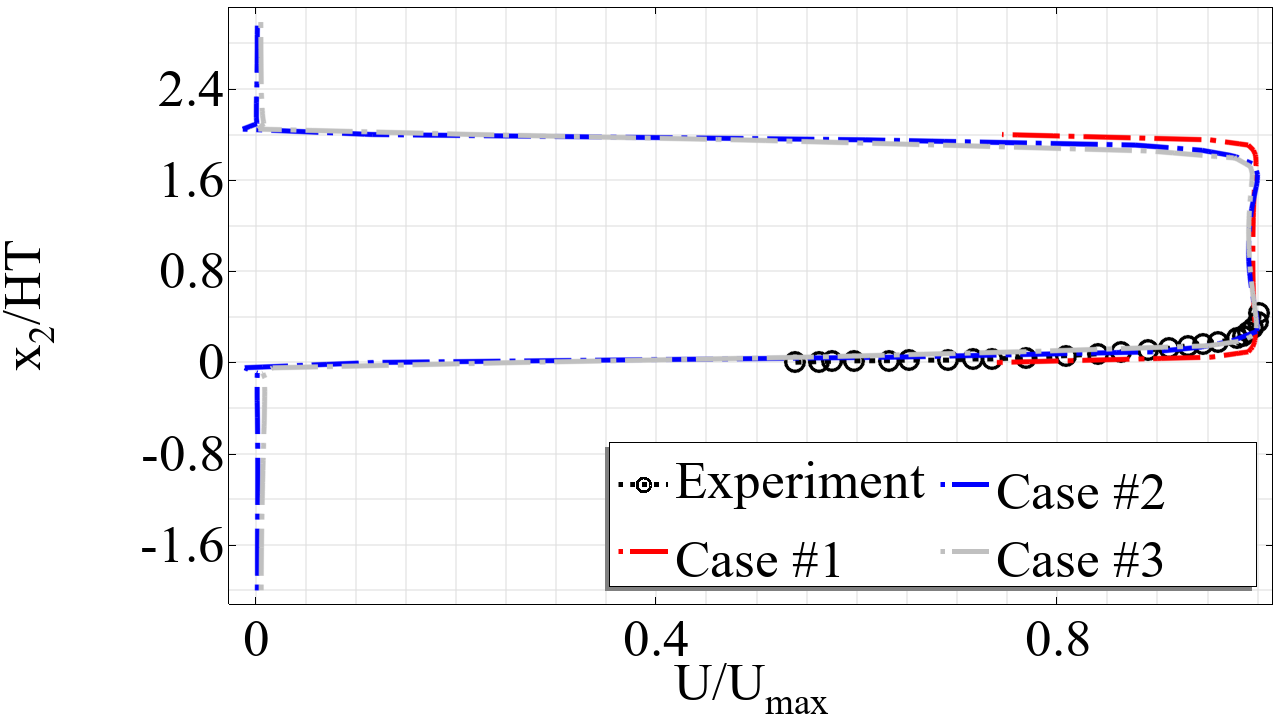}%
        \label{subfig:x_2_HT_minus1}
    }
    \hfill
    \subfloat[$x_1/HT = 0$]{%
        \includegraphics[width=0.4\textwidth]{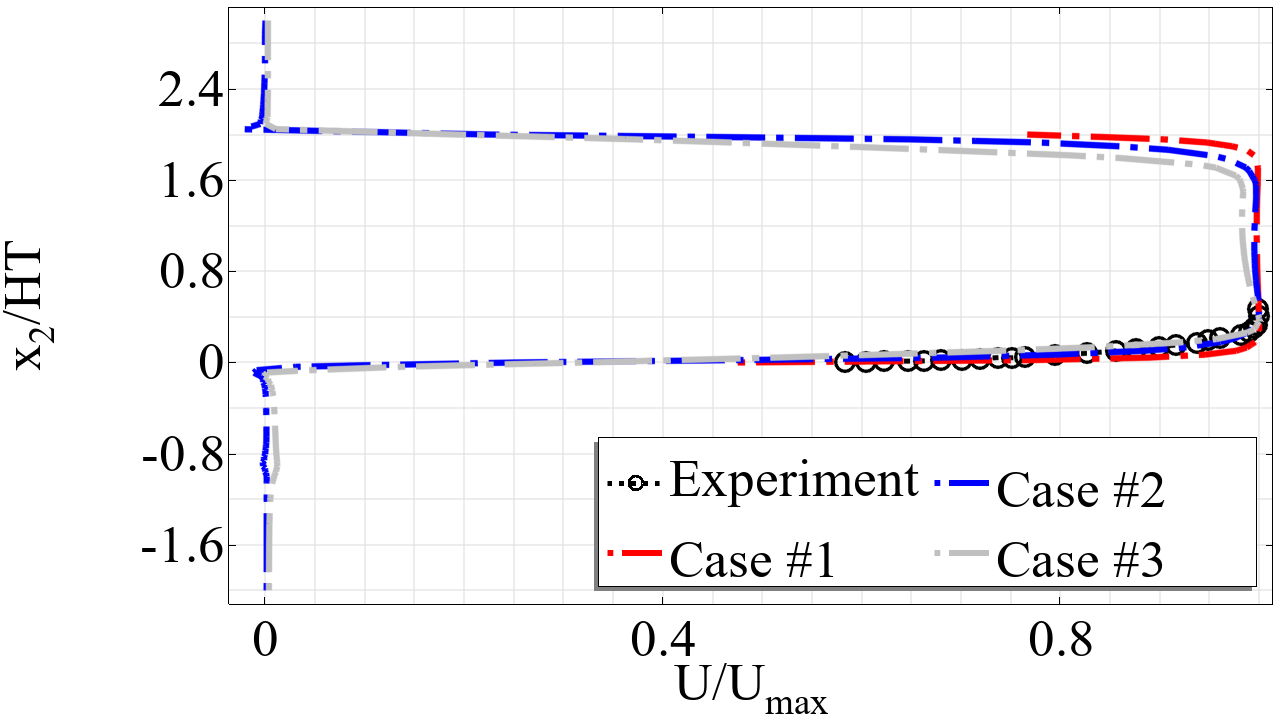}%
        \label{subfig:x_2_HT_0}
    }
    \\[0.2cm]
    \subfloat[$x_1/HT = 4/3$]{%
        \includegraphics[width=0.4\textwidth]{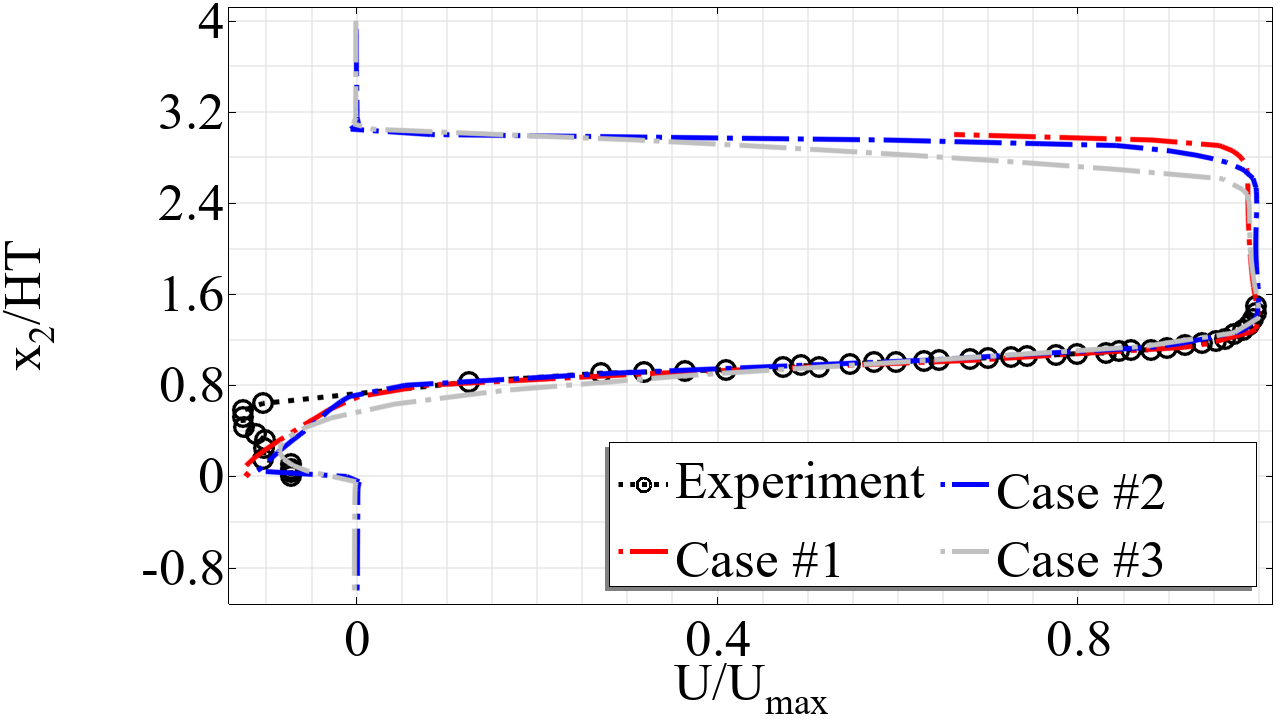}%
        \label{subfig:x_2_HT_4over3}
    }
    \hfill
    \subfloat[$x_1/HT = 8/3$]{%
        \includegraphics[width=0.4\textwidth]{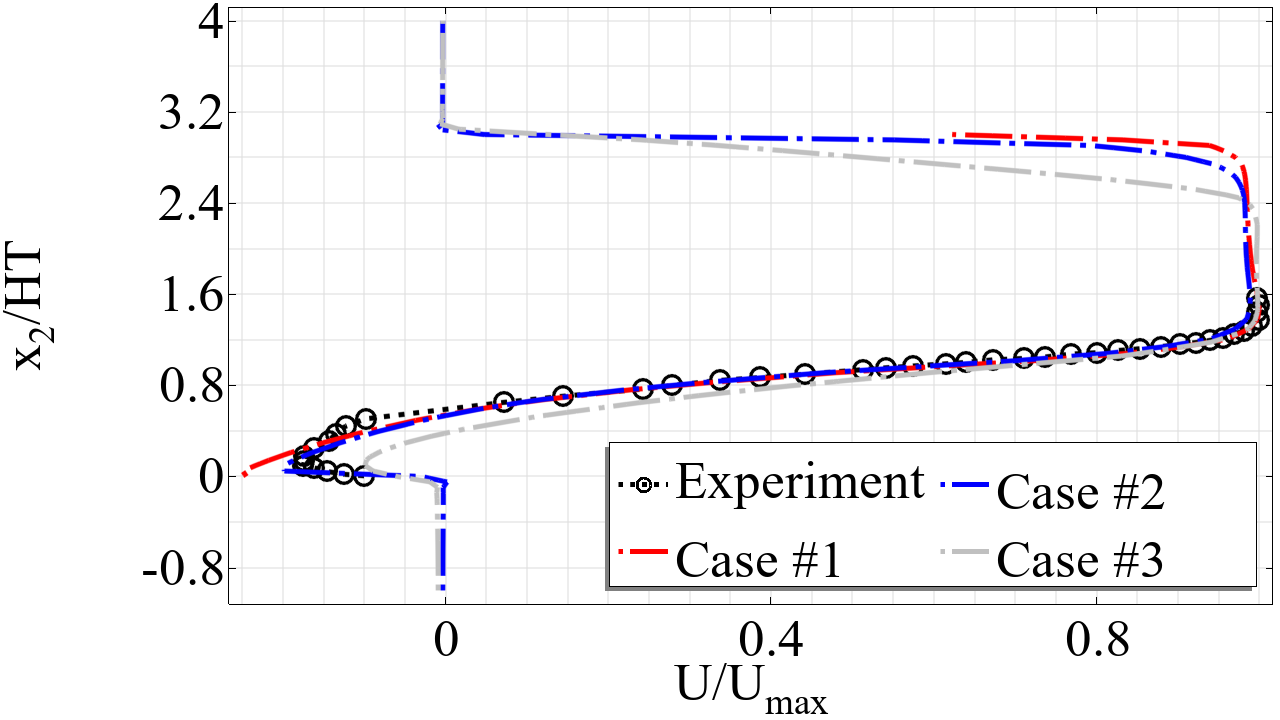}%
        \label{subfig:x_2_HT_8over3}
    }
    \\[0.2cm]
    \subfloat[$x_1/HT = 16/3$]{%
        \includegraphics[width=0.4\textwidth]{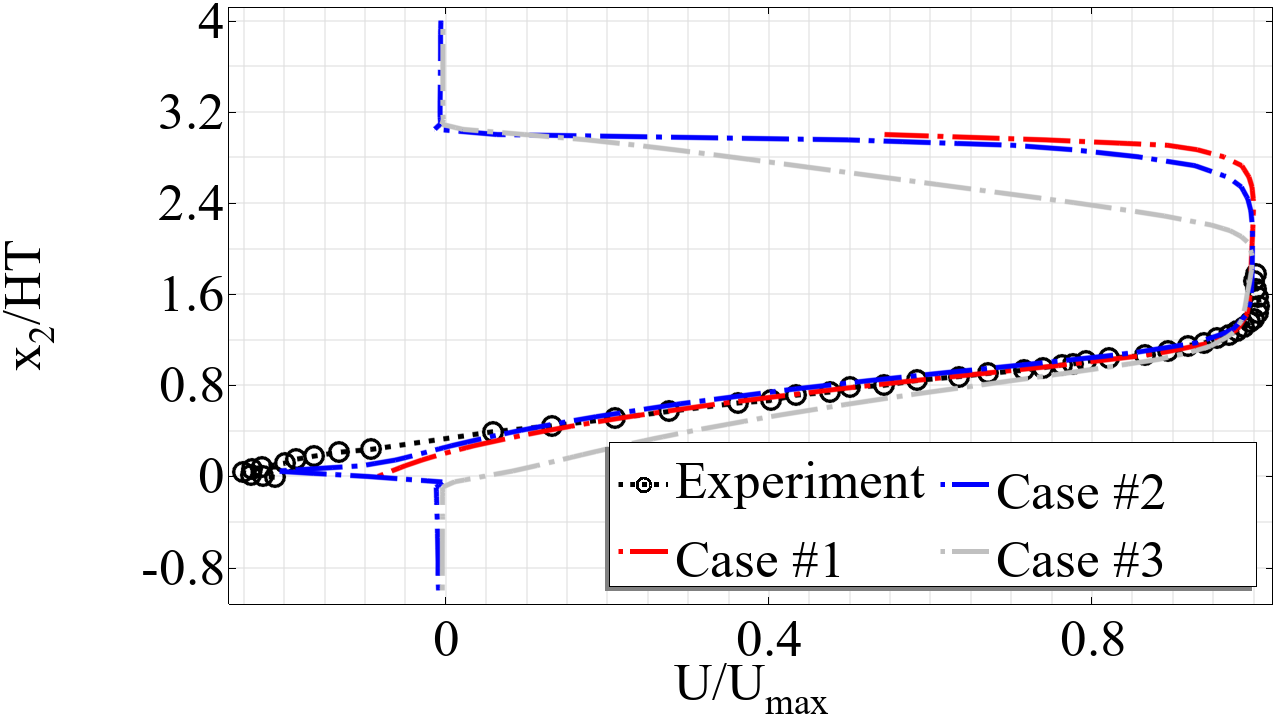}%
        \label{subfig:x_2_HT_16over3}
    }
    \hfill
    \subfloat[$x_1/HT = 24/3$]{%
        \includegraphics[width=0.4\textwidth]{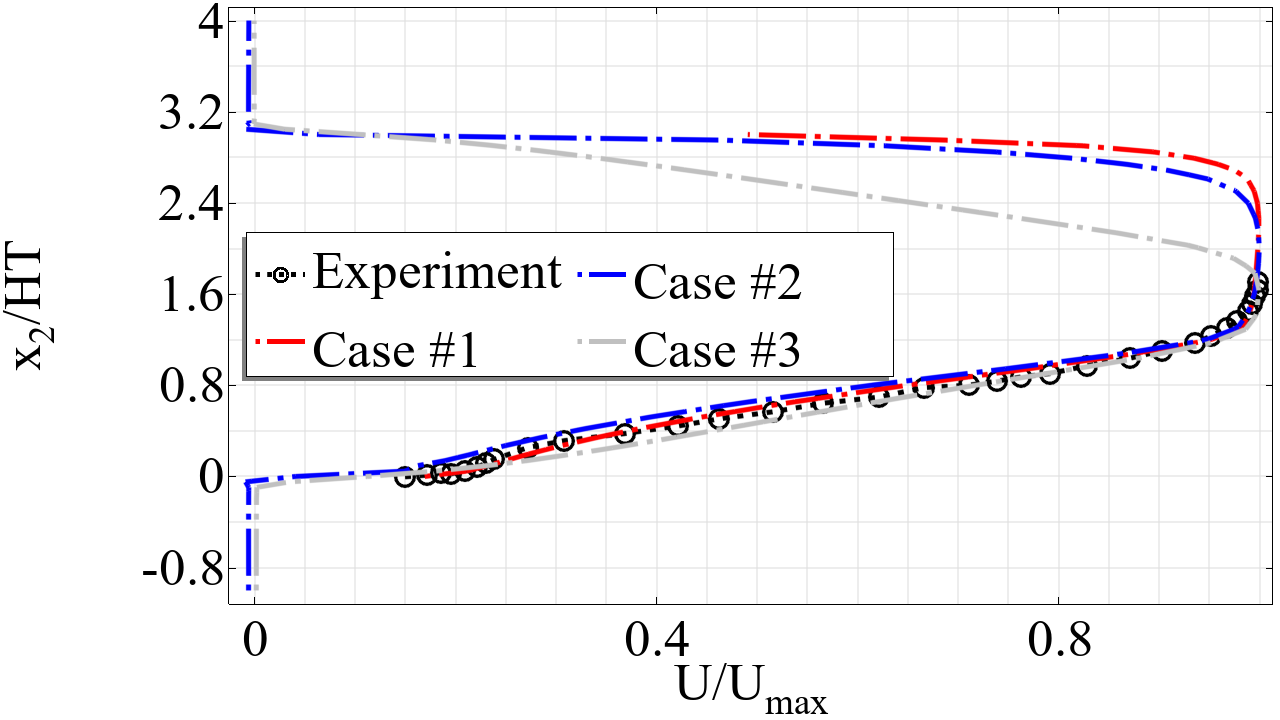}%
        \label{subfig:x_2_HT_24over3}
    }
    \\[0.2cm]
    \subfloat[$x_1/HT = 48/3$]{
        \includegraphics[width=0.4\textwidth]{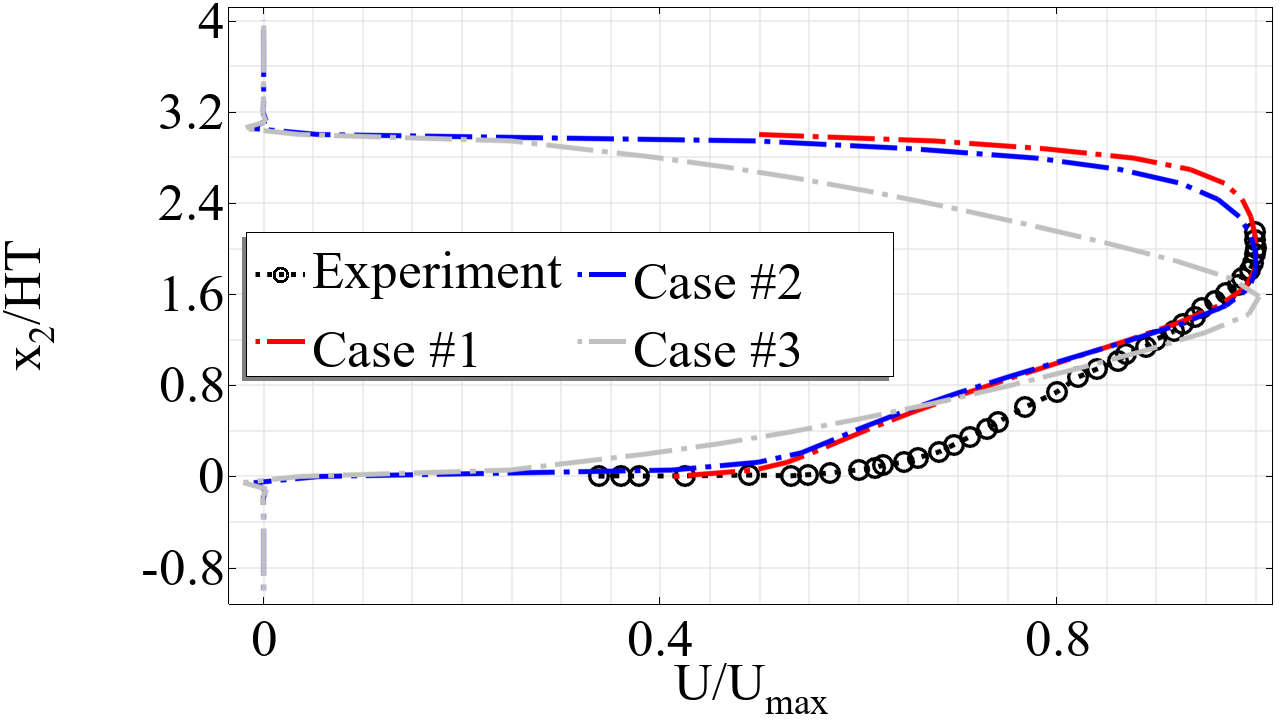}%
        \label{subfig:x_2_HT_48over3}
    }
    \caption{Schematics of the velocity profiles at various streamwise locations.
The experimental results are compared with three simulation cases: 
Case $\#1$ explicit wall model with explicit wall-function; 
Case $\#2$ the proposed implicit wall-function method; and 
Case $\#3$ ``conventional'' model with implicit wall.}
    \label{Fig:validation_cutting_line}
\end{figure}
\Cref{Fig:validation_cutting_line} shows the velocity profiles along different cutting lines at various distances from the inlet, for the three cases compared with the experimental results from Kim et al.\cite{kim1980investigation}. The results clearly indicate that the ``conventional'' method (case~\#3) fails to correctly capture the near-wall velocity distribution and deviates from the experimental real-life results, particularly at larger downstream locations. In contrast, the proposed implicit wall-function method (case~\#2) provides accurate predictions of the velocity field compared to the experimental data, as well as very close resemblance to the explicit wall model (case~\#1). 
It should be noted that all of the three cases do not sufficiently capture the velocity profile near the lower wall at the point furthest downstream, \Cref{subfig:x_2_HT_48over3}. This has nothing to do with the current implementation, but rather modeling errors introduced through the RANS and $k$-$\varepsilon$ models.

\begin{figure}
    \centering
    \subfloat[Velocity, Case~\#1]{%
        \includegraphics[width=0.32\textwidth]{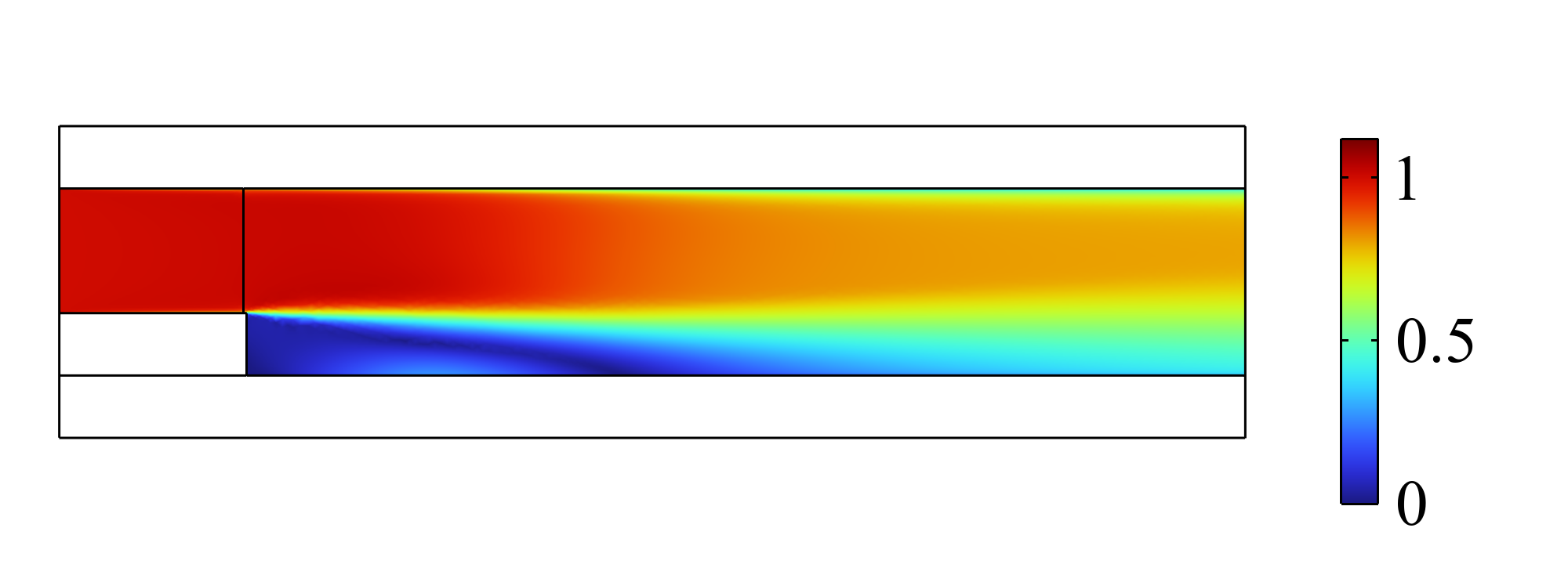}%
        \label{subfig:vel_case1}%
    }\hfill
    \subfloat[Turbulent kinetic energy $k$, Case~\#1]{%
        \includegraphics[width=0.32\textwidth]{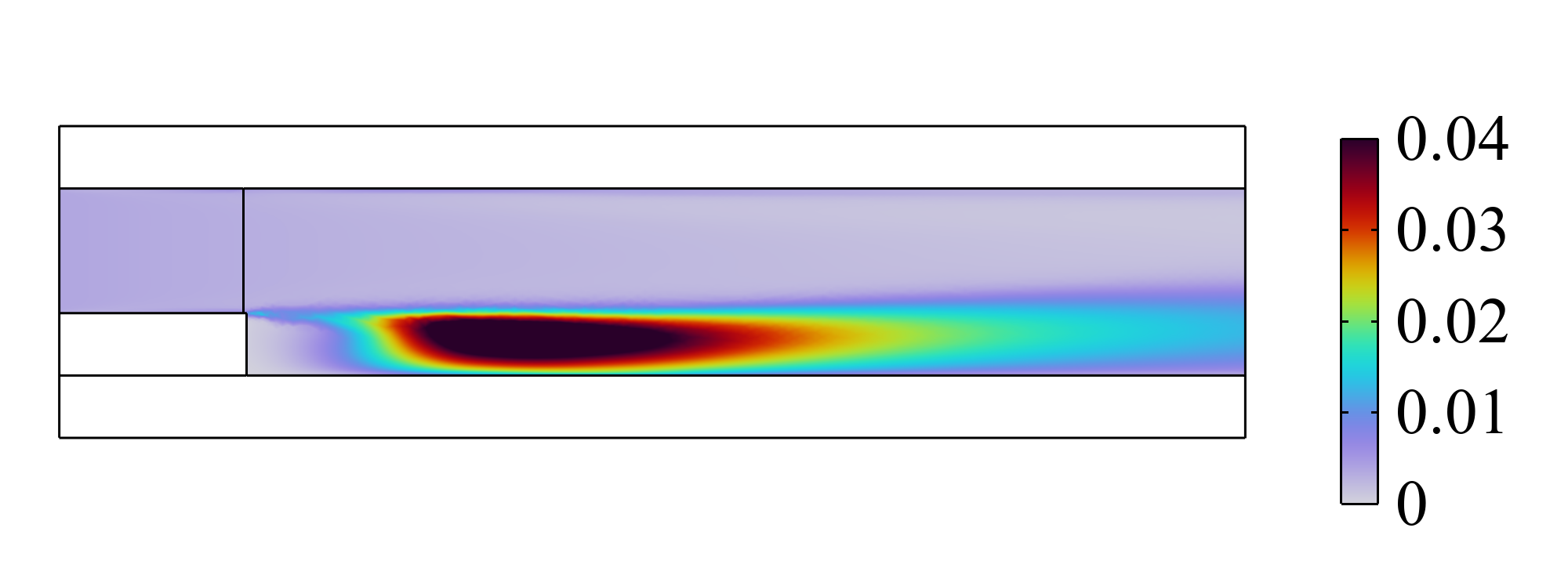}%
        \label{subfig:k_case1}%
    }\hfill
    \subfloat[Turbulent dissipation rate $\varepsilon$, Case~\#1]{%
        \includegraphics[width=0.32\textwidth]{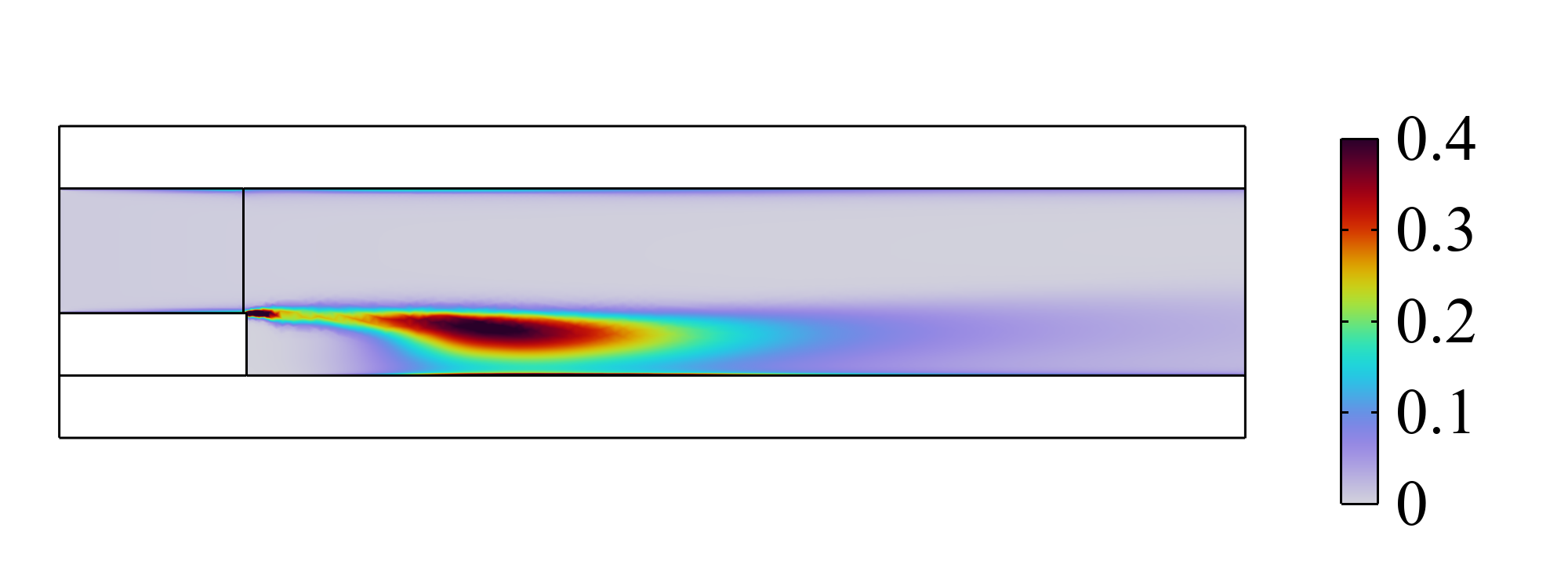}%
        \label{subfig:eps_case1}%
    }\\[2mm]
    
    \subfloat[Velocity, Case~\#2]{%
        \includegraphics[width=0.32\textwidth]{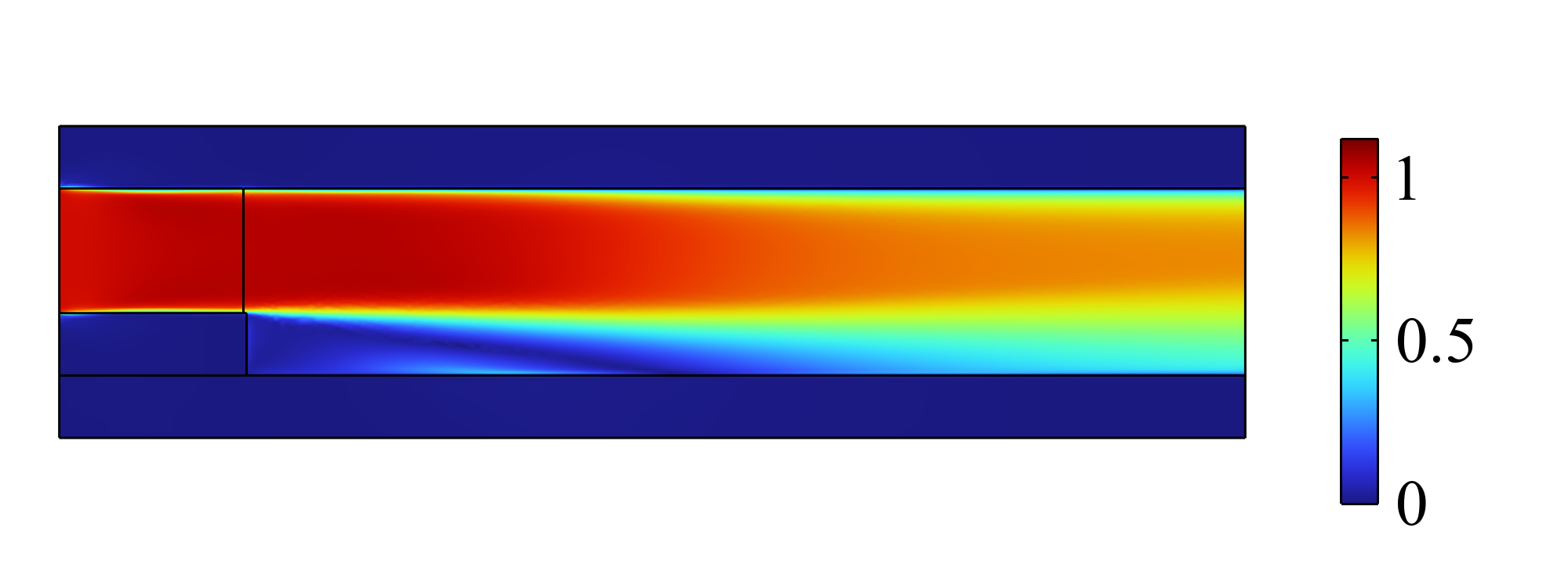}%
        \label{subfig:vel_case2}%
    }\hfill
    \subfloat[Turbulent kinetic energy $k$, Case~\#2]{%
        \includegraphics[width=0.32\textwidth]{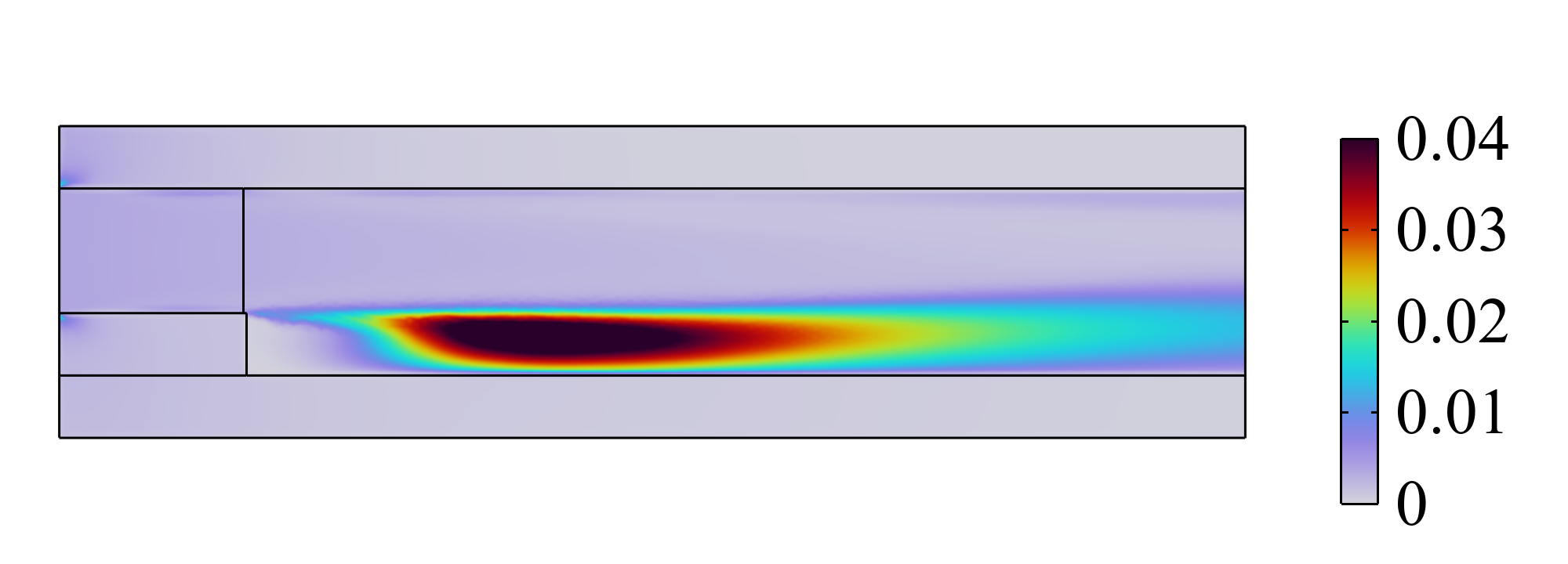}%
        \label{subfig:k_case2}%
    }\hfill
    \subfloat[Turbulent dissipation rate $\varepsilon$, Case~\#2]{%
        \includegraphics[width=0.32\textwidth]{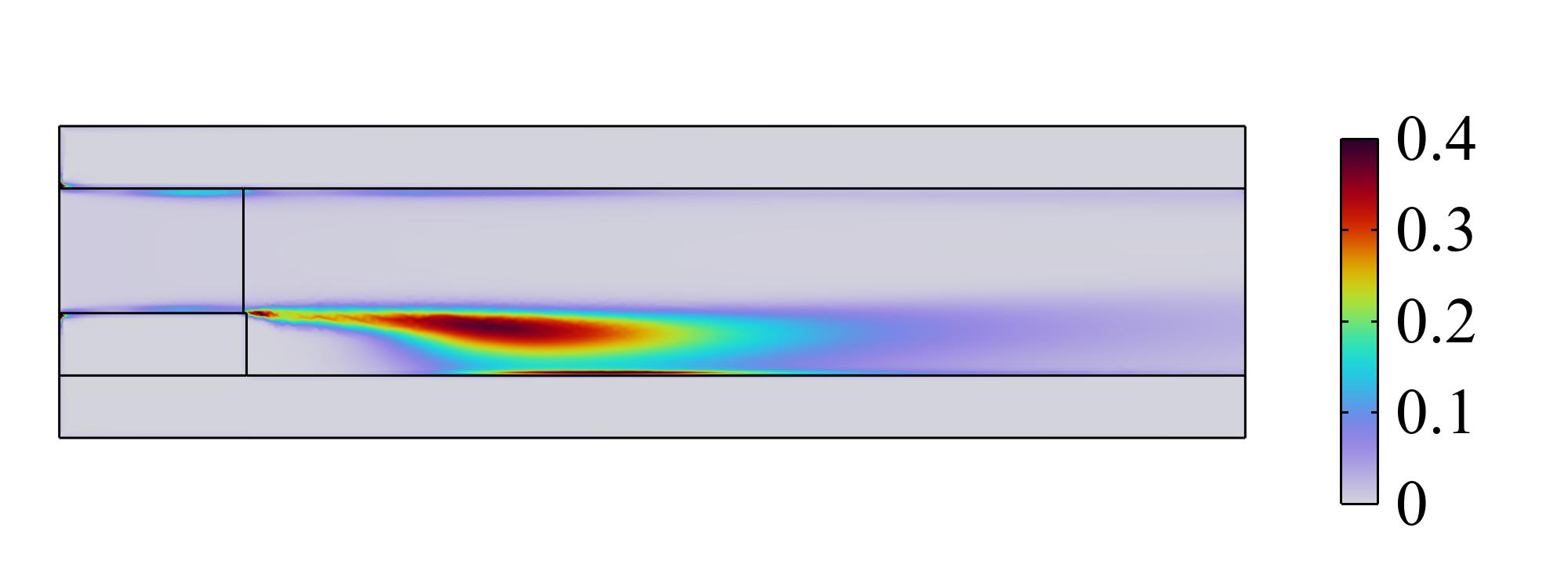}%
        \label{subfig:eps_case2}%
    }\\[2mm]
    
    \subfloat[Velocity, Case~\#3]{%
        \includegraphics[width=0.32\textwidth]{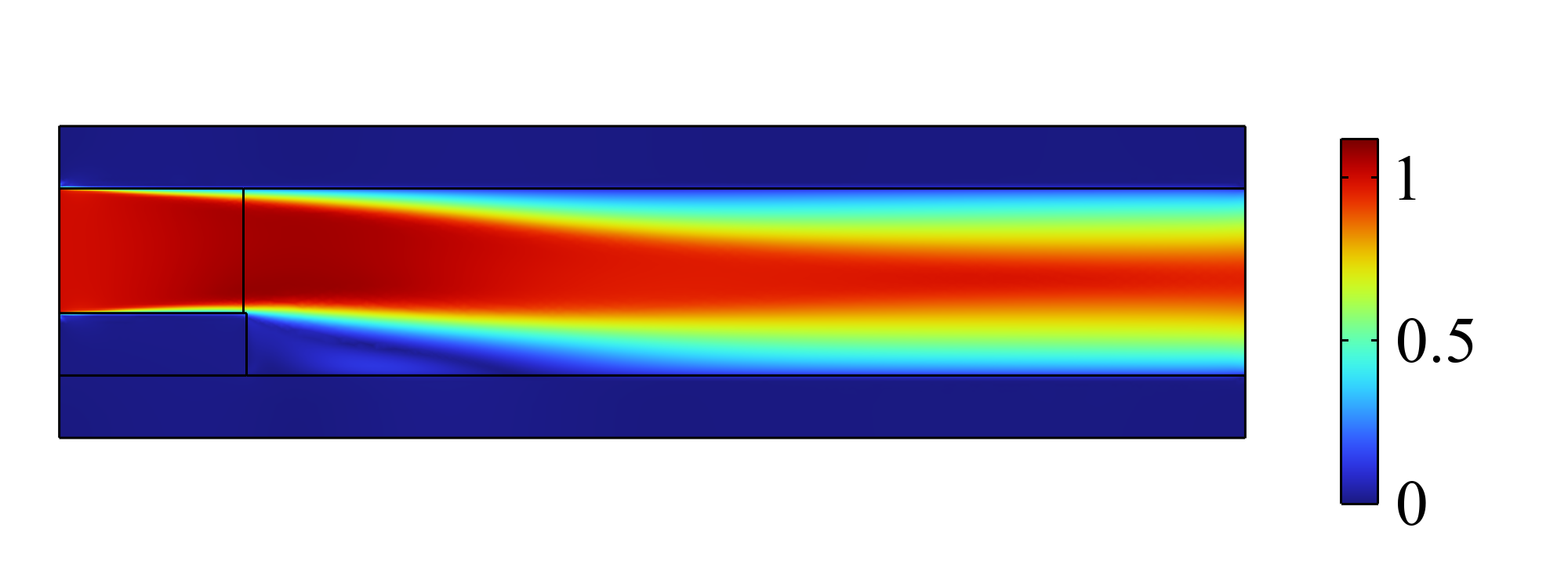}%
        \label{subfig:vel_case3}%
    }\hfill
    \subfloat[Turbulent kinetic energy $k$, Case~\#3]{%
        \includegraphics[width=0.32\textwidth]{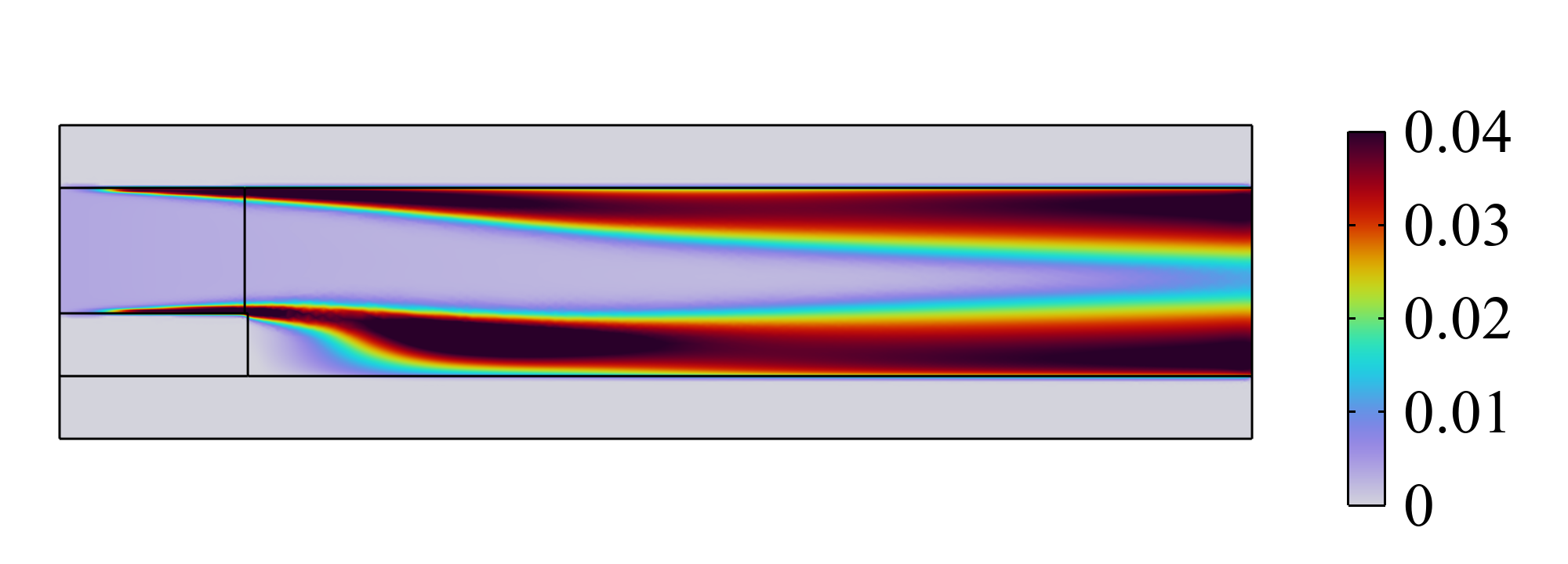}%
        \label{subfig:k_case3}%
    }\hfill
    \subfloat[Turbulent dissipation rate $\varepsilon$, Case~\#3]{%
        \includegraphics[width=0.32\textwidth]{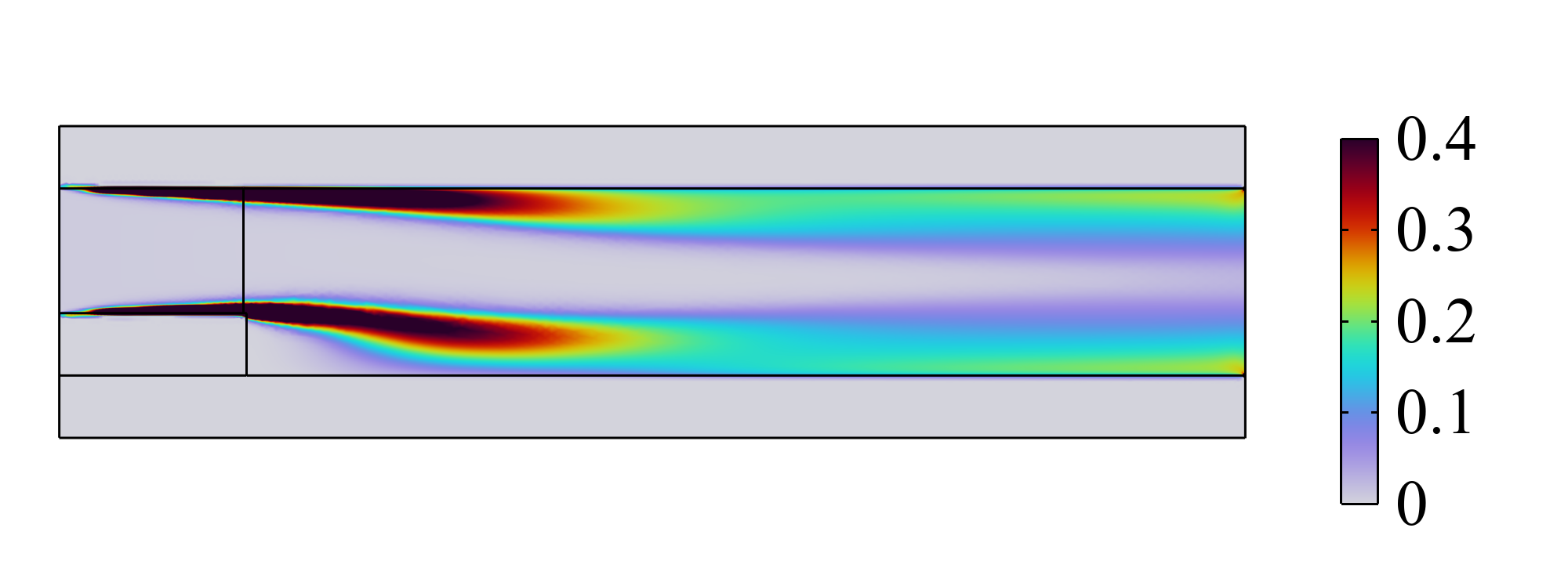}%
        \label{subfig:eps_case3}%
    }

    \caption{Visual summary of the comparison results for velocity $\boldsymbol{u}$, turbulent kinetic energy $k$, and turbulent dissipation rate $\varepsilon$ across different cases. Case~\#1 employs explicit wall-functions and is considered the reference accurate solution. Case~\#2 applies the proposed implicit wall-function method within a diffuse domain. Case~\#3 follows the ``conventional'' density-based turbulent flow modeling approach within a diffuse domain (applying volume forces on $k$ and $\varepsilon$~\cite{yoon2020topology}).}
    \label{fig:validation_summary}
\end{figure}
In~\Cref{fig:validation_summary}, the three cases are compared for the given mesh, demonstrating that the proposed implicit wall-function method (case~\#2) closely matches the explicit wall method (case~\#1), while differing significantly from the ``conventional'' method (case~\#3). It can be seen that the ``conventional'' approach predicts a significantly more curved flow path with boundary layers of a significant thickness further along the channel, with large turbulent quantities both at the top and bottom of the channel. This is because when applying no-slip velocity conditions and zero conditions on turbulent quantities, a coarse mesh will significantly over-predict the thickness of the boundary layer.
This is in stark contrast to the almost straight main flow path and very thin top boundary layer when using the wall-functions, both using explicit and implicit walls.

\section{Topology optimization test cases}\label{Section: Test cases}
\begin{figure}
    \centering
    \includegraphics[width=0.4\linewidth]{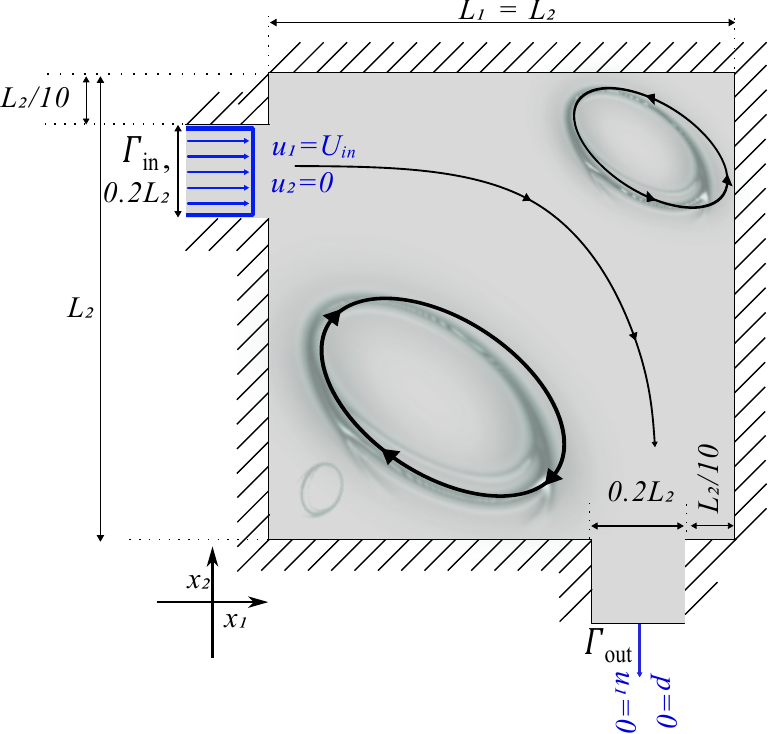}
    \caption{Pipe-bend schematic. The flow enters the square inlet region with a uniform velocity ($U_{\mathrm{in}}=1$), and the topology optimization seeks an optimal flow path that minimizes the average inlet pressure toward the outlet, where zero pressure and a normal-velocity outflow condition are prescribed.}
    \label{fig:Schematic_bending}
\end{figure}

\begin{figure}
    \centering
    \includegraphics[width=0.4\linewidth]{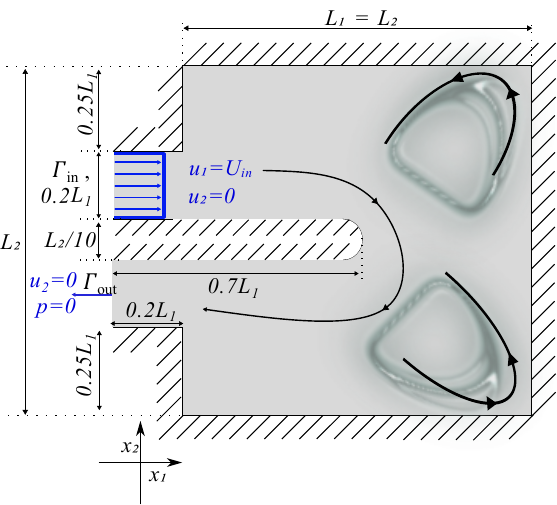}
    \caption{U-bend schematic. The flow enters the domain at the inlet with a uniform velocity ($U_{\mathrm{in}} = 1$) and, after following the optimized passage with the lowest average inlet pressure through the U-turn, exits at the outlet where a zero-pressure and normal-velocity boundary condition is imposed.}
    \label{fig:U_bending}
\end{figure}

\begin{figure}
    \centering
    \includegraphics[width=0.65\linewidth]{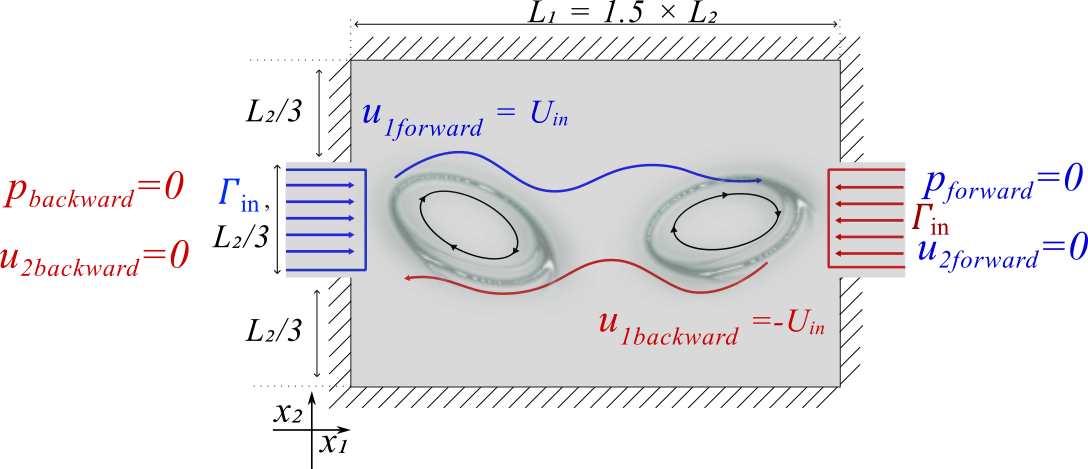}
    \caption{Schematic of the Tesla valve. Forward flow (blue) enters from the left with a uniform inlet velocity ($U_{\mathrm{in}} = 1$) and exits to the right under a zero-pressure, normal-velocity outlet condition. Backward flow (red) enters from the right with the same uniform inlet velocity and exits to the left. The topology optimization aims to find a geometry that maximizes the ratio of backward to forward average inlet pressure.}
    \label{fig:Tesla_valve}
\end{figure}

In this section, two common benchmark examples from the TuTO literature  are investigated using the proposed implicit wall-function  method and compared against the ``conventional'' method. We also treat the additional example of the turbulent Tesla valve problem. The considered cases are summarized in \Cref{tab:workbench_cases}, comprising three  problems: (i) the pipe-bend benchmark (\Cref{fig:Schematic_bending}), (ii) the U-bend benchmark (\Cref{fig:U_bending}), and (iii) the Tesla valve problem (\Cref{fig:Tesla_valve}).
\begin{table}[h!]
    \centering
    \caption{Problem configurations of the different examples are summarized, where cases employing the proposed implicit wall-function method are marked with “Yes” under the implicit wall-function column, while those following the ``conventional'' approach are indicated with “No”.}
    \label{tab:workbench_cases}
\begin{tabular}{lllll}
    \toprule
    \textbf{Problem} & \makecell{\textbf{Implicit}\\\textbf{wall-functions}} & \textbf{Vol. fraction} & \textbf{Reynolds} & \textbf{Example No.} \\
  
    Pipe-bend (\Cref{fig:Schematic_bending})    & No  & $V_f = 0.25$ & $5{,}000$ & Example \#1 \\
   
    Pipe-bend (\Cref{fig:Schematic_bending})    & Yes & $V_f = 0.25$ & $5{,}000$ & Example \#2 \\
    
    Pipe-bend (\Cref{fig:Schematic_bending})    & Yes & $V_f = 0.33 $ & $5{,}000$ & Example \#3 \\
   
    Pipe-bend (\Cref{fig:Schematic_bending})     & Yes & $V_f = 0.25$ & $200{,}000$ & Example \#4 \\
   
    U-bend (\Cref{fig:U_bending})    & No  & $V_f = 0.27$ & $5{,}000$ & Example \#5 \\

    U-bend (\Cref{fig:U_bending})    & Yes & $V_f = 0.27$ & $5{,}000$ & Example \#6 \\
  
    U-bend (\Cref{fig:U_bending})    & Yes & $V_f = 0.27$ & $200{,}000$ & Example \#7 \\
  
    Tesla valve (\Cref{fig:Tesla_valve})    & Yes & $V_f = 0.6$ & $5{,}000$ & Example \#8 \\
    \bottomrule
\end{tabular}
\end{table}

\subsection{Optimization problem formulations}\label{Subsection: Optimization problem}

For the pipe-bend and U-bend benchmarks, the objective functional is to minimize the average inlet pressure while ensuring that the fluid domain does not exceed the prescribed volume fraction. 
The optimization problem is formulated as:
\begin{equation}
\label{eq:optimization}
\begin{aligned}
{\min_{\gamma}}. \quad & \frac{\int_{\Gamma_{in}} p \, \mathrm{d}s}{{\int_{\Gamma_{in}} 1 \, \mathrm{d}s}} \\
\text{subject to} \quad 
& \frac{\int_{\Omega} \varphi \, \mathrm{d}V }{\int_{\Omega} 1 \, \mathrm{d}V}\leq V_f \\
& 0 \leq \gamma(\boldsymbol{x}) \leq 1 \quad \forall \boldsymbol{x} \in \Omega
\end{aligned}
\end{equation}

For the Tesla valve \cite{tesla1920valvular} case, the objective functional aims to minimize the ratio between the forward and backward average inlet pressures, thereby enhancing the so-called diodicity of the Tesla valve \cite{lin2015topology}. The maximum fluid volume fraction is limited to a value of $V_f=0.6$. 
The optimization problem is formulated as:
\begin{equation}
\label{eq:optimization2}
\begin{aligned}
{\min_{\gamma}}. \quad &  \int_{\Gamma_{in}} \frac{p_{\text{forward}}(\gamma(\boldsymbol{x}))}{p_{\text{backward}}(\gamma(\boldsymbol{x}))} \, \mathrm{d}s \\
    \text{subject to} \quad & \frac{\int_{\Omega} \varphi \, dV }{ \int_{\Omega} 1 \, \mathrm{~d}V }\leq V_f\\
& 0 \leq \gamma(\boldsymbol{x}) \leq 1 \quad \forall \boldsymbol{x} \in \Omega
\end{aligned}
\end{equation}

\subsection{Pipe-bend benchmark} \label{sec:bending_pipe_benchmark}
The pipe-bend is used as the first benchmark to compare optimized designs obtained with the proposed implicit wall-function method and the conventional approach. The problem setup is illustrated in \Cref{fig:Schematic_bending}, with a uniform inlet velocity $U_{\mathrm{in}}=1$ and a zero-pressure and a normal-velocity condition outlet.
\begin{table}[]
    \caption{Pipe-bend with proposed implicit wall-function method parameters.}
    \centering
    \label{tab:bending sweep_parameters}
    \begin{tabular}{@{}l l l l l l l l@{}}
        \toprule
        \textbf{Parameter} & \multicolumn{7}{l}{Value} \\
        \midrule
        $\beta$        & 4 & 6  & 9  & 13 & & & \\
        $q_a$          & 150 & 75 & 30 & 15 & & & \\
        $\psi_{\max}$  & 1000 & & & & & & \\
        $P_{con}$      & 4 & & & & & & \\
        $r_{1}$        & \multicolumn{7}{l}{$4 h_{\max}$} \\
        $V_f$          & \multicolumn{7}{l}{$\dfrac{1}{4}$ or $\dfrac{1}{3}$} \\
        $r_{2}$        & \multicolumn{7}{l}{$6 h_{\max}$} \\
        $Re_{inlet}$   & \multicolumn{7}{l}{$5000$} \\
        $h_{\max}$     & \multicolumn{7}{l}{$0.007$} \\
        $L_{1}$        & \multicolumn{7}{l}{$1$} \\
        $\alpha_{\max}$& \multicolumn{7}{l}{100} \\
        $U_{\mathrm{in}}$ & \multicolumn{7}{l}{1} \\
        \bottomrule
    \end{tabular}
\end{table}
The corresponding optimization parameters regarding the continuation approach are summarized in \Cref{tab:bending sweep_parameters}.
\begin{figure}
    \centering
    \includegraphics[width=0.4\linewidth]{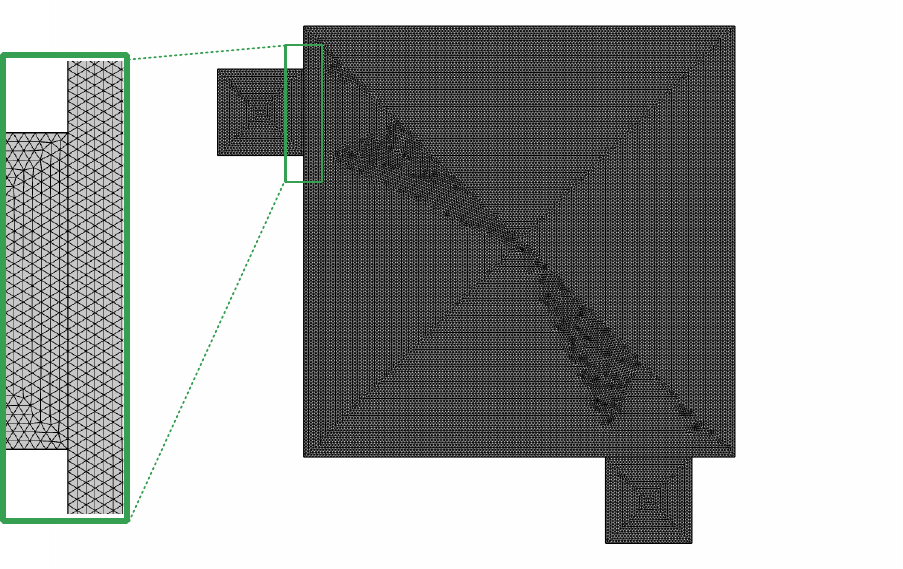}
    \caption{The mesh used for the pipe-bend benchmark is illustrated here for an approximate element size of $h_{max}=$ $0.007$, yielding 56,302 triangles, and 199,486 total DOFs.}
    \label{fig:bending_mesh}
\end{figure}
The computational mesh is presented in \Cref{fig:bending_mesh} with 56,302 triangles and a total number of degree of freedoms (DOFs) of 199,486.

\begin{figure}
    \centering
    \includegraphics[width=0.4\linewidth]{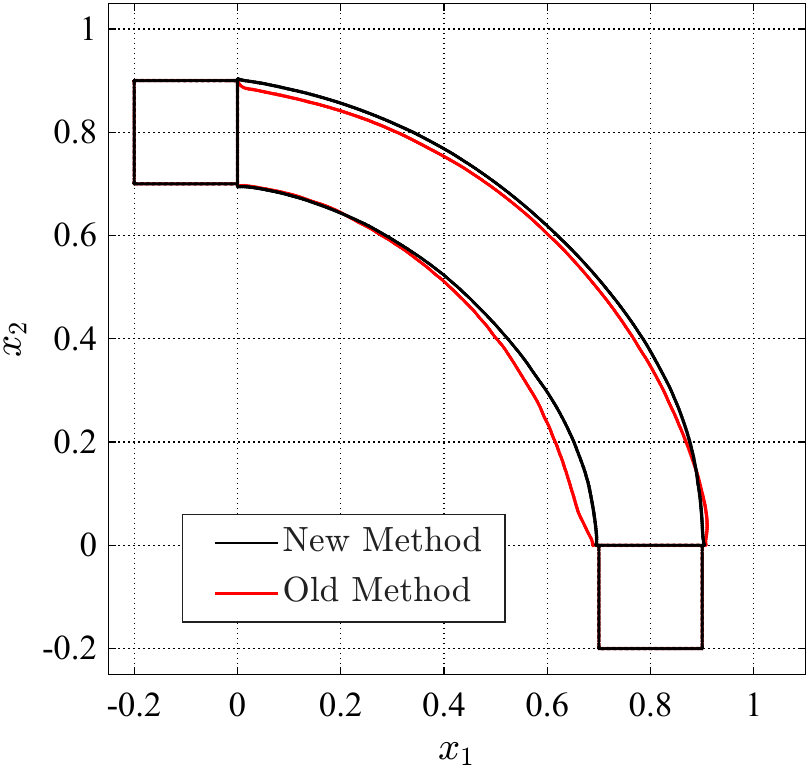}
    \caption{Comparison between optimized geometries resulted from proposed implicit wall-function method (Example \#2), and ``Conventional'' method (Example \#1, 
    with Brinkman penelaty on $k$ and $\varepsilon$, and without the implicit wall-functions), optimized result for $\text{Re}_{inlet}$ = 5000.}
    \label{fig:Comparison_old_new_methods}
\end{figure}
\begin{figure}
    \centering
    \subfloat[Proposed implicit wall-function method]{%
        \includegraphics[width=0.4\textwidth]{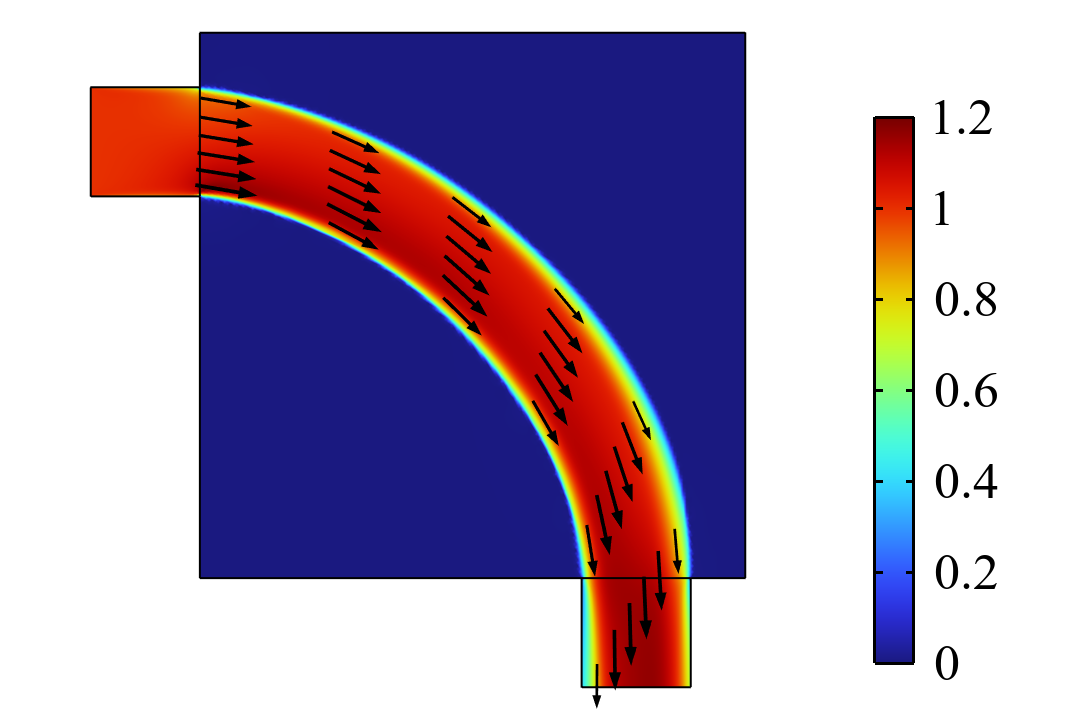}%
        \label{subfig:bending_tuto_velo}
    }
    \hspace{0.05\textwidth}
        \subfloat[``Conventional'' method]{%
        \includegraphics[width=0.4\textwidth]{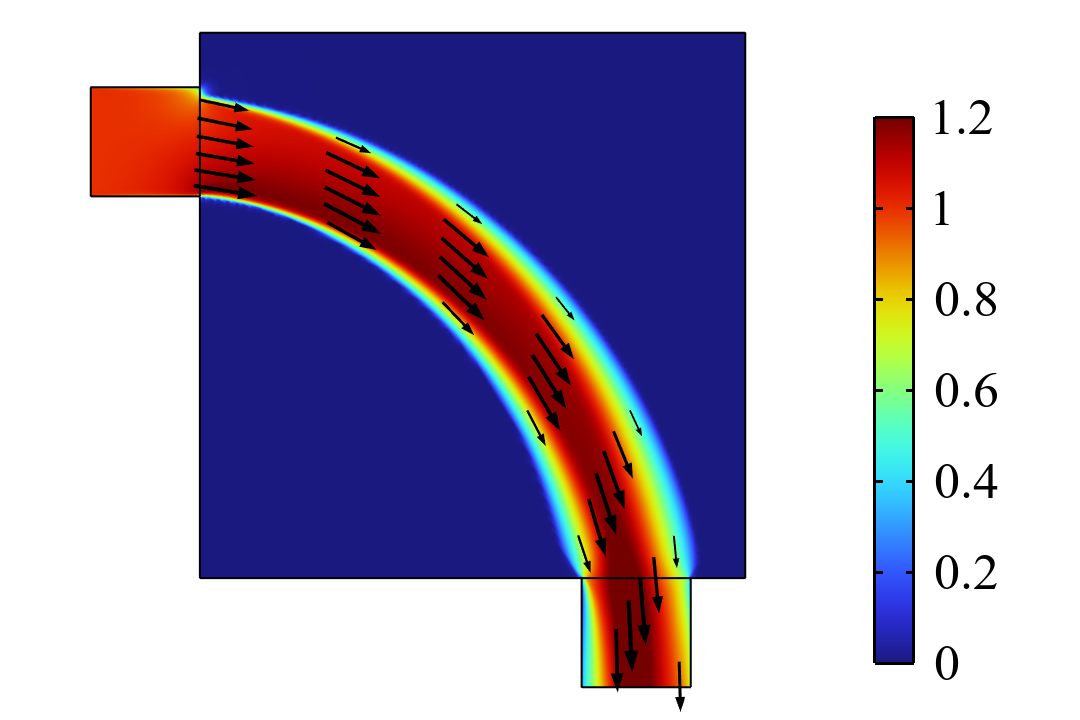}%
        \label{subfig:bending_conventional_velo}
    } 
    \caption{Comparison of the velocity field obtained with proposed implicit wall-function method and the ``conventional'' method (Examples \#1 and \#2). As illustrated, when the wall-functions are not employed, the boundary layer grows from the inlet and reaches its maximum thickness near the outlet. In contrast, when the implicit wall-function method is used, the velocity profile exhibits a more natural and physically accurate behavior, without the artificial boundary layer growth observed in the ``conventional'' approach.}
    \label{Fig:bending_velocity_comparison}
\end{figure}

As shown in \Cref{fig:Comparison_old_new_methods}, the geometry optimized with the proposed implicit wall-function approach differs systematically from the “conventional”~\Citep{yoon2020topology} formulation. The divergence between the designs appears already at the inlet, grows along the streamwise direction, and peaks near the outlet, where the two geometries deviate most. This behavior stems from the distinctly different velocity profiles predicted by the two models, as shown in \Cref{Fig:bending_velocity_comparison}: the proposed implicit wall-function method reproduces a physically consistent boundary layer that closely matches the explicit-wall reference, whereas the ``conventional'' penalization-based method causes an artificially thick and growing boundary-layer and increasingly erroneous velocity predictions, especially near the outlet.
As mentioned previously, the ``conventional'' method applies body-forces to the $k$–$\varepsilon$ equations to enforce $k=\varepsilon=0$ in solid regions, as well as no-slip velocity conditions. While these are the correct true boundary conditions, if the mesh is not sufficiently fine to capture the thin boundary layer, this method will vastly over-predict boundary layer thickness and growth.

\begin{figure}
    \centering

    \subfloat[Final topology optimized velocity field]{\includegraphics[width=0.4\textwidth]{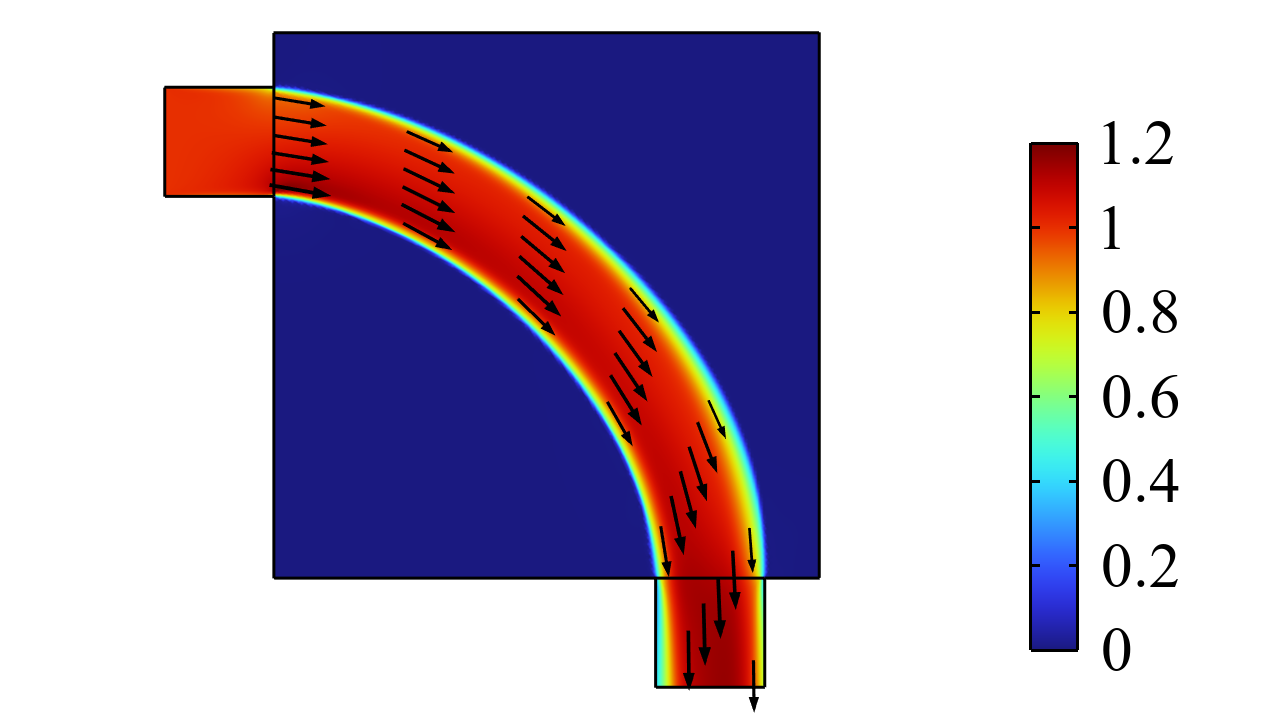}\label{subfig: im_vel}}
    \hspace{0.05\textwidth}
    \subfloat[Body fitted mesh velocity field]{\includegraphics[width=0.4\textwidth]{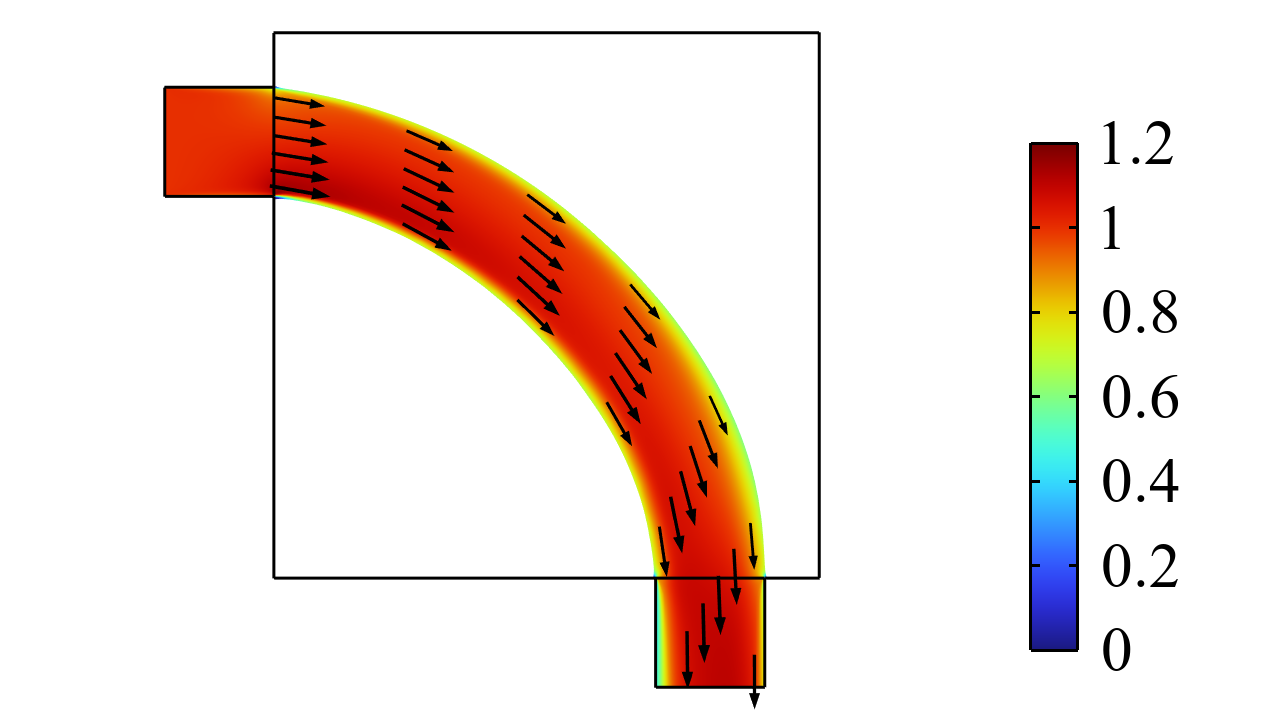}\label{subfig: ex_vel}}

    \vspace{2mm} 

    \subfloat[Final topology optimized pressure field]{\includegraphics[width=0.4\textwidth]{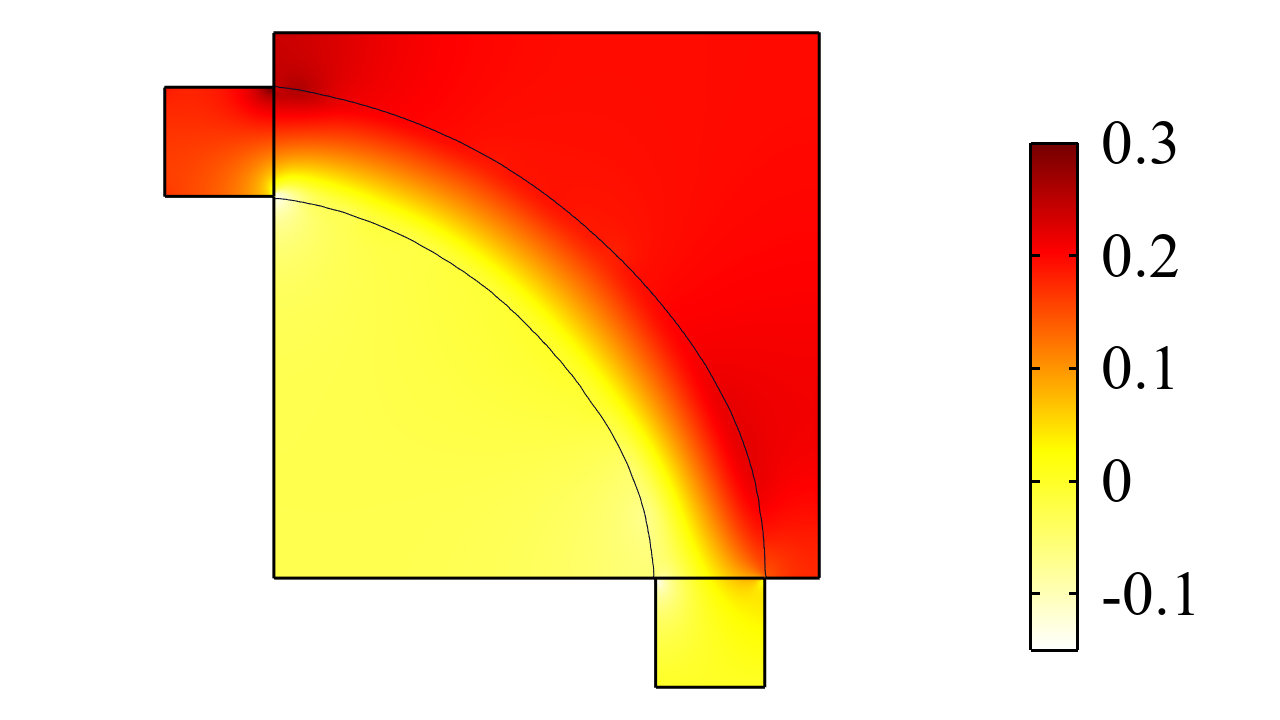}\label{subfig: im_per}}
    \hspace{0.05\textwidth}
    \subfloat[Body fitted mesh pressure field]{\includegraphics[width=0.4\textwidth]{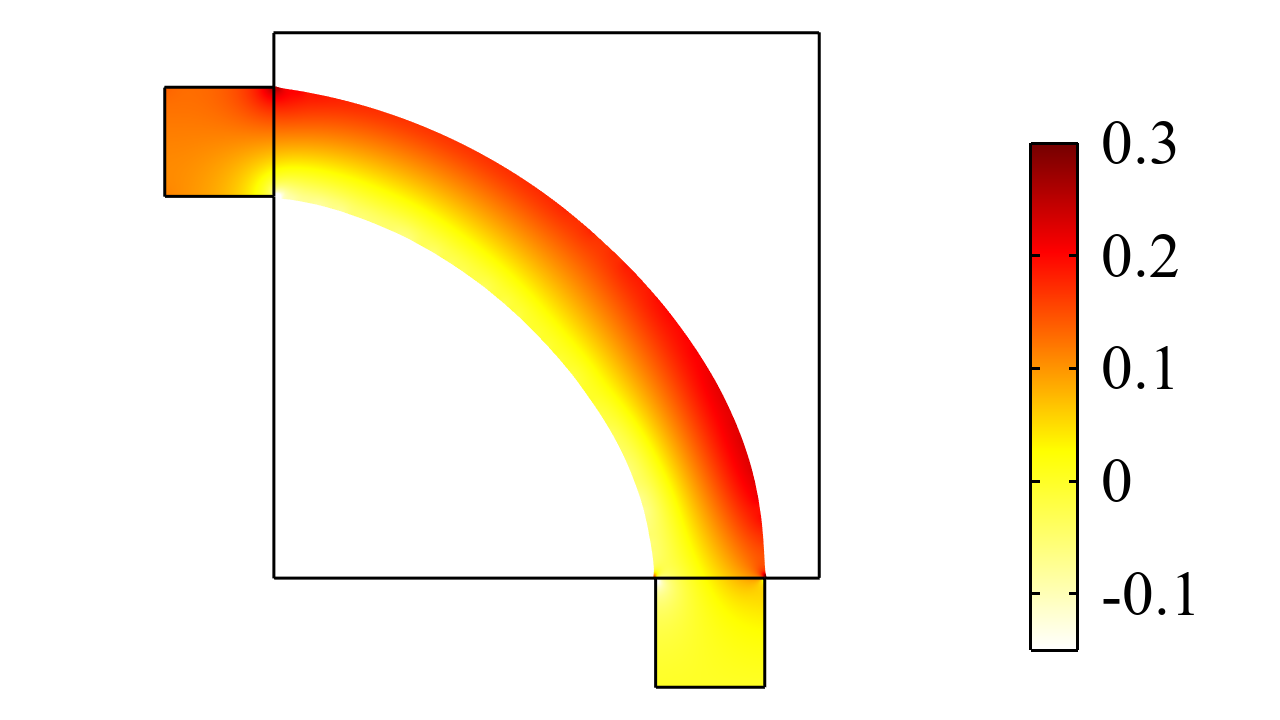}\label{subfig:ex_per}}

    \vspace{2mm} 

    \subfloat[Final topology optimized turbulent viscosity $\nu_t$]{\includegraphics[width=0.4\textwidth]{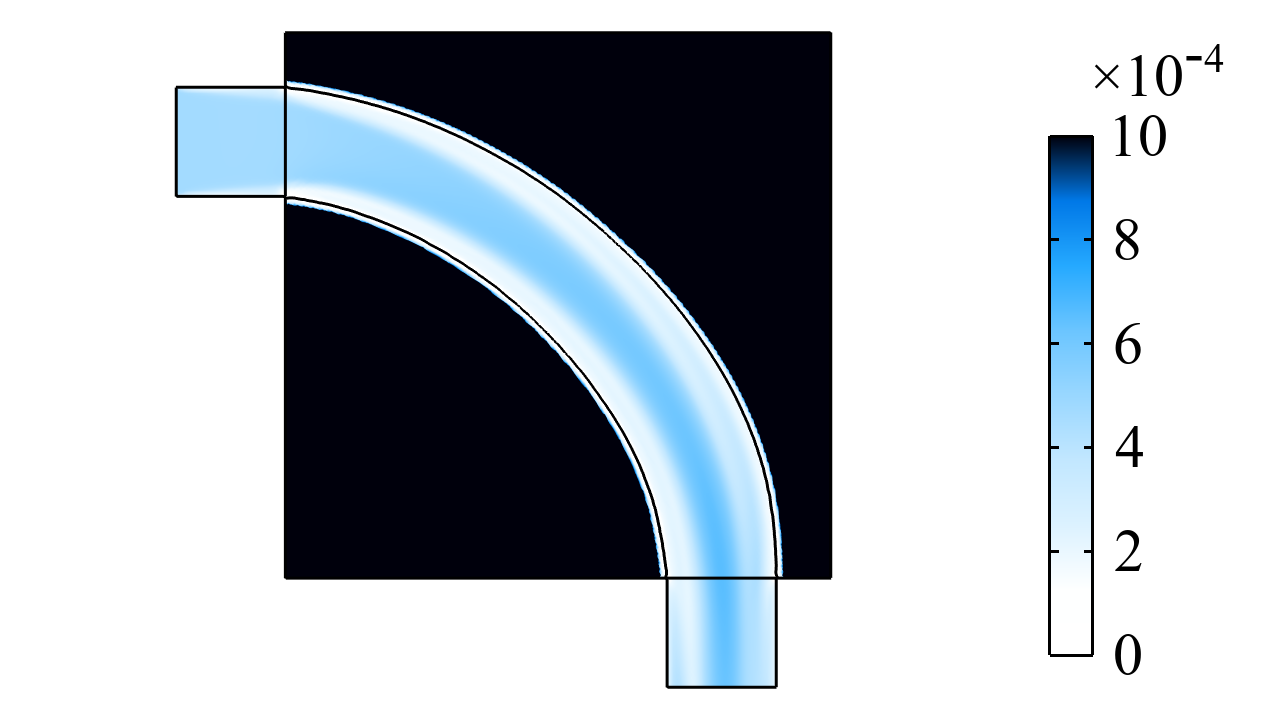}\label{subfig: muf}}
    \hspace{0.05\textwidth}
    \subfloat[Body fitted turbulent viscosity $\nu_t$]{\includegraphics[width=0.4\textwidth]{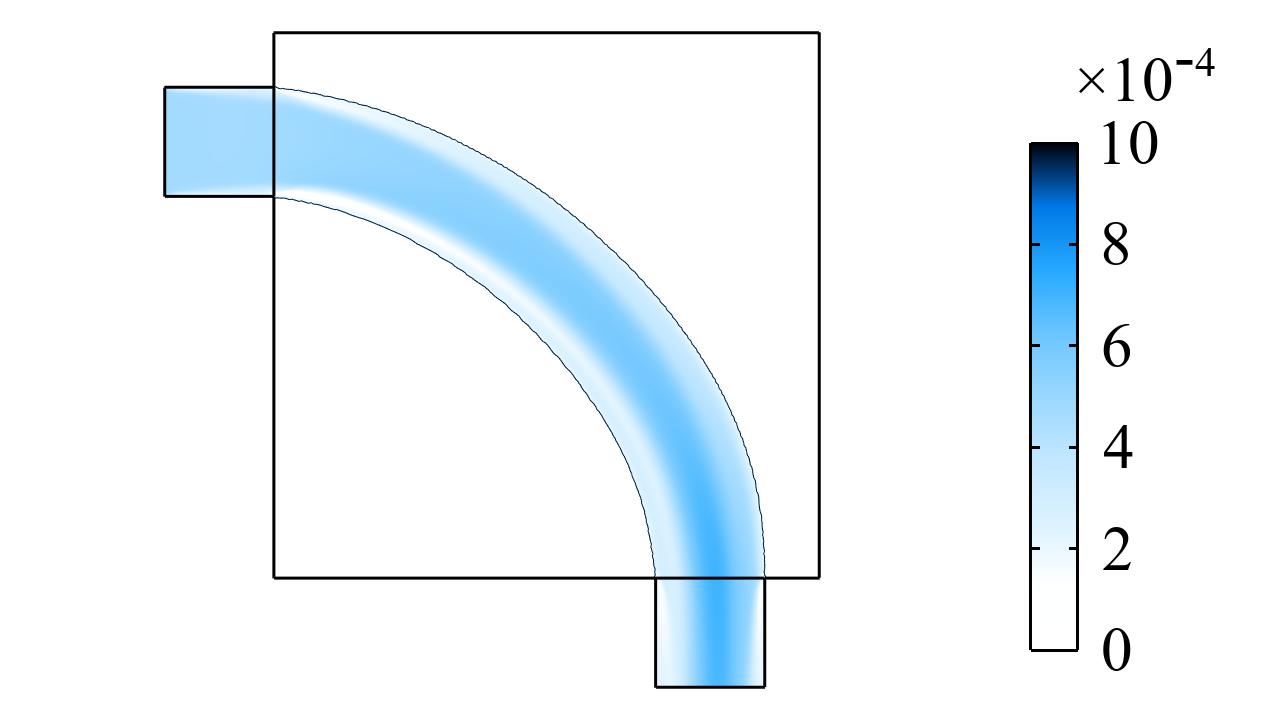}\label{subfig: ex_mut}}

    \caption{
Comparison between the optimized with implicit wall-function method pipe-bend geometry (Example \#2) and re-simulations with an explicit wall. 
The left column shows the topology-optimized results, while the right column presents the corresponding body-fitted re-simulation fields, facilitating direct visual comparison.
    }
    \label{fig:body_fitted_comparison_bending_pipe}
\end{figure}
To verify the accuracy of the proposed approach, the optimized design is re-simulated using an body-fitted representation with explicit wall-functions. As shown in \Cref{fig:body_fitted_comparison_bending_pipe}, close agreement is observed between the optimized result and the re-simulation of the body-fitted geometry. Excellent agreement is seen for the velocity $\boldsymbol{u}$, pressure $p$, and turbulent viscosity $\nu_t$ fields, confirming the accuracy of the proposed method.

When re-simulated using a body-fitted mesh and explicit wall-functions, Example~\#1 and \#2 yields a pressure drop of 0.1954 and 0.0924, respectively. This clearly shows that the optimized design using the proposed implict wall-function approach yields significantly better results.
It should be noted that there is an over-estimation in the pressure drop using the optimization model and the body-fitted re-simulations. This is common in TO of fluid flows and is to be expected due to the diffuse boundary. Overall, we often observe approximately double the pressure drop using the diffuse boundary than compared to the body-fitted equivalent. For Example~\#2, the optimization model predicts a pressure drop of 0.1731 and the body-fitted re-simulation predicts 0.0924.

\subsubsection{Pipe-bend with different volume fractions}
To assess how the prescribed volume fraction influences the optimized geometry, we applied the proposed method at a higher volume fraction setting \(V_f=\tfrac{1}{3}\) (Example~\#3) compared to the setting of \(V_f=\frac{1}{4}\) previously (Example~\#2). Other parameters are the same parameters as shown in \Cref{tab:bending sweep_parameters}.

\begin{figure}[bt]
    \centering
    \subfloat[$V_f=\frac{1}{4}$]{%
        \includegraphics[width=0.4\textwidth]{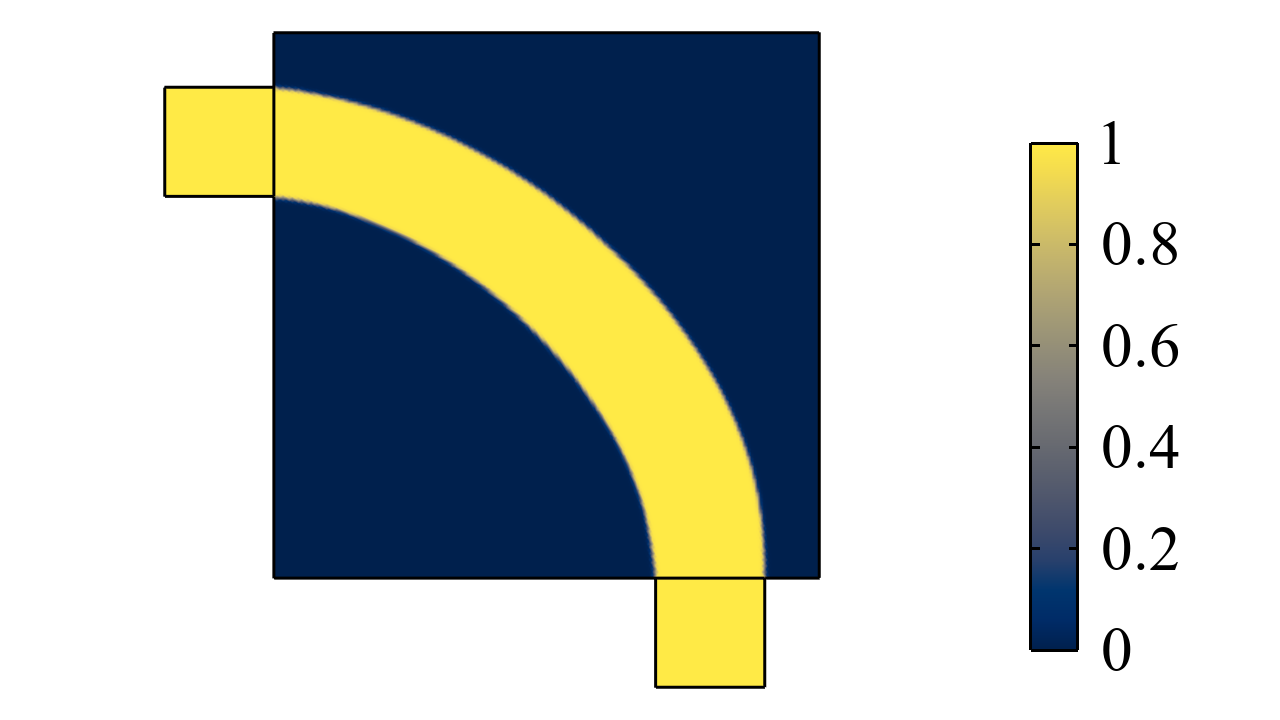}%
        \label{subfig:no_slip_final}
    }
    \hspace{0.05\textwidth}
        \subfloat[ $V_f=\frac{1}{3}$]{%
        \includegraphics[width=0.4\textwidth]{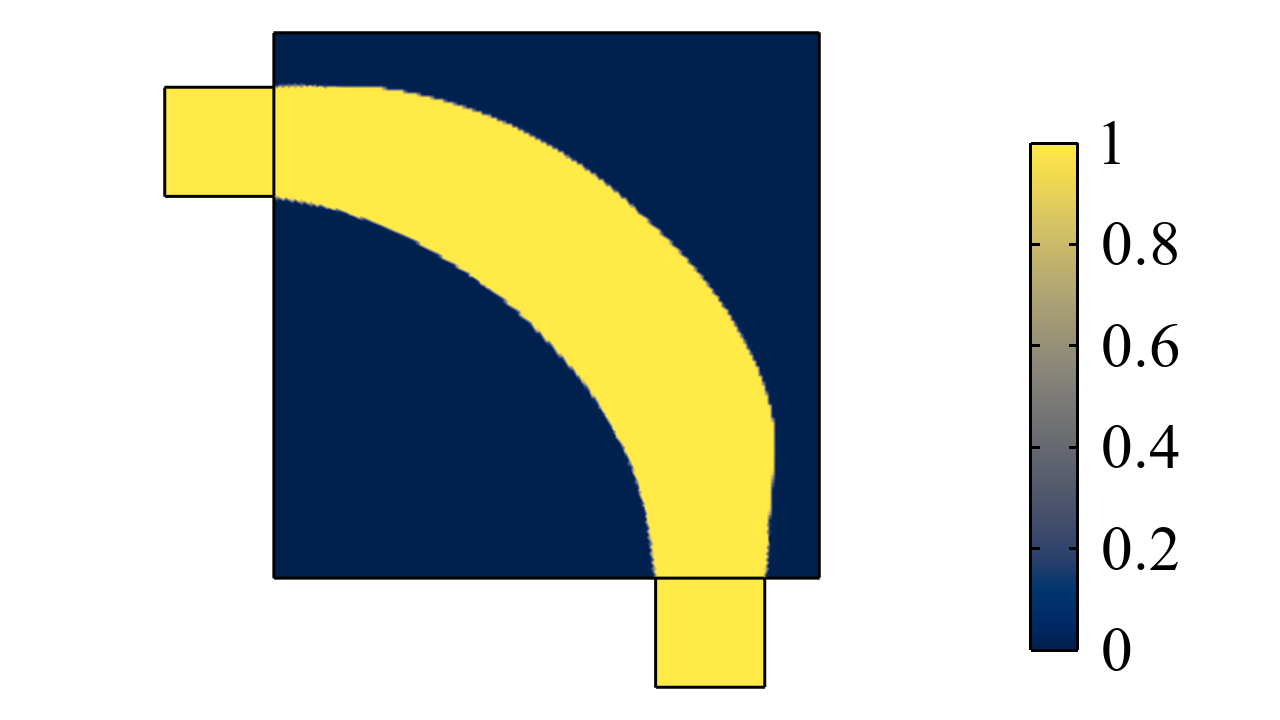}%
        \label{subfig:slip_final}
    } 
    \caption{Final topology-optimized, filtered--projected design variable \(\varphi\) using the proposed method. Left: \(V_f=1/4\)~(Example \#2). Right: \(V_f=1/3\)~(Example \#3).}
    \label{Fig:Bending_pipe_optimized_volume_fraction}
\end{figure}
Comparing the two optimized designs in \Cref{Fig:Bending_pipe_optimized_volume_fraction},
increasing the allowable fluid volume yields a visibly wider shaped channel as expected. The wider passage at \(V_f=\tfrac{1}{3}\) is consistent with
a lower hydraulic resistance (smaller average inlet pressure), whereas the thinner passage at
\(V_f=\tfrac{1}{4}\) is expected to increase losses but comes due to the requirement of lower fluid volume. Both designs indicate a clear curved channel approaching circular.

Furthermore, a re-simulation 
using an explicit wall representation is performed to further verify the accuracy of the proposed method for a different pipe geometry.
\begin{figure}[bt]
    \centering

    \subfloat[Optimized velocity field]{\includegraphics[width=0.4\textwidth]{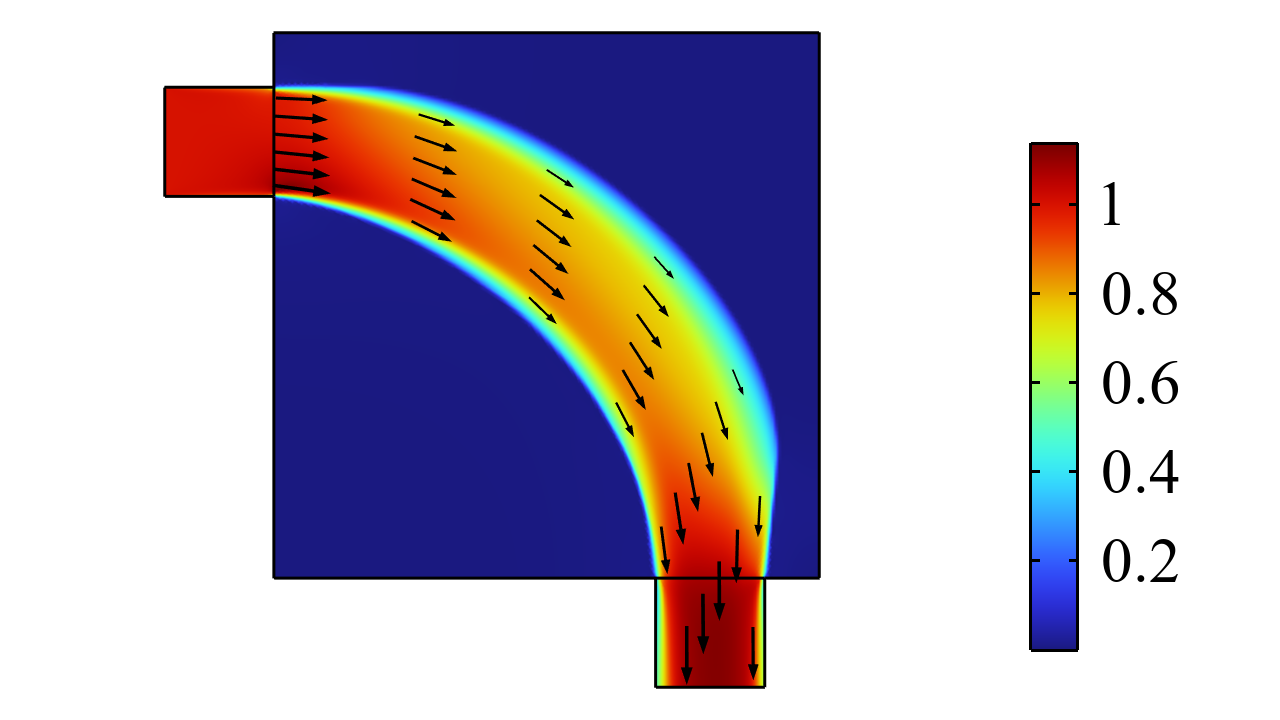}\label{subfig: im_vel_vf0.33}}
    \hspace{0.05\textwidth}
    \subfloat[Body-fitted velocity field]{\includegraphics[width=0.4\textwidth]{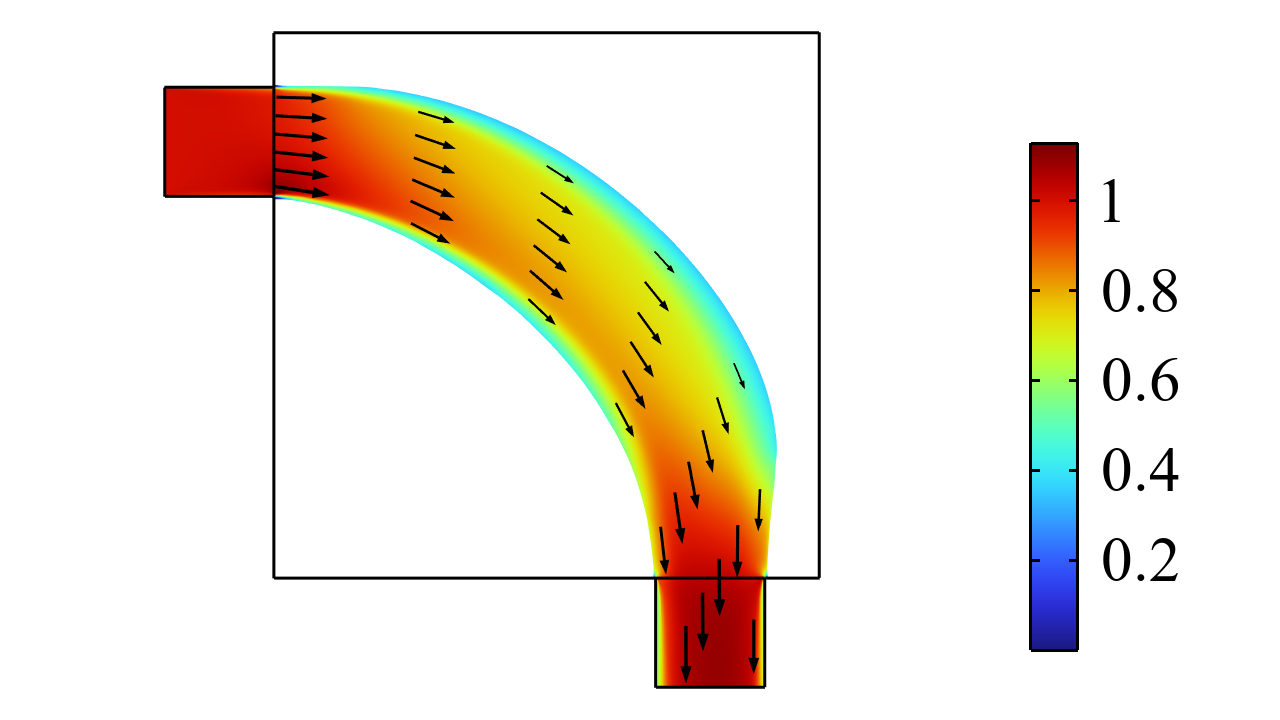}\label{subfig: ex_vel_vf0.33}}

    \vspace{2mm} 

    \subfloat[Optimized pressure field]{\includegraphics[width=0.4\textwidth]{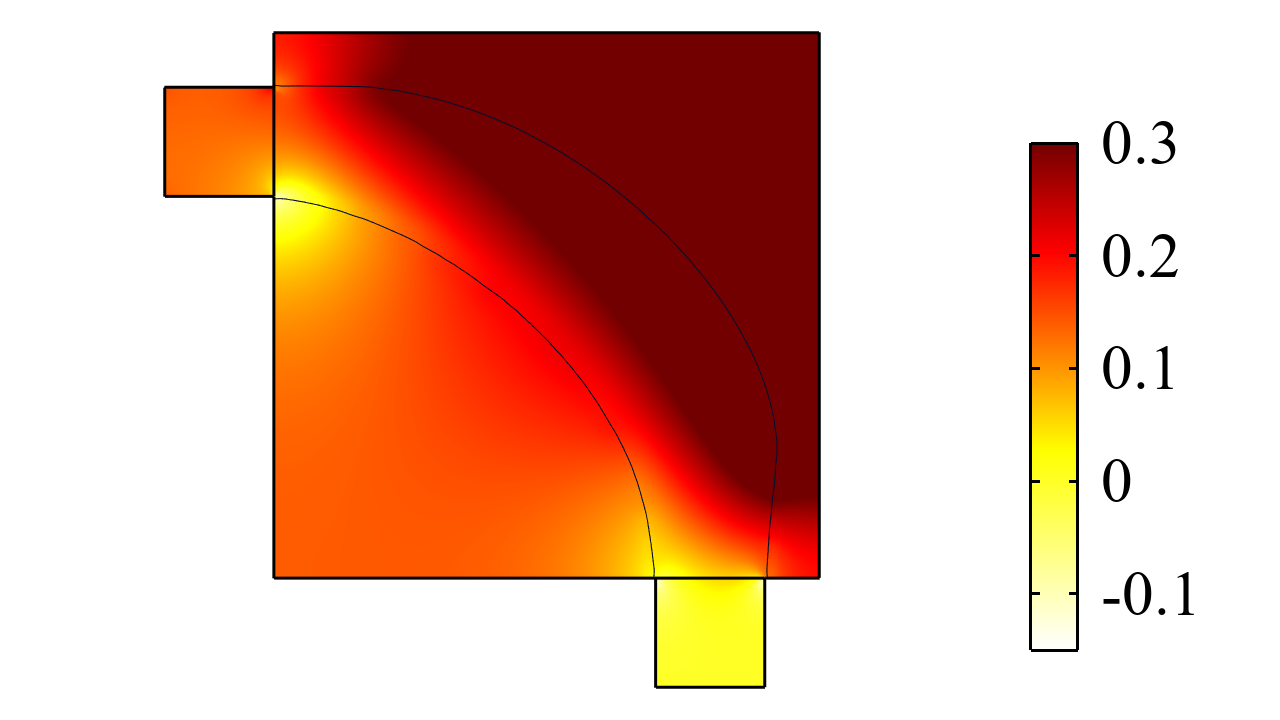}\label{subfig: im_per_vf0.33}}
    \hspace{0.05\textwidth}
    \subfloat[Body-fitted pressure field]{\includegraphics[width=0.4\textwidth]{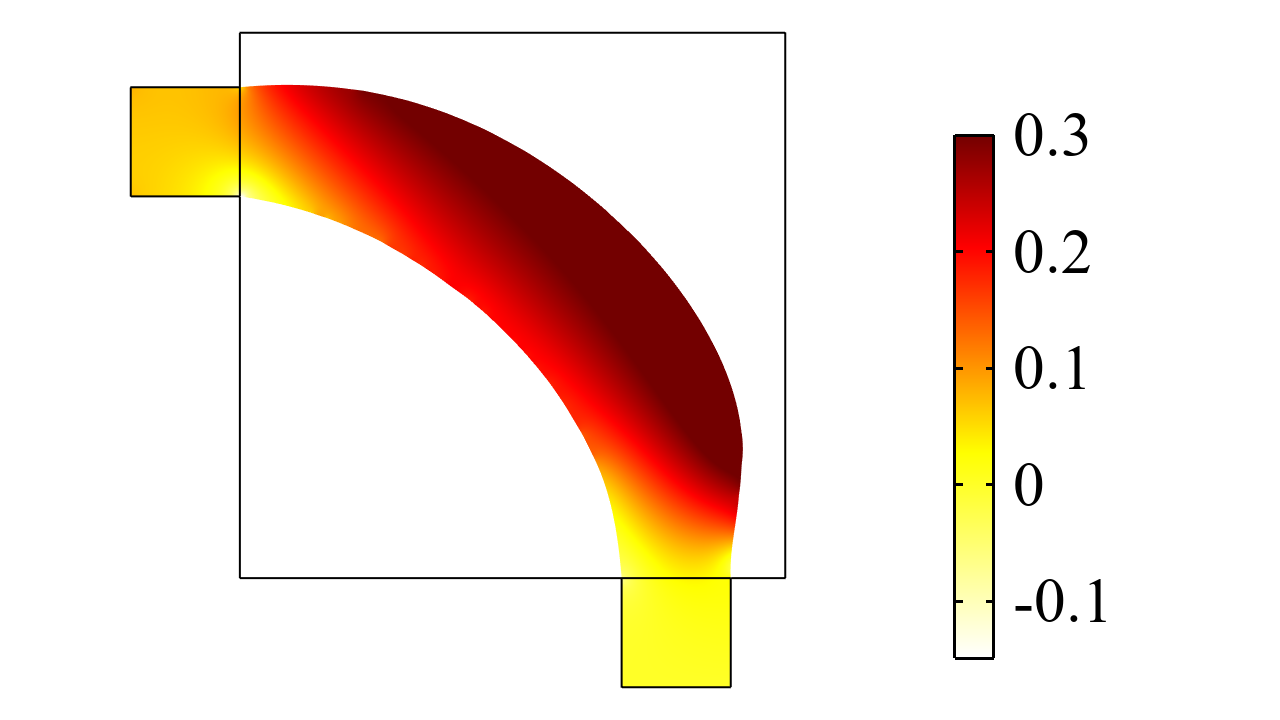}\label{subfig:ex_p_vf0.33}}

    \vspace{2mm} 

    \subfloat[Optimized turbulent viscosity $\nu_t$]{\includegraphics[width=0.4\textwidth]{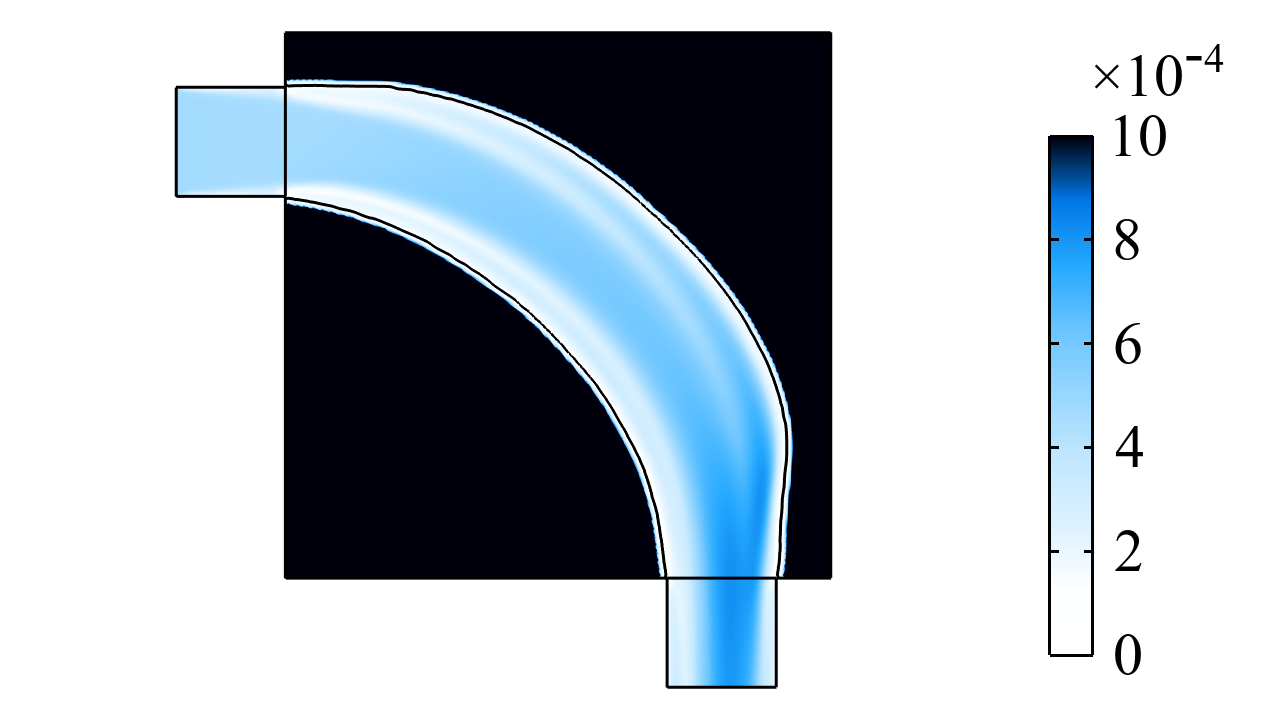}\label{subfig: im_mut_vf0.33}}
    \hspace{0.05\textwidth}
    \subfloat[Body-fitted turbulent viscosity $\nu_t$]{\includegraphics[width=0.4\textwidth]{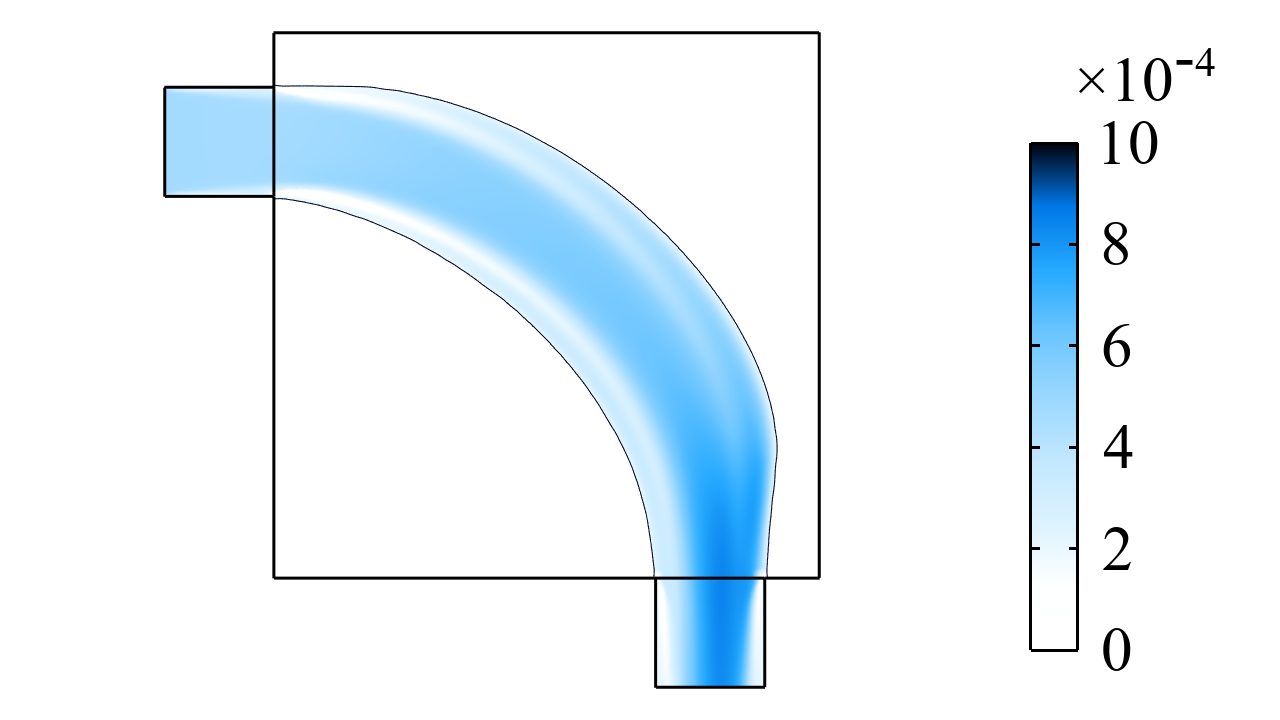}\label{subfig: ex_mut_vf0.33}}

    \caption{
Comparison of the results obtained using the proposed implicit wall-function method and the re-simulation with explicit wall-functions for a higher volume fraction ($V_f = \tfrac{1}{3}$, Example~\#3).}
    \label{fig:high_volfrac}
\end{figure}
The results are shown in \Cref{fig:high_volfrac} and it can be seen that the re-simulated results matches very well with those obtained from proposed method. It is important to note that the boundary layer here is thicker, because the channel becomes wider than the inlet in contrast to Example~\#2 (\Cref{fig:body_fitted_comparison_bending_pipe}). Thus, the turbulent viscosity distribution is also more complicated (\Cref{subfig: im_mut,subfig: im_mut_vf0.33}), but equally well captured.
 For Example~\#3, the optimization model predicts a pressure drop of 0.1303 and the body-fitted re-simulation predicts 0.0651.

\subsubsection{Pipe-bend with higher Reynolds number}
\begin{figure}
    \centering
    \includegraphics[width=0.4\linewidth]{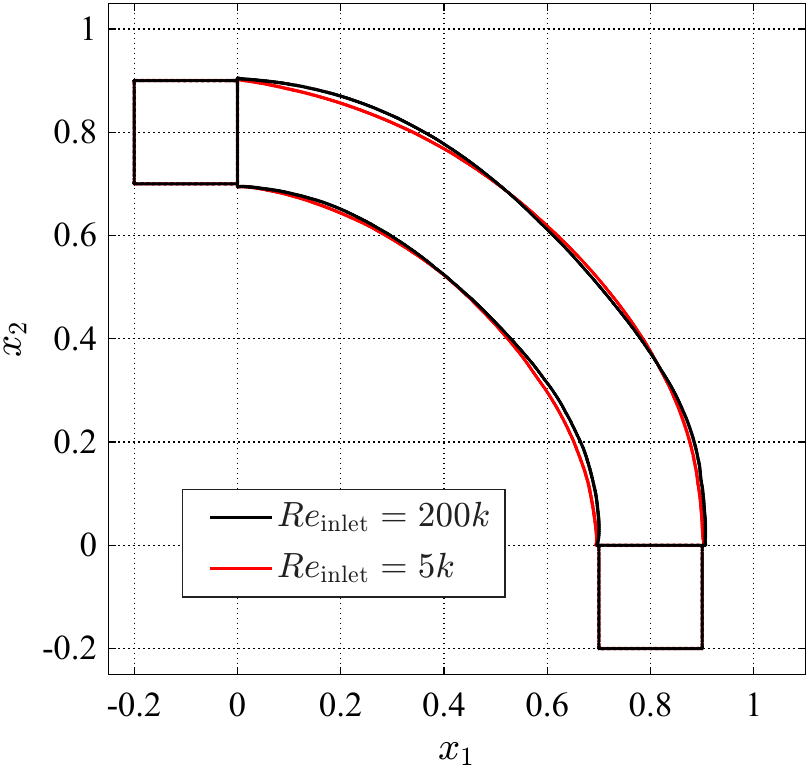}
    \caption{Comparison between optimized geometries for $Re=5k$~(Example \#2) and $Re=200k$~(Example \#4), using the proposed implicit wall-function approach.}
    \label{fig: Schematic_low_high_Reynolds}
\end{figure}

The proposed formulation enables accurate topology optimization at high Reynolds numbers on comparatively coarse meshes, thanks to the use of implicit wall-function method.
\Cref{fig: Schematic_low_high_Reynolds} shows the result for an increased Reynolds number of \(\mathrm{Re}=2\times10^{5}\)  (Example~\#4). 
The optimized geometries at low and high Reynolds numbers remain broadly similar, but with slightly larger curvature for the high Reynolds number case. This limited variation reflects that the boundary-layer characteristics change only modestly between these two regimes in our setup, which indicate low dependency of the optimized geometry in fully turbulent regime to the Reynolds number, leading to only minor differences in the final designs. Furthermore, the performance of the two are very similar at both Reynolds numbers when re-analyzed in a cross-check. For Example~\#4, the optimization model predicts a pressure drop of 0.1435 and the body-fitted re-simulation predicts 0.0470. 

\begin{figure}
    \centering
    \subfloat[Example~\#2, $Re=5{,}000$, $V_f = 0.25$]{
        \includegraphics[width=0.4\textwidth]{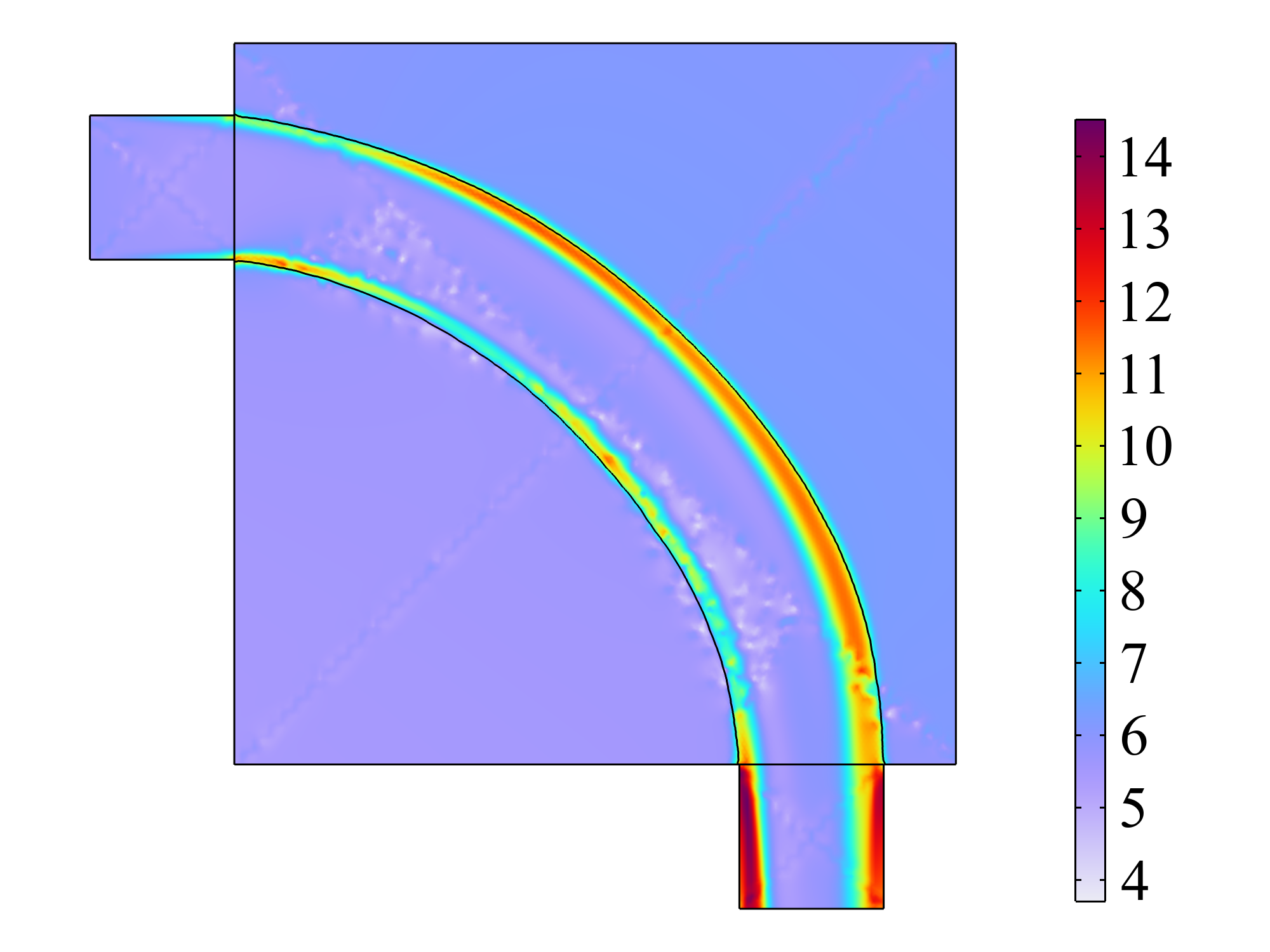}
        \label{subfig:yplus_bending_5k}
    }\hfill
    \subfloat[Example~\#3, $Re=5{,}000$, $V_f = 0.33$]{
        \includegraphics[width=0.4\textwidth]{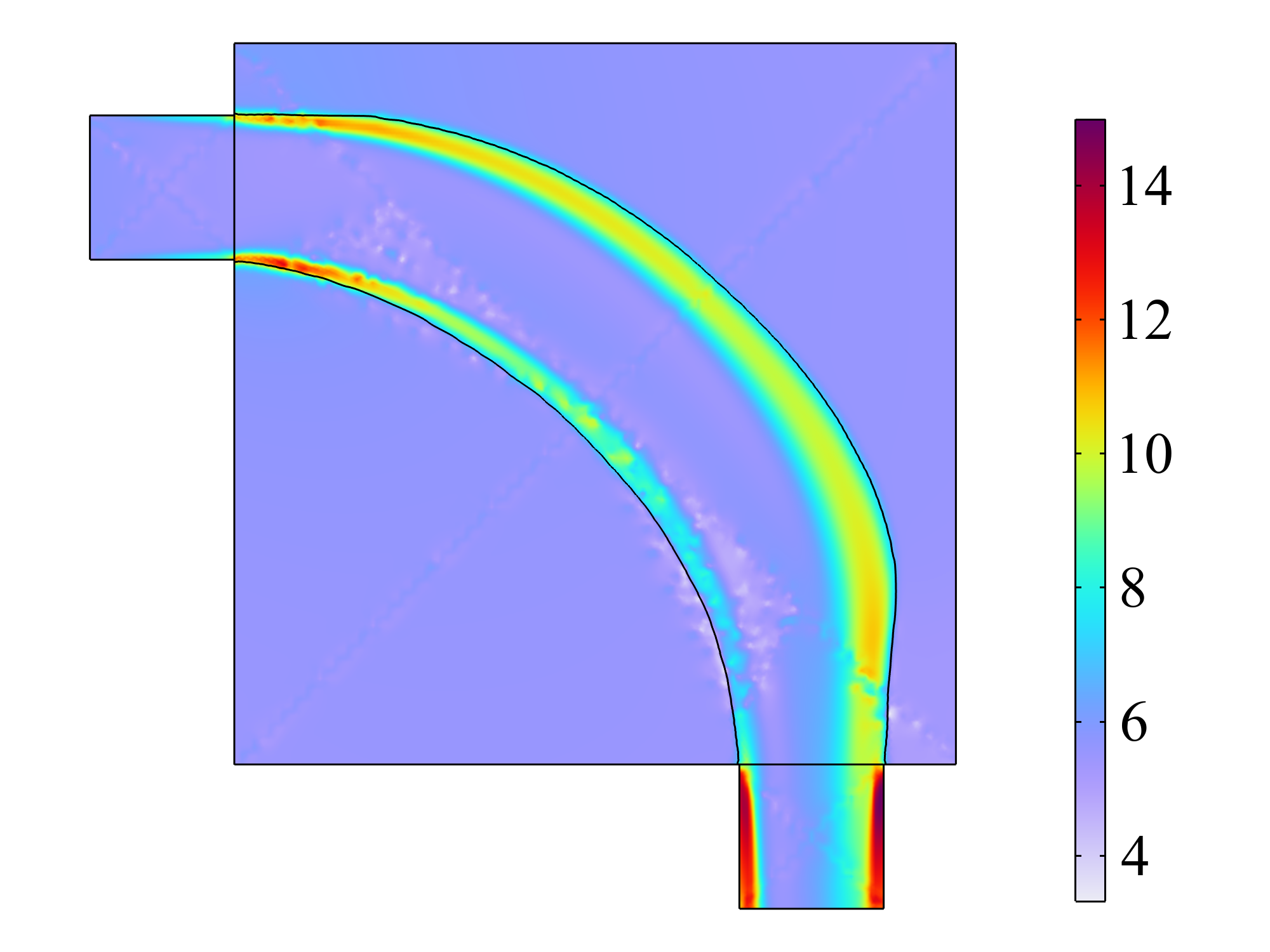}
        \label{subfig:yplus_bending_5k_vf0_33}
    }\hfill
    \subfloat[Example~\#4, $Re=200{,}000$, $V_f = 0.25$]{
        \includegraphics[width=0.4\textwidth]{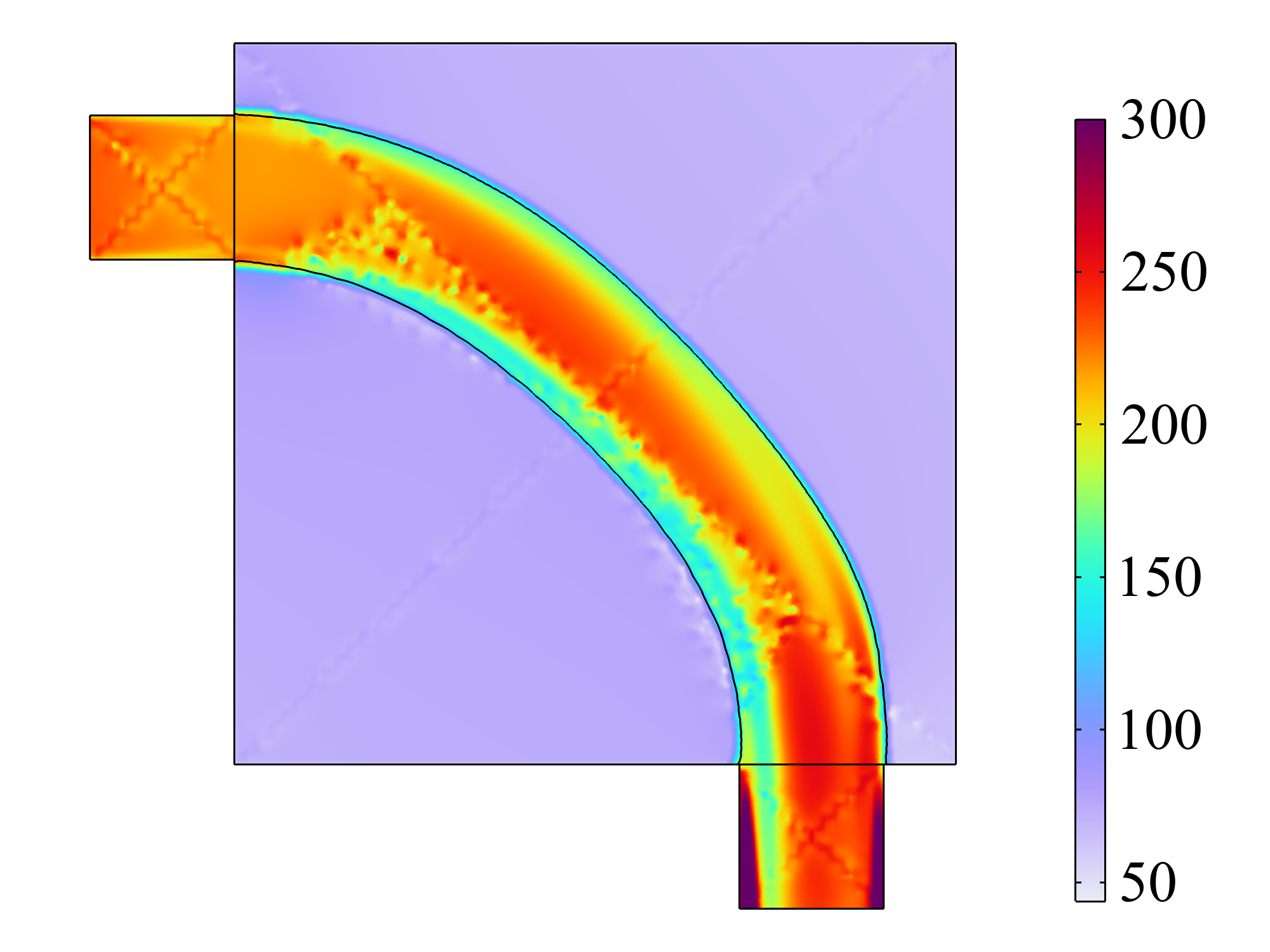}
        \label{subfig:yplus_bending_200k}
    }

    \caption{Implicit $y^{+}$ distributions for the optimized pipe-bend geometries in Examples~\#2–\#4, obtained using the proposed implicit wall-function method. In the vicinity of the walls, $y^{+}$ remains close to 11.06 for Examples~\#2 and \#3, and below 300 for Example~\#4, indicating that the logarithmic law of the wall is satisfied in all cases.}
    \label{fig:bending_yplus_all}
\end{figure}

\Cref{fig:bending_yplus_all} shows the implicit $y^{+}$ distribution for the pipe-bend (Examples~\#2--\#4). For the proposed implicit wall-function method, $y^+$ is defined implicitly through the domain. As illustrated in \Cref{fig:bending_yplus_all}, in the vicinity of the wall $y^{+}$ remains close to 11.06 and below 300. With the same mesh, when the Reynolds number increases from 5000 to 200,000, the corresponding $y^{+}$ values also increase, but remain within the allowable range.

\subsection{U-bend benchmark}
\begin{figure}[bt]
    \centering
    \includegraphics[width=0.4\linewidth]{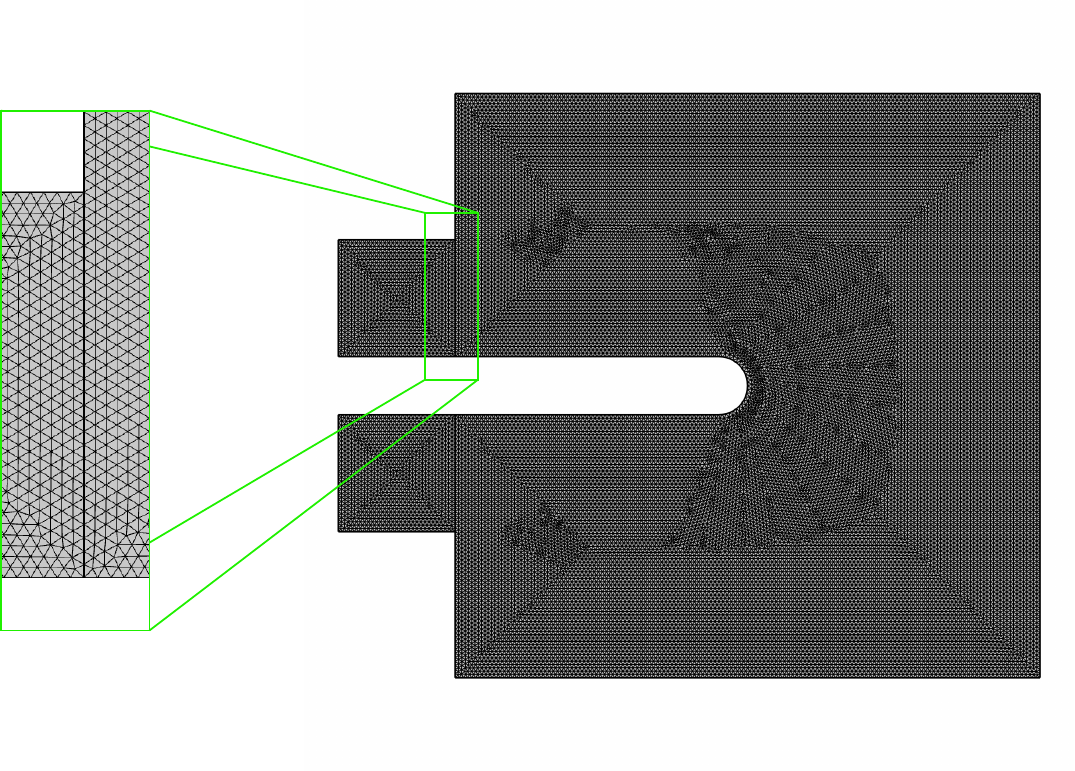}
    \caption{The mesh used for the U-bend benchmark is illustrated here for an approximate element size of $h_{max}=$ $0.007$, yielding 55,055 triangles, and 195,636 total DOFs.}
    \label{fig: Ubend_mesh}
\end{figure}

The second benchmark considered in this study is the U-bend case, illustrated in \Cref{fig:U_bending}. In this setup, a uniform inlet velocity enters the domain, passes through a $180^\circ$ bend, and is expected to leave the domain with a straight velocity profile and zero pressure at the outlet. The computational mesh is presented in \Cref{fig: Ubend_mesh} with 55,055 triangles and a total number of degree of freedoms (DOFs) of 195,636. The parameters and continuation approach is given in \Cref{tab:ubend sweep_parameters}.

\begin{table}[bt]
    \caption{U-bend with proposed implicit wall–functions method parameters.}
    \centering
    \label{tab:ubend sweep_parameters}
    \begin{tabular}{@{}l l l l l l l l@{}}
        \toprule
        \textbf{Parameter} & \multicolumn{7}{l}{Value} \\
        \midrule
        $\beta$        & 8 & 10 & 18 & 18 & & & \\
        $q_a$          & 150 & 75 & 35 & 12 & & & \\
        $\psi_{\max}$  & 1000 & & & & & & \\
        $P_{con}$      & 4 & & & & & & \\
        
        $r_{1}$        & \multicolumn{7}{l}{$4 h_{\max}$} \\
        $V_f$          & \multicolumn{7}{l}{0.27} \\
        $r_{2}$        & \multicolumn{7}{l}{$4 h_{\max}$} \\
        $Re_{inlet}$   & \multicolumn{7}{l}{$5000$} \\
        $h_{\max}$     & \multicolumn{7}{l}{$0.007$} \\
        $L_{1}$        & \multicolumn{7}{l}{$1$} \\
        $\alpha_{\max}$& \multicolumn{7}{l}{100} \\
        $U_{\mathrm{in}}$ & \multicolumn{7}{l}{1} \\
        \bottomrule
    \end{tabular}
\end{table}

\begin{figure}[bt]
    \centering
    \includegraphics[width=0.4\linewidth]{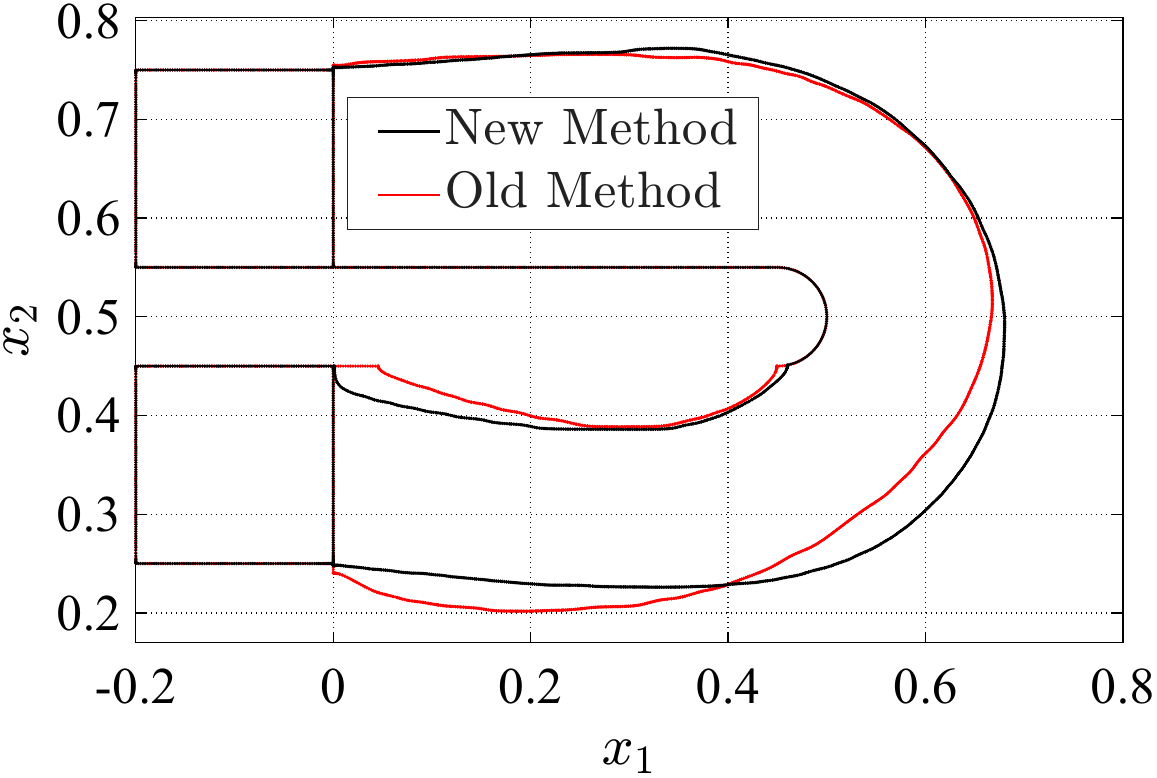}
    \caption{Comparison for the U-bend benchmark between the proposed implicit wall-function method~(Example \#6), and ``conventional'' method~(Example \#5) on the optimized geometry. The main difference happens close to outlet when the ``conventioanl'' method fails to accurately simulate boundary layer growth.}
    \label{fig: Ubend_optimized}
\end{figure}

The optimized geometries obtained using the proposed implicit wall-function method and the “conventional” method are shown in \Cref{fig: Ubend_optimized}. Similar to the pipe-bend benchmark, the proposed method produces geometries that differ from those of the conventional method, particularly in the downstream region and near the outlet. This difference arises from the more accurate boundary layer prediction achieved by the implicit wall-function formulation, whereas the conventional method introduces increasing errors in boundary layer representation. Although the design optimized using the ``conventional'' method may appear more streamline in some sense, because the curvature appears more accommodating, the final performance is significantly worse. When re-simulated using a body-fitted model with explicit wall-functions, the ``conventional'' design has a pressure drop of 0.9446 and the design from the proposed method has a pressure drop of only 0.3423.

\begin{figure}[bt]
    \centering

    \subfloat[Final topology optimized velocity field]{\includegraphics[width=0.4\textwidth]{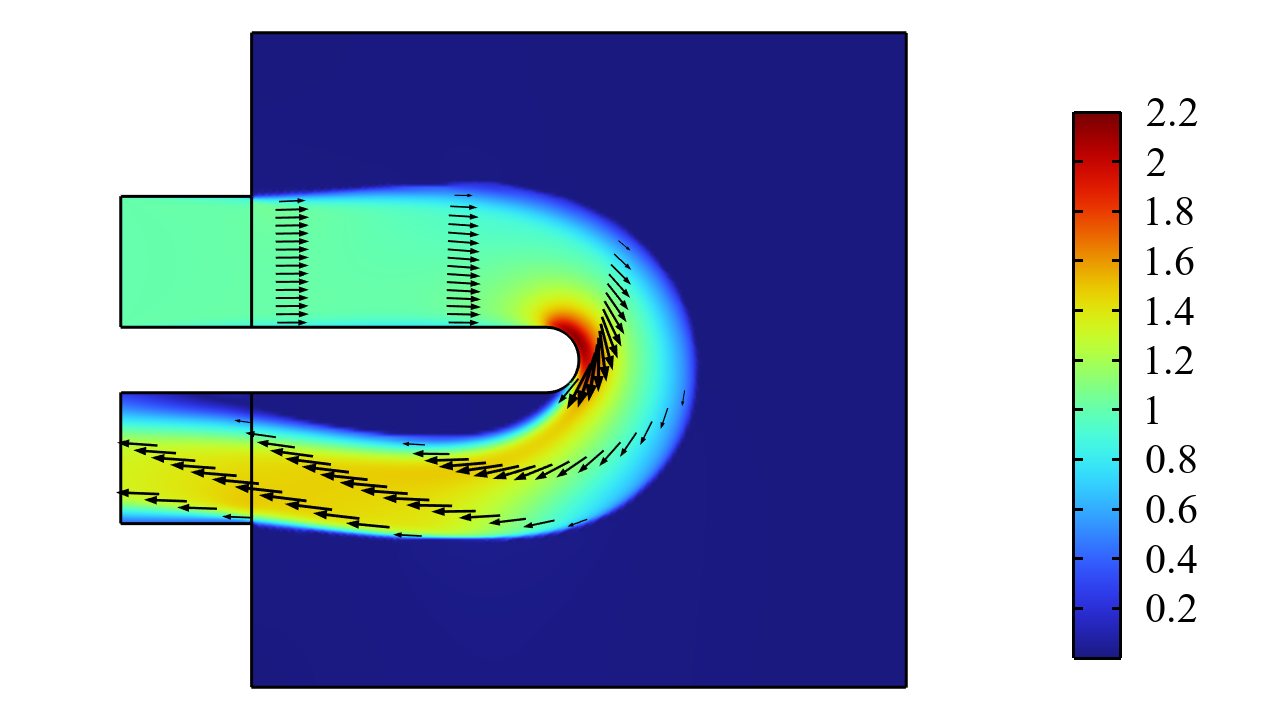}\label{subfig: im_vel}}
    \hspace{0.05\textwidth}
    \subfloat[Body-fitted mesh velocity field]{\includegraphics[width=0.4\textwidth]{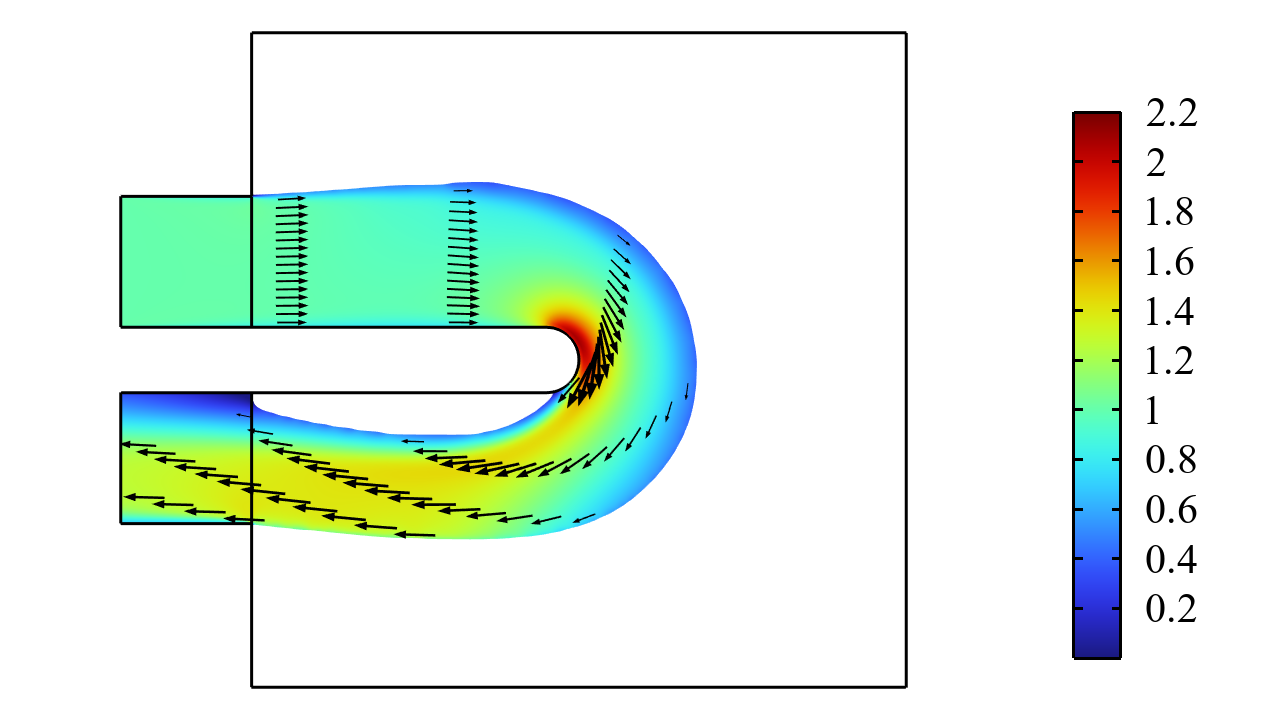}\label{subfig: ex_vel}}

    \vspace{2mm} 

    \subfloat[Final topology optimized pressure field]{\includegraphics[width=0.4\textwidth]{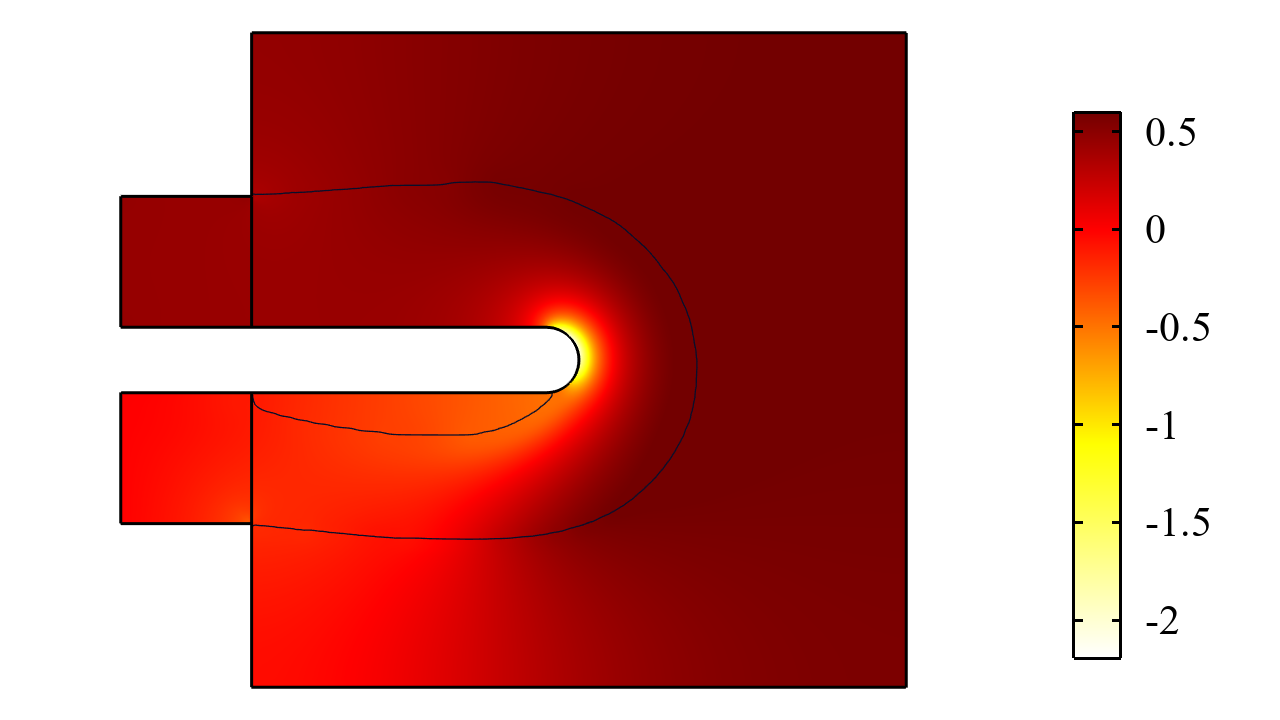}\label{subfig: im_per}}
    \hspace{0.05\textwidth}
    \subfloat[Body-fitted mesh pressure field]{\includegraphics[width=0.4\textwidth]{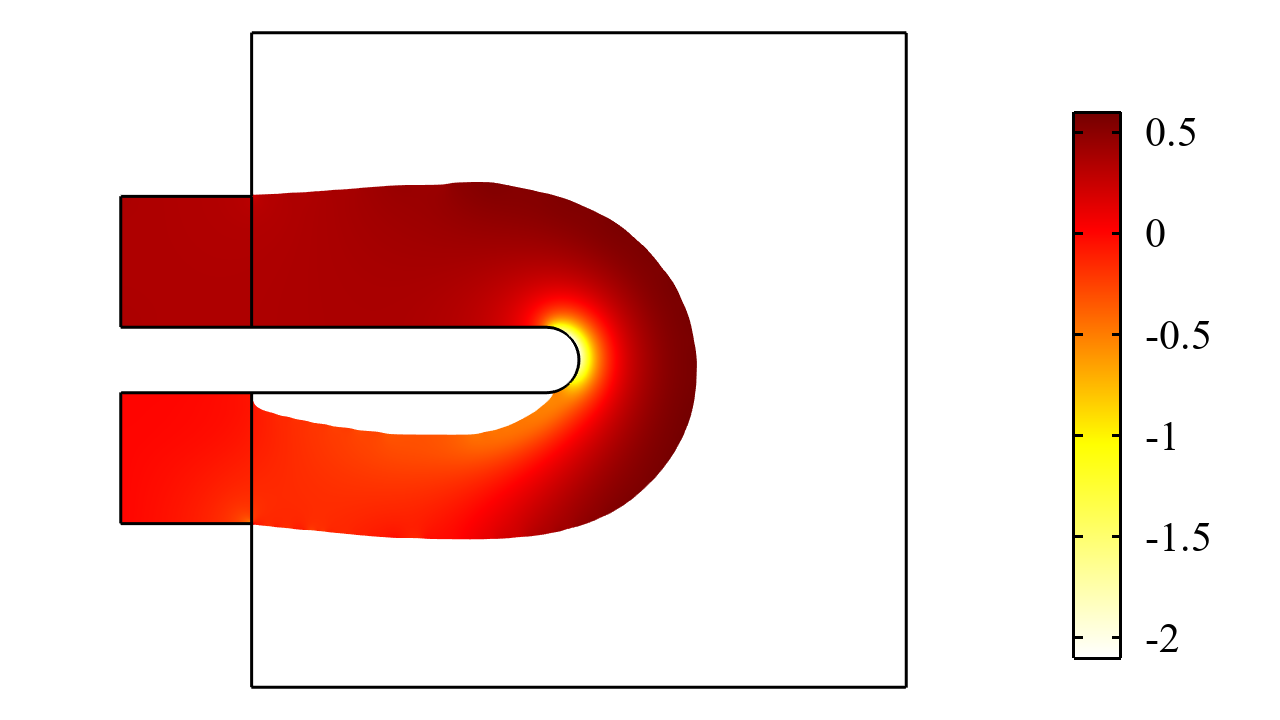}\label{subfig: ex_per}}

    \vspace{2mm} 

    \subfloat[Final topology optimized turbulent viscosity $\nu_t$]{\includegraphics[width=0.4\textwidth]{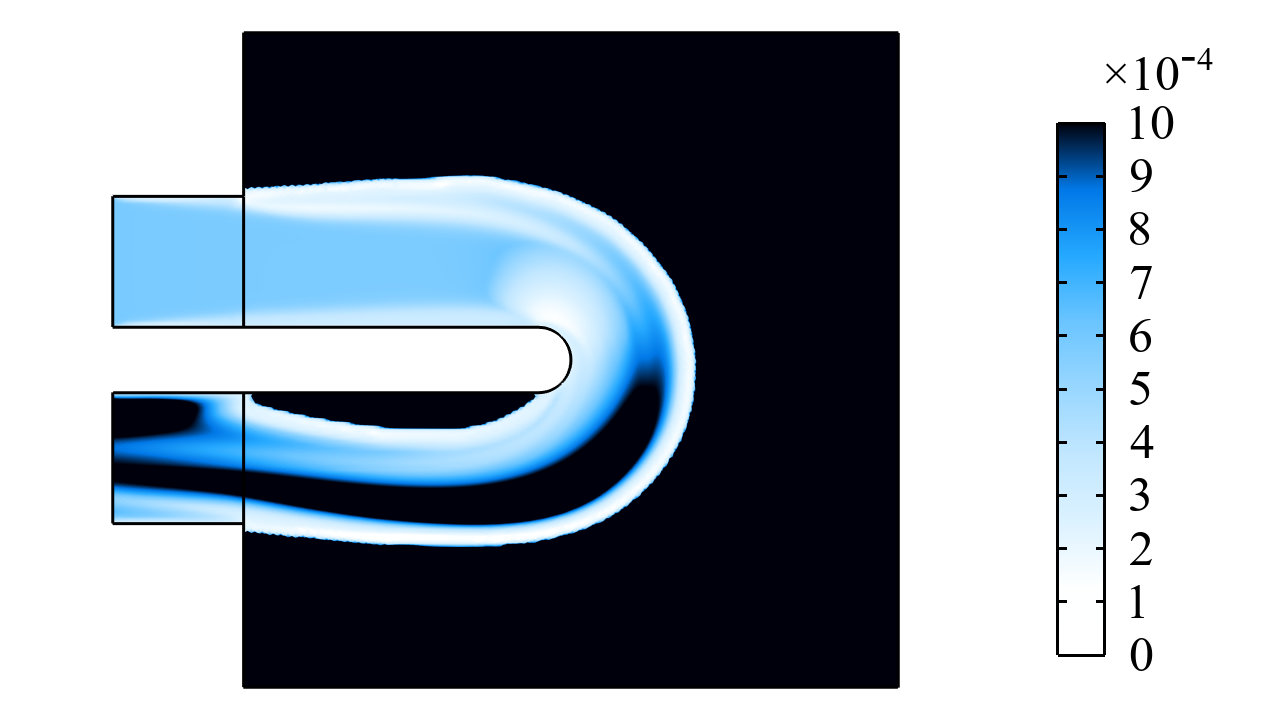}\label{subfig: im_mut}}
    \hspace{0.05\textwidth}
    \subfloat[Body-fitted turbulent viscosity $\nu_t$]{\includegraphics[width=0.4\textwidth]{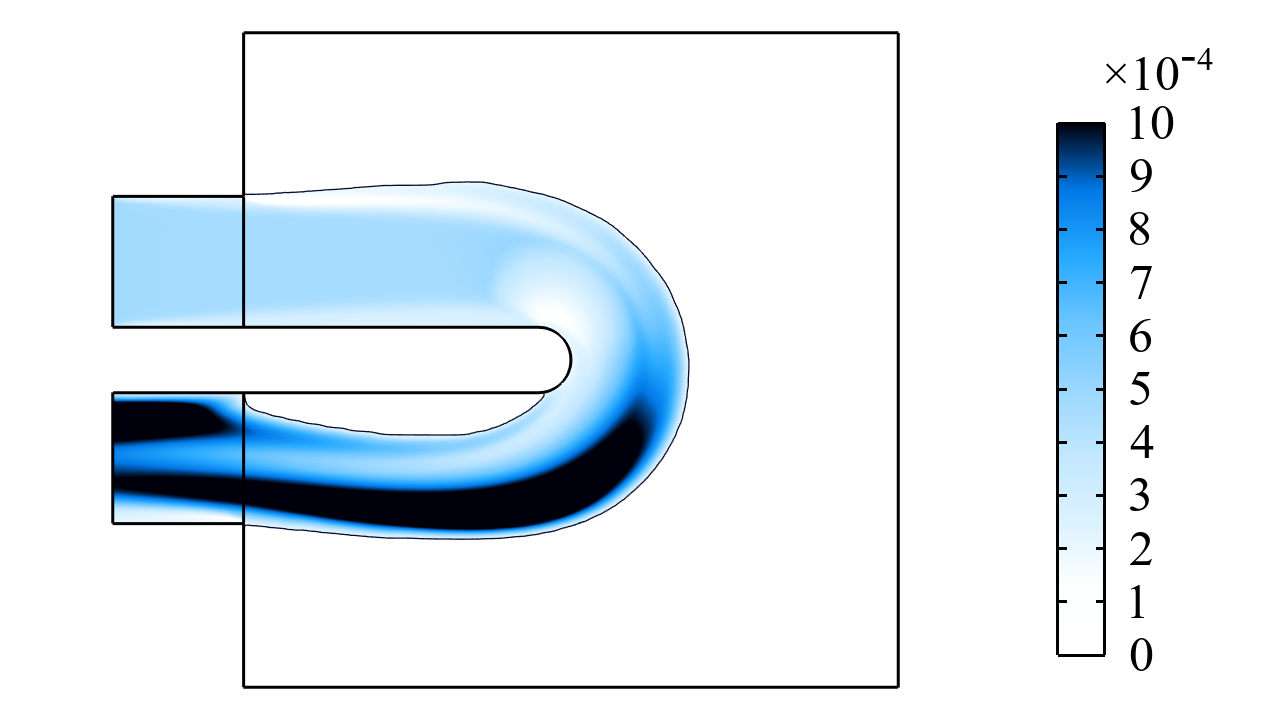}\label{subfig: ex_mut_U_bend}}

    \caption{
    Comparison between the optimized design and the body-fitted re-simulation for U-bend case (Example \#6).
    The left column shows the topology-optimized result with implicit wall-function, while the right column presents the corresponding body-fitted re-simulation fields.    
    }
    \label{fig: U_bending_body_fitted}
\end{figure}

As shown in \Cref{fig: U_bending_body_fitted}, the optimized diffuse-domain velocity, pressure, and turbulent viscosity fields closely match those obtained from the body-fitted mesh re-simulation. This demonstrates that the proposed implicit wall-function method can accurately reproduce the flow fields, yielding results highly consistent with explicit-wall representations.
As previously, there is a discrepancy between the pressure drop predicted by the optimization model and the re-simulation model. For Example~\#6, the optimization model predicts 0.4408 and the re-simulation model predicts 0.3423.

\subsubsection{U-bend benchmark with higher Reynolds number}
\begin{figure}[tb]
    \centering
        \includegraphics[width=0.4\textwidth]{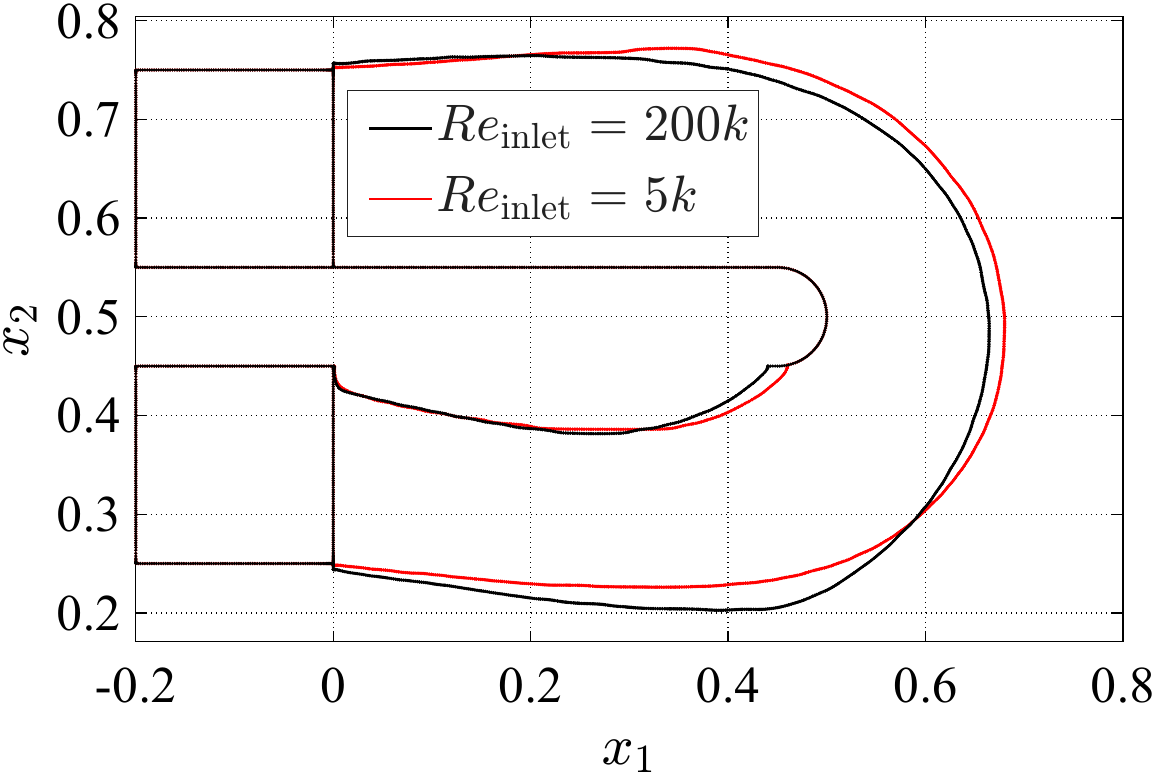}%
        \label{subfig:no_slip_final}

    \caption{$Re=5k$~(Example \#6) and $Re=200k$~(Example \#7), optimized geometry using implicit wall-function method.}
    \label{Fig:U_bending_comparison}
\end{figure}

Using wall-functions, pushing the Reynolds number to higher values (e.g., \(\operatorname{Re}=2\times10^{5}\)) is computationally easier, enabling accurate solutions even on comparatively coarse meshes. In the fully turbulent regime considered here \(\operatorname{Re} \in (5\times10^{3}, 2\times10^{5})\), the optimal topology is largely preserved, with only minor, localized adjustments, as illustrated in \Cref{Fig:U_bending_comparison}.
Increasing \(\operatorname{Re}\) from \(5\times10^{3}\) to \(2\times10^{5}\) leads to slight geometric deviations and minor changes in pressure drop as predicted by the optimization model. However, when re-simulated using a body-fitted model with explicit wall-functions, a large difference in pressure drop is observed with 0.3423 and 0.2118 for \(\text{Re}=5k\) and \(\text{Re}=200k\), respectively.
In both cases, the geometry obtained with \(\text{Re}=200k\) performs slightly better (around \(1\%\)), which indicates that the geometric changes are local optima, rather than true optima for the given conditions and the same continuation approach.

\begin{figure}[bt]
    \centering
    \subfloat[Example~\#6, U-bend, $Re = 5{,}000$]{
        \includegraphics[width=0.4\textwidth]{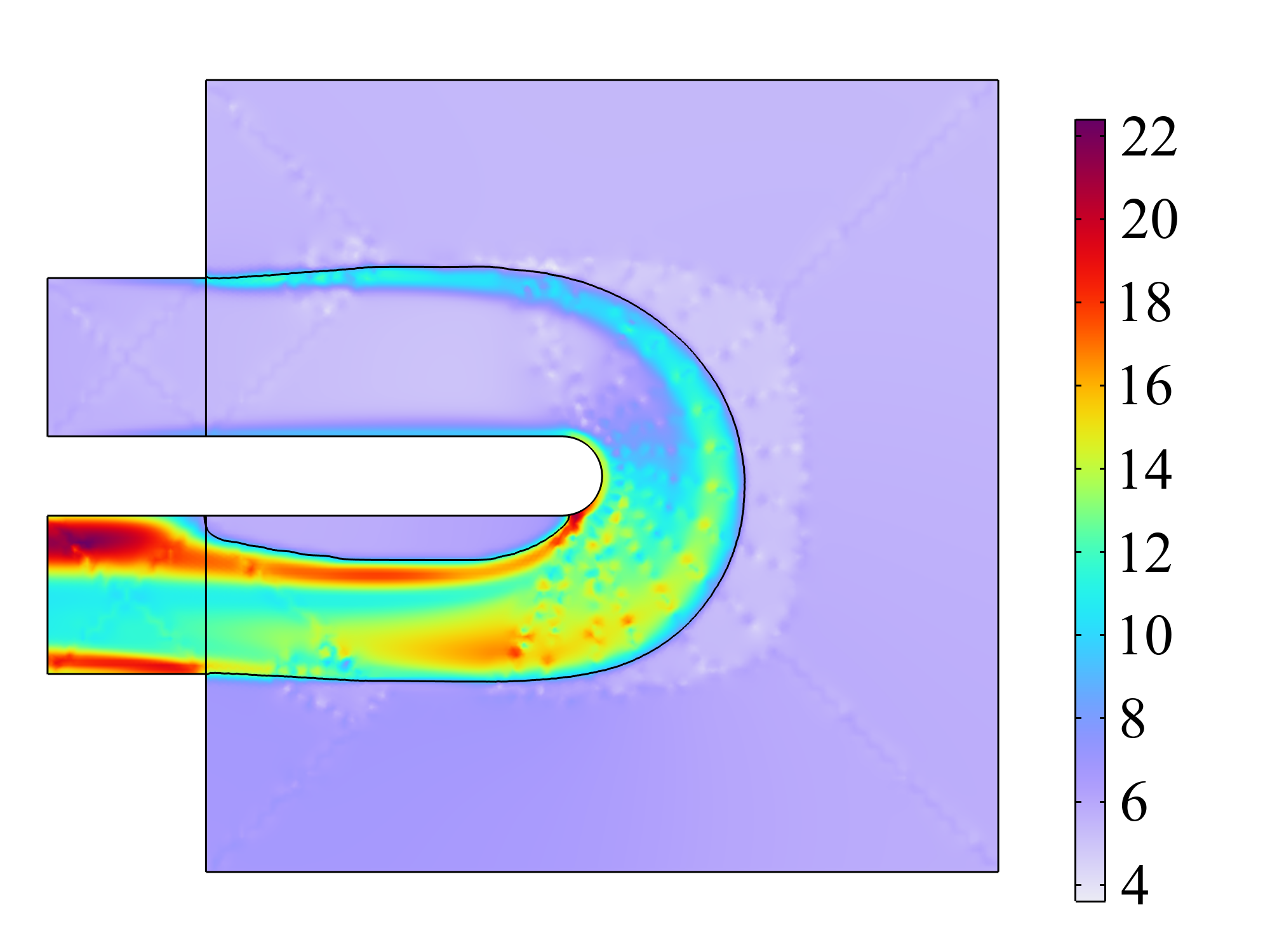}
        \label{subfig:yplus_ubend_5k}
    }\hfill
    \subfloat[Example~\#7, U-bend, $Re = 200{,}000$]{
        \includegraphics[width=0.4\textwidth]{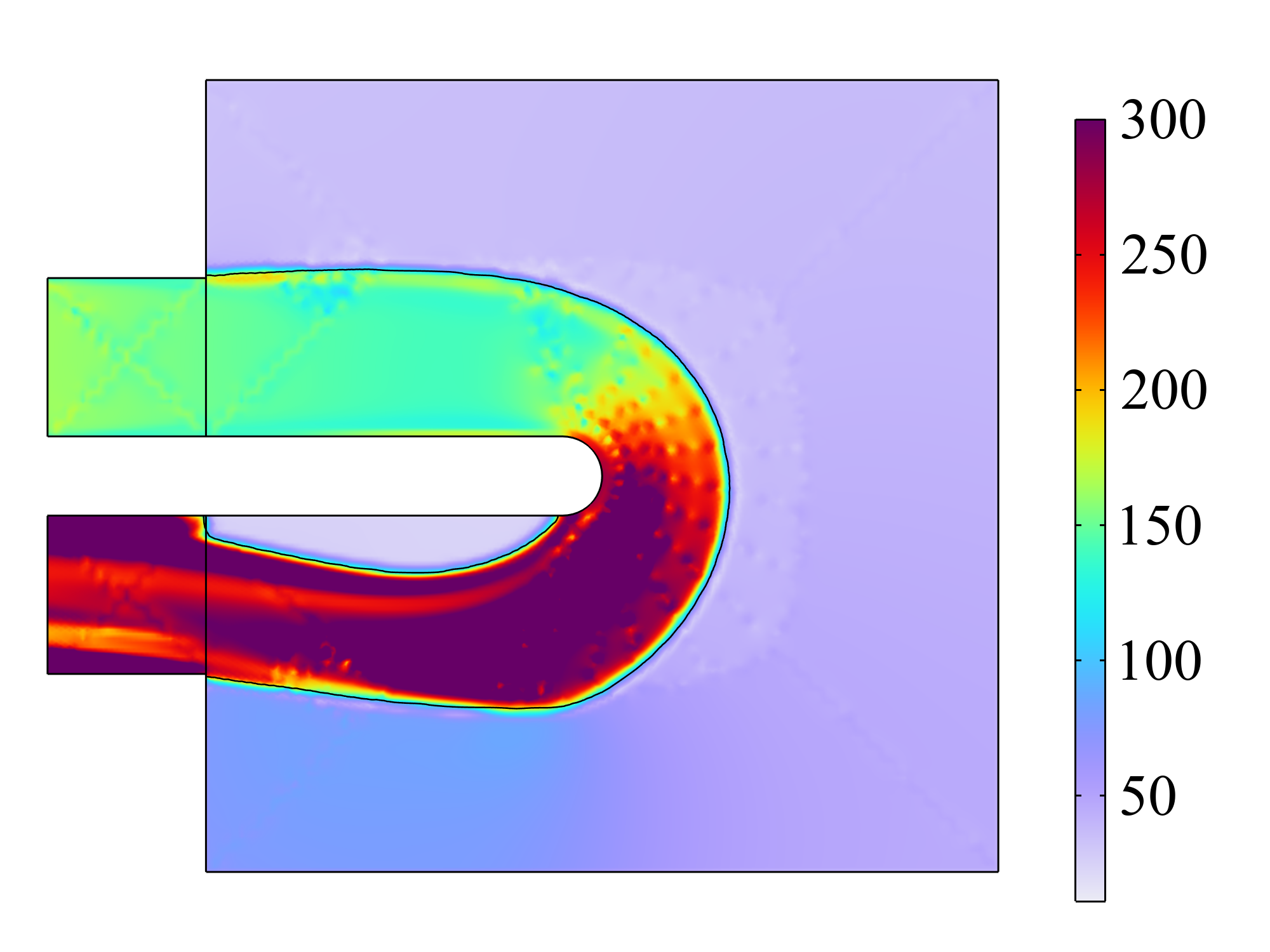}
        \label{subfig:yplus_ubend_200k}
    }

    \caption{Implicit $y^{+}$ distributions for the optimized U-bend geometries in Examples~\#6 and \#7, obtained using the proposed implicit wall-function method. In the vicinity of the walls, $y^{+}$ remains close to 11.06 for Example~\#6 and below 300 for Example~\#7, indicating that the logarithmic law of the wall is satisfied in both cases.}
    \label{fig:ubend_yplus_all}
\end{figure}
\Cref{fig:ubend_yplus_all} shows the implicit $y^{+}$ distribution for the U-bend (Examples~\#6 and \#7). In the area close to the wall, $y^{+}$ remains close to 11.06 for Example~\#6, and below 300 for Example~\#7, confirming that the logarithmic law of the wall holds.

\subsubsection{Comparison with literature and verification with LES}

Having compared the proposed method to the ``conventional'' method, we seek to compare our design to the literature using wall-functions.
The only reference in the literature with a topology-optimized design using wall-functions for the k-$\varepsilon$ RANS model, is the work by \cite{picelli2022topology} using the TOBS-GT method --- a method that utilizes body-fitted meshes with explicit walls. The resulting design is extracted from \cite[Fig. 16a]{picelli2022topology} and loaded into COMSOL to ensure simulation under the exact same conditions as our results. It should be noted that this design is not obtained using average inlet pressure directly as objective functional, but instead has used the dissipated energy functional \cite{borrvall2003topology,picelli2022topology} --- so the comparison is not entirely fair, although the two objectives should be proportional. The optimized designs are compared using a body-fitted mesh and both RANS $k-\varepsilon$ and LES models.
\begin{figure}[tb]
    \centering
    \subfloat[Proposed approach, velocity]{\includegraphics[width=0.30\textwidth]{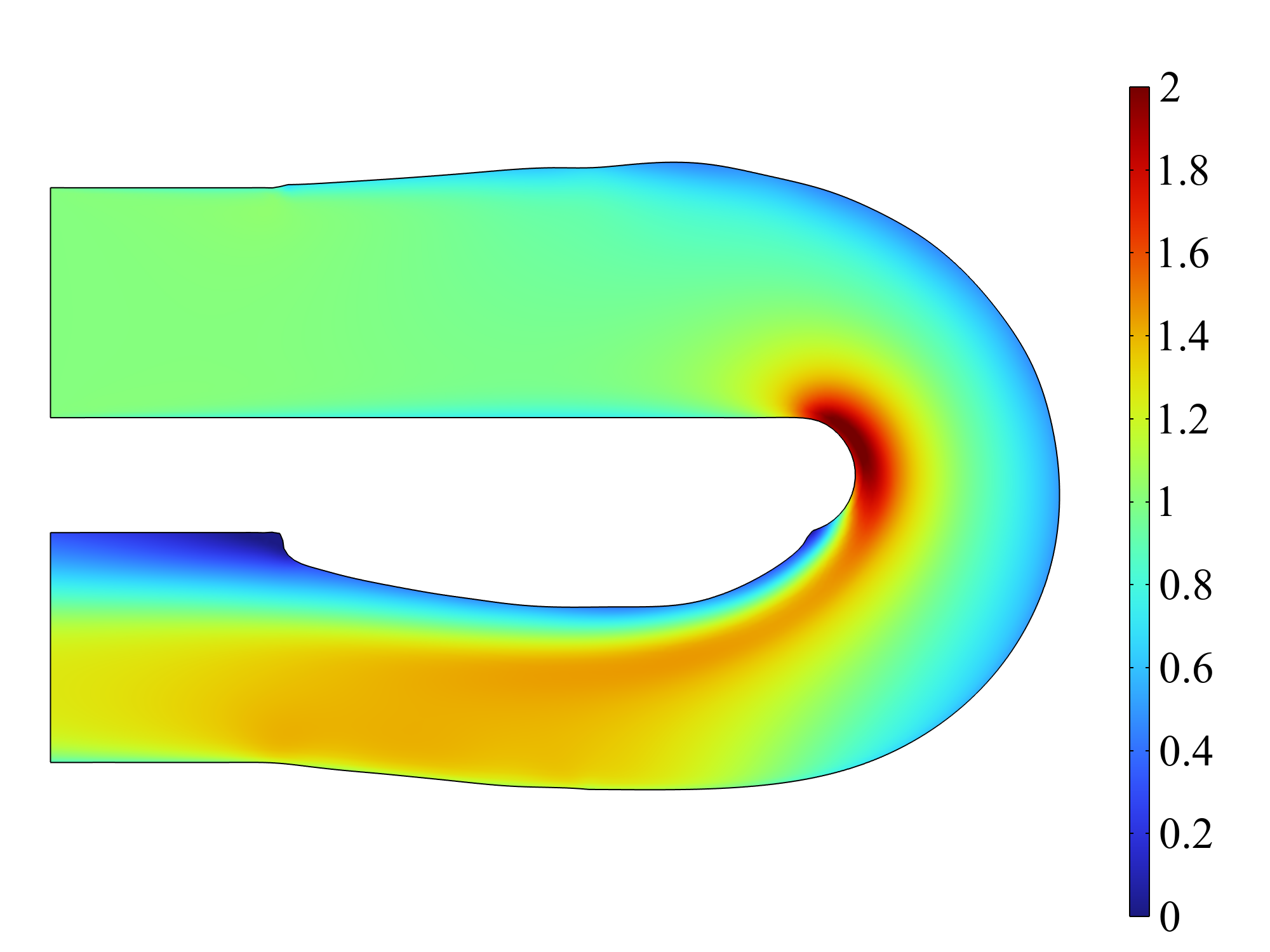}\label{subfig:comparison_RANS_velocityField_ours}}
    \hspace{0.04\textwidth}
    \subfloat[TOBS-GT method \cite{picelli2022topology}, velocity]{\includegraphics[width=0.30\textwidth]{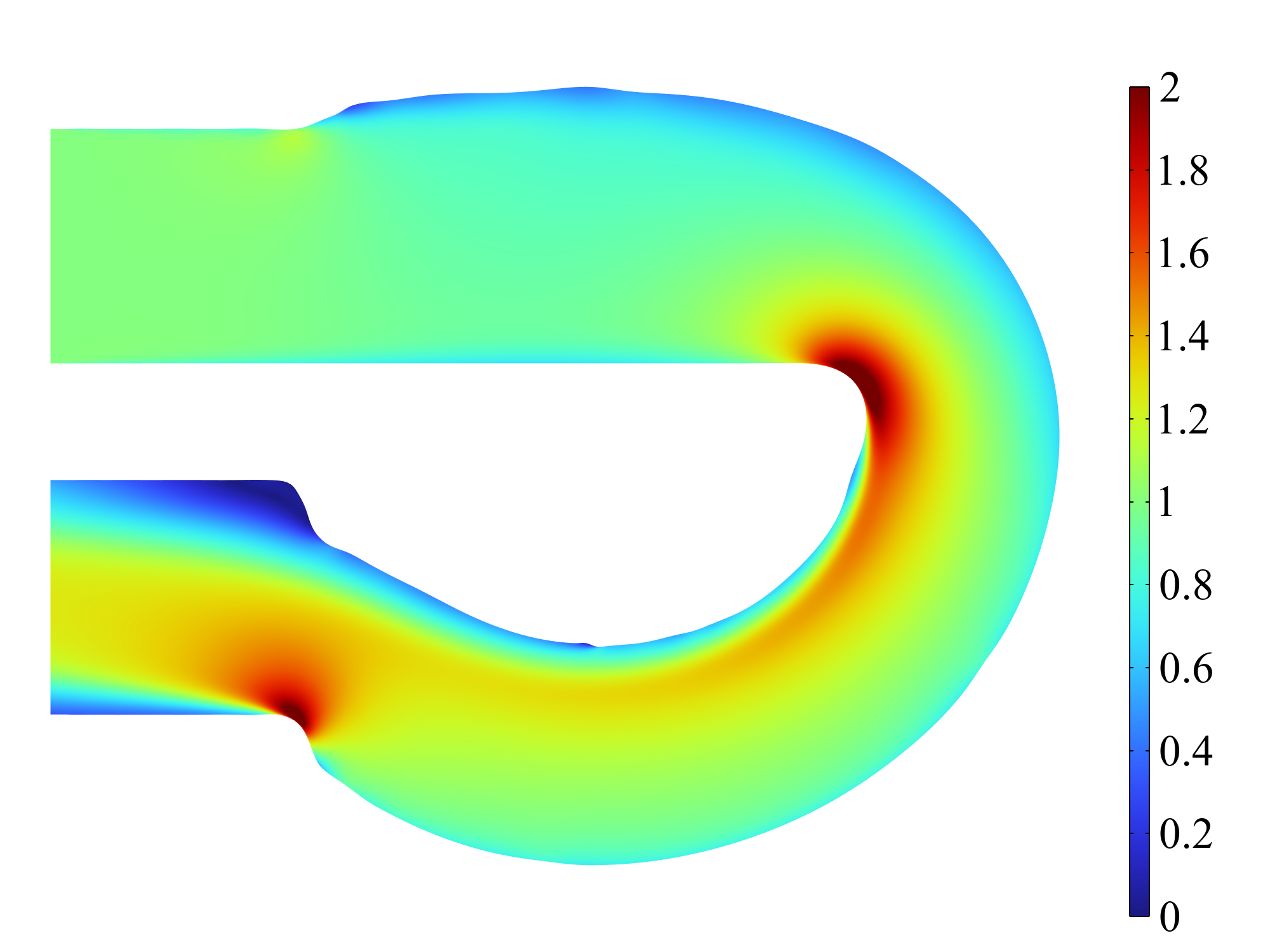}\label{subfig:comparison_RANS_velocityField_picelli}}
    \hspace{0.04\textwidth}
    \subfloat[``Conventional'' approach, velocity]{\includegraphics[width=0.30\textwidth]{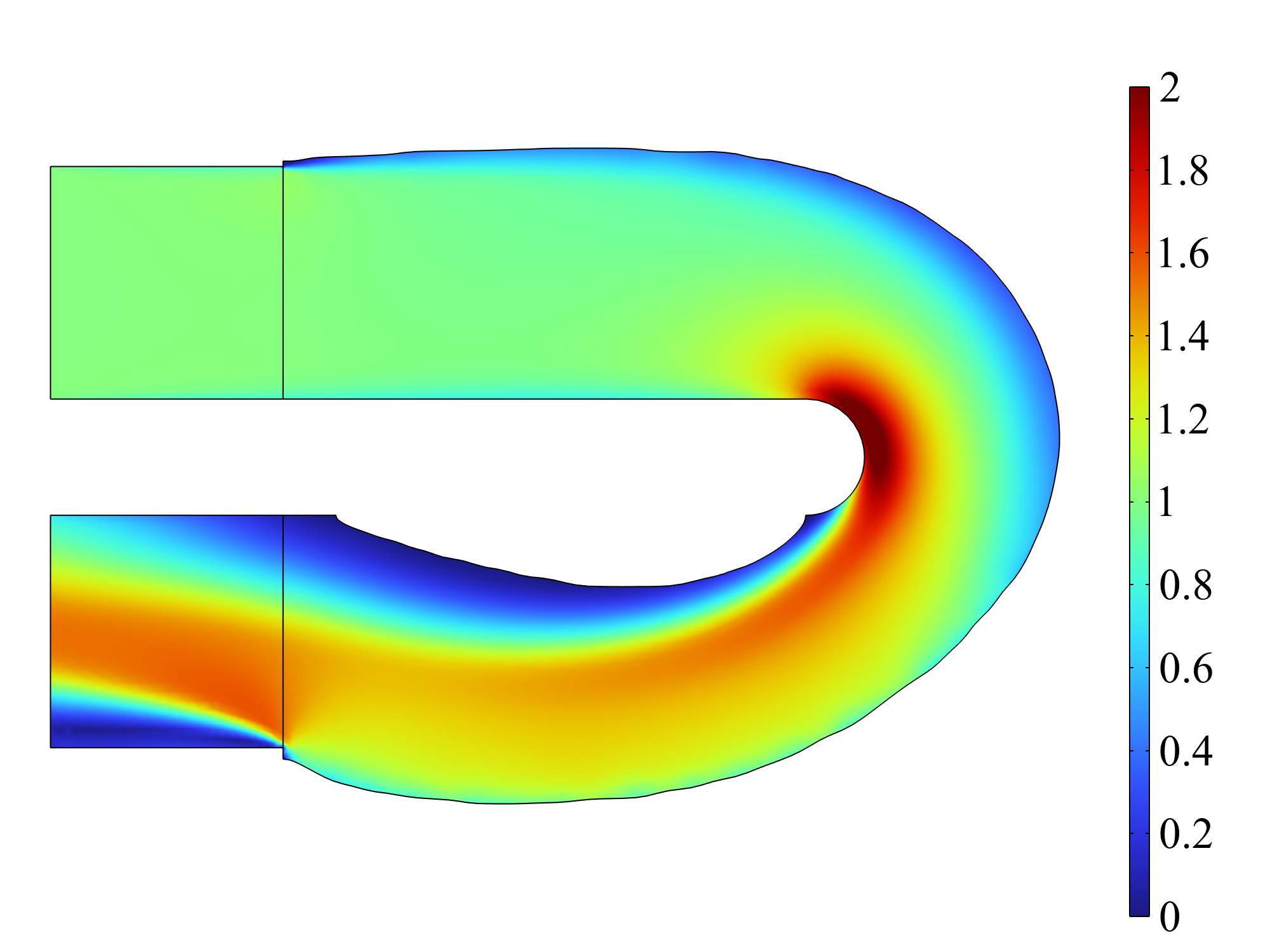}\label{subfig:comparison_RANS_velocityField_trad}}
    \\
    \subfloat[Proposed approach, pressure]{\includegraphics[width=0.30\textwidth]{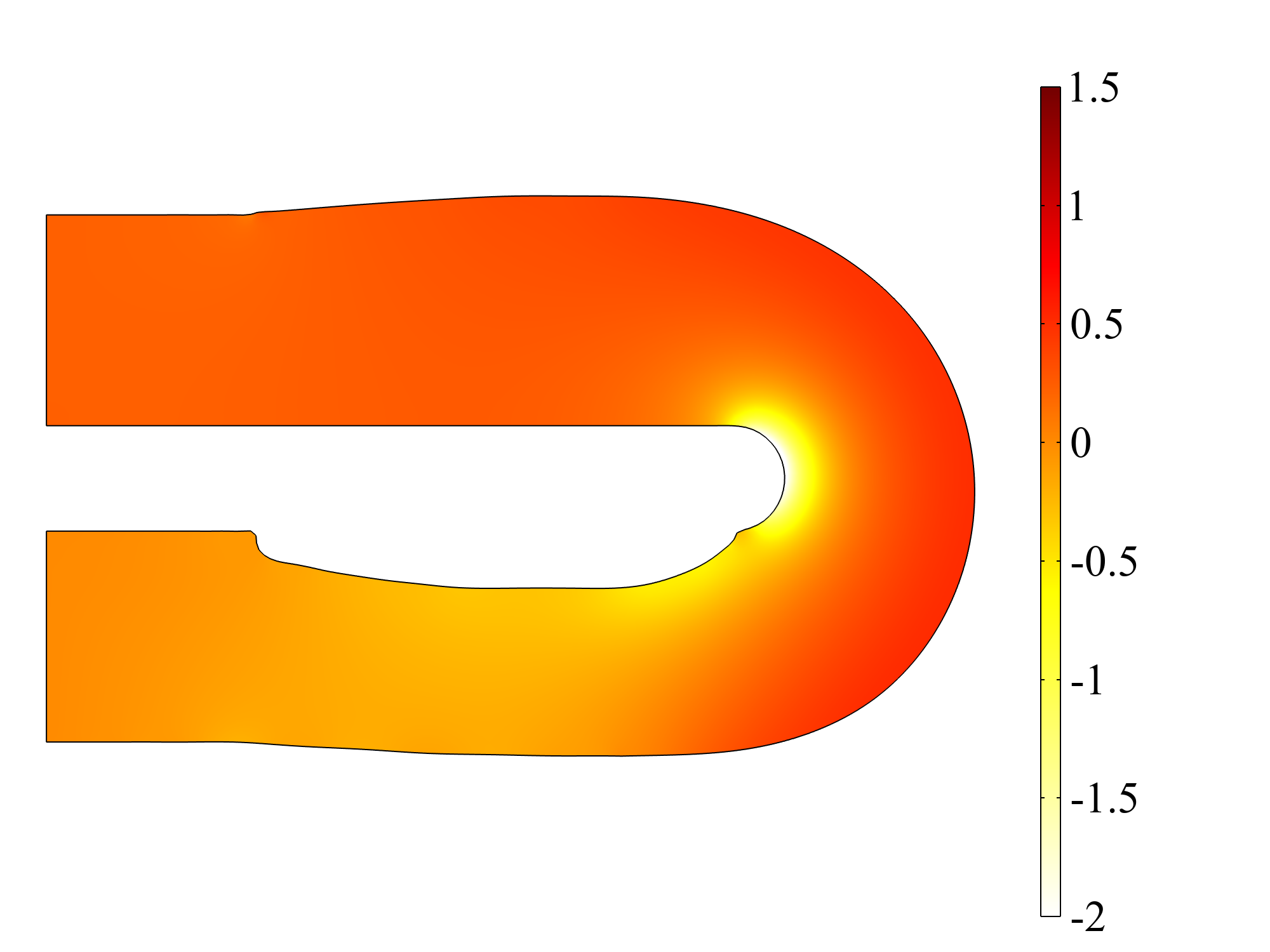}\label{subfig:comparison_RANS_pressureField_ours}}
    \hspace{0.04\textwidth}
    \subfloat[TOBS-GT method \cite{picelli2022topology}, pressure]{\includegraphics[width=0.30\textwidth]{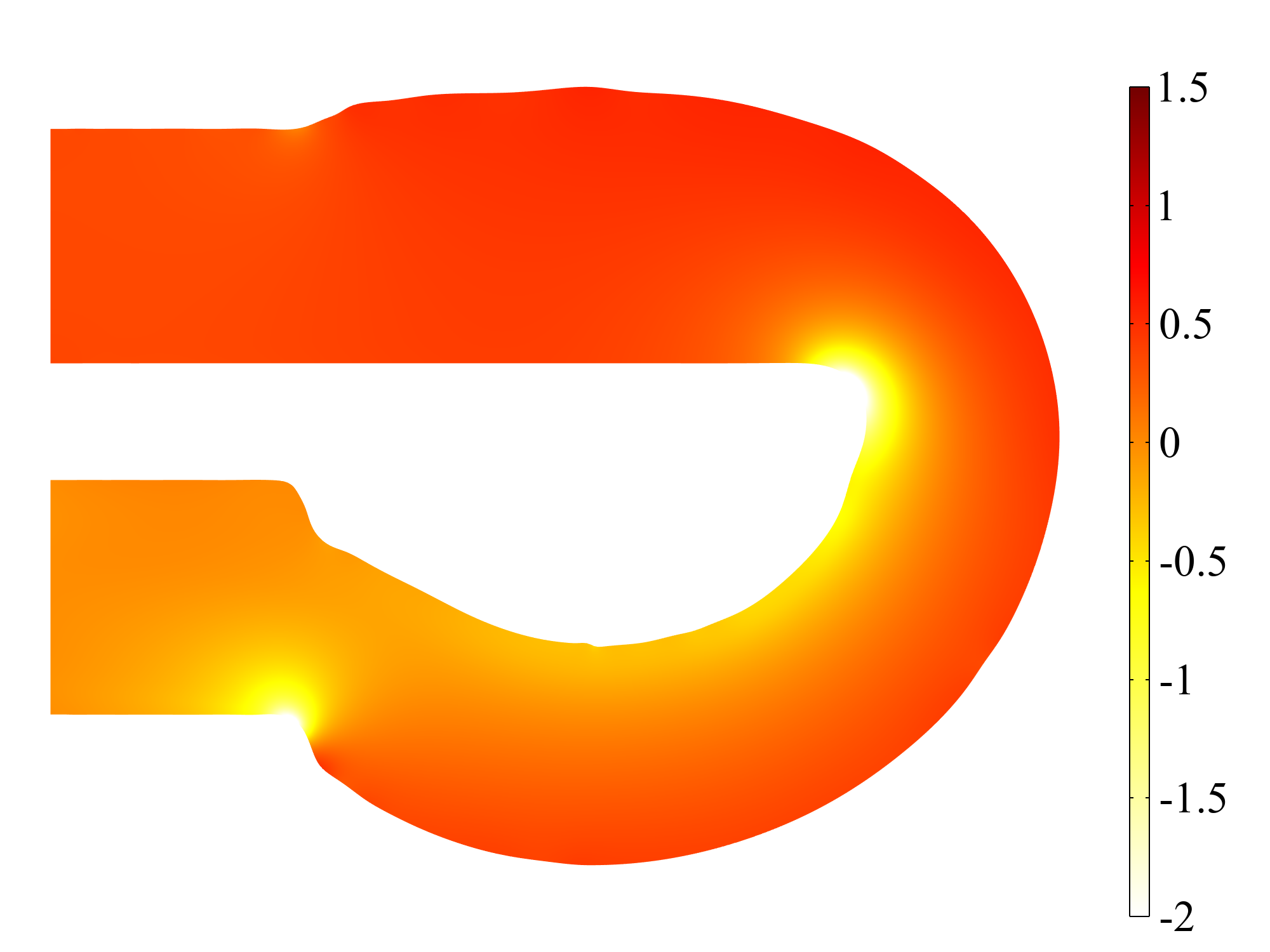}\label{subfig:comparison_RANS_pressureField_picelli}}
    \hspace{0.04\textwidth}
    \subfloat[``Conventional'' approach, pressure]{\includegraphics[width=0.30\textwidth]{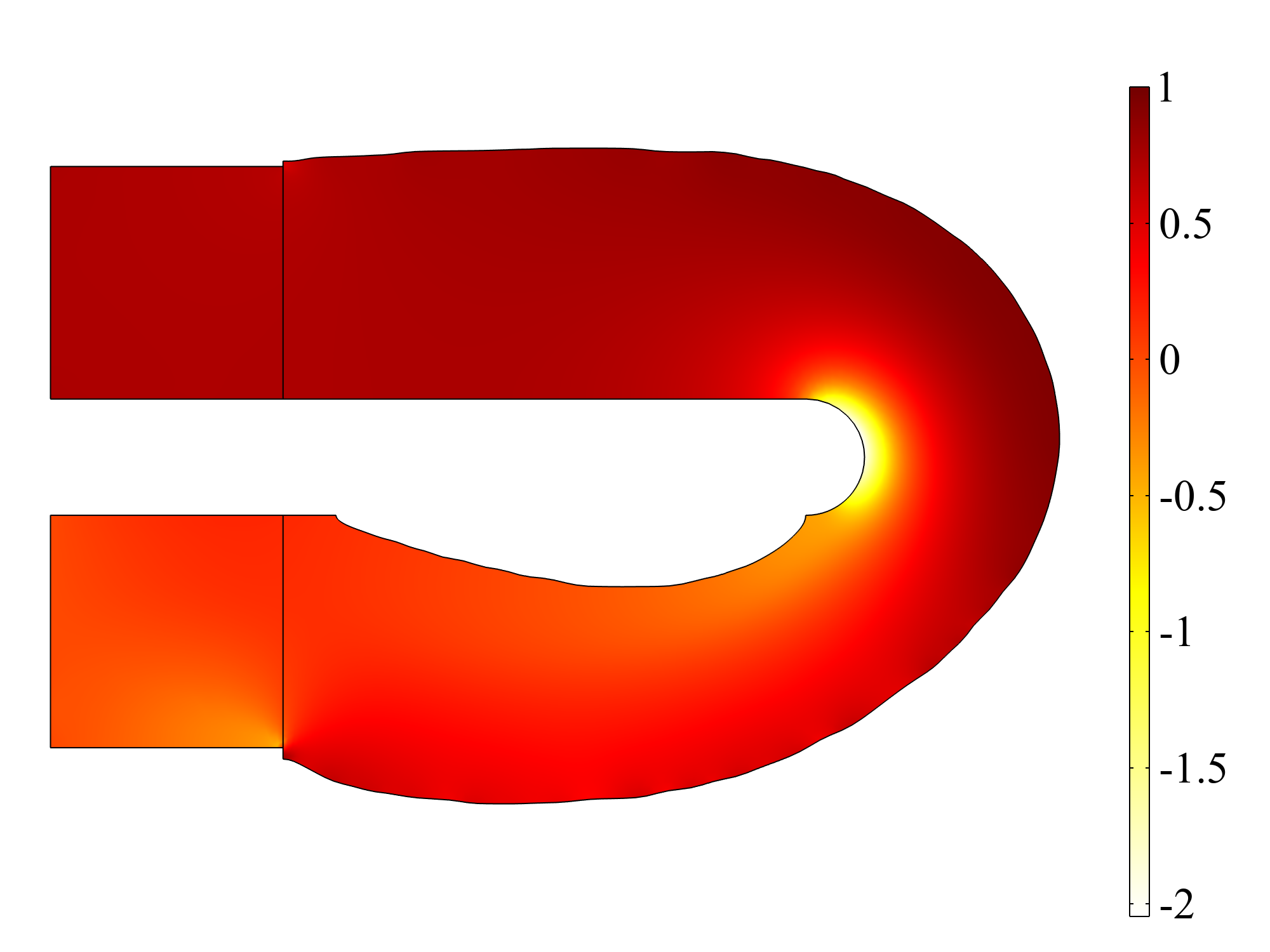}\label{subfig:comparison_RANS_pressureField_trad}}
    \caption{Velocity and pressure fields of the three different designs as calculated using a body-fitted mesh with explicit wall-functions with standard $k$-$\varepsilon$ model.}\label{fig:comparison_RANS}
\end{figure}
\Cref{fig:comparison_RANS} shows the pressure fields using similar body-fitted meshes with explicit wall-functions applied for three different results: the proposed method; the paper by \citet{picelli2022topology}; and the ``conventional'' approach.

\begin{table}[tb]
    \centering
    \begin{tabular}{lccc}
        \toprule
         & \multicolumn{3}{c}{Design} \\
        Model & Proposed   &    TOBS \cite{picelli2022topology}   &    ``Conventional'' \\
        \midrule
        RANS $k-\varepsilon$ & 0.3423 & 0.4103 & 0.9446 \\
        LES                  & 0.4014 & 0.6332 & 1.2806 \\
        \bottomrule
    \end{tabular}
    \caption{Inlet pressure for the three different designs as calculated using body-fitted meshes and either RANS $k-\varepsilon$ or LES, computed from the fields shown in \Cref{fig:comparison_RANS} and \Cref{fig:comparison_LES}.}
    \label{tab:comparison_pressureDrop}
\end{table}
The first row of \Cref{tab:comparison_pressureDrop} shows the pressure drops for the three different designs as computed using a body-fitted mesh using explicit wall-functions for the RANS $k-\varepsilon$ model.
The pressure drop is evaluated to be $0.3423$ for our design and $0.4103$ for the design of \citet{picelli2022topology}, with the ``convectional'' design all the way up at $0.9446$. This shows that the proposed method generates the best performing geometry for this problem.  

In order to further verify the performance of the optimized designs, higher fidelity simulations are run in lieu of physical experiments. We use the next level of fidelity in turbulence models and set up a large eddy simulation (LES) in COMSOL. The details of the LES study are given in Appendix \ref{app:LESdetails} and only the results are given here for brevity. 
\begin{figure}[bt]
    \centering
    \subfloat[Proposed approach, velocity]{\includegraphics[width=0.3\textwidth]{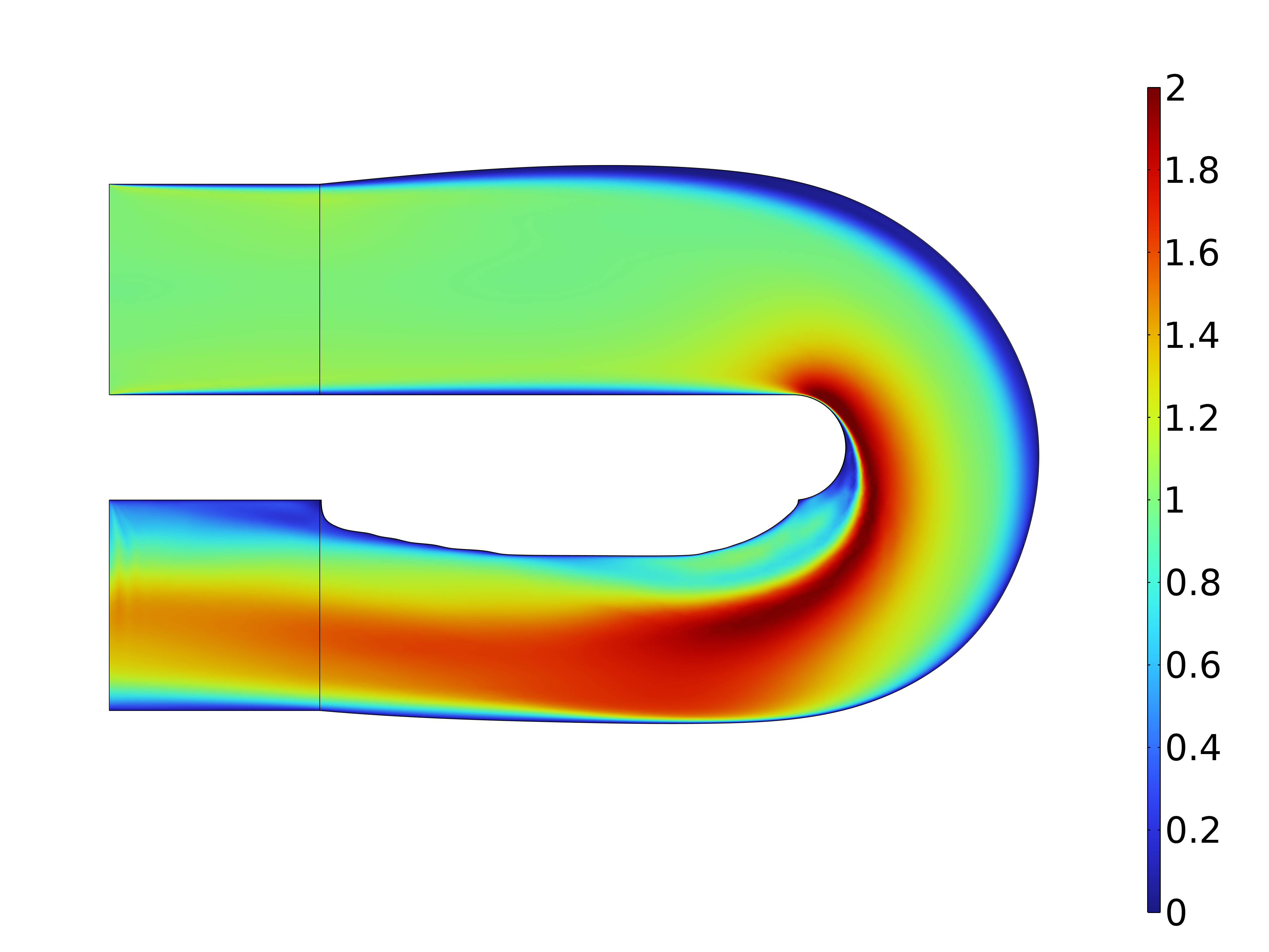}\label{subfig:comparison_LES_velocityField_ours}}
    \hfill
    \subfloat[TOBS-GT method \citep{picelli2022topology}, velocity]{\includegraphics[width=0.3\textwidth]{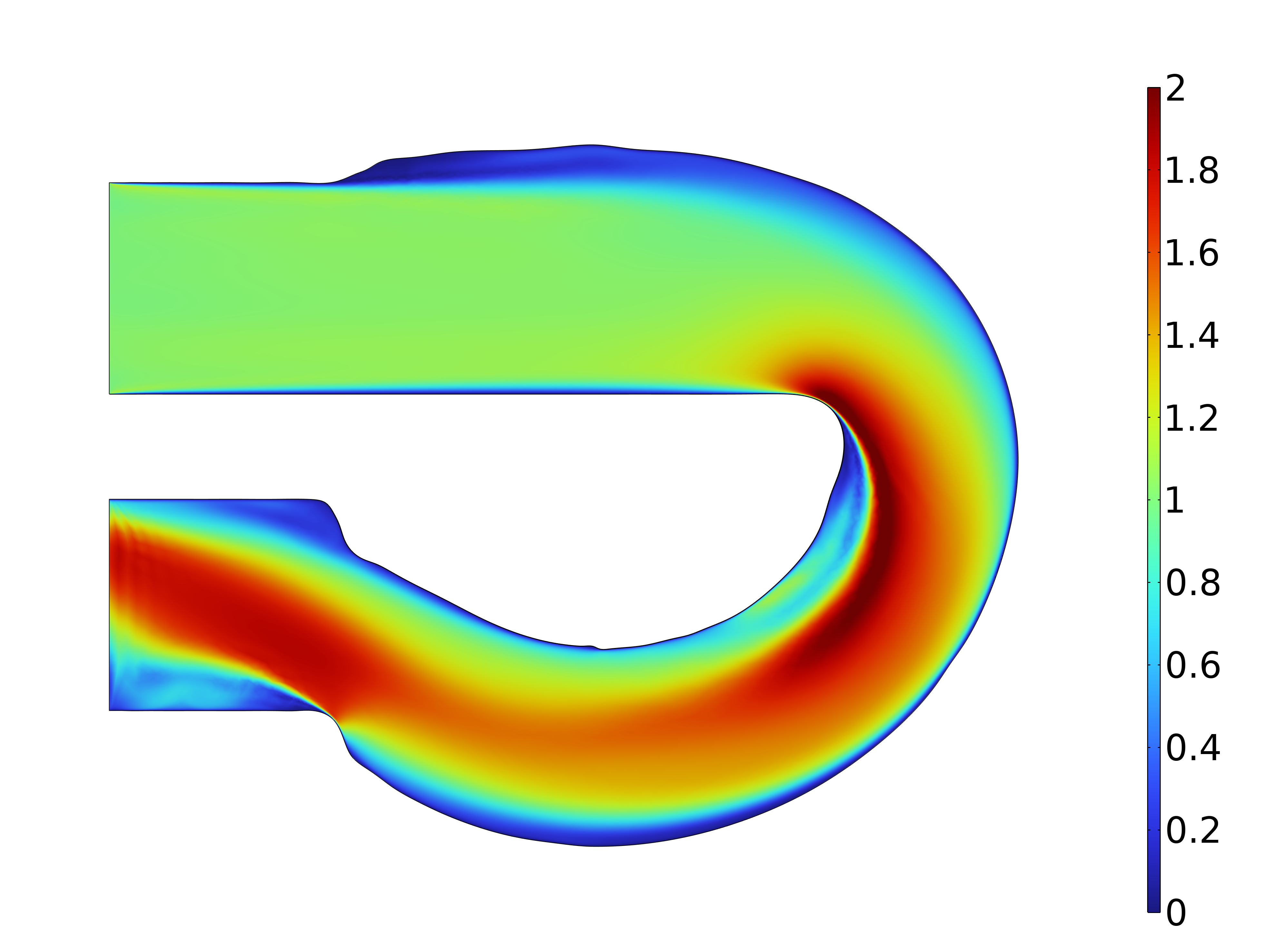}\label{subfig:comparison_LES_velocityField_picelli}}
    \hfill
    \subfloat[``Conventional'' approach, velocity]{\includegraphics[width=0.3\textwidth]{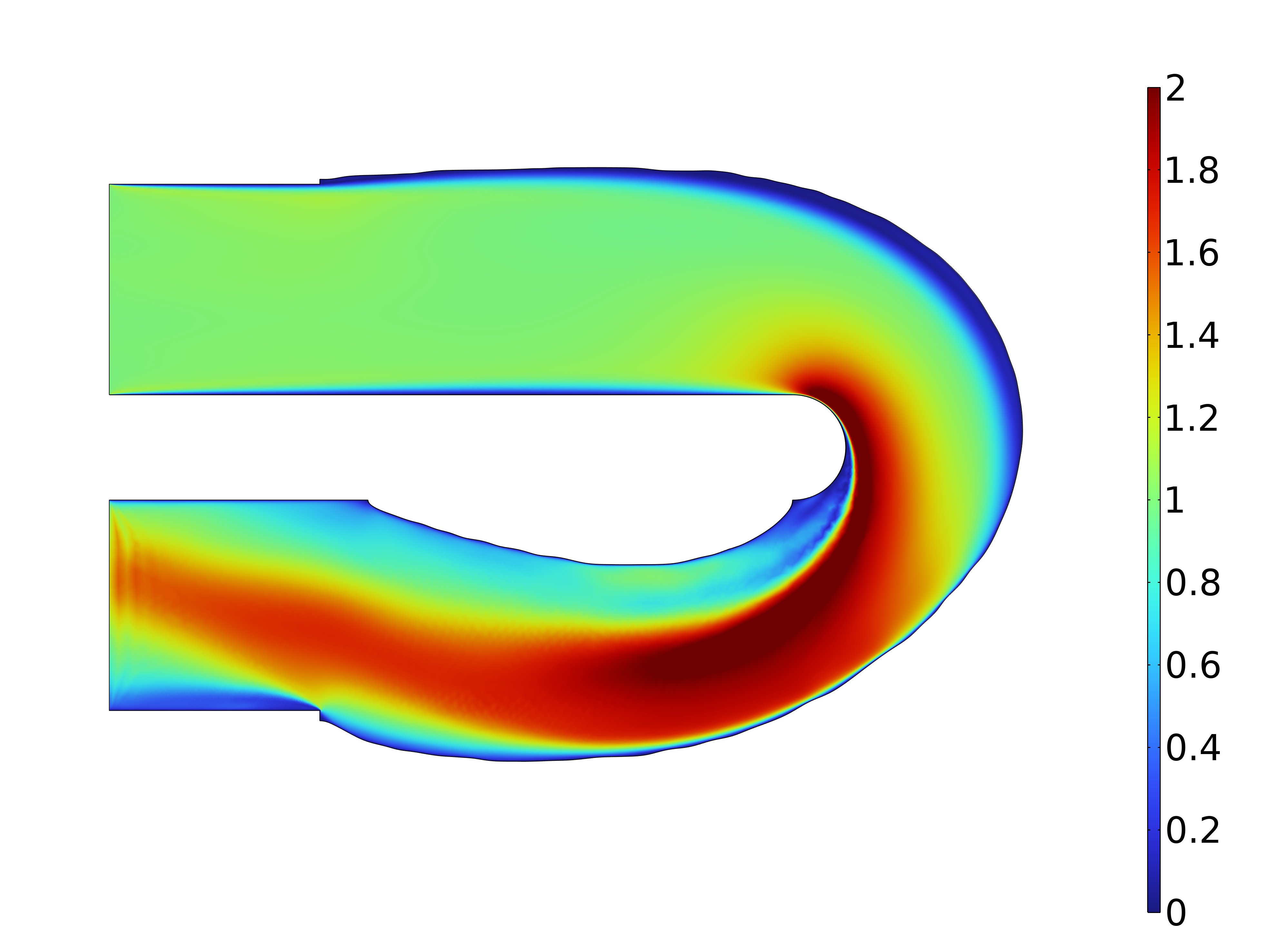}\label{subfig:comparison_LES_velocityField_trad}}
    \\
    \subfloat[Proposed approach, pressure]{\includegraphics[width=0.3\textwidth]{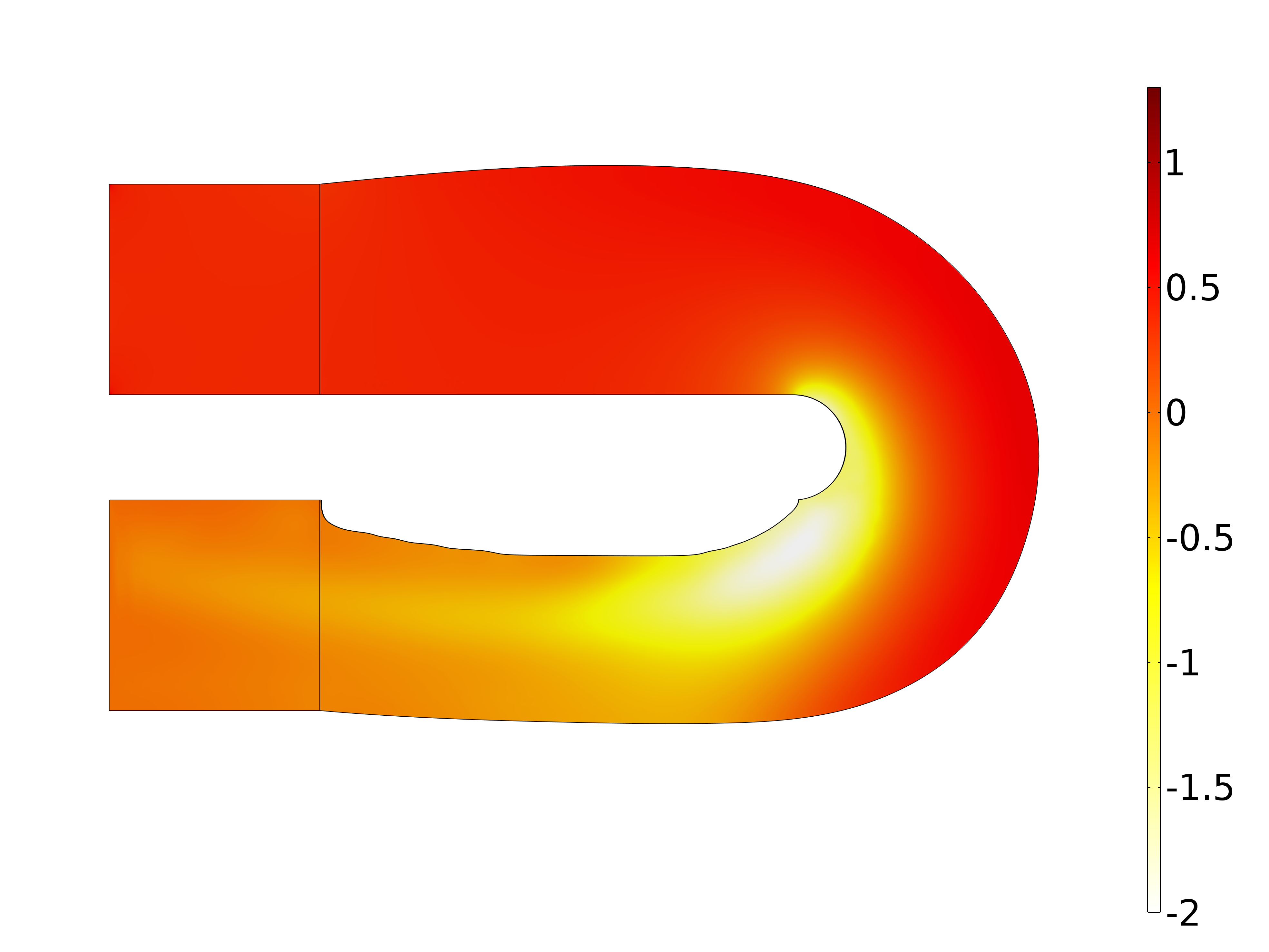}\label{subfig:comparison_LES_pressureField_ours}}
    \hfill
    \subfloat[TOBS-GT method \citep{picelli2022topology}, pressure]{\includegraphics[width=0.3\textwidth]{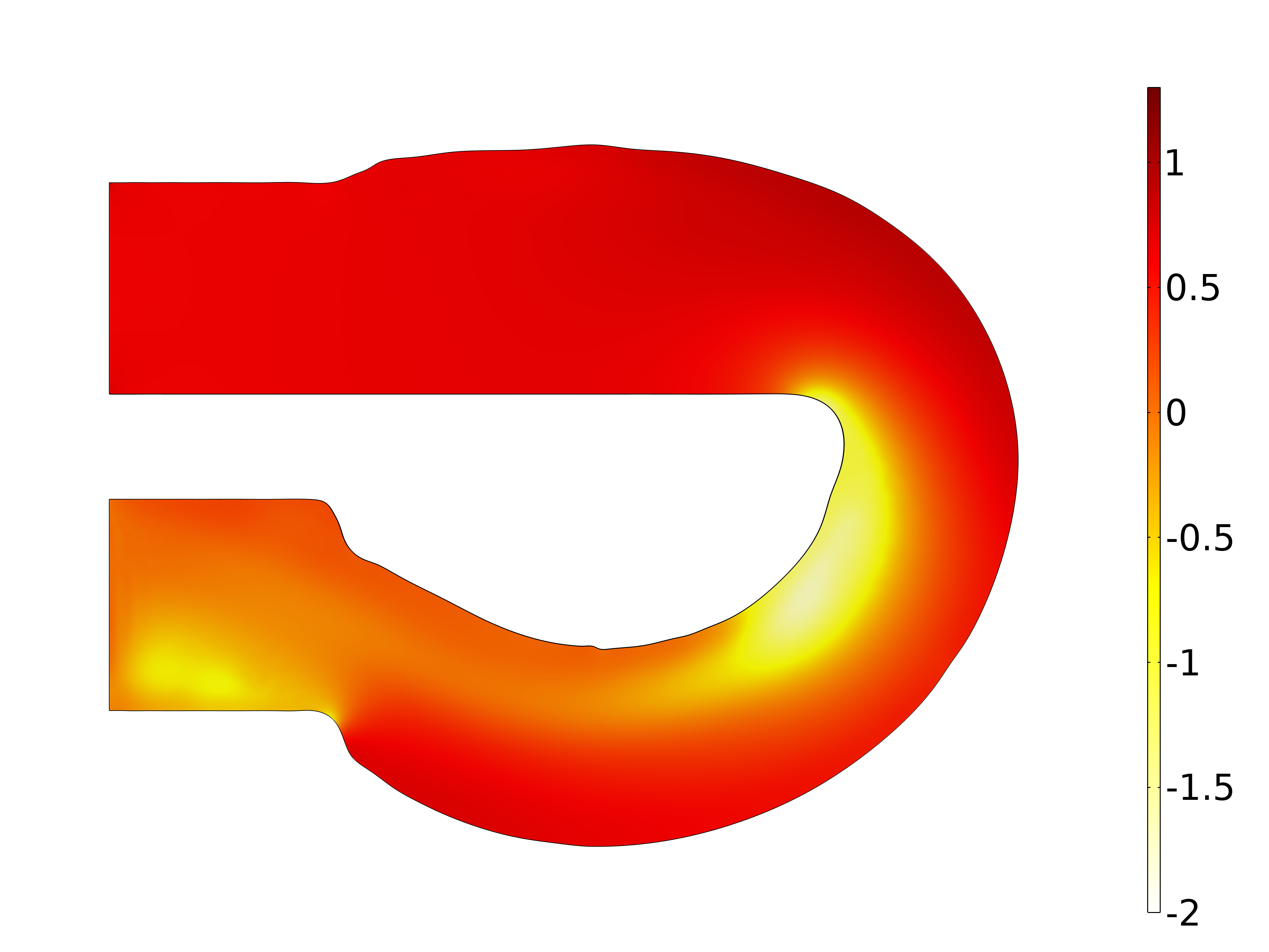}\label{subfig:comparison_LES_pressureField_picelli}}
    \hfill
    \subfloat[``Conventional'' approach, pressure]{\includegraphics[width=0.3\textwidth]{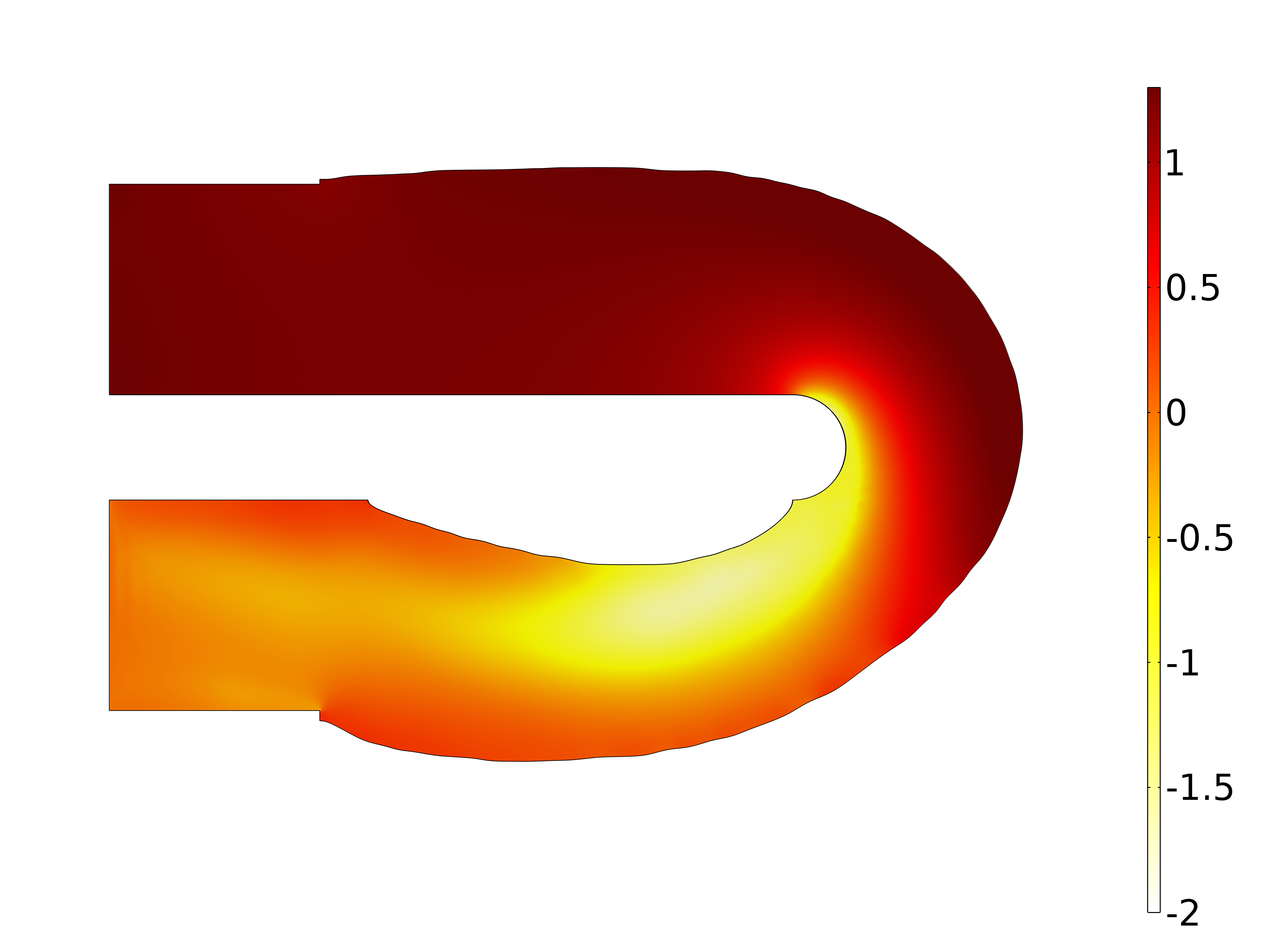}\label{subfig:comparison_LES_pressureField_trad}}
    \caption{Time-averaged velocity and pressure fields of the three different designs as calculated using a body-fitted mesh with LES model.}\label{fig:comparison_LES}
\end{figure}
\Cref{fig:comparison_LES} shows the time-averaged velocity and pressure fields for the three designs: the proposed implicit wall-function method; the result by \cite{picelli2022topology}; the ``conventional'' method.
It can be seen that the overall trends are similar, but there are important differences between the $k-\varepsilon$ model with explicit wall-functions and the LES model.

The second row of \Cref{tab:comparison_pressureDrop} shows the pressure drops for the three different designs as computed using a body-fitted mesh using the LES model.
The time-averaged pressure drop is computed to be $0.4014$, $0.6332$, and $1.2806$, respectively, following the same trend observed using the RANS $k-\varepsilon$ model above.
Despite seeming more accommodating to inertia and having a smoother directional change, the design generated by TOBS-GT \cite{picelli2022topology} appears to have a $57\%$ larger average pressure drop than the proposed implicit wall-function approach. However, both designs generated using wall-functions are significantly better than the design generated using the ``conventional'' approach without wall-functions, with a pressure drop 3 and 2 times larger than the proposed and TOBS approaches, respectively. This is to be expected, since the ``conventional'' approach would require a significantly finer mesh than currently used (and commonly used in the literature) to capture the thin boundary layers and near wall behavior.

\FloatBarrier

\subsection{Tesla valve}
The ``Tesla valve''~\cite{tesla1920valvular} is a passive, no-moving-parts flow rectifier that leverages geometry to induce direction-dependent hydraulic losses. Functionally, a Tesla valve is designed to minimize average inlet pressure in the forward direction, while inducing a significantly larger average inlet pressure in the reverse direction --- thereby acting as a fluid diode.  At low Reynolds numbers, the response is primarily viscosity-dominated, whereas at high Reynolds numbers, inertial effects become the governing mechanism. Because the architecture features sharp bends and sudden expansions, highly turbulent regimes occur, and accurate treatment of near-wall physics becomes decisive, making the implementation of wall-functions particularly important on the coarser meshes usually used for topology optimization.  
\begin{figure}
    \centering
    \includegraphics[width=0.5\linewidth]{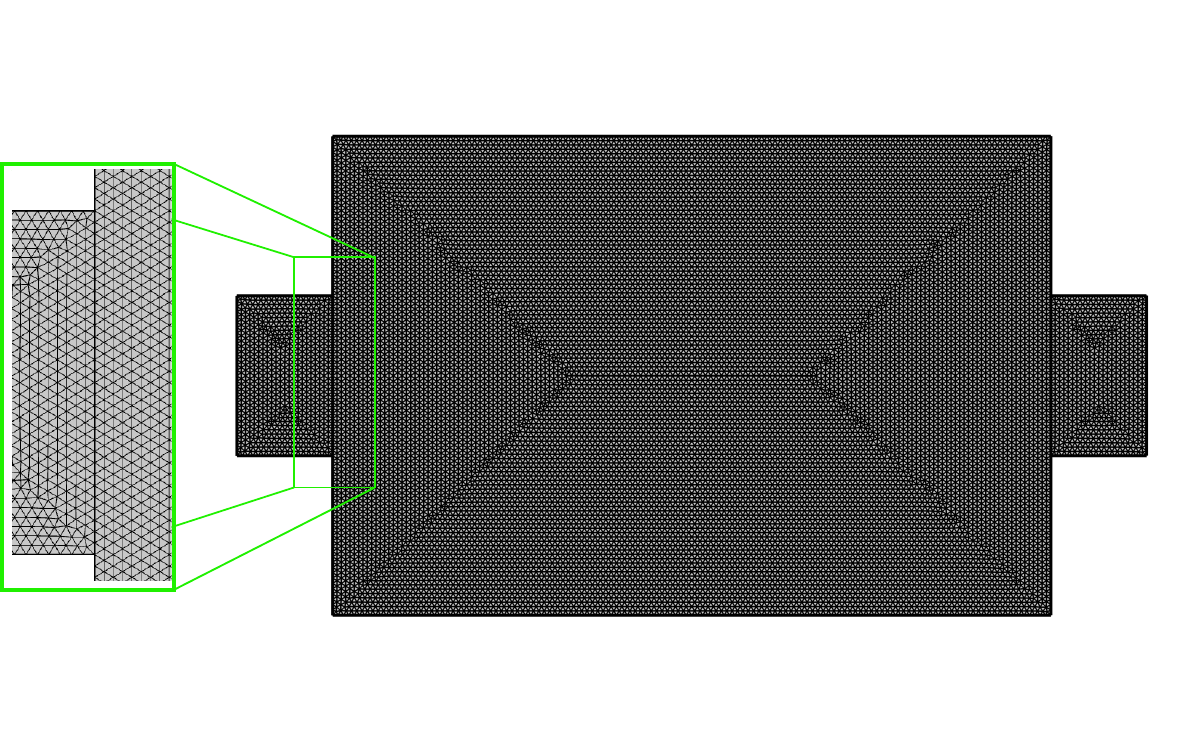}
    \caption{The mesh used for the Tesla valve benchmark is illustrated here for an approximate element size of $h_{max}=$ $0.0125$, yielding 32,516 triangles, and 198,228 total DOFs.}
    \label{fig: Tesla_valve_mesh}
\end{figure}
The schematic of the Tesla valve is shown in \Cref{fig:Tesla_valve}. The flow enters with a uniform inlet velocity ($U_{\mathrm{in}} = 1$), and a zero pressure together with a straight-out velocity condition at the outlet. The computational mesh is presented in \Cref{fig: Tesla_valve_mesh} with 32,516 triangles and a total number of degree of freedoms (DOFs) of 198,228. The parameters and continuation scheme used in the topology optimization are summarized in \Cref{tab:Tesla_valve_sweep_parameters}.
\begin{table}[t!]
    \caption{Tesla-valve with implicit wall-function method parameters.}
    \centering
    \label{tab:Tesla_valve_sweep_parameters}
    \begin{tabular}{@{}l l l l l l l l@{}}
        \toprule
        \textbf{Parameter} & \multicolumn{7}{l}{Value} \\
        \midrule
        $\beta$        & 8 & 12 & 18 & 24 & 30 & 36 & 45 \\
        $q_a$          & 100 & 33 & 11 & 3 & 1 & 1 & 1 \\
        $\alpha_{\max}$& 250 & 270 & 300 & 350 & 350 & 350 & 400 \\
        $h_{\max}$     & 0.01125  \\
        $\psi_{\max}$  & \multicolumn{7}{l}{1000} \\
        $P_{con}$            & \multicolumn{7}{l}{4} \\
        $r_{1}$     & \multicolumn{7}{l}{$4 h_{\max}$} \\
        $V_f$          & \multicolumn{7}{l}{0.6} \\
        $r_{2}$     & \multicolumn{7}{l}{$4 h_{\max}$} \\
        $Re_{inlet}$   & \multicolumn{7}{l}{5000} \\
        $L_{1}$        & \multicolumn{7}{l}{1} \\
        $U_{\mathrm{in}}$       & \multicolumn{7}{l}{1} \\
        \bottomrule
    \end{tabular}
\end{table}

\begin{figure}
    \centering
    \renewcommand{\arraystretch}{1.2}
    \setlength{\tabcolsep}{-2pt}
    \caption{Tesla valve (Example \#8) with implicit wall-function method, optimization history. 
    figures (a) to (d) show the forward velocity fields; 
    figures (e) to (h) show the backward velocity fields;
    figures (i) to (l) illustrate the forward pressure fields; 
    figures (m) to (p) illustrate the backward pressure fields; 
   figures (q) to (t) display the filtered-projected design variables, $\varphi$; 
    figures (u) to (x) present the wall intensity fields.}
    \label{fig:Optimization_history_of_the_tesla_valve}

    \begin{tabular}{cccc}
        \text{Iteration = 5} & \text{Iteration = 15} & \text{Iteration = 100} & \text{Iteration = 200} \\
        \subfloat[]{\includegraphics[width=0.26\textwidth]{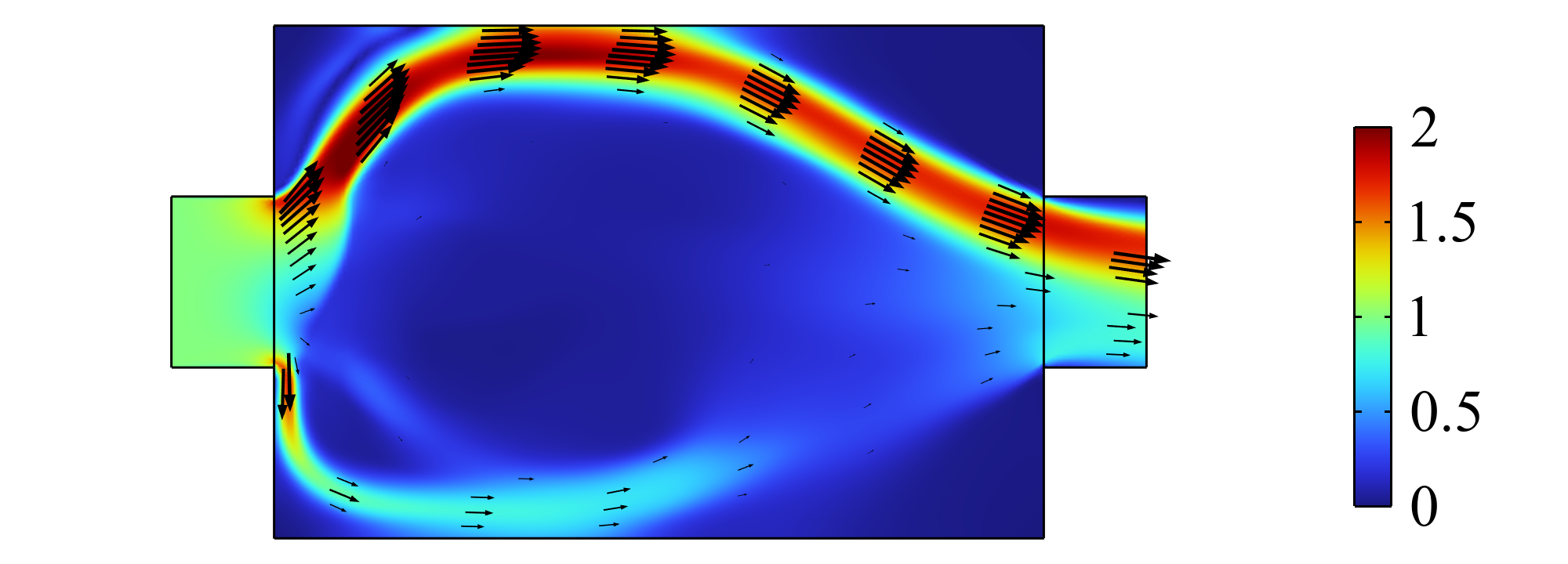}\label{subfig:iter5_v}} &
        \subfloat[]{\includegraphics[width=0.26\textwidth]{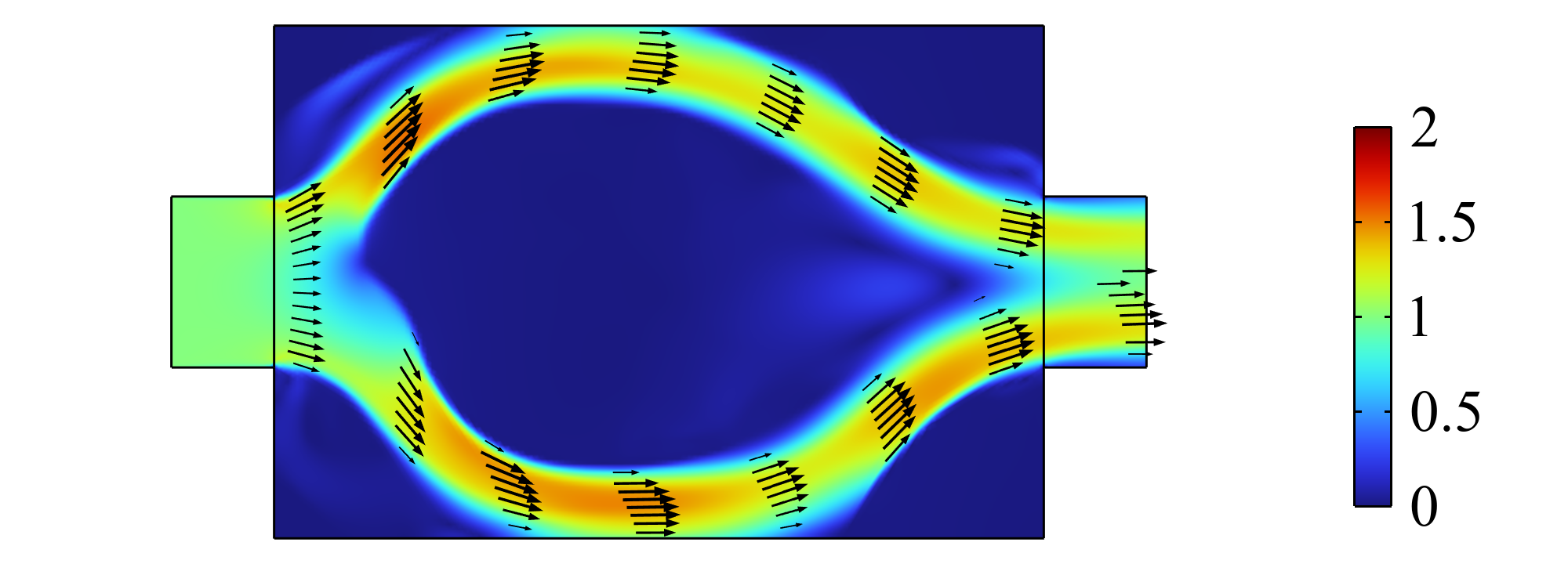}\label{subfig:iter15_v}} &
        \subfloat[]{\includegraphics[width=0.26\textwidth]{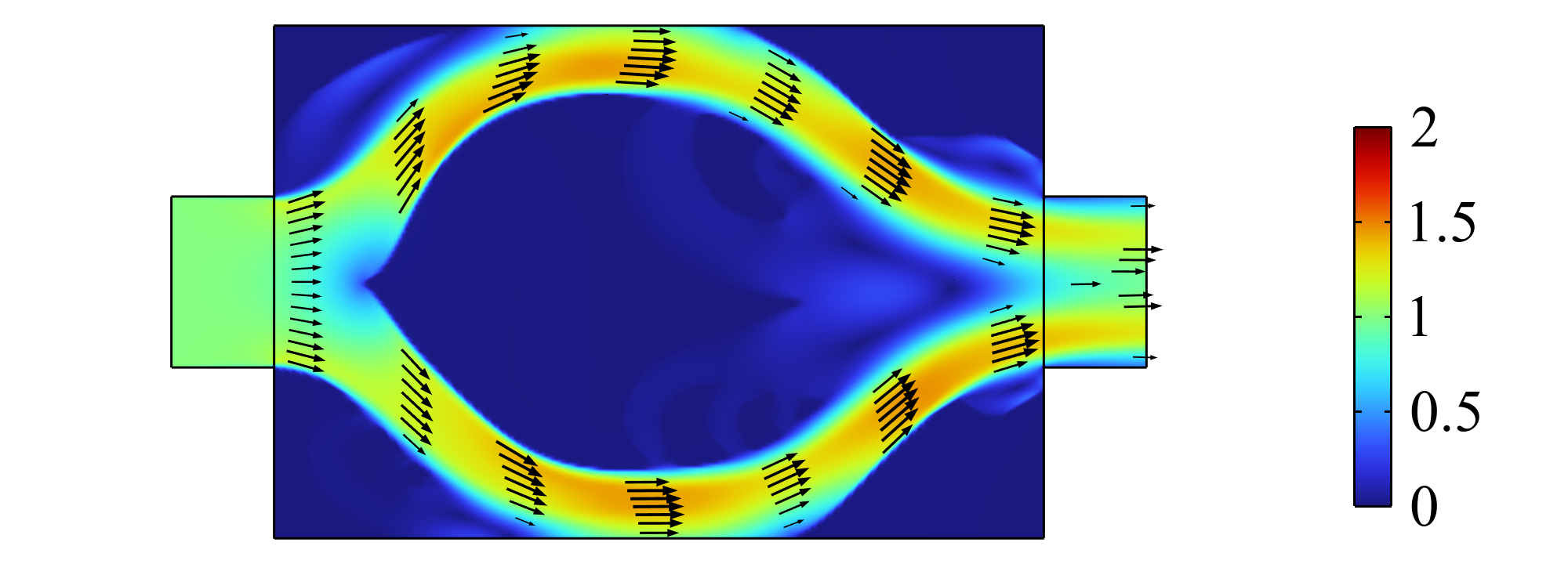}\label{subfig:iter100_v}} &
        \subfloat[]{\includegraphics[width=0.26\textwidth]{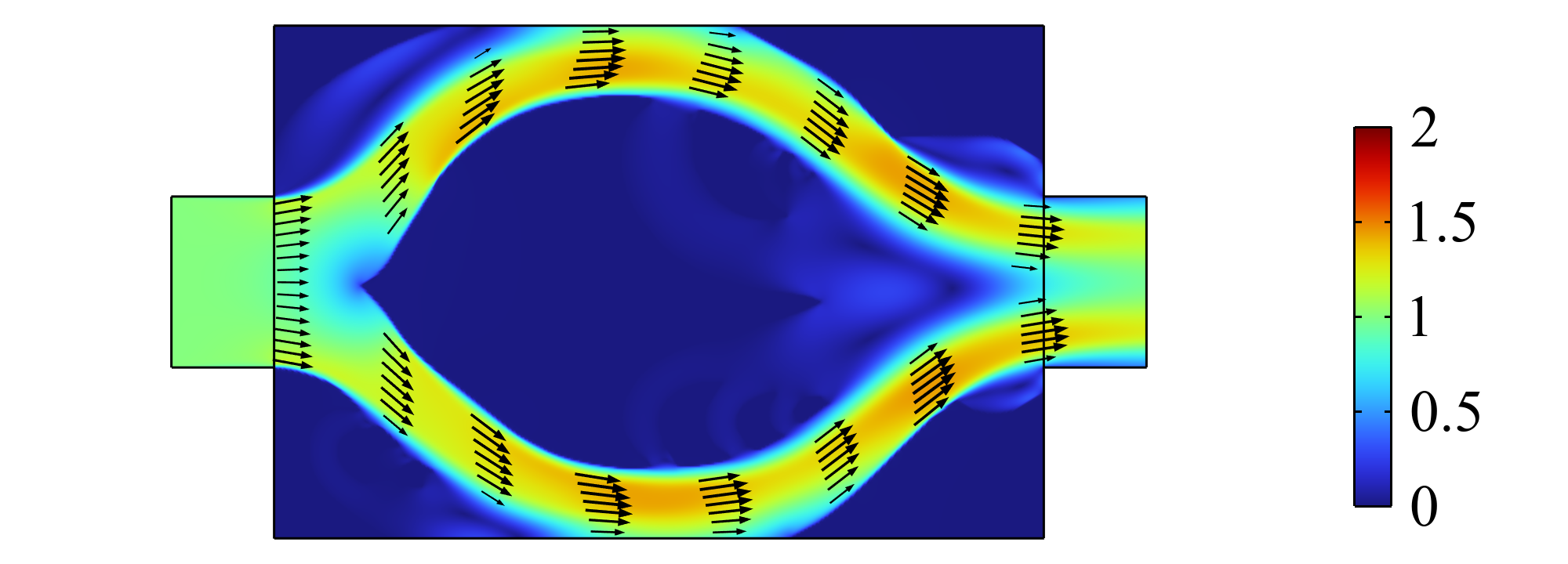}\label{subfig:iter200_v}} \\
        \addlinespace[2mm]

        \subfloat[]{\includegraphics[width=0.26\textwidth]{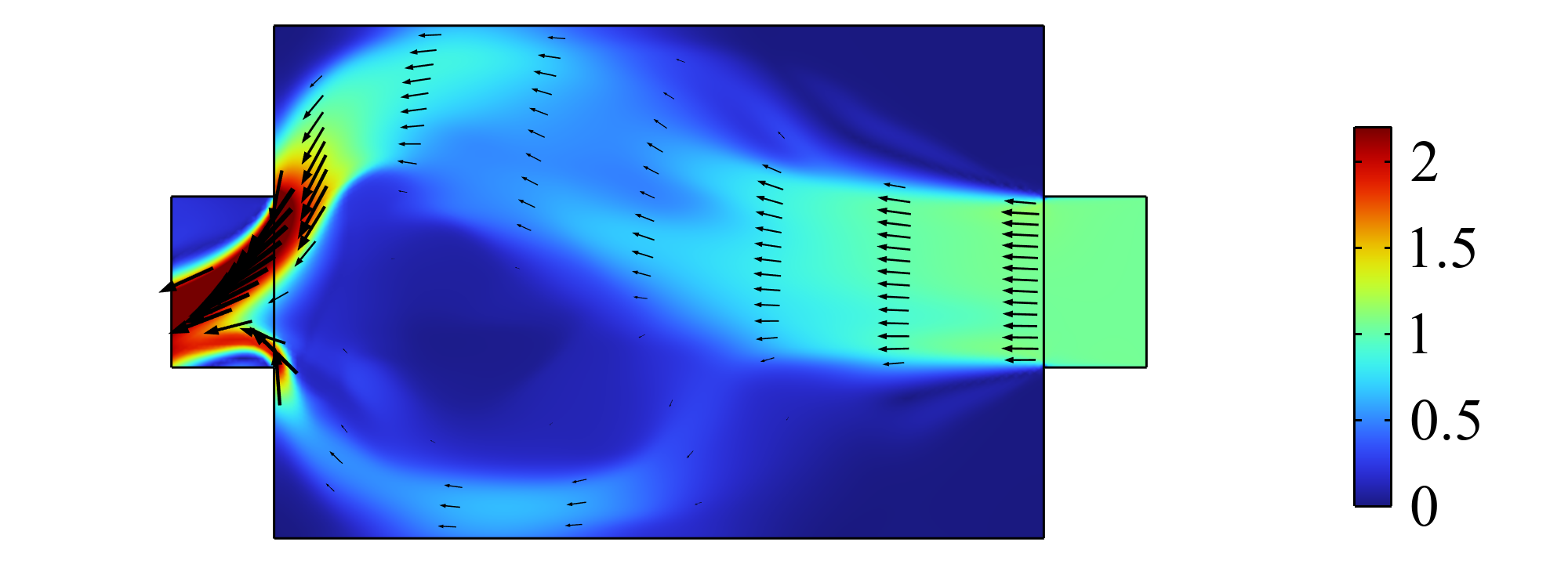}\label{subfig:iter5_v_b}} &
        \subfloat[]{\includegraphics[width=0.26\textwidth]{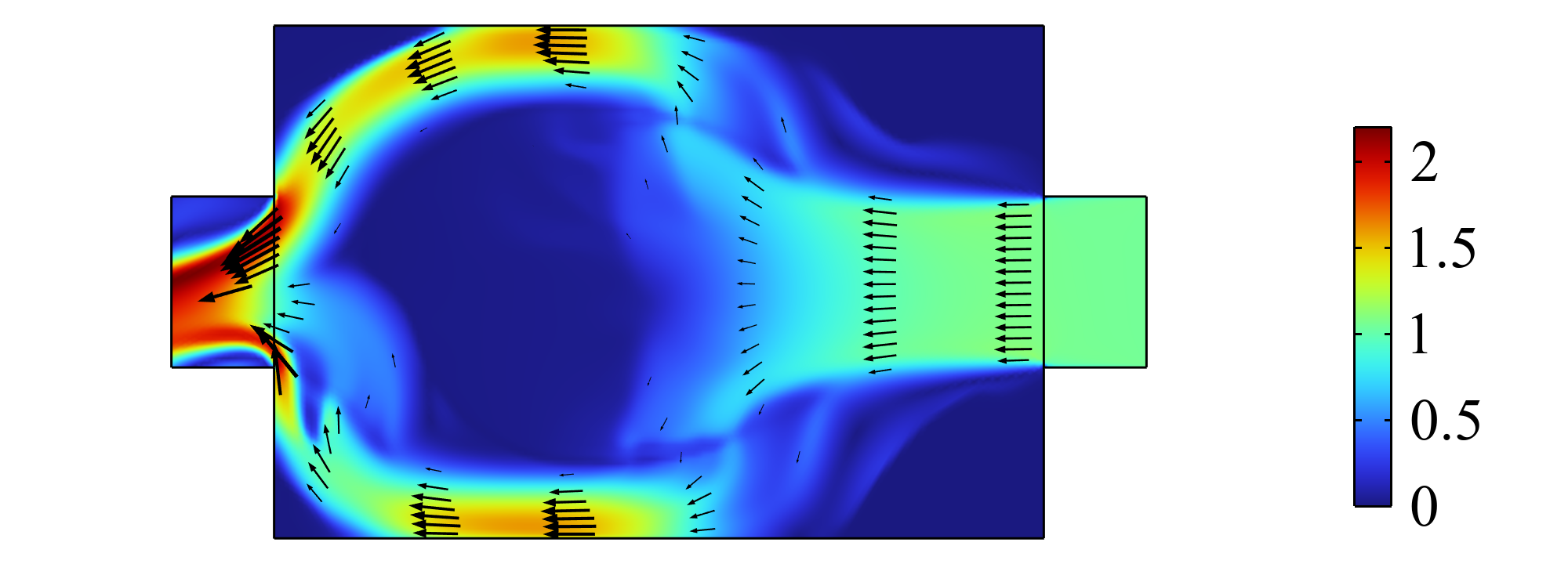}\label{subfig:iter15_v_b}} &
        \subfloat[]{\includegraphics[width=0.26\textwidth]{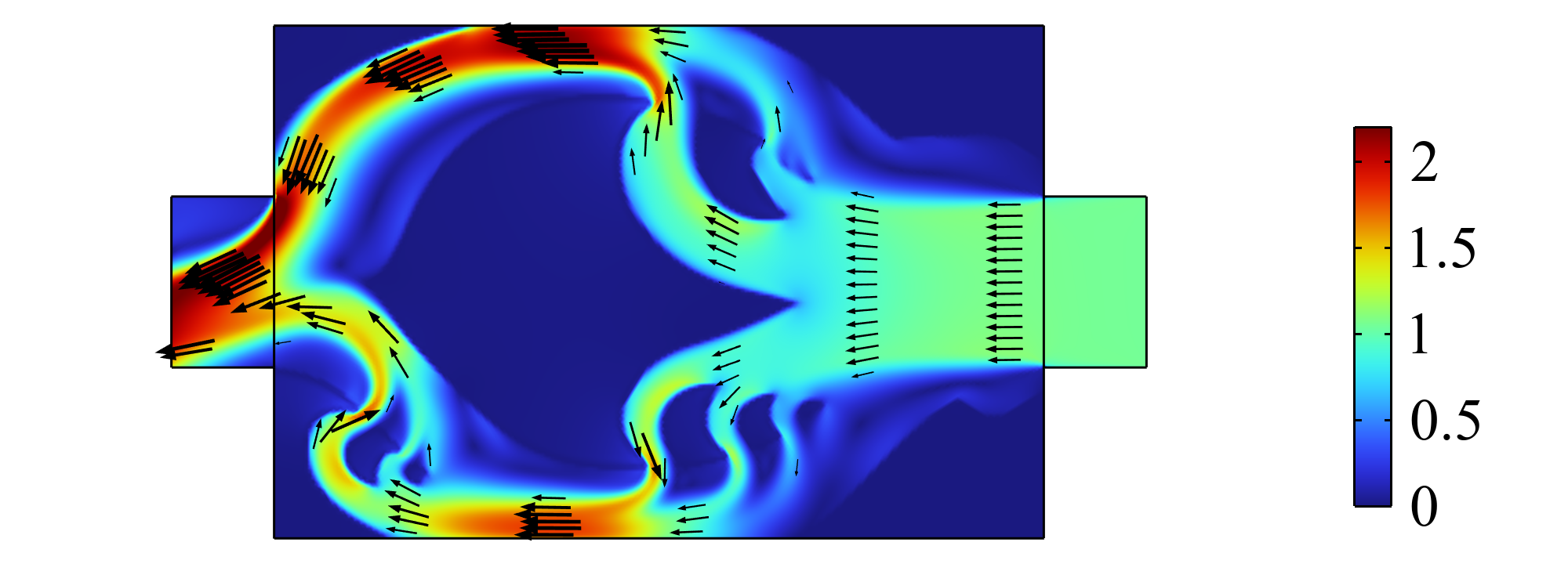}\label{subfig:iter100_v_b}} &
        \subfloat[]{\includegraphics[width=0.26\textwidth]{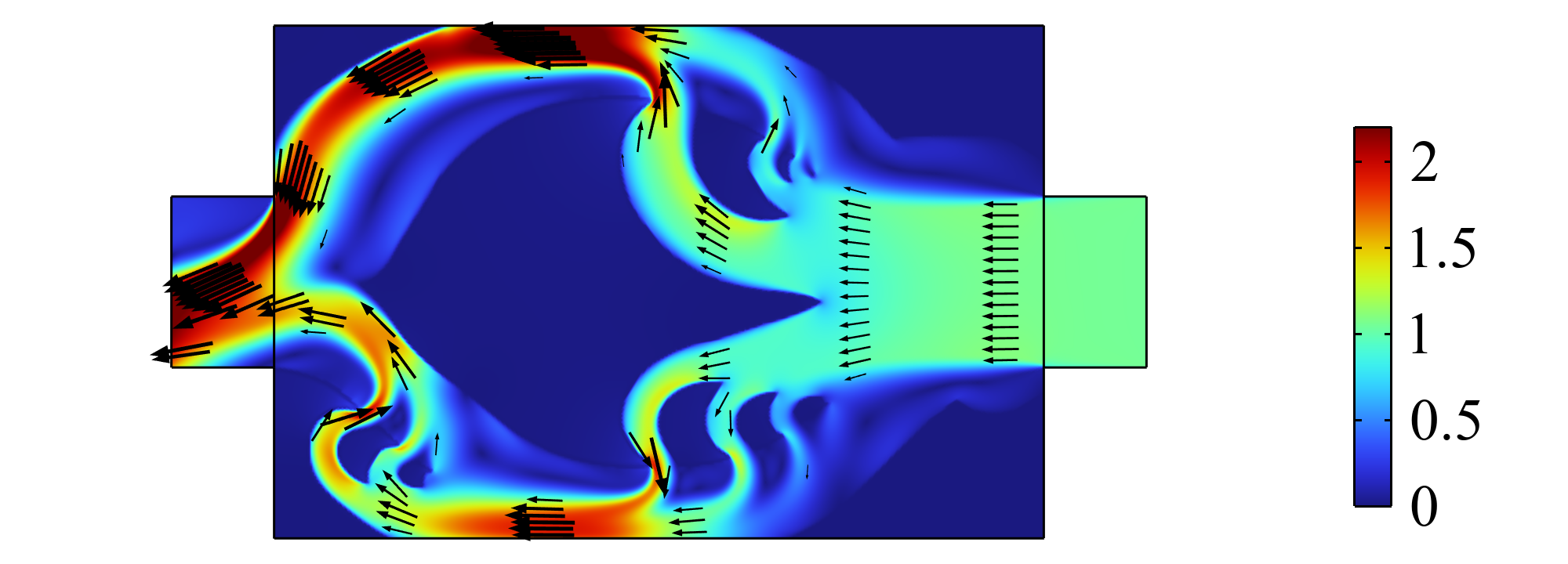}\label{subfig:iter200_v_b}} \\
        \addlinespace[2mm]

        \subfloat[]{\includegraphics[width=0.26\textwidth]{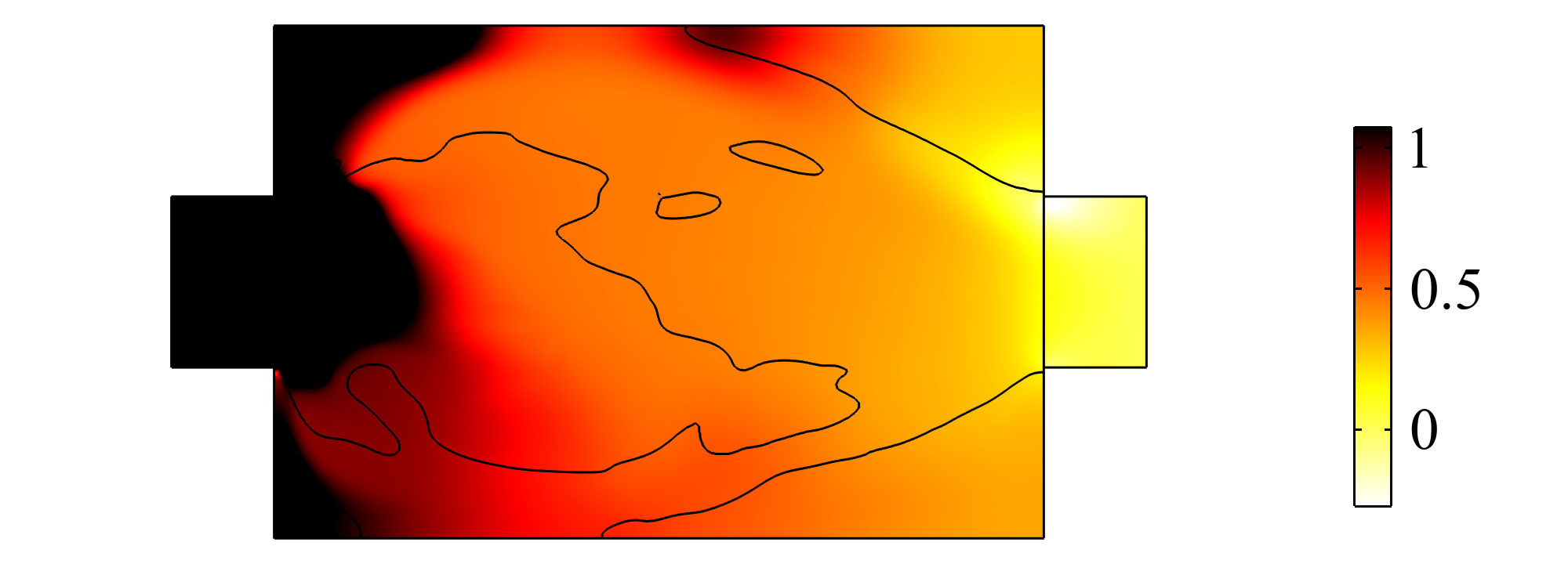}\label{subfig:iter5_p}} &
        \subfloat[]{\includegraphics[width=0.26\textwidth]{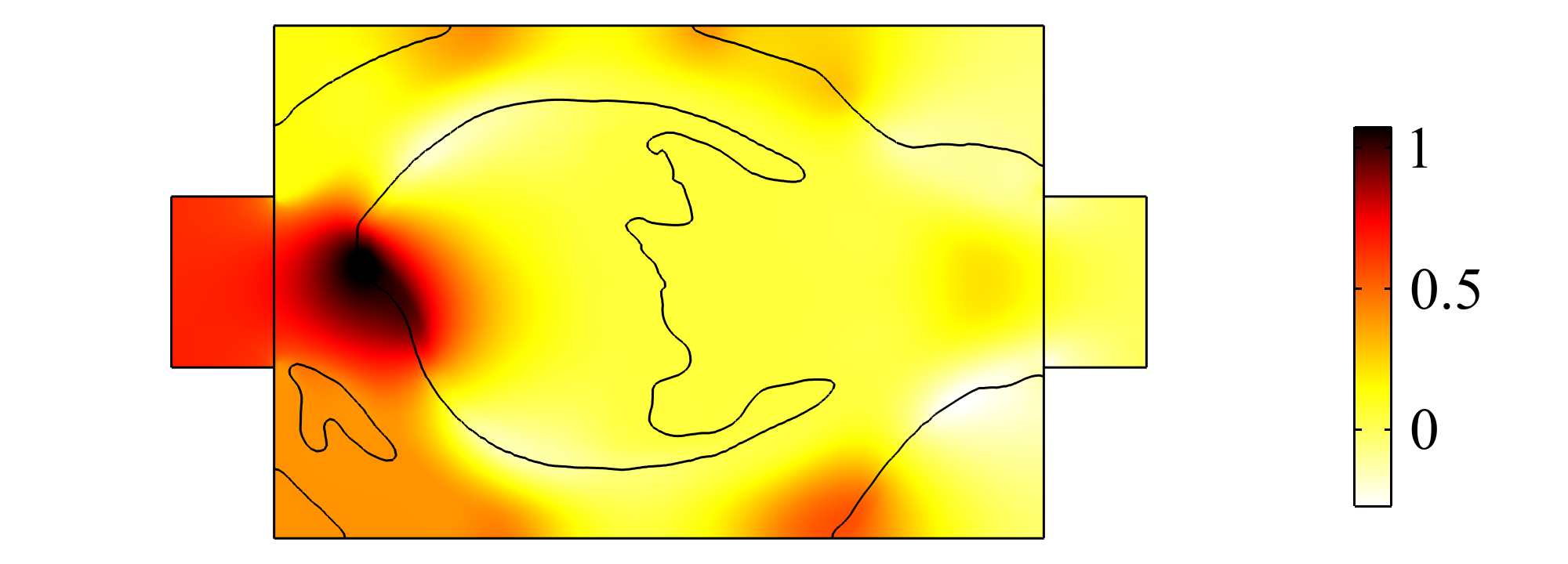}\label{subfig:iter15_p}} &
        \subfloat[]{\includegraphics[width=0.26\textwidth]{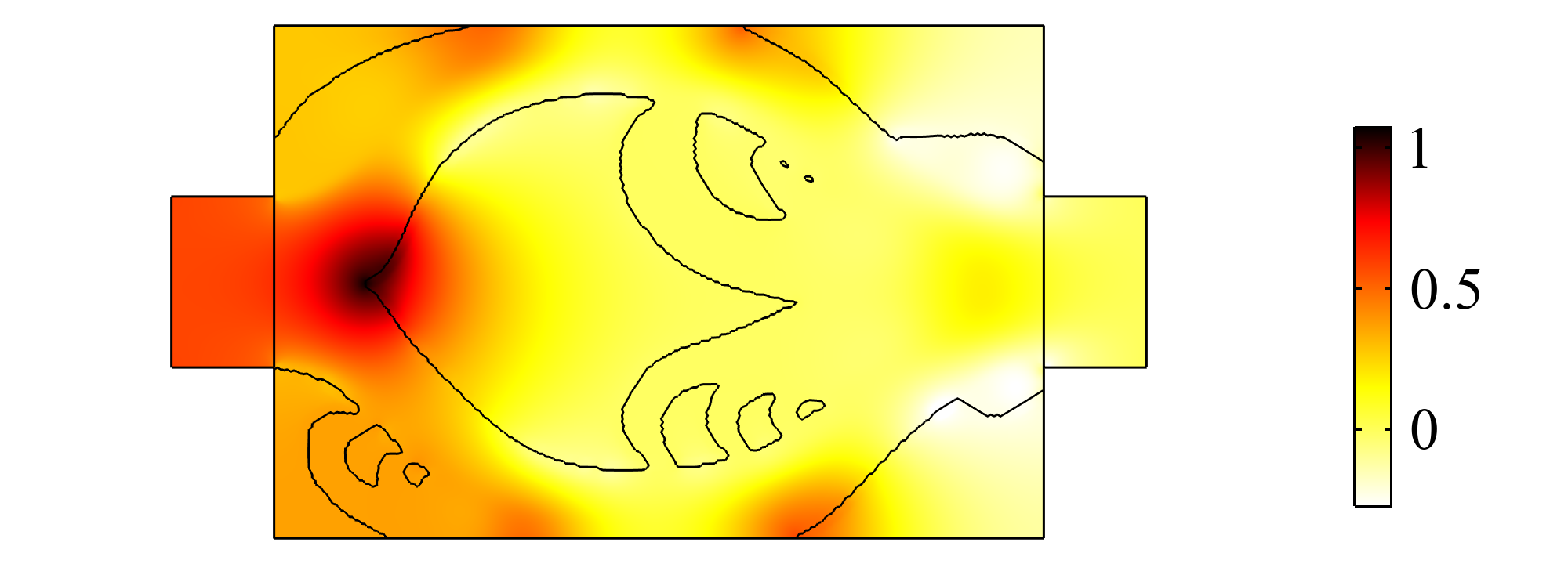}\label{subfig:iter100_p}} &
        \subfloat[]{\includegraphics[width=0.26\textwidth]{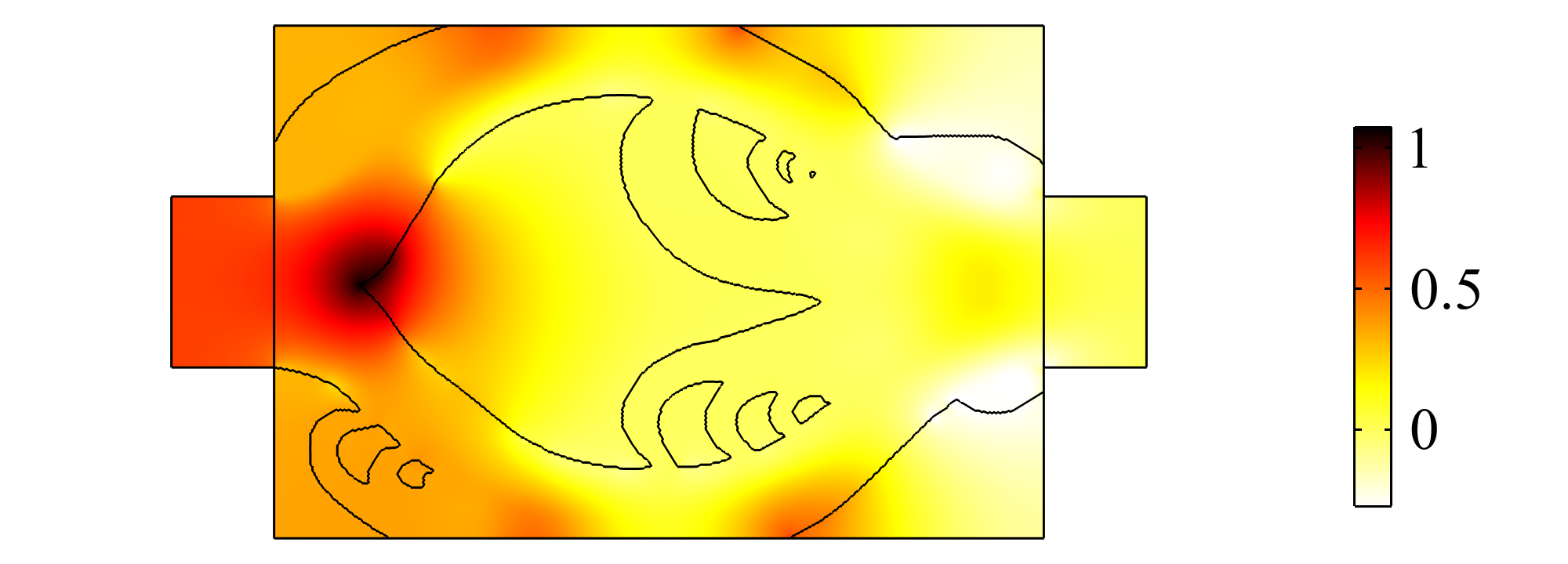}\label{subfig:iter200_p}} \\
        \addlinespace[2mm]

        \subfloat[]{\includegraphics[width=0.26\textwidth]{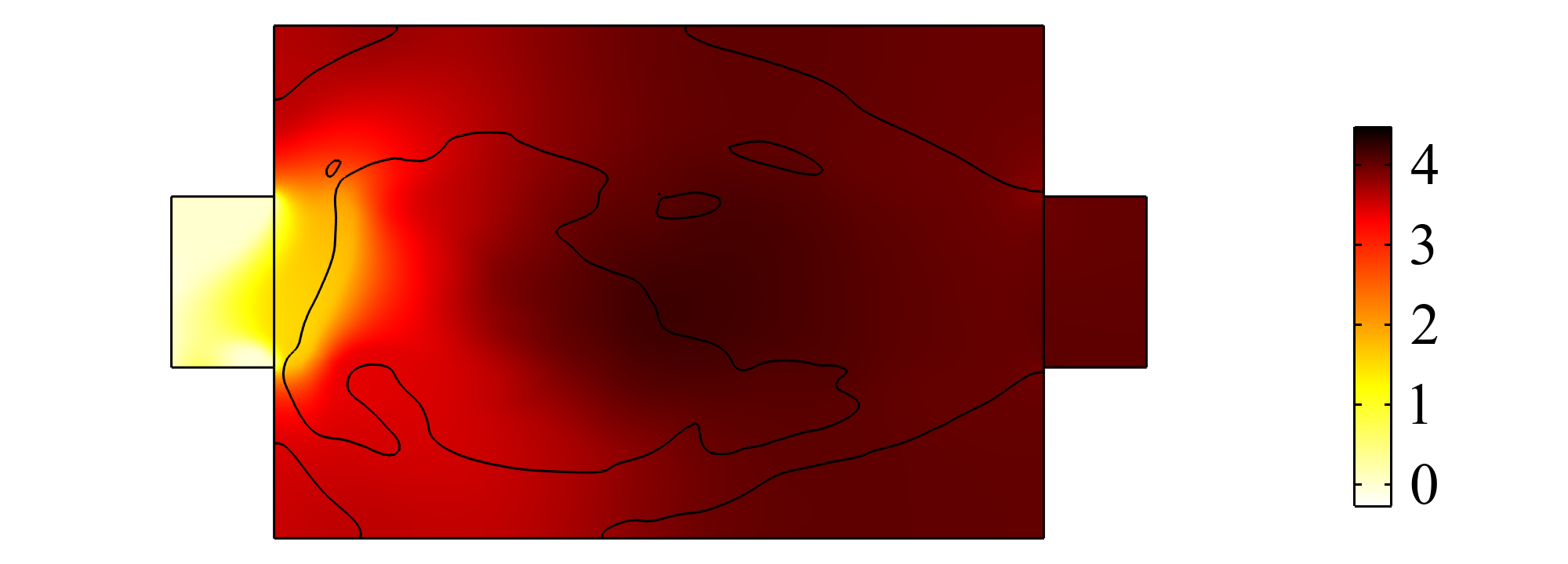}\label{subfig:iter5_p_b}} &
        \subfloat[]{\includegraphics[width=0.26\textwidth]{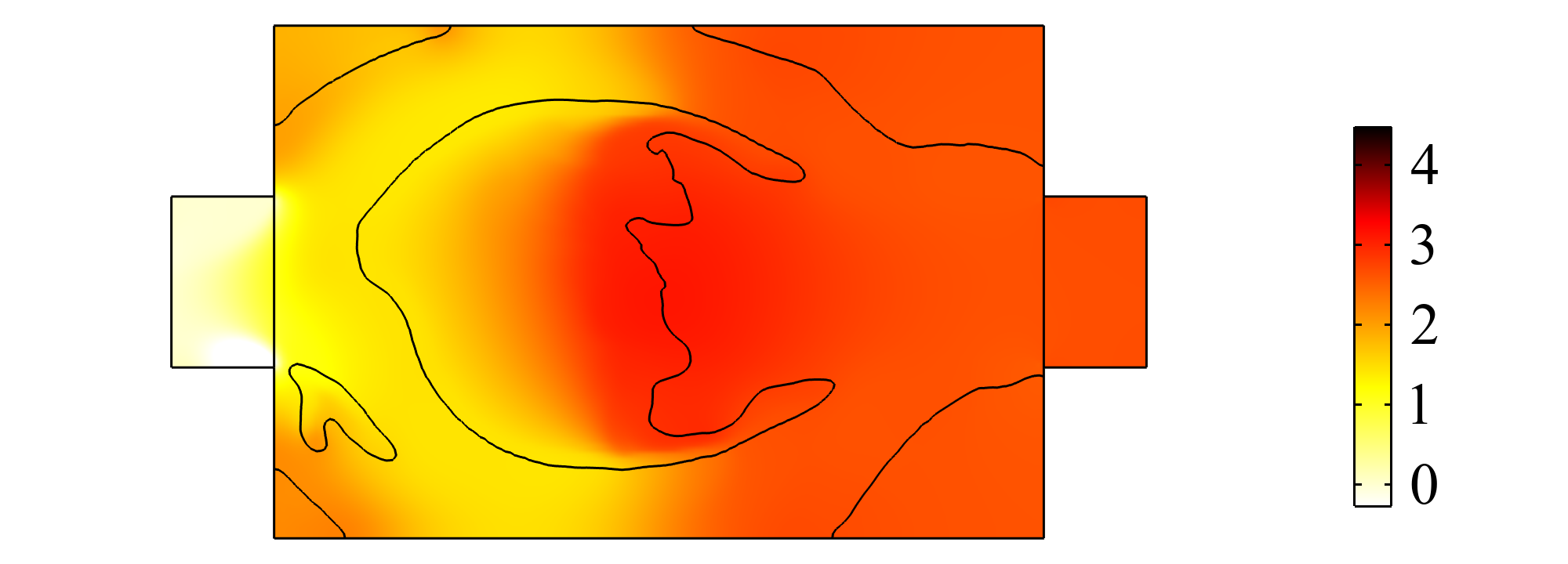}\label{subfig:iter15_p_b}} &
        \subfloat[]{\includegraphics[width=0.26\textwidth]{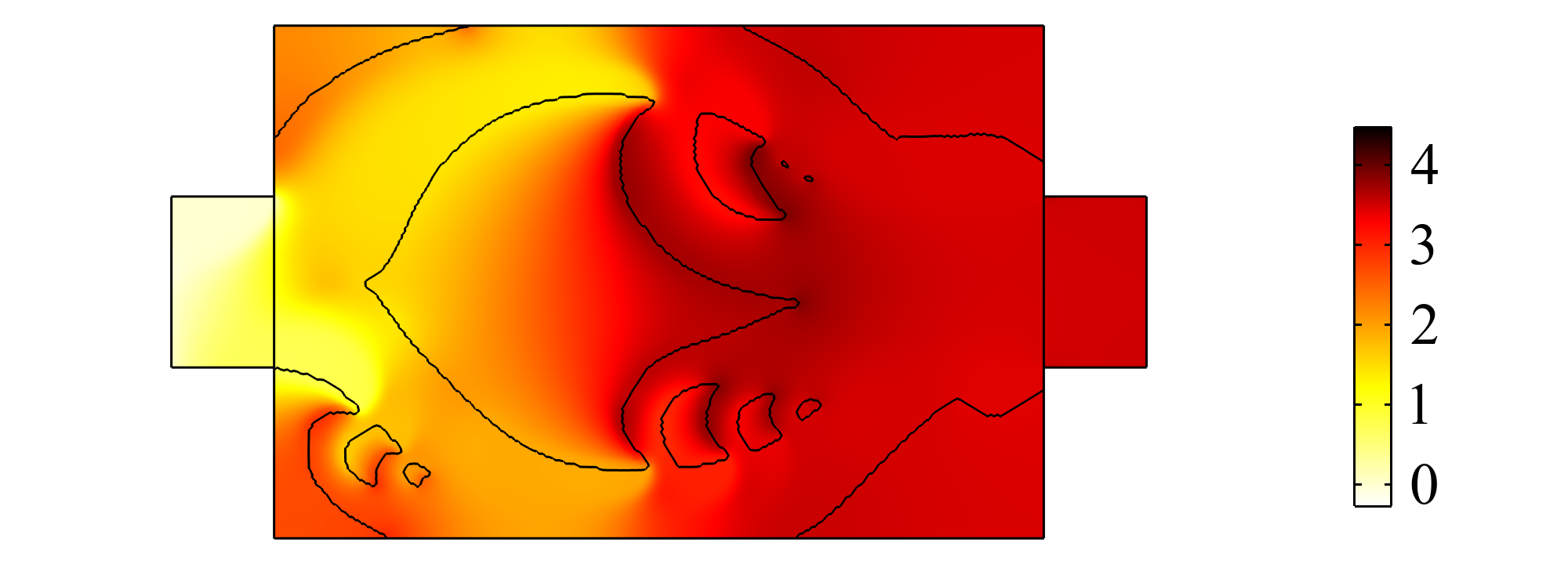}\label{subfig:iter100_p_b}} &
        \subfloat[]{\includegraphics[width=0.26\textwidth]{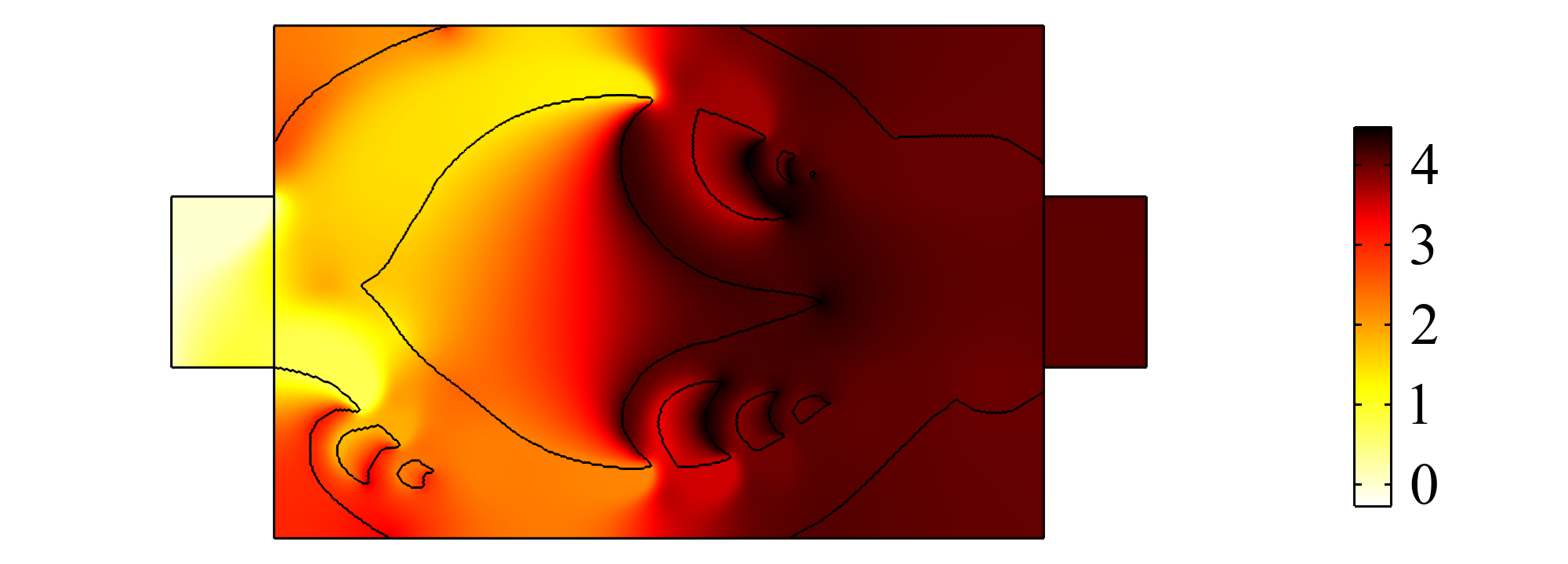}\label{subfig:iter200_p_b}} \\
        \addlinespace[2mm]

        \subfloat[]{\includegraphics[width=0.26\textwidth]{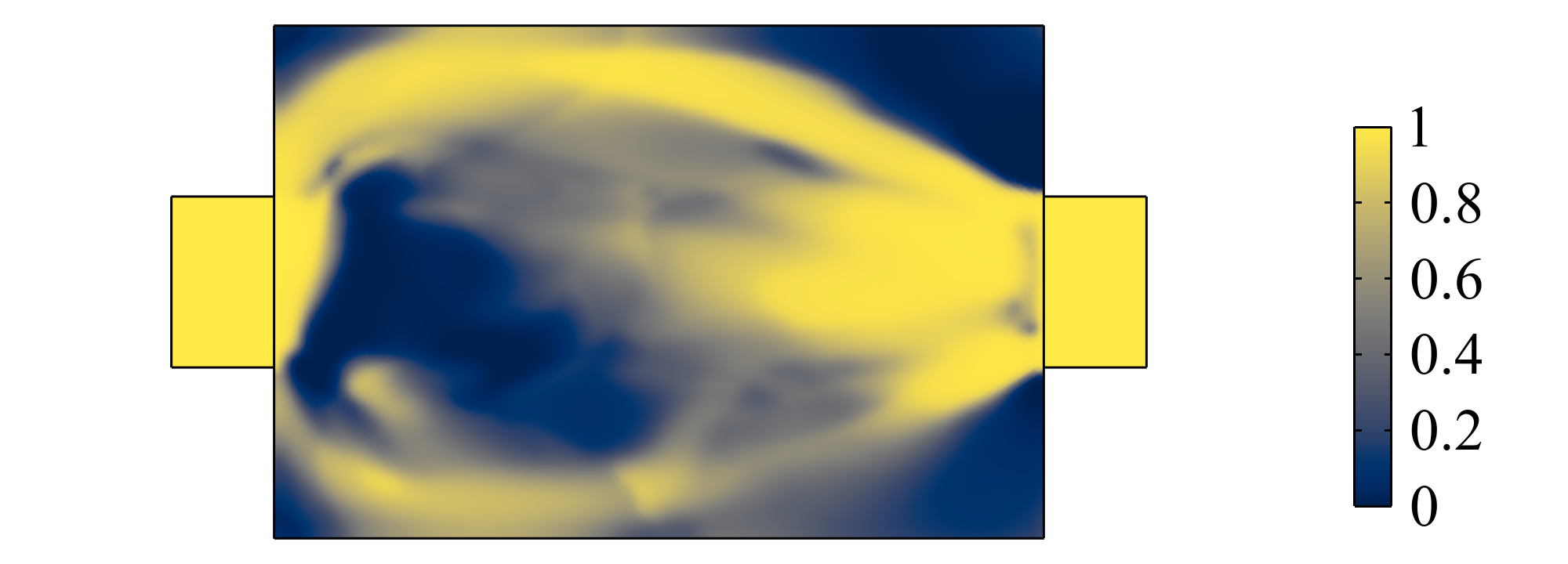}\label{subfig:iter5_dv}} &
        \subfloat[]{\includegraphics[width=0.26\textwidth]{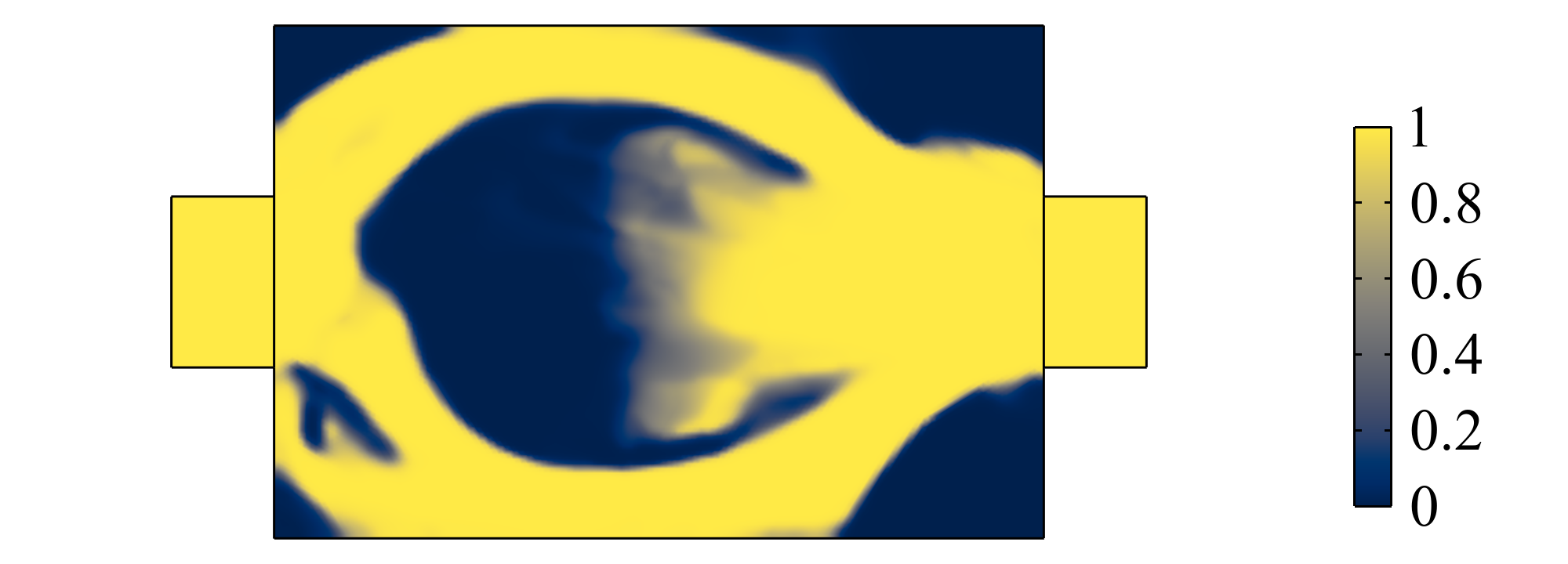}\label{subfig:iter15_dv}} &
        \subfloat[]{\includegraphics[width=0.26\textwidth]{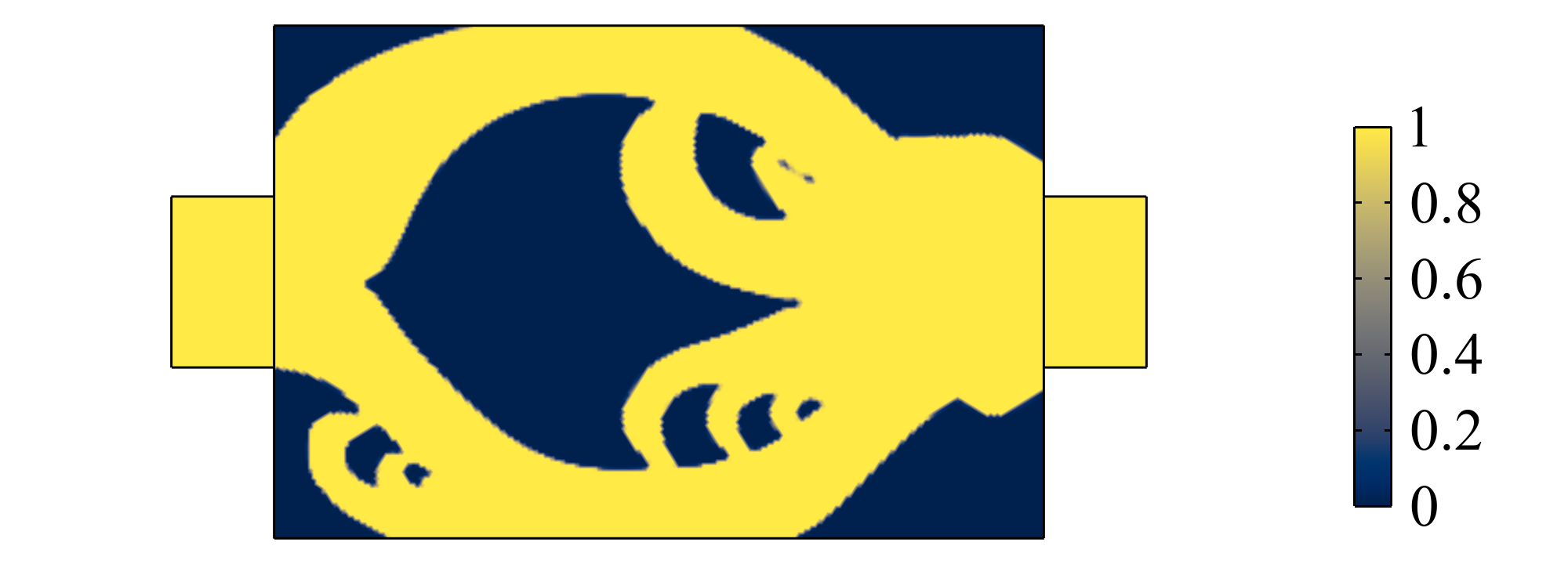}\label{subfig:iter100_dv}} &
        \subfloat[]{\includegraphics[width=0.26\textwidth]{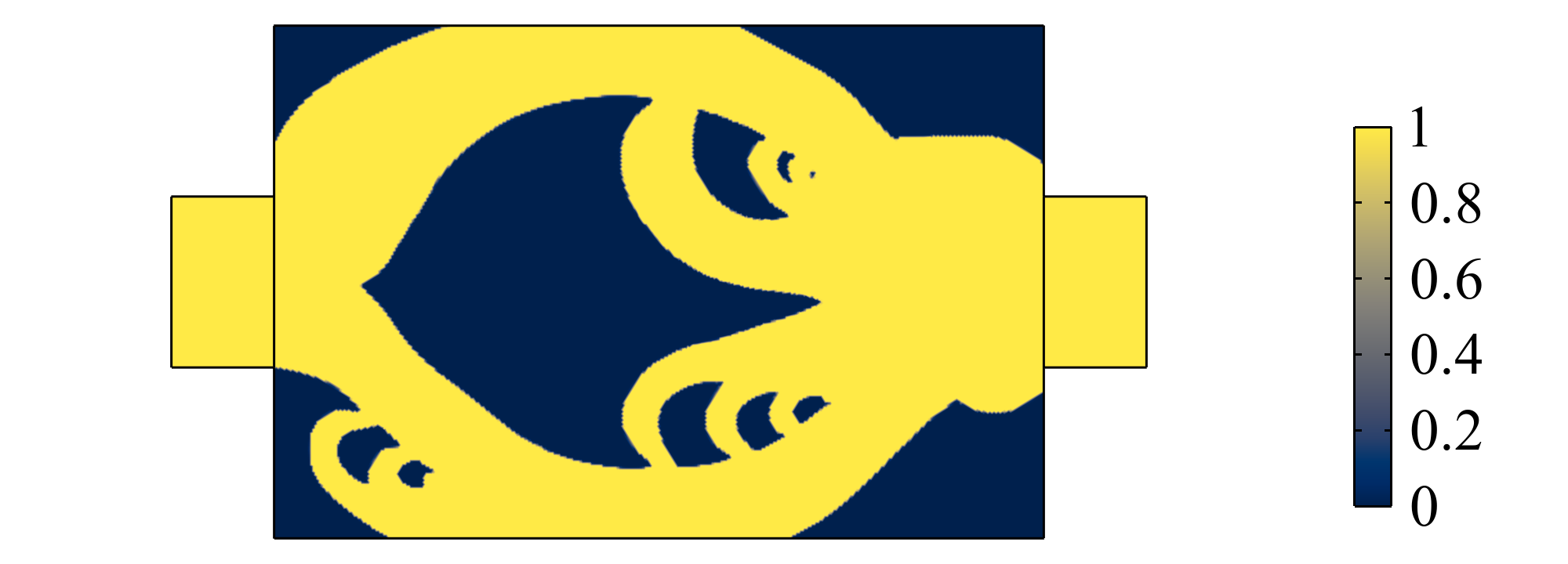}\label{subfig:iter200_dv}} \\
        \addlinespace[2mm]

        \subfloat[]{\includegraphics[width=0.26\textwidth]{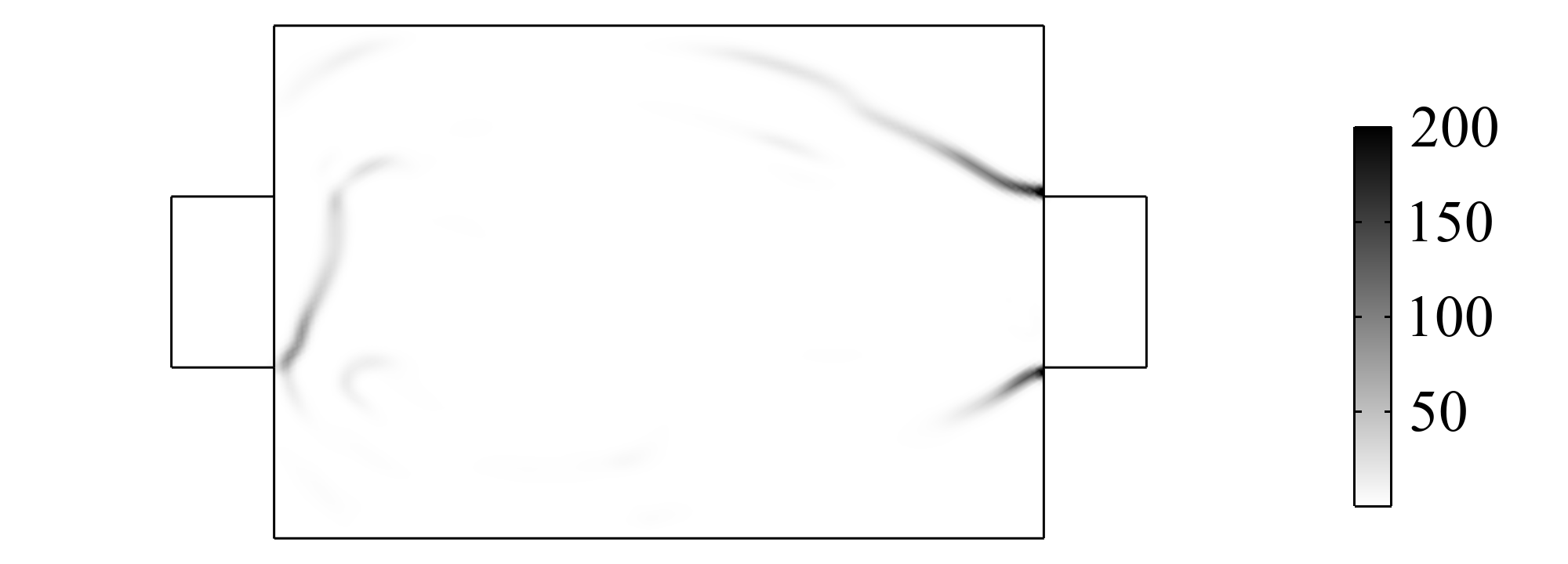}\label{subfig:iter5_w}} &
        \subfloat[]{\includegraphics[width=0.26\textwidth]{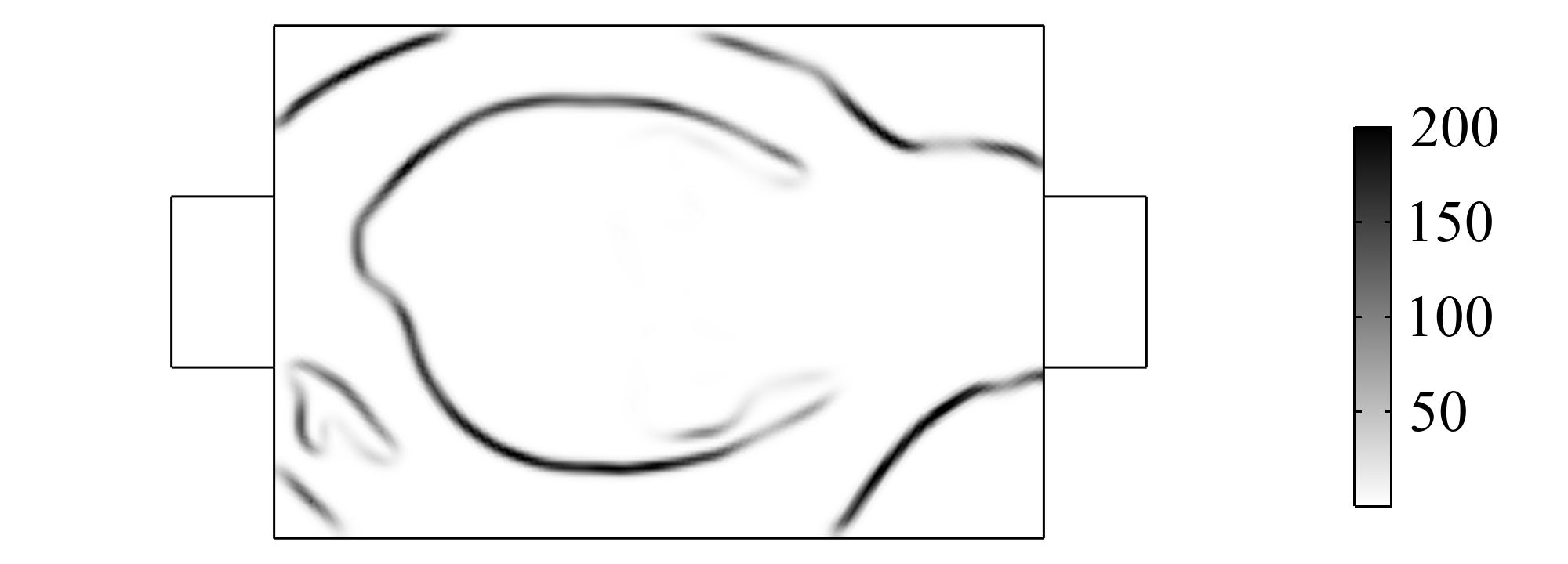}\label{subfig:iter15_w}} &
        \subfloat[]{\includegraphics[width=0.26\textwidth]{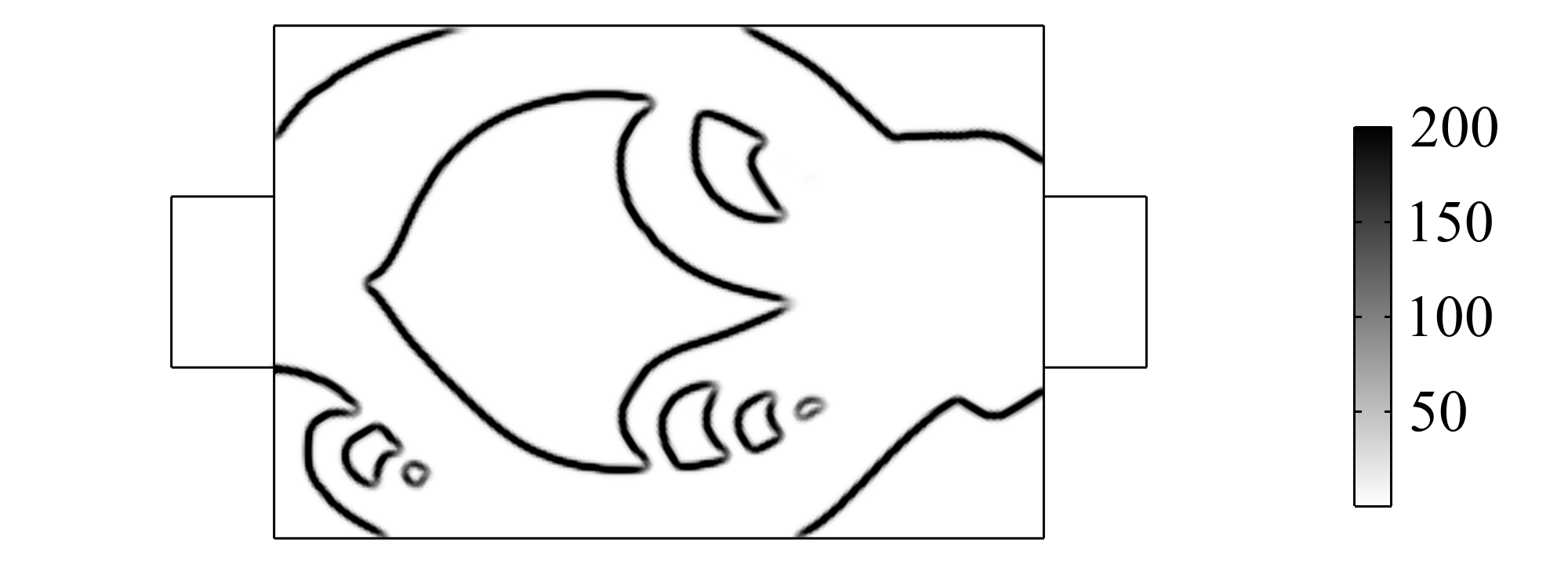}\label{subfig:iter100_w}} &
        \subfloat[]{\includegraphics[width=0.26\textwidth]{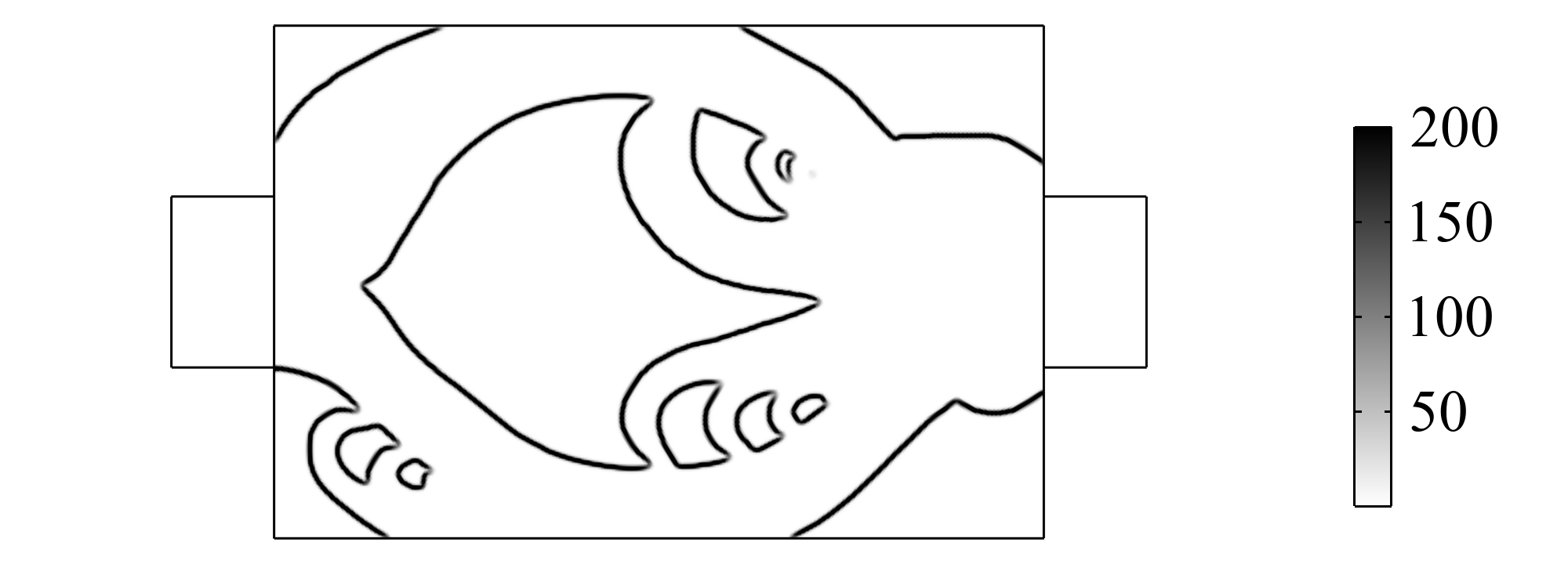}\label{subfig:iter200_w}} \\
    \end{tabular}
    \label{fig:Optimization_history_of_the_tesla_valve}
\end{figure}

The optimization history of the Tesla valve, illustrating the evolution of the design throughout the optimization process, is shown in \Cref{fig:Optimization_history_of_the_tesla_valve}. It can be seen that the overall topology slowly develops (\Cref{subfig:iter5_dv} - \ref{subfig:iter200_dv}) and that the wall-functions are gradually imposed through an equivalently developing wall intensity field (\Cref{subfig:iter5_w} - \ref{subfig:iter200_w}). As the topology develops, the velocity fields (\Cref{subfig:iter5_v} - \ref{subfig:iter200_v}) become more clear and especially the backwards flow (\Cref{subfig:iter5_v_b} - \ref{subfig:iter200_v_b}) becomes more complicated since the topology gets more complicated. For the forward flow direction (\Cref{subfig:iter200_v}), the optimized design seems to be somewhat streamlined to simply reroute the flow around the top and bottom of the main features. For the backwards flow direction (\Cref{subfig:iter200_v_b}), the optimized design has many small channels and obstacles with a curvature to change the direction of the flow, thereby inducing energy losses. This is clearly seen from the pressure fields, where for the forward flow (\Cref{subfig:iter200_p}) a single stagnation region is the main minimal driver of energy loss, but for the backward flow (\Cref{subfig:iter200_p_b}) the small individual channels cause a large energy loss.

\begin{figure}
    \centering
    \subfloat[Forward flow: Implicit wall]
    {\includegraphics[width=0.4\textwidth]{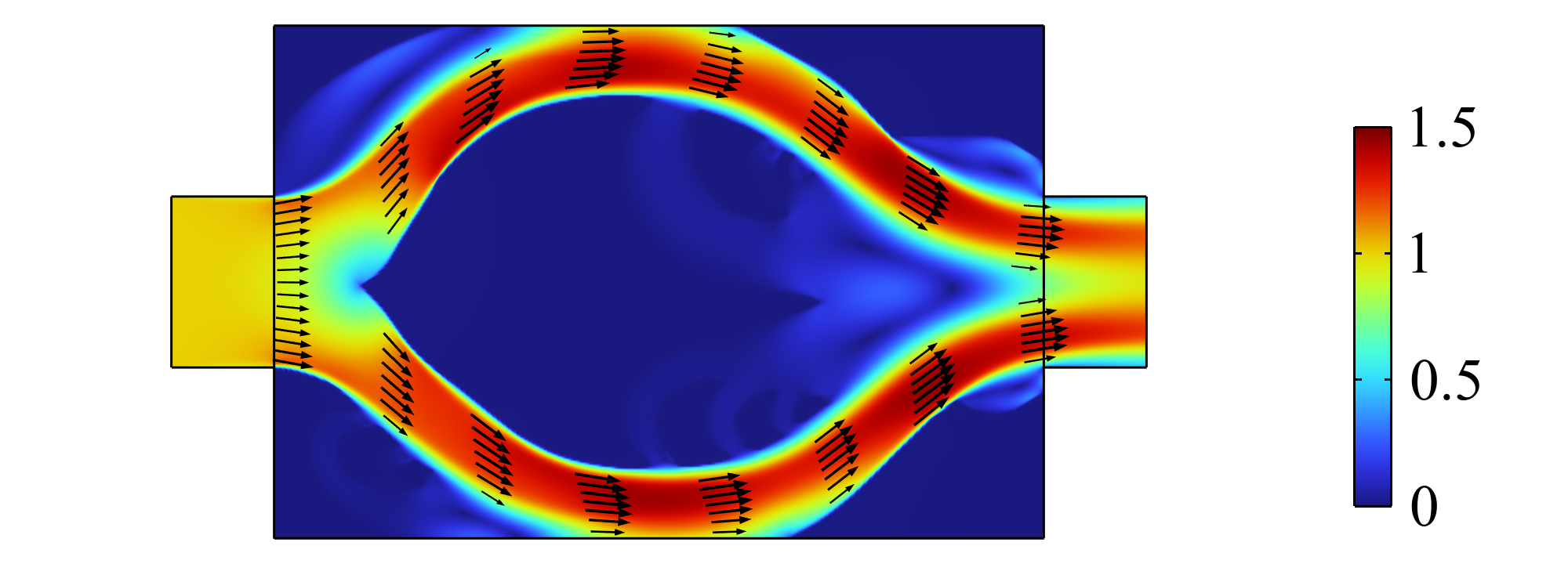}\label{subfig:manuscript_Springer/Tesla_velocity_it_200}}
    \hspace{0.05\textwidth}
    \subfloat[Forward flow: Explicit wall]
    {\includegraphics[width=0.4\textwidth]{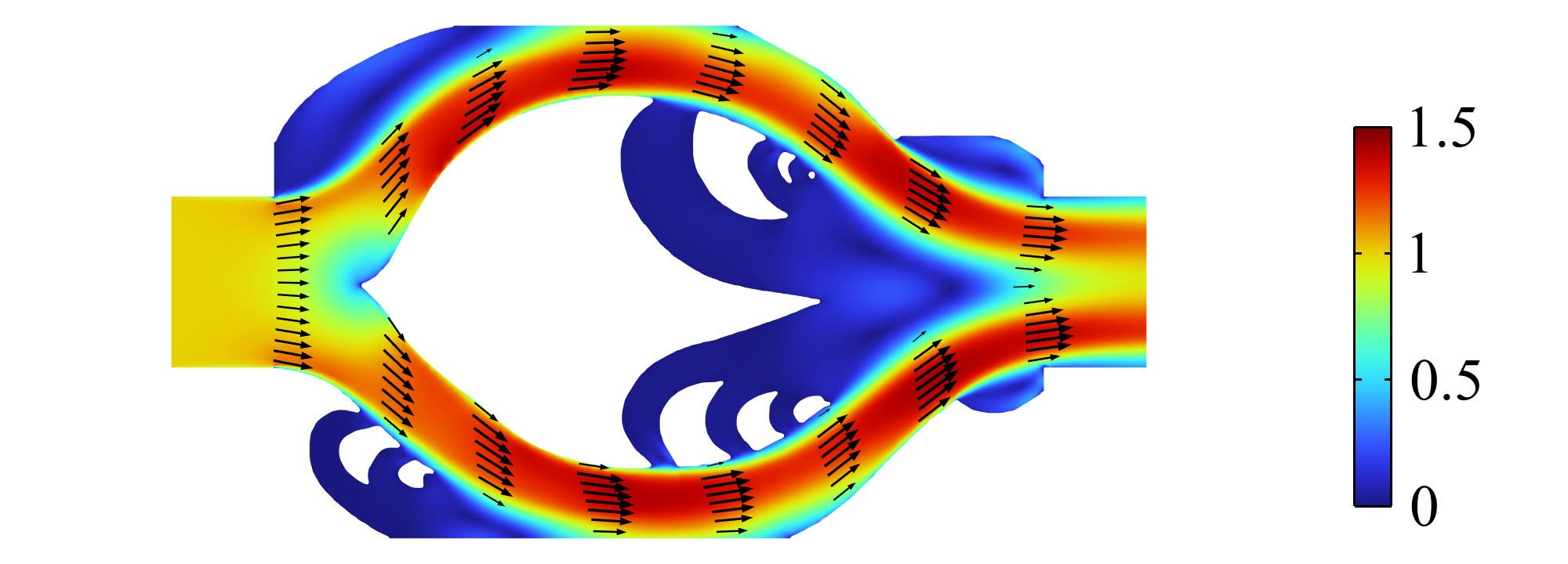}\label{subfig:Tesla_backward_exvelo}} \\

    \subfloat[Backward flow: Implicit wall]
    {\includegraphics[width=0.4\textwidth]{Tesla_velocity_backward_it_200.png}\label{subfig:Tesla_velocity_backward_it_200}}
    \hspace{0.05\textwidth}
    \subfloat[Backward flow: Explicit wall]
    {\includegraphics[width=0.4\textwidth]{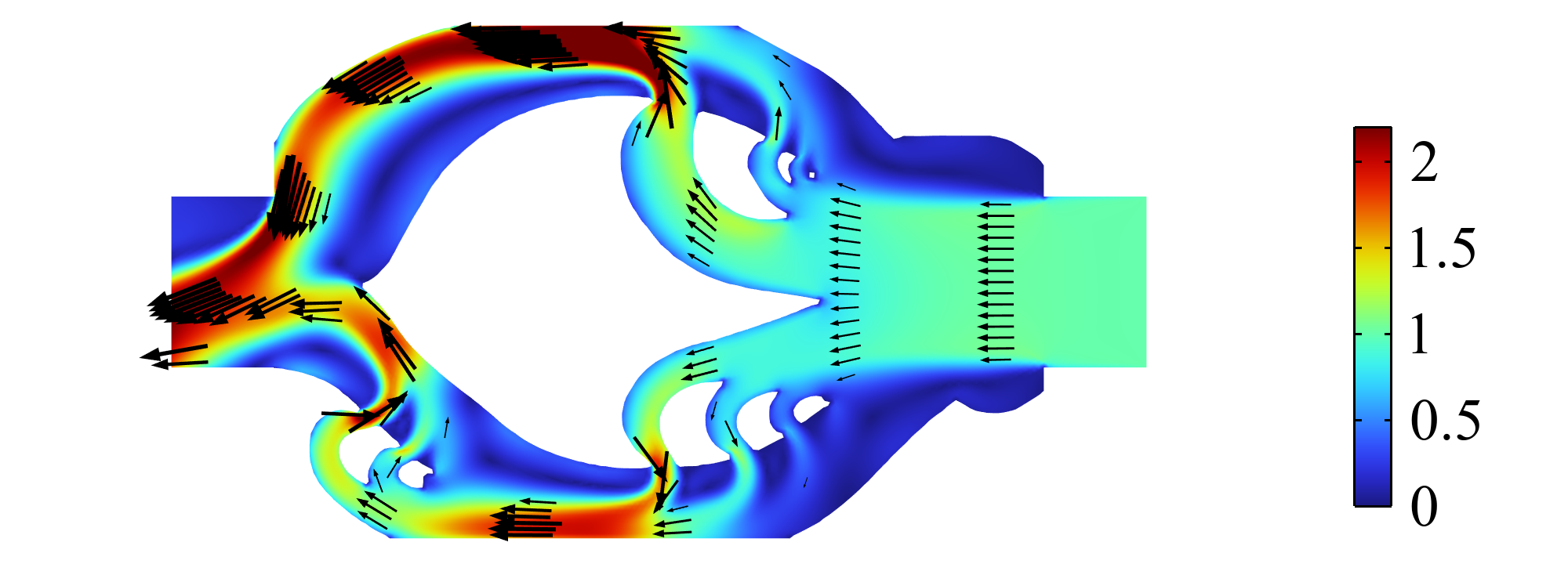}\label{subfig:Tesla_pressure_exvelo}} \\

    \subfloat[Forward turbulent viscosity $\nu_t$ : Implicit wall]
    {\includegraphics[width=0.4\textwidth]{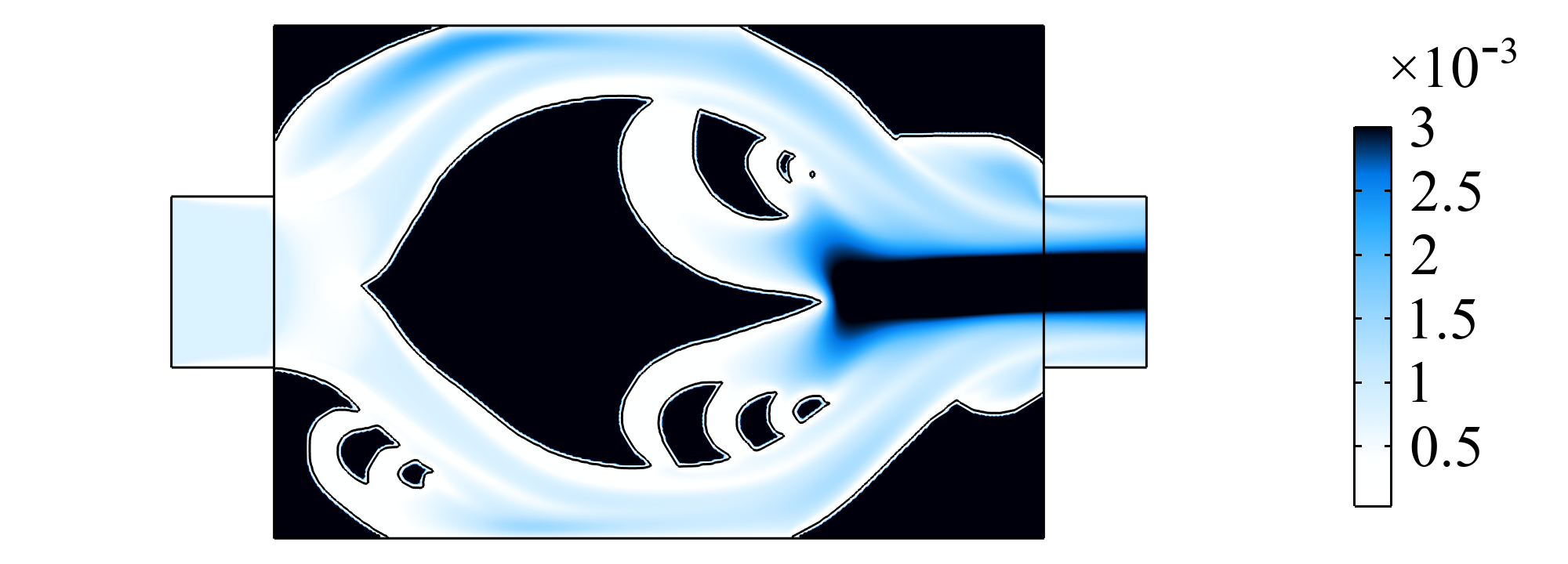}\label{subfig:forward_imp}}
    \hspace{0.05\textwidth}
    \subfloat[Forward turbulent viscosity $\nu_t$ : Explicit wall]
    {\includegraphics[width=0.4\textwidth]{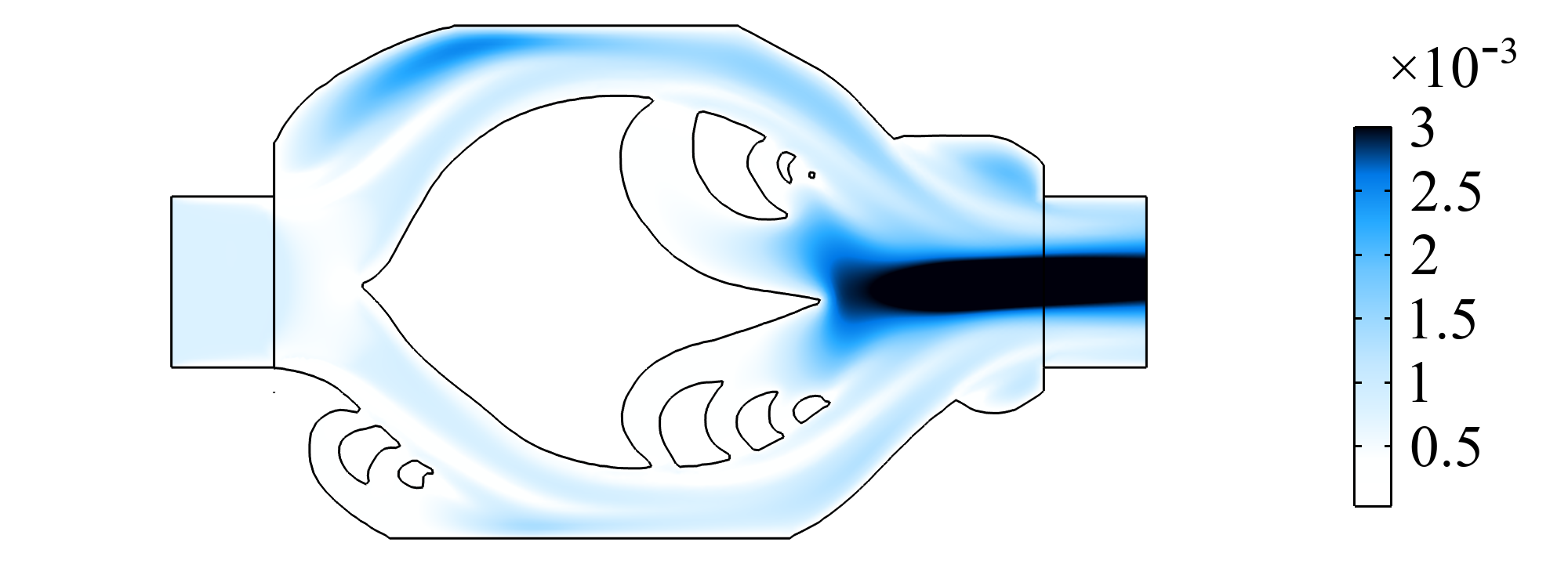}\label{subfig:forward_exp}} \\
    
    \subfloat[Backward turbulent viscosity $\nu_t$ : Implicit wall]
    {\includegraphics[width=0.4\textwidth]{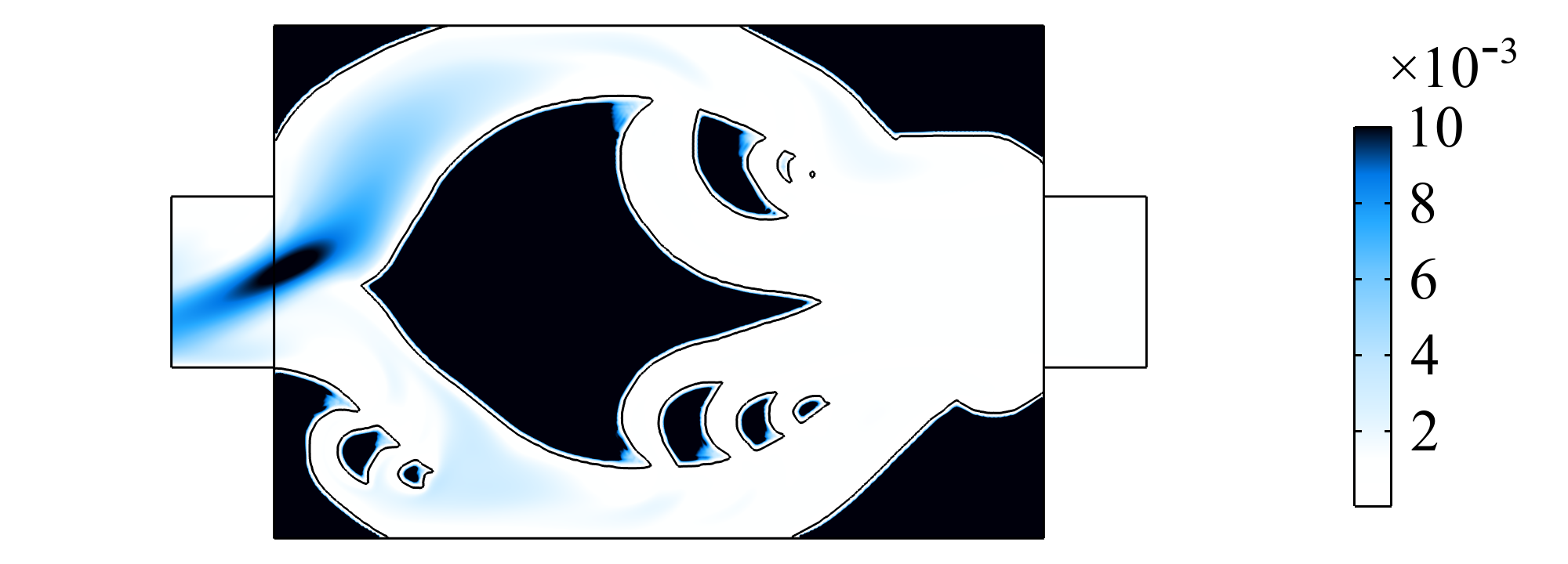}\label{subfig:backward_imp}}
    \hspace{0.05\textwidth}
    \subfloat[Backward turbulent viscosity $\nu_t$ : Explicit wall]
    {\includegraphics[width=0.4\textwidth]{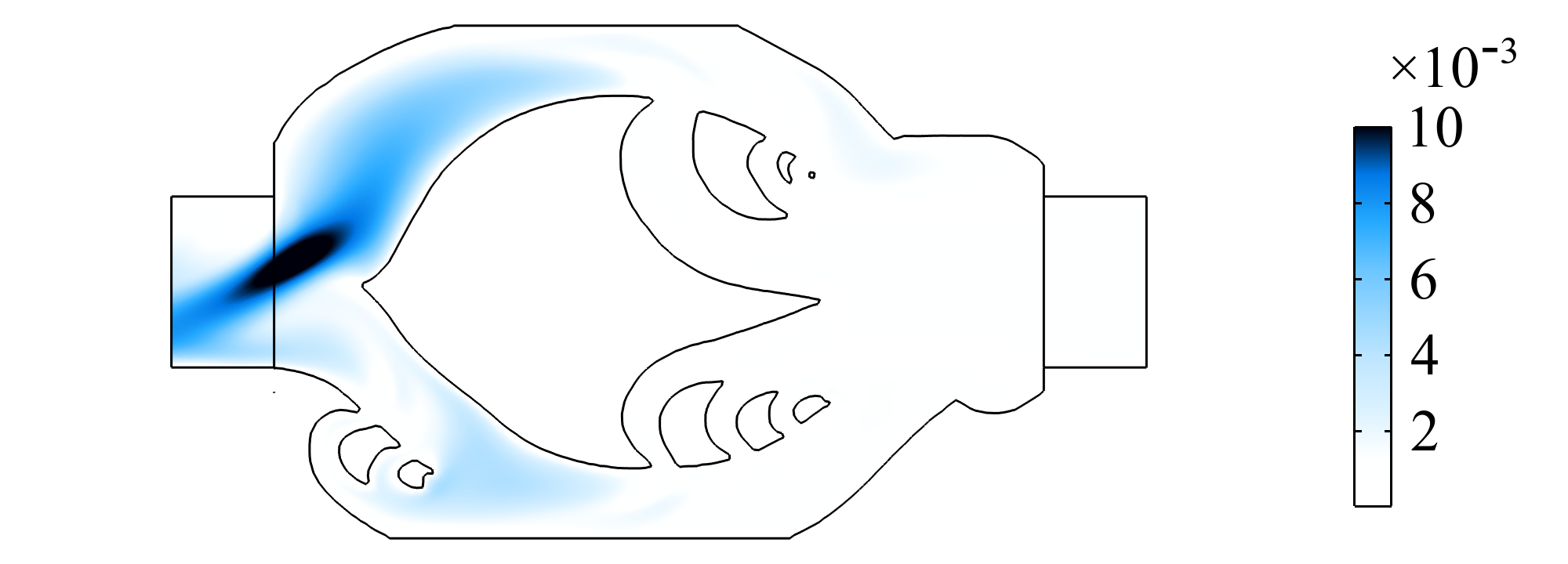}\label{subfig:backward_exp}}

    \subfloat[Forward pressure profile: Implicit wall]
    {\includegraphics[width=0.4\textwidth]{Tesla_pressure_forward_it_200.png}\label{subfig:Tesla_pressure_forward_it_200}}
    \hspace{0.05\textwidth}
    \subfloat[Forward pressure profile: Explicit wall]
    {\includegraphics[width=0.4\textwidth]{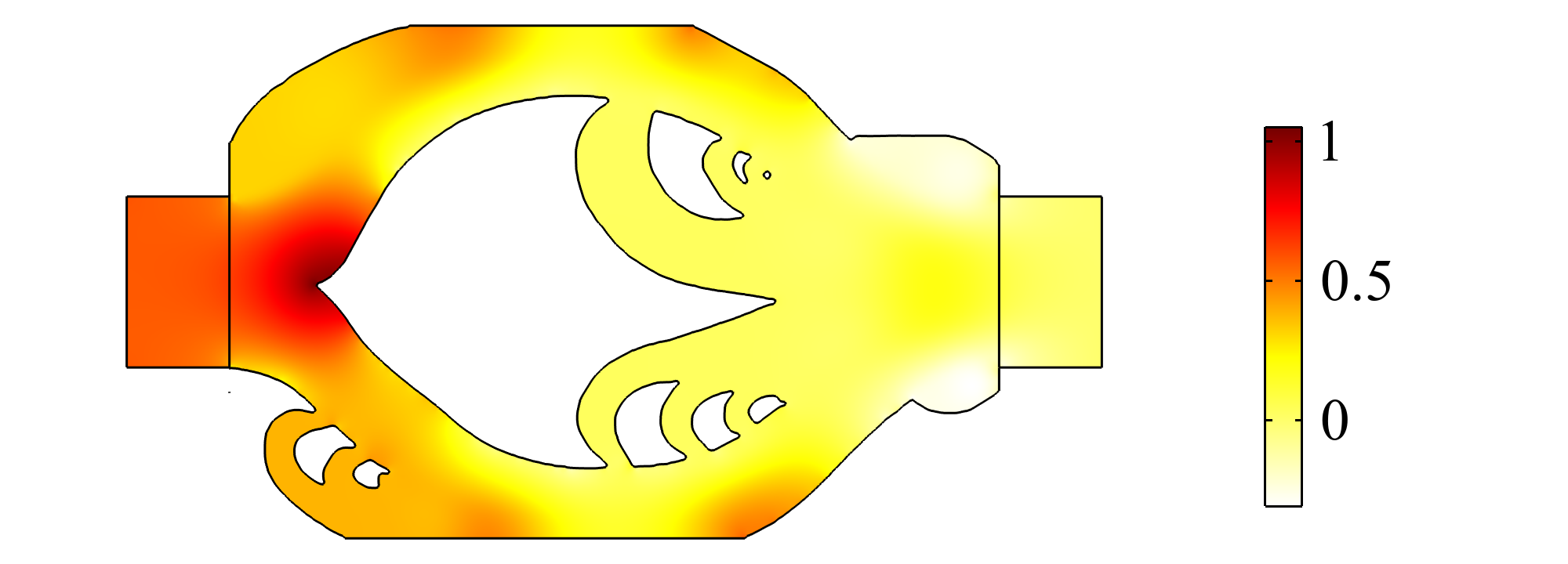}\label{subfig:Tesla_expressure_forward}} \\
    
    \subfloat[Backward pressure profile: Implicit wall]
    {\includegraphics[width=0.4\textwidth]{Tesla_pressure_backward_it_200.png}\label{subfig:Tesla_pressure_backward_it_200}}
    \hspace{0.05\textwidth}
    \subfloat[Backward pressure profile: Explicit wall]
    {\includegraphics[width=0.4\textwidth]{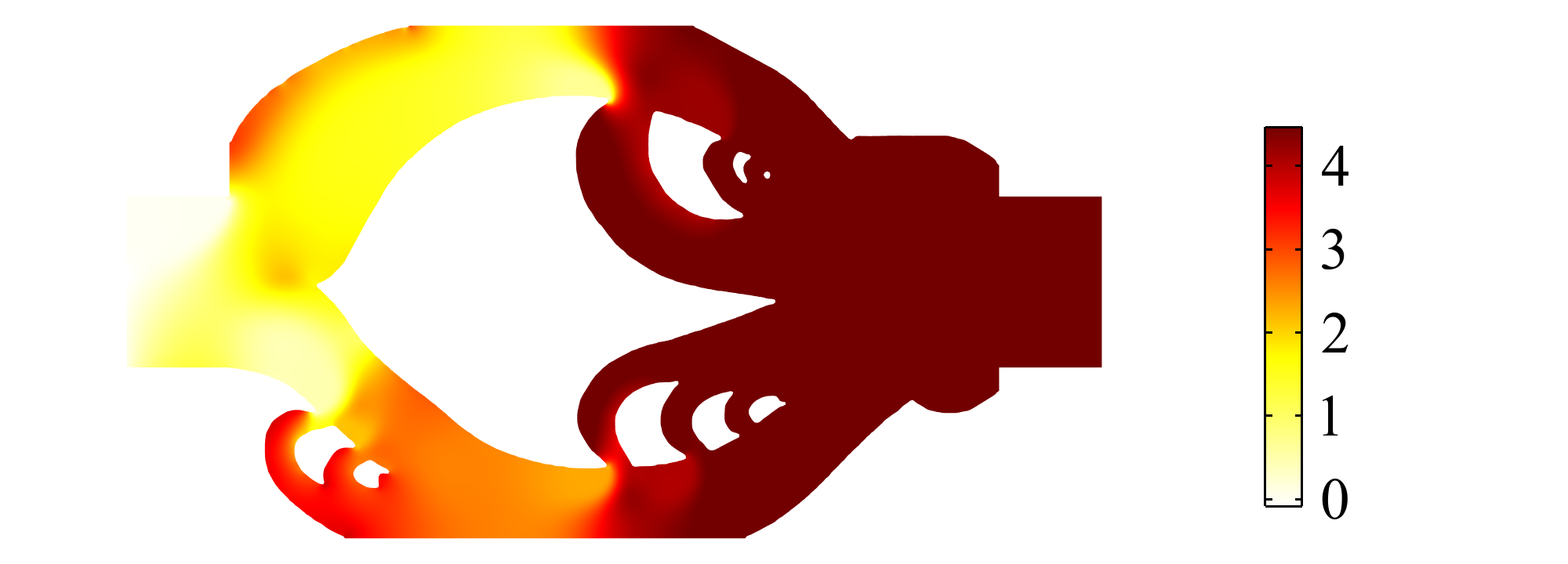}\label{subfig:Tesla_expressure_backward}}
    
    \caption{Optimized Tesla valve features obtained using the proposed implicit wall-function method (left) and corresponding results from the body-fitted re-simulation with explicit wall-functions (right) for the Tesla valve benchmark (Example~\#8).}
    \label{fig:explicit_vs_implicit}
\end{figure}

As shown in \Cref{fig:explicit_vs_implicit}, the forward and backward velocity, turbulent viscosity ($\nu_T$), and pressure fields obtained from the proposed implicit wall-function method are compared with the corresponding results from the re-simulation using a body-fitted mesh with explicit wall-functions. The figure demonstrates that even for complex topologies, the final optimized fields with the proposed implicit wall-function method closely match those obtained from the body-fitted geometry simulation with explicit wall-functions, highlighting both the accuracy of the proposed method and the consistency of the chosen parameters employed in the present study. 
In contrast to the first two benchmarks, the observed difference in pressure drop, between the optimization and body-fitted models, is significantly smaller. For the forward flow direction, the optimization model predicts 0.5962 and the body-fitted model predicts 0.5710. For the backward flow direction, the optimization model predicts 4.0812 and the body-fitted model predicts 4.4473. The smaller discrepancy is most likely due to the higher final $\beta$ value of the continuation scheme (\Cref{tab:Tesla_valve_sweep_parameters}).

\begin{figure}
    \centering
    \subfloat[Forward flow]{
        \includegraphics[width=0.4\textwidth]{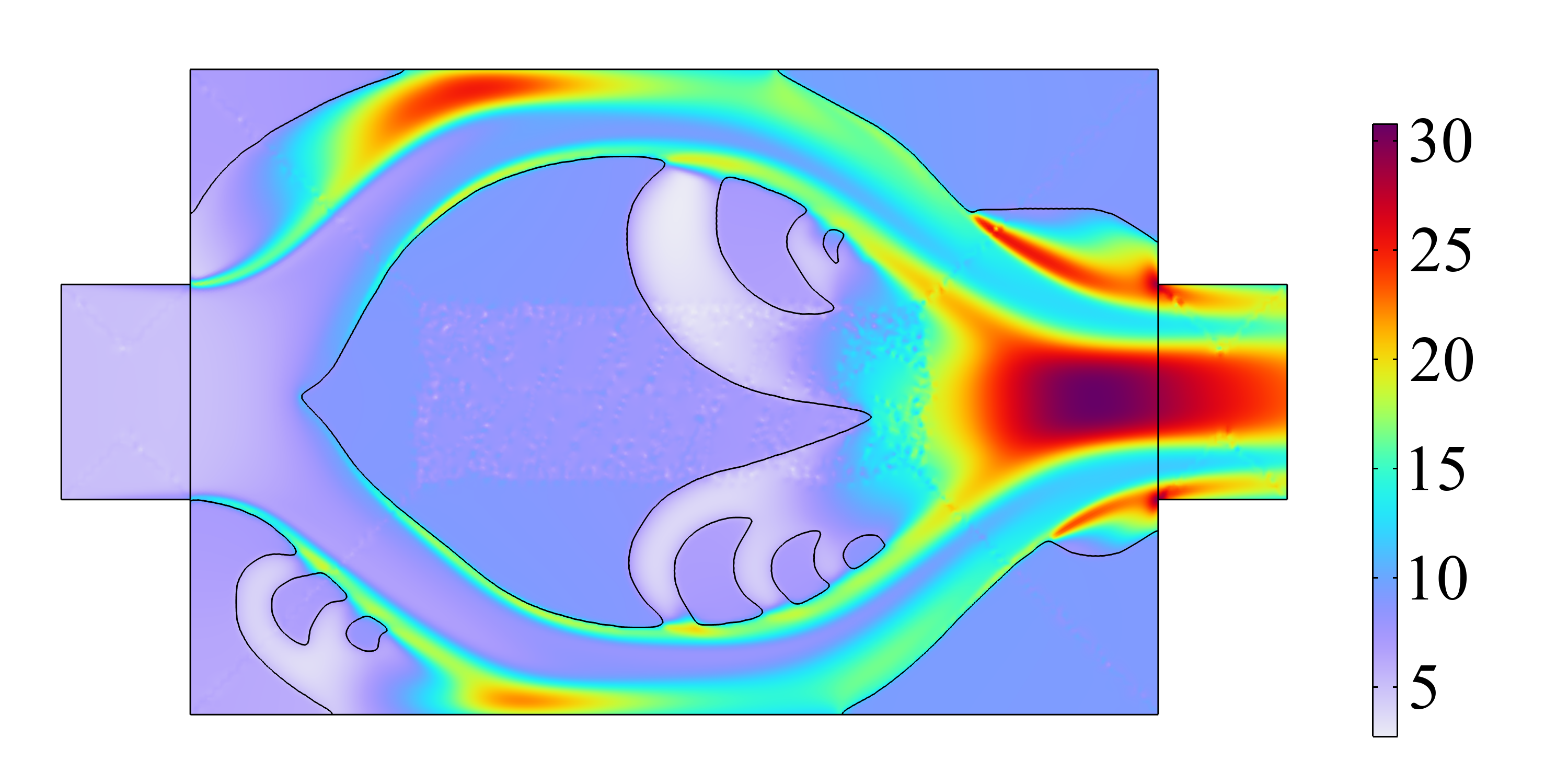}
        \label{subfig:yplus_forward}
    }\hfill
    \subfloat[Backward flow]{
        \includegraphics[width=0.4\textwidth]{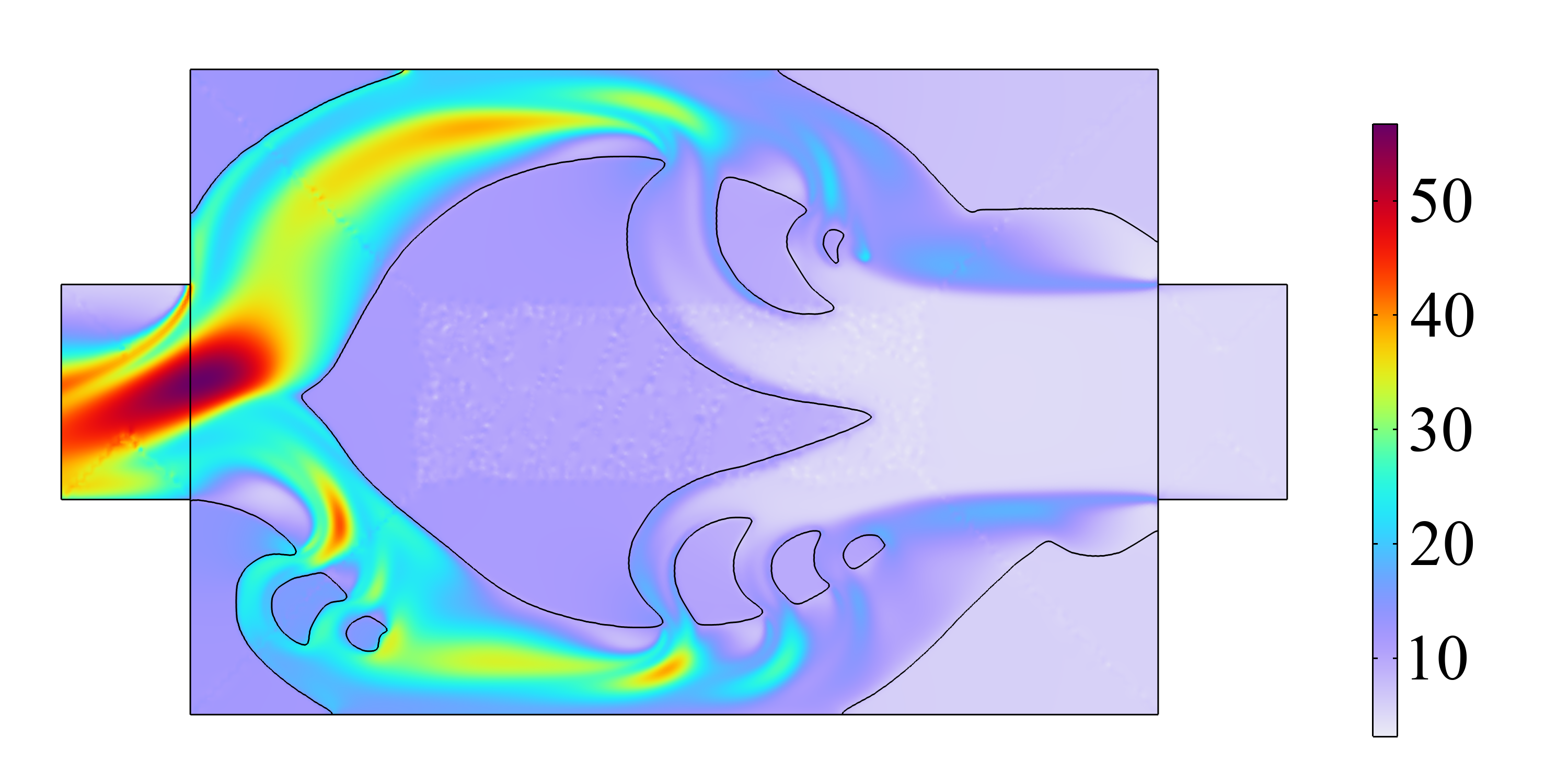}
        \label{subfig:yplus_backward}
    }

    \caption{Implicit $y^{+}$ distributions for the optimized geometry of Tesla valve~(Example \#8) under forward and backward flow conditions, obtained using the proposed implicit wall-function method. In the vicinity of the walls, $y^{+}$ remains within the logarithmic law region.}
    \label{fig:yplus_forward_backward}
\end{figure}

\Cref{fig:yplus_forward_backward}, shows the implicitly defined $y^+$ for the Tesla valve benchmark inside the domain and for forward and backward flow directions. It can be seen that the $y^+$ values remains in the logarithmic sub-domain as assumed by the wall-functions.

\subsubsection{Tesla valve in different scenarios}
\begin{figure}
    \centering
    \subfloat[Proposed implicit wall-function method optimized geometry\label{subfig:tes_pic1}]{
        \includegraphics[width=0.4\textwidth]{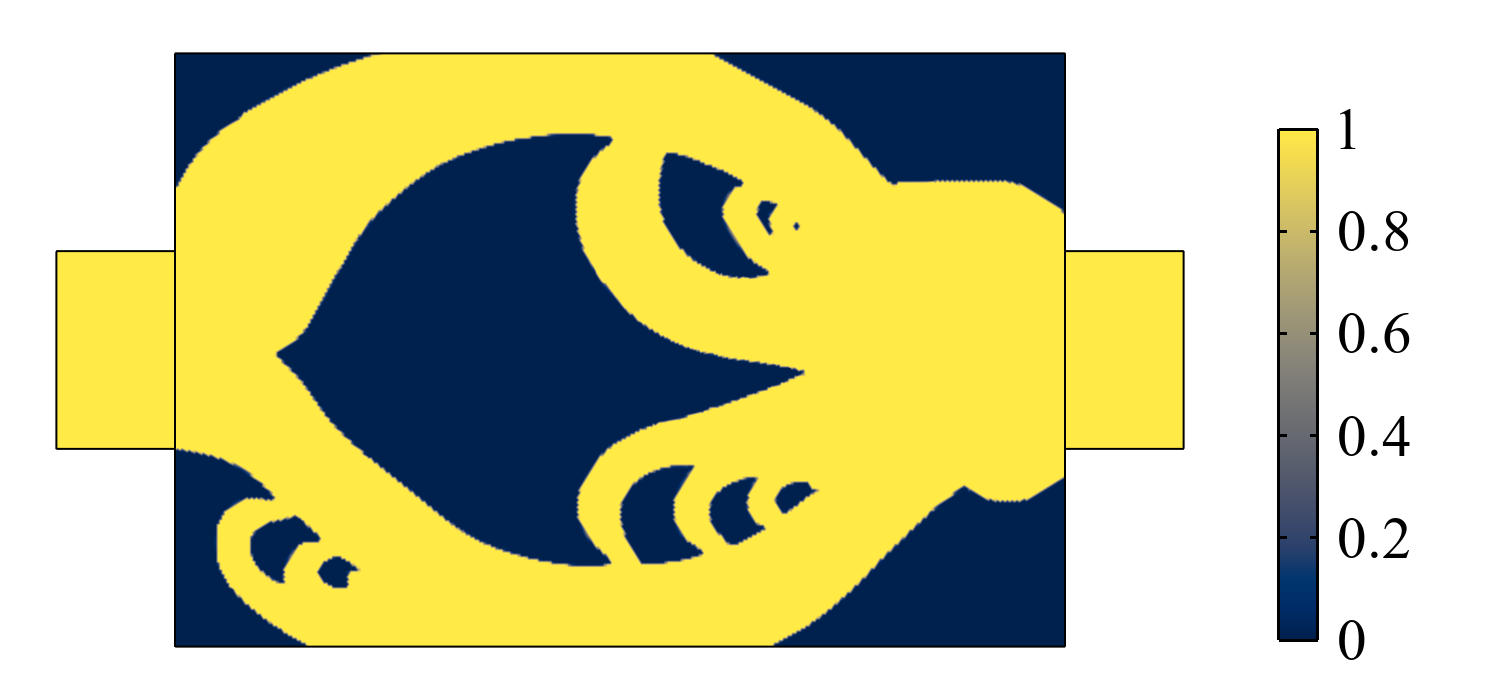}
    }\\[1ex] 

    \subfloat[Laminar free-slip optimized geometry\label{subfig:tes_pic2}]{
        \includegraphics[width=0.4\textwidth]{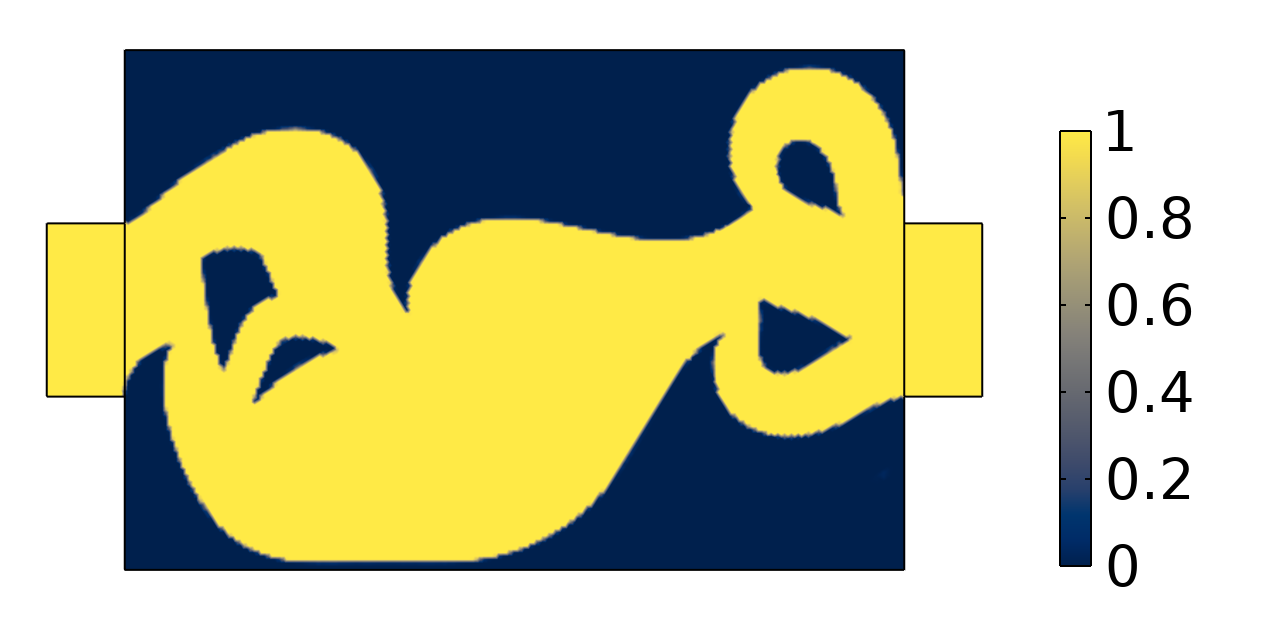}
    }\\[1ex]

    \subfloat[Laminar no-slip optimized geometry\label{subfig:tes_pic3}]{
        \includegraphics[width=0.4\textwidth]{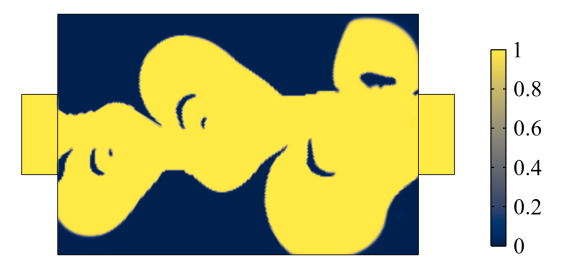}
    }

    \caption{Final topology-optimized geometries of the Tesla valve under three different flow assumptions and boundary conditions: 
    (a) turbulent flow with the proposed implicit wall-function method at $\operatorname{Re}_{\text{inlet}}=5000$, 
    (b) laminar flow with a free-slip boundary condition $\operatorname{Re}_{\text{inlet}}=33$, and 
    (c) laminar flow with a no-slip boundary condition $\operatorname{Re}_{\text{inlet}}=33$.}
    \label{fig:three_pics}
\end{figure}

The Tesla valve operates across flow regimes where either viscous forces (low Reynolds number) or inertia (high Reynolds number) dominate. Because the boundary-layer treatment in the Tesla valve strongly influences the topology produced by the optimization, we investigate this effect by performing optimizations in different regimes (both laminar and turbulent) and, in the laminar case, with different wall boundary conditions (both no-slip and free-slip). In total, we optimize the geometry in three representative cases: (a) a fully turbulent setting ($\text{Re}_{inlet} = 5000$) using the proposed implicit wall-function method; (b) a laminar setting ($\text{Re}_{inlet} = 33$) with free-slip walls~\cite{bayat2025density}; and (c) a laminar setting ($\text{Re}_{inlet} = 33$) with no-slip walls. The corresponding optimized geometries are shown in \Cref{fig:three_pics}.

In the turbulent case (\Cref{subfig:tes_pic1}), energy losses due to inertial forces and turbulent variables are captured. The optimizer therefore favors configurations that guide the forward flow along relatively direct paths with only smooth turns, while recirculation-promoting features appear and penalize the reverse direction. The resulting design exhibits locally constricted throats that induce form losses primarily in backward operation, achieving strong rectification with only a modest impact on forward resistance.

The laminar no-slip case (\Cref{subfig:tes_pic3}) is dominated by viscous forces both at the boundary and in the bulk domain. In contrast, for the laminar free-slip case (\Cref{subfig:tes_pic2}), tangential viscous stresses at the wall are suppressed and wall viscous dissipation disappears, while viscous stresses in the bulk domain remains due to the low Reynolds number. Thus, in the laminar no-slip case, the optimizer straightens the main conduits and reduces excessive constrictions to limit wall related energy dissipation. While, in the free-slip case, the optimizer is able to conduct more re-directions using more walls, without having energy losses due to the wall viscous stresses.

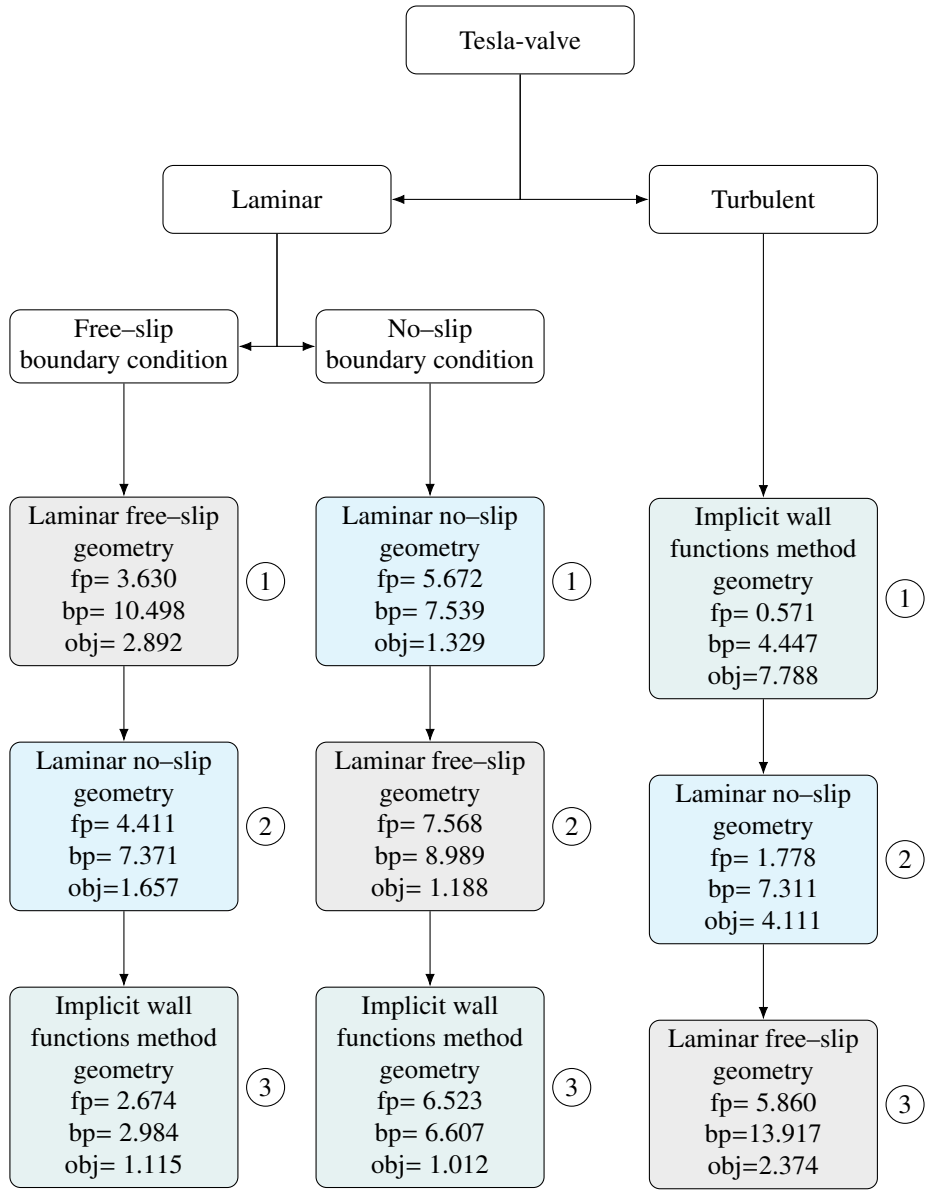
\begin{figure}
\centering
\begin{tikzpicture}[
  node distance=10mm and 18mm,
  box/.style={draw, rounded corners, minimum width=30mm, minimum height=9mm, align=center},
  circ/.style={draw, circle, inner sep=1pt, minimum size=5mm},
  conn/.style={-Latex}
]

\node[box] (root) {Tesla-valve};

\node[box, below=12mm of root, xshift=-32mm] (lam) {Laminar};
\node[box, below=12mm of root, xshift= 32mm] (turb) {Turbulent};

\draw[conn] (root.south) |- (lam.east);
\draw[conn] (root.south) |- (turb.west);

\node[box, below left =10mm and -10mm of lam] (slip)   {Free--slip\\boundary condition};
\node[box, below right=10mm and -10mm of lam] (noslip) {No--slip \\boundary condition};

\draw[conn] (lam.south) |- (slip.east);
\draw[conn] (lam.south) |- (noslip.west);

\node[box,fill=darkgray!10, below=15mm of slip] (s1) {Laminar free--slip \\geometry\\fp= 3.630\\bp= 10.498 \\obj= 2.892};
\node[box,fill=cyan!10, below=10mm of s1]  (s2) {Laminar no--slip\\ geometry\\fp= 4.411\\bp= 7.371\\obj=1.657};
\node[box, fill=teal!10, below=10mm of s2]  (s3) {Implicit wall \\functions method\\ geometry\\fp= 2.674\\bp= 2.984\\obj= 1.115};
\draw[conn] (slip.south) -- (s1.north);
\draw[conn] (s1.south)   -- (s2.north);
\draw[conn] (s2.south)   -- (s3.north);
\node[circ, right=1mm of s1] {1};
\node[circ, right=1mm of s2] {2};
\node[circ, right=1mm of s3] {3};

\node[box,fill=cyan!10, below=15mm of noslip] (n1) {Laminar no--slip\\ geometry\\fp= 5.672\\bp= 7.539\\obj=1.329};
\node[box,fill=darkgray!10, below=10mm of n1]     (n2) {Laminar free--slip \\geometry\\fp= 7.568\\bp= 8.989\\obj= 1.188};
\node[box, fill=teal!10, below=10mm of n2]     (n3) {Implicit wall \\functions method\\ geometry\\fp= 6.523\\bp= 6.607\\obj= 1.012};
\draw[conn] (noslip.south) -- (n1.north);
\draw[conn] (n1.south)     -- (n2.north);
\draw[conn] (n2.south)     -- (n3.north);
\node[circ, right=1mm of n1] {1};
\node[circ, right=1mm of n2] {2};
\node[circ, right=1mm of n3] {3};

\node[box, fill=teal!10, below=35mm of turb] (t1) {Implicit wall \\functions method\\ geometry\\fp= 0.571\\bp= 4.447\\obj=7.788};
\node[box,fill=cyan!10, below=10mm of t1]   (t2) {Laminar no--slip\\ geometry\\fp= 1.778\\bp= 7.311\\ obj= 4.111};
\node[box,fill=darkgray!10, below=10mm of t2]   (t3) {Laminar free--slip \\geometry\\fp= 5.860\\ bp=13.917\\obj=2.374};
\draw[conn] (turb.south) -- (t1.north);
\draw[conn] (t1.south)   -- (t2.north);
\draw[conn] (t2.south)   -- (t3.north);
\node[circ, right=1mm of t1] {1};
\node[circ, right=1mm of t2] {2};
\node[circ, right=1mm of t3] {3};

\end{tikzpicture}
\caption{Tesla valve case study, illustrating the laminar with free-slip~(gray) and no-slip~(blue), and turbulent flow with implicit wall-functions~(green) optimized geometries, re-simulated with three different conditions (Laminar with free-slip and no-slip boundary conditions, and turbulent with no-slip boundary conditions). The notations are defined as follows: fp denotes the forward average inlet pressure, bp the backward average inlet pressure, and obj the objective function.}
\label{fig:the_comparison_different_tesla_valves}
\end{figure}
To assess cross-regime robustness, a cross-check is performed, where each geometry is re-simulated across three scenarios --- turbulent flow, laminar free-slip, and laminar no-slip. As summarized in \Cref{fig:the_comparison_different_tesla_valves}, two clear trends emerge. First, each design performs best in the regime and boundary condition setting for which it was optimized. Second, performance deteriorates markedly when transferred across regimes: the turbulent-optimized geometry exhibits low diodicity under laminar conditions, and conversely, the laminar-optimized geometries show low diodicity in turbulent flow. Finally, it is worth noting that the turbulent-optimized geometry achieves a significantly higher diodicity of $7.788$, compared to only $4.111$ for the laminar-optimized geometry, highlighting the substantial improvement in the objective function when employing turbulent topology optimization. The significantly higher diodicity in the turbulent optimization arises because the direction-dependency of the flow due to inertial forces and separation points can be captured with turbulent models, but not in laminar regimes (low $\text{Re}$), in which viscous forces dominate and the design is less sensitive to flow direction (forward and backward flows), as losses due to inertial forces are neglected.

\subsection{Summary of the benchmarks}
A detailed summary of the final objective values for all cases considered in this study, together with the corresponding body-fitted re–simulation average inlet pressures, is provided in \Cref{tab: All_cases_table}. The table reports the final average inlet pressures values for Examples~\#1–\#8 as well as for the literature comparison and the LES re-simulations.
\begin{table}[]
    \centering
    \caption{Final objective function values and body-fitted re-simulation average inlet pressures for the different examples. The symbol ‘---’ indicates that the corresponding quantity was not evaluated in the present work.}
    \label{tab:final_objectives}
    \begin{tabular}{lccc}
        \toprule
        Example &
        Pressure drop from optimization &
        \multicolumn{2}{c}{Pressure drop from re-simulation } \\
        & &
        $k$-$\varepsilon$  &
        LES  \\
        \midrule
        Example~\#1               & 0.4023  & 0.1954  & ---   \\
        Example~\#2               & 0.1731  & 0.0924  & ---   \\
        Example~\#3               & 0.1303  & 0.0651  & ---   \\
        Example~\#4               & 0.1435  & 0.0470  & ---   \\
        Example~\#5               & 0.8327  & 0.9446  & 1.2806\\
        Literature comparison \cite{picelli2022topology}     & ---     & 0.4103  & 0.6332  \\
        Example~\#6               & 0.4408  & 0.3423  & 0.4014\\
        Example~\#7               & 0.4108  & 0.2118  & ---   \\
        Example~\#8, forward flow  & 0.5962  & 0.5710  & ---   \\
        Example~\#8, backward flow & 4.0812  & 4.4473  & ---   \\
        \bottomrule
    \end{tabular}
    \label{tab: All_cases_table}
\end{table}
\FloatBarrier

\section{Conclusions}\label{Section: Conclusion}
We presented a turbulent topology optimization framework that couples a density-based formulation with an implicit wall–function treatment of near–wall physics for the standard $k$-$\varepsilon$ RANS model. The approach delivers accurate designs in complex, turbulence–dominated flows while keeping costs tractable. The key findings are:

\begin{enumerate}
  \item \textbf{Better accuracy on coarser meshes.} By operating on meshes that target $y^+\!\in[11.06,\,300]$ instead of $y^+\!\approx 1$, the method avoids fully resolving the viscous sub-layer, reducing the number of degrees of freedom and lightening the computational cost.
  \item \textbf{Correct boundary–layer behavior.} The formulation accurately predicts boundary–layer growth without the common over–thickening in conventional approaches, producing velocity and pressure profiles that agree with expectations and explicit wall re-simulations.
  \item \textbf{High–$\operatorname{Re}$ robustness.} The framework sustains stable optimization runs at Reynolds numbers up to the hundreds of thousands (for some problems), maintaining convergence of the non-linear RANS system while retaining mesh economy. Although at high Reynolds numbers convergence may still be an issue, it is much less problematic than when not using the implicit wall-function approach.
  \item \textbf{Accuracy in highly non-linear physics and complex geometries.} The method remains reliable for the strongly non-linear $k$-$\varepsilon$ equations and in intricate layouts (e.g., Tesla valve channels with sharp bends and narrow channels).
  \item \textbf{Ease of integration.} The approach is relatively simple to implement in existing density–based TO: it requires only additional operations that are already common in density-based TO.
\end{enumerate}

The present formulation is based on the $k$-$\varepsilon$ standard model and the wall treatment is based on log-law of the wall, thus the approach does carry all the limitations of both $k$-$\varepsilon$ method and the log-law of the wall. In regions of strong separation, reattachment or stagnation, when the $k$-$\varepsilon$ model or log-law fails to predict the flow actions correctly, the accuracy of the proposed implicit wall-function model is expected to deteriorate. The authors suggest extending the present topology optimization with implicit wall-functions framework to alternative RANS closures, such as $k$-$\omega$ model, or to higher-fidelity approaches like LES, which can likewise benefit from implicitly defined wall-functions. Furthermore, there are two parameters in the present approach, related to the implicitly defined wall-functions, that must be tuned carefully to capture the wall effects: the wall intensity $\psi$ and the second filter radius $r_2$. The parameter $\psi$ controls how strongly the implicit wall condition is applied, while $r_2$ controls the effective wall thickness; both must be selected accurately. Rigorous studies and dimensional analysis \cite{Theulings2023,Theulings2025} may be promising directions to ease the selection of these in the future.

\subsection*{Supplementary Information}\label{subsection: Supplementary information}
No supplementary information is used in this paper.

\subsection*{Acknowledgments}\label{subecton: Acknoledgement}
Part of the computation done for this project was performed on the UCloud interactive HPC system, which is managed by the eScience Center at the University of Southern Denmark. The second author received financial support from the European Union through the Marie Sklodowska Curie Actions Postdoctoral Fellowship 22 (Grant No. 101106842). \par

\section*{Declarations}\label{Section: Declarations}

\subsection*{Conflict of interest}\label{Subsecton: Declaration of Competing interest}
The authors declare that they have no competing financial interests for this paper.

\subsection*{Replication of results}\label{Subsection: Replication of results}
A detailed implementation and algorithm of the proposed method have been presented in~\Cref{Section: Formulation} and one can follow them and reproduce the results. In case of further queries, please contact the corresponding author(s). \par

\subsection*{Author contributions}\label{Subsection: Author contribution}
Amirhossein Bayat (\url{abay@sdu.dk}): Conceptualization, Methodology, Software, Investigation, Validation, Writing -- original draft. 
Hao Li (\url{hli@sdu.dk}): Validation, Resources, Supervision, Writing -- Review \& Editing, Project administration. 
Joe Alexandersen (\url{joal@sdu.dk}): Conceptualization, Methodology, Software, Validation, Resources, Supervision, Writing -- Review \& Editing, Funding acquisition. 

\bibliographystyle{elsarticle-num-names}

\bibliography{main}

\appendix
\renewcommand{\theequation}{A.\arabic{equation}}
\setcounter{equation}{0}
\renewcommand{\thefigure}{A.\arabic{figure}}
\setcounter{figure}{0}

\section{Details of LES verification studies} \label{app:LESdetails}

The LES verification studies for the U-bend benchmark are set up in COMSOL using their Residual-Based Variational Multiscale (RBVM) implementation \cite{COMSOL_LES}. Although LES inherently requires a three-dimensional domain, we have run two-dimensional approximations by having a thin domain in the third-direction with symmetry boundary conditions on the lateral sides.

\subsection{Meshes}

\begin{figure}
    \centering
    \subfloat[Present approach]{\includegraphics[width=0.4\textwidth]{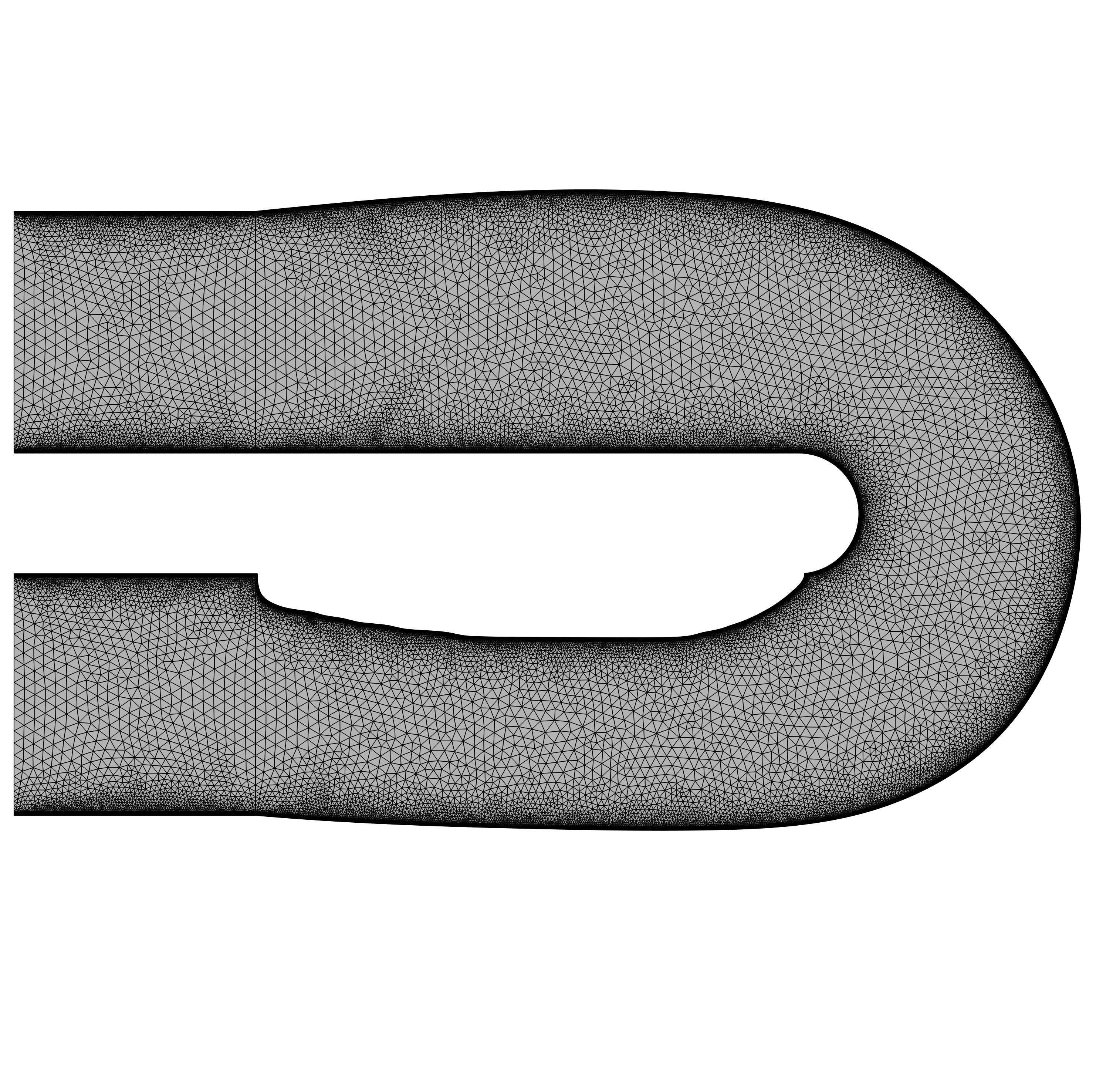}\label{subfig:app_LES_mesh_ours}}
    \hfill
    \subfloat[Present approach - zoom in near boundary]{\includegraphics[width=0.4\textwidth]{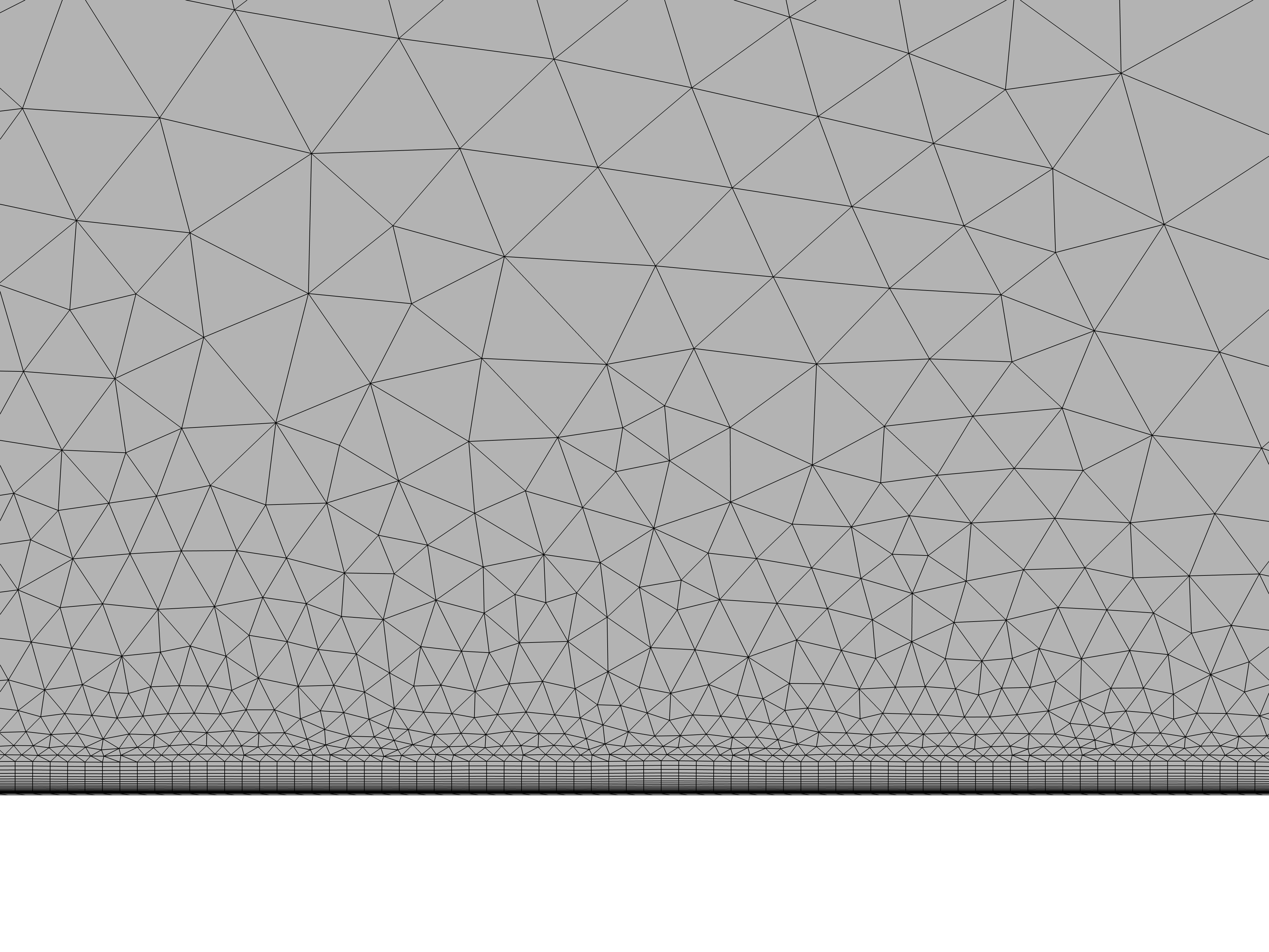}\label{subfig:app_LES_mesh_zoom}}
    \\
    \subfloat[``Conventional'' approach]{\includegraphics[width=0.4\textwidth]{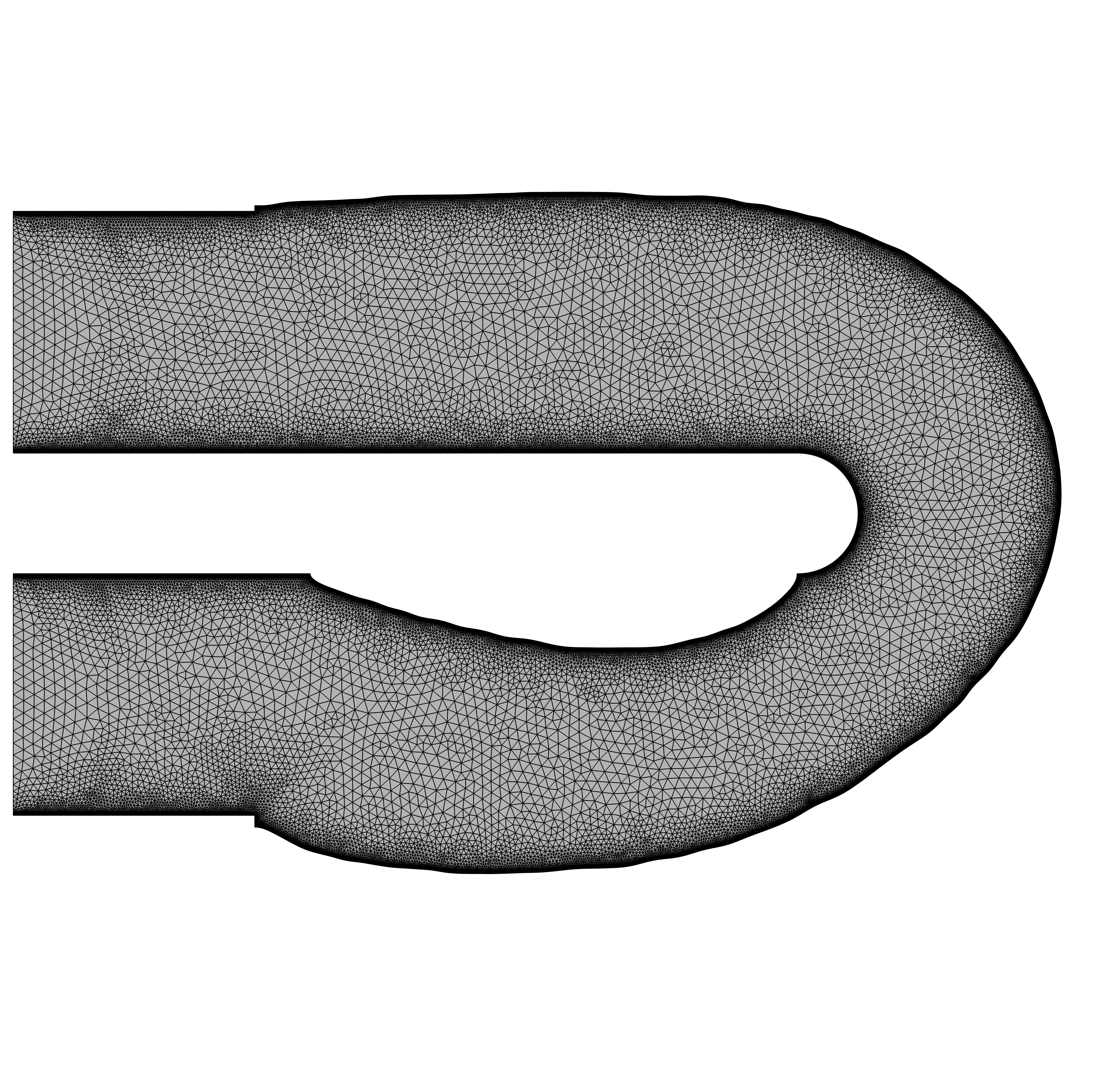}\label{subfig:app_LES_mesh_trad}}
    \hfill
    \subfloat[TOBS result from \citet{picelli2022topology}]{\includegraphics[width=0.4\textwidth]{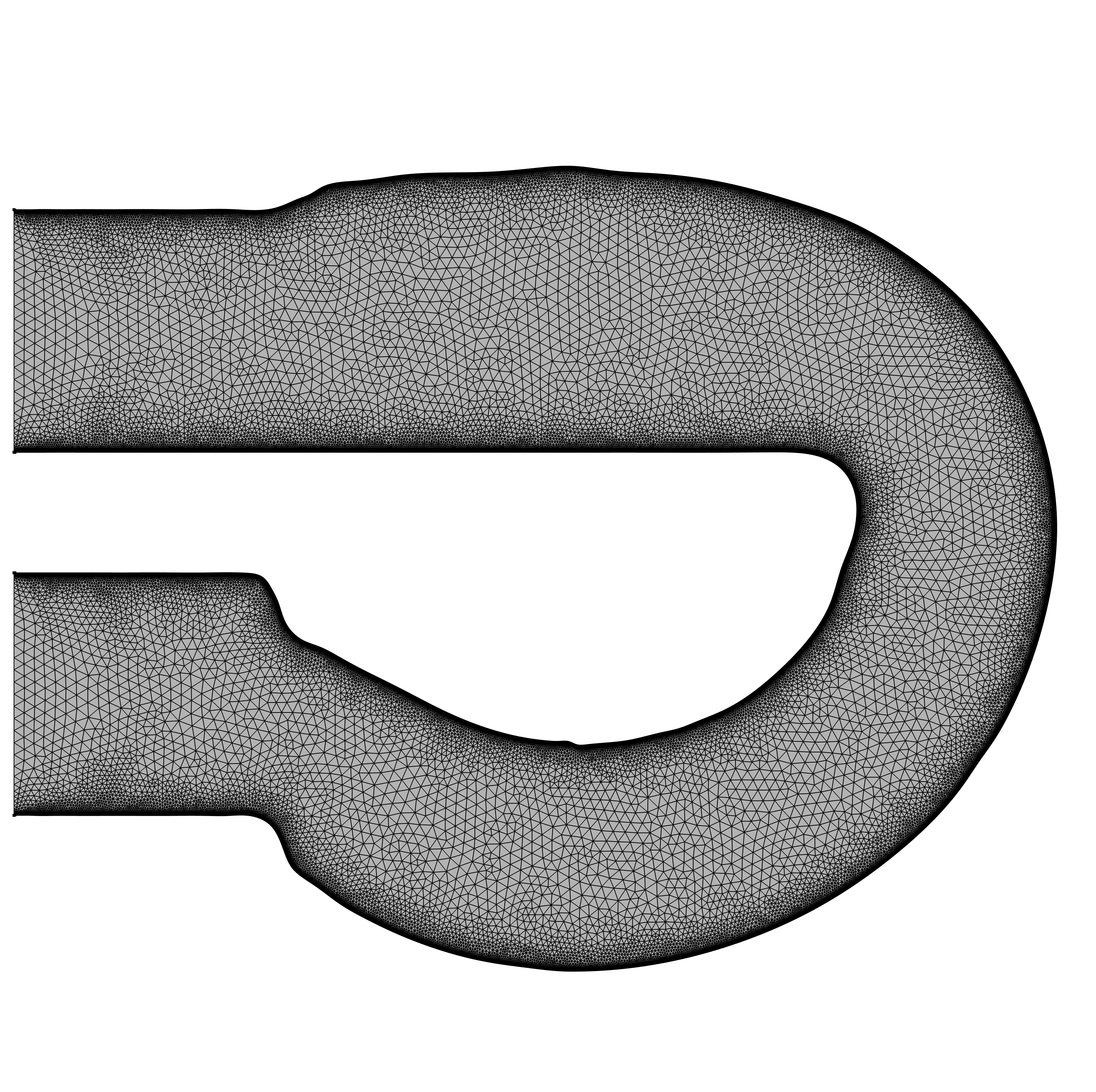}\label{subfig:app_LES_mesh_trad}}
    \caption{Meshes used for the LES verification of performance.}
    \label{fig:app_LES_meshes}
\end{figure}
The meshes used for the three cases are shown in \Cref{fig:app_LES_meshes}.
The initial unstructured triangular meshes are created on one lateral surface based on requiring: a maximum and minimum element size of 0.01 and 0.001, respectively, in the flow domain; a maximum and minimum element size of 0.001 and 0.00005, respectively, on the wall. Boundary layer meshing is then imposed along the walls using 18 layers of quadrilateral elements with a stretching factor of 1.15 and a thickness adjustment factor of 0.5. Finally, the mesh is extruded through the third dimension such that only one element exists in the through-thickness direction, making triangular elements into prisms and quadrilateral elements into hexahedrons.
The final meshes consist of:
\begin{itemize}
    \item Design using proposed method: 55,287 triangular prisms, 64,098 boundary layer elements, 748,840 total DOFs
    \item Design using ``conventional'' method: 55,542 triangular prisms, 63,288 boundary layer elements, 743,184 total DOFs
    \item Design using TOBS from \cite{picelli2022topology}: 58,060 triangular prisms, 66,780 boundary layer elements, 781,888 total DOFs
\end{itemize}

The computational time are above 48 hours for all three designs on 14 cores (AMD Ryzen Threadripper PRO 5975WX). The resolution used is severely restricted by limitations on computational time, rather than specific requirements to $y^+$ or formal mesh convergence studies.

\subsection{Computational details}

The boundary conditions are set as specified in the original problem formulation in \Cref{sec:bending_pipe_benchmark}. The inlet velocity is imposed in an average fashion \cite{COMSOL_LES} and the same goes for the outlet pressure. The inlet is seeded with synthetic turbulence with the following parameters: turbulent intensity medium (0.05), geometry-based turbulence length scale, 100 Fourier modes, and a random seed of 113013.

To speed up the computations, the simulations are first run on a very coarse mesh (build based on $4\times$ coarsening of the final mesh). The results from the final time are then used as initial conditions on a coarse mesh (build based on $2\times$ coarsening of the final mesh). Finally, the results from the final time are then used as initial conditions on the final mesh.

On the two coarse meshes, the simulated time span is set to 2 units of time, where it seems the flow has reached a quasi-repetitive state. For the final mesh, the simulated time span is set to 5 units of time, since it is seen to be sufficient, but is also strongly restricted by available computational time. The flow and pressure fields are saved at every 0.01 after 2 units of time. These saved states are subsequently used to compute time-averaged velocity and pressure fields for comparison with each other and the k-$\varepsilon$ model.


\end{document}